\documentclass[11pt,showpacs,preprint,preprintnumbers,nofootinbib,
superscriptaddress,amsmath,amssymb]{revtex4}

\usepackage{graphicx}
\usepackage{dcolumn}
\usepackage{bm}
\usepackage{amssymb}
\usepackage{amsmath}
\usepackage{epsfig}    
\usepackage{color}
\usepackage{slashed}
\begin{document} 

\title{
%Radiative corrections to the Higgs boson coupling constants
%in two Higgs doublet models 
Fingerprinting the extended Higgs sector using one-loop corrected\\ Higgs boson couplings and 
future precision measurements
}

\preprint{UT-HET 099}
\author{Shinya Kanemura}
\email{kanemu@sci.u-toyama.ac.jp}
\affiliation{Department of Physics, University of Toyama, \\3190 Gofuku, Toyama 930-8555, Japan}
\author{Mariko Kikuchi}
\email{kikuchi@jodo.sci.u-toyama.ac.jp}
\affiliation{Department of Physics, University of Toyama, \\3190 Gofuku, Toyama 930-8555, Japan}
\author{Kei Yagyu}
\email{K.Yagyu@soton.ac.uk}
\affiliation{School of Physics and Astronomy, University of Southampton, Southampton, SO17 1BJ, United Kingdom}

\begin{abstract}

We calculate radiative corrections to a full set of 
coupling constants for the 125 GeV Higgs boson at the one-loop level in two Higgs doublet models with four types of Yukawa interaction under the 
softly-broken discrete $Z_2$ symmetry.
The renormalization calculations are performed in the on-shell scheme, in which the gauge dependence in the mixing parameter which appears in the previous calculation 
is consistently avoided. 
We first show the details of our renormalization scheme, and present the complete set of the analytic formulae of the renormalized couplings. 
We then numerically demonstrate how the inner parameters of the model can be extracted by the future precision measurements of these couplings at the high luminosity LHC and 
the International Linear Collider. 

\end{abstract}
\maketitle

\section{Introduction}

The LHC Run-I has confirmed the existence of a 
Higgs boson ($h$)~\cite{LHC_Higgs_ATLAS,LHC_Higgs_CMS}, whose properties are in agreement with those of the standard model (SM) 
within the uncertainties of the current data~\cite{ATLAS_Coupling1,ATLAS_Coupling2,ATLAS_Coupling3,CMS_Coupling0,CMS_Coupling,LHC_spin}.
Thanks to the discovery of the Higgs boson, 
the SM was established as an effective theory to describe physics at the scale of electroweak symmetry breaking.
In spite of the success of the SM, 
there are many motivations to consider new physics beyond the SM 
such as to solve the gauge hierarchy problem and to explain phenomena 
like neutrino oscillation, dark matter and baryon asymmetry of the Universe.
There have been various new physics models proposed, some of which predict new particles at the electroweak to TeV scales.
However, currently none of such new particles has been discovered yet.
Their discovery is one of the main tasks of the LHC Run-II, which will start its operation in 2015. 
    
Even though the Higgs boson shows SM like properties, 
the Higgs sector can be extended from the minimal form with only an isospin doublet field. 
Indeed, there is no theoretical reason for the hypothesis of the minimal structure for the Higgs sector. 
Thus there are possibilities for extended Higgs sectors such as those with additional iso-singlets, doublets, and/or triplets. 
These extended Higgs sectors can also be consistent with all the current LHC data in some portions of their parameter space.  

Extended Higgs sectors are often introduced in various new physics models.
For example, the Minimal Supersymmetric SM (MSSM) requires the Higgs sector with two doublet fields~\cite{MSSM,Higgs_hunters}.
Multi Higgs structures are also studied in the context of 
additional CP violating phases~\cite{CPV} and also realization of the strong first order phase transition~\cite{1opt}, 
both of which are required for successful electroweak baryogenesis~\cite{ewbg}.
Models with the Type-II seesaw scenario are motivated to generate tiny neutrino masses 
by introducing a triplet field~\cite{HTM}.  
An additional singlet is required in the Higgs sector of the models with spontaneous breakdown 
of the $U(1)_{B-L}$ symmetry~\cite{B-L,B-L_c,B-L_dm}, which may be related to the mechanism of neutrino mass generation~\cite{B-L_rad}.
Introduction of an additional unbroken symmetry into an extended Higgs sector, such as a discrete $Z_2$ symmetry~\cite{IDM,Ma-Deshpande}
or a global $U(1)$ symmetry~\cite{Ko}, can provide candidates of dark matter.   
Under the $Z_2$ or the global $U(1)$ symmetry, if some of the scalar fields are assigned to be odd or to be charged, respectively,  
they cannot decay into a pair of SM particles so that the lightest one is stable. 
Such an unbroken symmetry can also be embedded into models with a radiative generation of neutrino masses~\cite{B-L_rad,radseesaw-original, radseesaw-susy,radseesaw-dirac,radseesaw-dm,radseesaw_dim7,aks}, 
where the existence of tiny neutrino masses and dark matter can be explained by the same origin of the symmetry. 
%In Ref.~\cite{aks}, in addition to tiny neutrino masses and dark matter, 
%the baryon asymmetry of the Universe can also be simultaneously explained based on the electroweak baryogenesis mechanism in the TeV scale physics. 
%
Therefore, a characteristic Higgs sector appears in each new physics model.

There are several important properties which characterize the structure of the Higgs sector. 
First of all, it is important to know the number of scalar multiplets and their representations.
Second, does it respect new symmetries (global or discrete/exact or softly-broken)?
Third, the mass of the second Higgs boson generally contains information of the new scale which does not appear in the SM. 
Fourth, the strength of the coupling constants among extra Higgs bosons 
provides information of the dynamics of the Higgs potential 
which is essentially important to understand nature of electroweak symmetry breaking. 
Finally, the decoupling property~\cite{decoupling} of extra Higgs bosons is closely connected to physics beyond the SM.
Therefore, by future measurements of these properties, 
the Higgs sector can be reconstructed, and the direction of new physics beyond the SM can be determined.

The direct search of extra Higgs bosons can provide a clear evidence to a non-minimal Higgs sector.
The current data accumulated from previous collider experiments such as LEP~\cite{LEP1,LEP2} and Tevatron~\cite{Tevatron1,Tevatron2,Tevatron3,Tevatron4,Tevatron5,Tevatron6} 
have already given lower bounds for masses of the extra Higgs bosons.  
At the LHC Run-I, in spite of the discovery of a Higgs boson with the mass of 125 GeV, 
no extra Higgs boson has been found, and the parameter space for additional light Higgs bosons 
has been constrained to the considerable extent in regions with relatively smaller masses of the extra Higgs bosons~\cite{tautau-ATLAS,tautau-CMS,HWW-ATLAS,HA-THDM-CMS,LHC_Extra3,LHC_Extra5,LHC_Extra6,LHC_Extra7,LHC_Extra8,LHC_Extra9,LHC_Extra10,LHC_Extra11,LHC_Extra12}.
At the LHC Run-II, with the energy of 13-14 TeV and the integrated luminosity of 300 fb$^{-1}$, 
wider regions of masses of the extra Higgs bosons will be surveyed.

In addition to direct searches, 
new physics models beyond the SM have also been indirectly investigated 
by utilizing precision measurements of various physics observables such as the oblique parameters at LEP/SLC experiments~\cite{LEP_Indirect}. 
Flavour experiments have also been used to constrain the mass of charged Higgs bosons which appears in  extended Higgs sectors~\cite{bsg,Misiak}.
Now that the measured couplings of the Higgs boson $h$ with the SM particles are consistent with the 
predictions in the SM within the uncertainties, 
it is time to consider fingerprinting of extended Higgs sectors~\cite{fingerprint,Kanemura_HPNP} 
by calculating radiative corrections to the predictions of those observables which will be  measured with more precision at future experiments 
such as the LHC Run-II, the high luminosity (HL)-LHC~\cite{HLLHC_ATLAS,HLLHC_CMS,HLLHC_Rep} with the integrated luminosity of 3000 fb$^{-1}$ 
and future lepton colliders like the International Linear Collider (ILC)~\cite{ILC_TDR,ILC_white}.
In new physics models with extended Higgs sectors, the coupling constants of $h$ with the SM particles are generally predicted 
with deviations from the SM predictions due to field mixing and loop contributions of non-SM particles.
Although no deviation has been found up to now in the Higgs boson couplings within the uncertainty of the current data, 
a deviation could be found in future experiments where more precise measurements will be attained. 
We then are able to indirectly obtain information of the second Higgs boson from these deviations. 
Furthermore, a pattern of these deviations strongly depends on the structure of the Higgs sector, 
so that by comparing theoretical predictions of the Higgs couplings 
in various new physics models with future experimental data the shape of the Higgs sector can be determined indirectly.
In order to compare the theory predictions to future precision data at the HL-LHC and also the ILC, where
coupling constants are expected to be measured typically by a few percent or better accuracy, 
evaluations of the Higgs boson couplings including radiative corrections are inevitable.  

There are many studies for radiative corrections in extended Higgs sectors in the literature.  
Radiative corrections to the electroweak gauge boson two point functions (oblique corrections)
have been studied in extended Higgs sectors in Refs.~\cite{delrho_THDM,Blank_Hollik,Chen-Dawson-Jackson,Kanemura-Yagyu}.  
Loop induced vertices $hgg$~\cite{hgg_sm}, $h\gamma\gamma$~\cite{hgamgam_sm,hgamgam_thdm,hgamgam_Zgam_thdm,hgamgam_HTM,hgamgam_Zgam_HTM,Chiang-Yagyu-gamgam} 
and $hZ\gamma$~\cite{hZgam_sm,hgamgam_Zgam_thdm,hgamgam_Zgam_HTM,hZgam_HTM,Chiang-Yagyu-gamgam}
have been evaluated in extended Higgs sectors. 
Those to the Higgs boson couplings have been investigated in the two Higgs doublet model (THDM)
in Refs.~\cite{thdm_rad_susy,KKOSY,KOSY,KKY} and in the Higgs triplet model in Refs.~\cite{AKKY_Lett,AKKY_Full}.

In this paper, we study electroweak radiative corrections to the coupling constants of the 125 GeV Higgs boson $h$ 
in the THDM~\cite{THDM_rev} with the softly-broken $Z_2$ symmetry~\cite{GW}. 
Under the $Z_2$ symmetry, four types of Yukawa interactions~\cite{Barger,Grossman,Akeroyd,typeX} are possible
depending on the assignment of the $Z_2$ charges into quarks and leptons.
We investigate radiative corrections to the full set of 
Higgs boson couplings ($hWW$, $hZZ$, $htt$, $hbb$, $h\tau\tau$, $hhh$, 
$h\gamma\gamma$, $hZ\gamma$ and $hgg$) at the one-loop level in all types of the THDMs. 
We employ an improved on-shell renormalization scheme in our renormalization calculation 
where the gauge dependence in the calculation of the mixing angle in the previous studies 
is eliminated\footnote{According to Ref.~\cite{gauge_depend}, the gauge dependence exists in a renormalization of a mixing angle.}. 
We then evaluate deviations in these coupling constants from the SM predictions under the constraint of current experimental data 
and theoretical bounds such as vacuum stability and perturbative unitarity.  

Furthermore, 
we investigate how we can extract information of the inner parameters such as the mass of the second Higgs boson and mixing angles 
when the scale factors $\kappa_X^{}$ are 
experimentally determined with the expected uncertainties at the HL-LHC and the ILC, 
where $\kappa_X^{}$ are the ratios of the measured couplings $hXX$ from the SM predictions.  
Evaluating $\kappa_X^{}$ at the one-loop level in the THDMs,  
we discuss the possibility to measure properties of the Higgs sector using the future precision data by fingerprinting, 
and finally we determine the structure of the Higgs sector. 

This paper is organized as follows. 
In Sec.~\ref{sec model}, we define the Lagrangian of THDMs, and give formulae for the Higgs boson masses and the Higgs boson couplings at the tree level.
After that, we discuss constraints from vacuum stability and perturbative unitarity as the theoretical bounds. 
We then discuss the bounds from the electroweak oblique parameters, flavour experiments, direct searches of extra Higgs bosons at the LHC and the 
measurements of Higgs boson couplings at the LHC Run-I. 
In addition, we shortly summarize future prospects for extra Higgs boson searches and  precision measurements of the Higgs boson $h$ at the LHC Run-II, the HL-LHC and the ILC. 
In Sec.~\ref{renormalization}, we explain renormalization in the electroweak sector, the Yukawa sector, and the Higgs sector in the THDMs.
We also discuss the modified renormalization scheme. 
In Sec.~\ref{reno_couplings}, we give formulae of renormalized Higgs couplings and loop induced decay rates.
We numerically estimate decoupling properties and non-decoupling effects of our one-loop calculations in the section.
In Sec.~\ref{determination}, we demonstrate how we can extract inner parameters by using future precision data.   
Discussions and conclusions are given in Sec. VI.

\section{Two Higgs doublet models}\label{sec model}

\subsection{Lagrangian}
 
\begin{table}[t]
\begin{center}
{\renewcommand\arraystretch{1.2}
\begin{tabular}{c|ccccccc|ccc}\hline\hline
&
\multicolumn{7}{c|}{$Z_2$ charge}
&\multicolumn{3}{c}{Mixing factor}\\     \cline{2-11}
&$\Phi_1$&$\Phi_2$&$Q_L$&$L_L$&
$u_R$&$d_R$&$e_R$
&$\xi_u$ &$\xi_d$&$\xi_e$ \\\hline
Type-I &$+$&
$-$&$+$&$+$&
$-$&$-$&$-$&$\cot\beta$&$\cot\beta$&$\cot\beta$ \\\hline
Type-II&$+$&
$-$&$+$&$+$&
$-$
&$+$&$+$& $\cot\beta$&$-\tan\beta$&$-\tan\beta$ \\\hline
Type-X &$+$&
$-$&$+$&$+$&
$-$
&$-$&$+$&$\cot\beta$&$\cot\beta$&$-\tan\beta$ \\\hline
Type-Y &$+$&
$-$&$+$&$+$&
$-$
&$+$&$-$& $\cot\beta$&$-\tan\beta$&$\cot\beta$ \\\hline\hline
\end{tabular}}
\caption{Charge assignment of the softly-broken $Z_2$ symmetry and the mixing factors in Yukawa interactions given in Eq.~(\ref{yukawa_thdm}).}
\label{yukawa_tab}
\end{center}
\end{table}

In this section, we define the Lagrangian in the THDM with the softly-broken $Z_2$ symmetry, where 
the Higgs sector is composed of two isospin doublet scalar fields $\Phi_1$ and $\Phi_2$.  
The charge assignment for the $Z_2$ symmetry is shown in Table~\ref{yukawa_tab}. 
The following Lagrangian is modified from the SM:  
\begin{align}
\mathcal{L}_{\text{THDM}} = \mathcal{L}_{\text{kin}} + \mathcal{L}_{Y} - V, 
\end{align}
where $\mathcal{L}_{\text{kin}}$, $\mathcal{L}_Y$ and $V$ are respectively the kinetic Lagrangian, the Yukawa Lagrangian and the scalar potential. 
Throughout the paper, we assume the CP invariance in the Higgs sector. 

First, the kinetic Lagrangian is given by
\begin{align}
\mathcal{L}_{\text{kin}} &=|D_\mu\Phi_1|^2+|D_\mu\Phi_2|^2,
\end{align}
where $D_\mu$ is the covariant derivative: 
\begin{align}
D_\mu = \partial_\mu-\frac{i}{2}g \tau^a W_\mu^a-\frac{i}{2}g'B_\mu, 
\end{align}
with $W_\mu^a$ ($a=$1-3) and $B_\mu$ being the $SU(2)_L$ and $U(1)_Y$ gauge bosons, respectively. 
The two doublet fields can be parameterized as 
\begin{align}
\Phi_i=\left[\begin{array}{c}
w_i^+\\
\frac{1}{\sqrt{2}}(v_i+h_i+iz_i)
\end{array}\right],\hspace{3mm}(i=1,2), 
\end{align}
where $v_1$ and $v_2$ are the vacuum expectation values (VEVs) for $\Phi_1$ and $\Phi_2$, 
which satisfy $v\equiv\sqrt{v_1^2+v_2^2}=(\sqrt{2}G_F)^{-1/2}$. 
The ratio of the two VEVs is defined as $\tan\beta=v_2/v_1$.  
The mass eigenstates for the scalar bosons are obtained by the following orthogonal transformations as
\begin{align}
\left(\begin{array}{c}
w_1^\pm\\
w_2^\pm
\end{array}\right)&=R(\beta)
\left(\begin{array}{c}
G^\pm\\
H^\pm
\end{array}\right),\quad 
\left(\begin{array}{c}
z_1\\
z_2
\end{array}\right)
=R(\beta)\left(\begin{array}{c}
G^0\\
A
\end{array}\right),\quad
\left(\begin{array}{c}
h_1\\
h_2
\end{array}\right)=R(\alpha)
\left(\begin{array}{c}
H\\
h
\end{array}\right), \notag\\
\text{with}~R(\theta) &= 
\left(
\begin{array}{cc}
\cos\theta & -\sin\theta\\
\sin\theta & \cos\theta
\end{array}\right),
\label{mixing}
\end{align}
where $G^\pm$ and $G^0$ are the Nambu-Goldstone bosons absorbed by the longitudinal component of $W^\pm$ and $Z$, respectively.  
The mixing angle $\alpha$ is expressed in terms of the mass matrix elements for the CP-even scalar states as shown in Eqs.~(\ref{tan2a})-(\ref{m12}). 
As the physical degrees of freedom, 
we have a pair of singly-charged Higgs boson $H^\pm$, a CP-odd Higgs boson $A$ and two CP-even Higgs bosons $h$ and $H$. 
We define $h$ as the observed Higgs boson with the mass of about 125 GeV. 

In terms of the mass eigenbasis of the Higgs fields, the interaction terms among the Higgs bosons and the weak gauge bosons are given by 
\begin{align}
\mathcal {L}_{\text{kin}}
=&
[\sin(\beta-\alpha) h+ \cos(\beta-\alpha) H] 
\Big(\frac{m_W^2}{v} W^{+\mu} W^-_\mu +  \frac{m_Z^2}{2v} Z^\mu Z_\mu \Big)\notag\\
&+ g_{\phi_1\phi_2 V}^{}(\partial^\mu\phi_1\phi_2-\phi_1\partial^\mu\phi_2) V_\mu 
+g_{\phi_1\phi_2 V_1V_2}^{}\, \phi_1\phi_2V_1^\mu V_{2\mu}, 
\end{align}
where coefficients of the Scalar-Scalar-Gauge vertex $g_{\phi_1\phi_2 V}^{}$ and those of 
the Scalar-Scalar-Gauge-Gauge vertex $g_{\phi_1\phi_2 V_1V_2}^{}$ are listed in Appendix A. 

Next, we discuss the Yukawa Lagrangian. The most general form under the $Z_2$ symmetry is given by  
\begin{align}
-{\mathcal L}_Y =
&Y_{u}{\overline Q}_Li\sigma_2\Phi^*_uu_R^{}
+Y_{d}{\overline Q}_L\Phi_dd_R^{}
+Y_{e}{\overline L}_L\Phi_e e_R^{}+\text{h.c.},
\end{align}
where $\Phi_{u,d,e}$ are either $\Phi_1$ or $\Phi_2$. 
Depending on the $Z_2$ charge assignment, there are four types of Yukawa interactions~\cite{Barger,Grossman}, which we call as 
Type-I, Type-II, Type-X and Type-Y~\cite{typeX}. 
The interaction terms are expressed in terms of the mass eigenstates
of the Higgs bosons as
\begin{align}
-{\mathcal L}_Y^{\text{int}}=&
\sum_{f=u,d,e}\frac{m_f}{v}\left( \xi_h^f{\overline
f}fh+\xi_H^f{\overline f}fH-2iI_f \xi_f{\overline f}\gamma_5fA\right)\notag\\
&+\frac{\sqrt{2}}{v}\left[V_{ud}\overline{u}
\left(m_d\xi_d\,P_R-m_u\xi_uP_L\right)d\,H^+
+m_e\xi_e\overline{\nu^{}}P_Re^{}H^+
+\text{h.c.}\right],  \label{yukawa_thdm}
\end{align}
where $\xi_h^f$ and $\xi_H^f$ are defined by
\begin{align}
\xi_h^f &= \sin(\beta-\alpha)+\xi_f \cos(\beta-\alpha), \\
\xi_H^f &= \cos(\beta-\alpha)-\xi_f \sin(\beta-\alpha), 
\end{align}
and $\xi_f$ in each type of Yukawa interactions are given in Table~\ref{yukawa_tab}. 
In Eq.~(\ref{yukawa_thdm}), $I_f$ represents the third component of the isospin of a fermion $f$; i.e., $I_f=+1/2$ $(-1/2)$ for $f=u~(d,e)$. 

The Higgs potential under the softly-broken $Z_2$ symmetry and the CP invariance is given by  
\begin{align}
V &=m_1^2|\Phi_1|^2+m_2^2|\Phi_2|^2-m_3^2(\Phi_1^\dagger \Phi_2 +\text{h.c.})\notag\\
& +\frac{1}{2}\lambda_1|\Phi_1|^4+\frac{1}{2}\lambda_2|\Phi_2|^4+\lambda_3|\Phi_1|^2|\Phi_2|^2+\lambda_4|\Phi_1^\dagger\Phi_2|^2
+\frac{1}{2}\lambda_5\left[(\Phi_1^\dagger\Phi_2)^2+\text{h.c.}\right]. \label{pot_thdm2}
\end{align}
The tadpole terms for $h_1$ and $h_2$ are respectively calculated as  
\begin{align}
\frac{T_1}{v\cos\beta}&=-m_1^2+M^2\sin^2\beta-\frac{v^2}{2}(\lambda_1\cos^2\beta+\bar{\lambda}\sin^2\beta),\\
\frac{T_2}{v\sin\beta}&=-m_2^2+M^2\cos^2\beta-\frac{v^2}{2}(\lambda_2\sin^2\beta+\bar{\lambda}\cos^2\beta),
\end{align}
where $\bar{\lambda}\equiv\lambda_3+\lambda_4+\lambda_5$, 
and $M$ describes the soft breaking scale of the $Z_2$ symmetry: 
\begin{align}
M^2=\frac{m_3^2}{\sin\beta\cos\beta}. \label{bigm}
\end{align}
We note that $M^2$ can be taken to be both positive and negative values. 
By requiring the tree level tadpole conditions; i.e., $T_1=T_2=0$, 
$m_1^2$ and $m_2^2$ can be eliminated in the Higgs potential.   
 
The squared masses of $H^\pm$ and $A$ are calculated as 
\begin{align}
m_{H^\pm}^2=M^2-\frac{v^2}{2}(\lambda_4+\lambda_5),\quad m_A^2&=M^2-v^2\lambda_5.  \label{mass1}
\end{align}%
Those for the CP-even Higgs bosons and the mixing angle $\alpha$ are given by 
\begin{align}
&m_H^2=\cos^2(\alpha-\beta)M_{11}^2+\sin^2(\alpha-\beta)M_{22}^2+\sin2(\alpha-\beta)M_{12}^2,\\
&m_h^2=\sin^2(\alpha-\beta)M_{11}^2+\cos^2(\alpha-\beta)M_{22}^2-\sin2(\alpha-\beta)M_{12}^2,\\
&\tan 2(\alpha-\beta)=\frac{2M_{12}^2}{M_{11}^2-M_{22}^2}, \label{tan2a}
\end{align} 
where $M_{ij}^2$ ($i,j=1,2$) are the mass matrix elements for the CP-even scalar states in the basis of $(h_1,h_2)R(\beta)$: 
\begin{align}
M_{11}^2&=v^2(\lambda_1\cos^4\beta+\lambda_2\sin^4\beta)+\frac{v^2}{2}\bar{\lambda}\sin^22\beta,  \label{m11}  \\
M_{22}^2&=M^2+v^2\sin^2\beta\cos^2\beta(\lambda_1+\lambda_2-2\bar{\lambda}), \label{m22}  \\
M_{12}^2&=\frac{v^2}{2}\sin2\beta(-\lambda_1\cos^2\beta+\lambda_2\sin^2\beta)+\frac{v^2}{2}\sin2\beta\cos2\beta\bar{\lambda}.  \label{m12}
\end{align}

Thus, ten parameters in the potential ($v_{1,2}$, $m_{1\text{-}3}^2$ and $\lambda_{1\text{-}5}$)
can be described by the eight physical parameters 
$m_h$, $m_H^{}$, $m_A^{}$, $m_{H^\pm}^{}$, $\alpha$, $\beta$, $v$ and $M^2$, and two tadpoles $T_1$ and $T_2$ which are taken to be zero at the tree level.  
The quartic couplings $\lambda_1$-$\lambda_5$ in the potential are then rewritten in terms of the physical parameters as
\begin{align}
\lambda_1v^2 &= (m_H^2\tan^2\beta + m_h^2)\sin^2(\beta-\alpha) 
            +(m_H^2 + m_h^2\tan^2\beta)\cos^2(\beta-\alpha) \notag\\
             &~~~+2(m_H^2-m_h^2) \sin(\beta-\alpha)\cos(\beta-\alpha)\tan\beta -M^2\tan^2\beta, \notag\\
\lambda_2v^2 &= (m_H^2\cot^2\beta + m_h^2)\sin^2(\beta-\alpha) 
            +(m_H^2 + m_h^2\cot^2\beta)\cos^2(\beta-\alpha) \notag\\
             &~~~-2(m_H^2-m_h^2) \sin(\beta-\alpha)\cos(\beta-\alpha) \tan\beta -M^2\cot^2\beta, \notag\\
\lambda_3v^2 &= (m_H^2- m_h^2)[\cos^2(\beta-\alpha) -\sin^2(\beta-\alpha) +(\tan\beta-\cot\beta)\sin(\beta-\alpha)\cos(\beta-\alpha)]\notag\\
             &~~~ +2m_{H^\pm}^2-M^2, \notag\\
\lambda_4v^2&=M^2+m_A^2-2m_{H^\pm}^2,\notag\\
\lambda_5v^2&=M^2-m_A^2. \label{physical}
\end{align}

We here define the so-called scaling factors to describe deviations in the Higgs boson couplings from the SM prediction as follows: 
\begin{align}
&\kappa_V^{} \equiv \frac{g_{hVV}^{\text{THDM}}}{g_{hVV}^{\text{SM}}},~~\text{for}~~V=Z,~W,\quad
\kappa_f^{} \equiv \frac{y_{hff}^{\text{THDM}}}{y_{hff}^{\text{SM}}},\quad
\kappa_h^{} \equiv \frac{\lambda_{hhh}^{\text{THDM}}}{\lambda_{hhh}^{\text{SM}}}, \label{scaling1}
\end{align} 
where $g_{hVV}^{\text{SM}}$,~$y_{hff}^{\text{SM}}$ and $\lambda_{hhh}^{\text{SM}}$ 
are the $hVV$, $hf\bar{f}$ and $hhh$ coupling constants in the SM, respectively, and 
those with THDM in the superscript are corresponding predictions in the THDM. 
The scaling factors for loop induced couplings can also be defined by 
\begin{align}
\kappa_\gamma^2 \equiv \frac{\Gamma(h\to \gamma\gamma)_{\text{THDM}}}{\Gamma(h\to \gamma\gamma)_{\text{SM}}},\quad 
\kappa_{Z\gamma}^2 \equiv \frac{\Gamma(h\to Z\gamma)_{\text{THDM}}}{\Gamma(h\to Z\gamma)_{\text{SM}}},\quad
\kappa_{g}^2 \equiv \frac{\Gamma(h\to gg)_{\text{THDM}}}{\Gamma(h\to gg)_{\text{SM}}}, \label{scaling2} 
\end{align}
where $\Gamma(h\to XY)_{\text{SM}}$ and $\Gamma(h\to XY)_{\text{THDM}}$ 
are respectively the decay rates of the $h\to XY$ mode in the SM and in the THDM. 
At the tree level, the scaling factors are given by 
\begin{align}
\kappa_V^{} &= \sin(\beta-\alpha),\\ 
\kappa_f^{} &= \xi_h^f=\sin(\beta-\alpha)+\xi_f\cos(\beta-\alpha),\\
\kappa_h^{} &=  \sin(\beta-\alpha) -\frac{2(M^2-m_h^2)}{m_h^2}\sin(\beta-\alpha)\cos^2(\beta-\alpha)\notag\\
&-\frac{M^2-m_h^2}{m_h^2}\cos^3(\beta-\alpha)(\cot\beta-\tan\beta). 
\label{scaling3}
\end{align}
We can see that all the scaling factors become unity when $\sin(\beta-\alpha)=1$ is taken, so that 
we call this limit as the SM-like limit~\cite{Gunion-Haber}. 

It is convenient to introduce a parameter $x$ defined as
\begin{align}
x \equiv \frac{\pi}{2}- (\beta-\alpha),  \label{xpara}
\end{align}
where $x\to 0$ corresponds to the SM-like limit. 
We note that in the MSSM, the sign of $x$ is determined to be negative due to supersymmetric relations~\cite{Higgs_hunters}. 
Because the current LHC data suggest that the observed Higgs boson is SM-like, the case with $|x|\ll 1$ describes such a situation. 
In this case, we obtain 
\begin{align}
\kappa_V^{} & =   1-\frac{x^2}{2} + \mathcal{O}(x^3),\label{kv}\\
\kappa_f^{} & =   1+\xi_f\, x-\frac{x^2}{2} + \mathcal{O}(x^3), \label{ku} \\
\kappa_h^{} & =   1+\left(\frac{3}{2}-\frac{2M^2}{m_h^2}\right)\,x^2 + \mathcal{O}(x^3). 
\end{align}
As it has already been pointed out in Ref.~\cite{fingerprint}, 
looking at the correlation between $\kappa_f^{}$ and $\kappa_{f'}^{}$ $(f\neq f')$ 
is quite useful to distinguish the four types of Yukawa interactions.  

\begin{figure}[t]
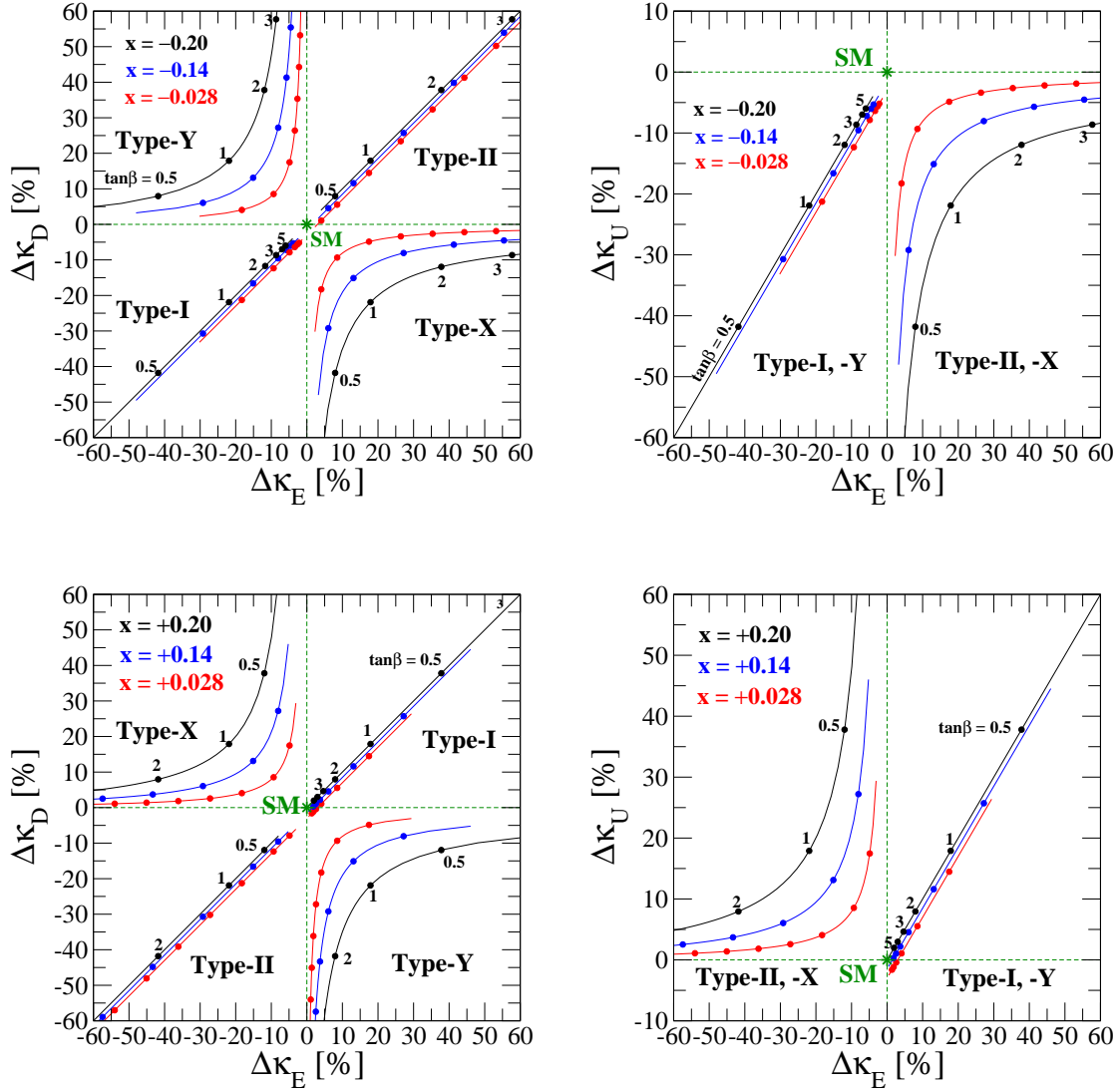

%========================================
\centering
\includegraphics[width=7cm]{FP_tree.eps} \hspace{5mm} 	
\includegraphics[width=7cm]{FP_tree_eu.eps} \\ \vspace{10mm}
\includegraphics[width=7cm]{FP_tree_plus.eps} \hspace{5mm} 	
\includegraphics[width=7cm]{FP_tree_eu_plus.eps} \\ \vspace{5mm}
\caption{Tree level predictions on the $\Delta\kappa_E$-$\Delta \kappa_D$ (left panel) and $\Delta\kappa_E$-$\Delta\kappa_U$ (right panel) 
plane in the four types of Yukawa interactions. 
The black, blue and red curves respectively show the case of $|x|=0.20$ [$\sin(\beta-\alpha)\simeq 0.98$], $|x|=0.14$  [$\sin(\beta-\alpha)\simeq 0.99$] 
and $|x|=0.028$  [$\sin(\beta-\alpha)\simeq 0.996$]. 
The sign of $x$ is taken to be negative in the upper figures and positive in the lower figures.  }
\label{FP_tree}
%========================================
\end{figure}

In Fig.~\ref{FP_tree}, we show the tree level predictions on the $\Delta\kappa_E$-$\Delta\kappa_D$ plane (left panels) 
and $\Delta\kappa_E$-$\Delta\kappa_U$ plane (right panels) 
in the four types of Yukawa interactions, where $\Delta\kappa_X=\kappa_X - 1$. 
The subscripts $E,~D$ and $U$ respectively represent the flavour independent charged leptons, down-type quarks and up-type quarks. 
In this plot, we take $|x|=0.2$, $0.14$ and $0.028$, and the sign of $x$ is set to be negative (positive) for upper (lower) panels.  
As it can be seen, the predictions for the four types of Yukawa interacitons appear in different quadrants of the $\Delta\kappa_E$-$\Delta\kappa_D$ plane. 
Therefore, at least from the tree level result, 
we can discriminate the type of Yukawa interaction in the THDM by looking at the measured values of $\Delta \kappa_E$ and $\Delta \kappa_D$. 

In Ref.~\cite{KKY}, one-loop corrected Yukawa couplings have been calculated in the four types of Yukawa interactions in the THDM. 
It has been clarified that the predictions in the four types of Yukawa interactions are well separated on the $\Delta\kappa_E$-$\Delta\kappa_D$ plane
at the one-loop level 
even if we scan the inner parameters under the constraints from perturbative unitarity and vacuum stability. 

\subsection{Vacuum stability and perturbative unitarity}

A set of quartic coupling constants in the Higgs potential $\lambda_1$-$\lambda_5$ is constrained by taking into account 
vacuum stability and perturbative unitarity as follows.

First, we require that the Higgs potential is bounded from below in any direction with a large scalar field value. 
The sufficient condition to keep such a stability of the vacuum is given by~\cite{Ma-Deshpande,VS_THDM,VS_THDM2} %% Add explanations
\begin{align}
\lambda_1>0, \quad \lambda_2>0,\quad \sqrt{\lambda_1\lambda_2}+\lambda_3+\text{MIN}(0,\lambda_4+\lambda_5,\lambda_4-\lambda_5)>0.
\end{align}

Second, the perturbative unitarity bound~\cite{Uni-2hdm1,Uni-2hdm2,Uni-2hdm3,Uni-2hdm4} is given by requiring that 
all the independent eigenvalues of the $T$ matrix $a_{i,\pm}^0$ ($i=1$-6) for the $S$-wave amplitude of the elastic scatterings of 2-body boson states
are satisfied as
\begin{align}
|a_{i,\pm}^0|\leq\frac{1}{2}, 
\end{align}
where each of $a_{i,\pm}^0$ is given by~\cite{Uni-2hdm2,Uni-2hdm3,Uni-2hdm4} 
\begin{align}
a_{1,\pm}^0 &=  \frac{1}{32\pi}
\left[3(\lambda_1+\lambda_2)\pm\sqrt{9(\lambda_1-\lambda_2)^2+4(2\lambda_3+\lambda_4)^2}\right],\\
a_{2,\pm}^0 &=
\frac{1}{32\pi}\left[(\lambda_1+\lambda_2)\pm\sqrt{(\lambda_1-\lambda_2)^2+4\lambda_4^2}\right],\\
a_{3,\pm}^0 &= \frac{1}{32\pi}\left[(\lambda_1+\lambda_2)\pm\sqrt{(\lambda_1-\lambda_2)^2+4\lambda_5^2}
\right],\\
a_{4,\pm}^0 &= \frac{1}{16\pi}(\lambda_3+2\lambda_4\pm 3\lambda_5),\\
a_{5,\pm}^0 &= \frac{1}{16\pi}(\lambda_3\pm\lambda_4),\\
a_{6,\pm}^0 &= \frac{1}{16\pi}(\lambda_3\pm\lambda_5).
\end{align}

\begin{figure}[t]
%========================================
\centering
\includegraphics[width=9cm]{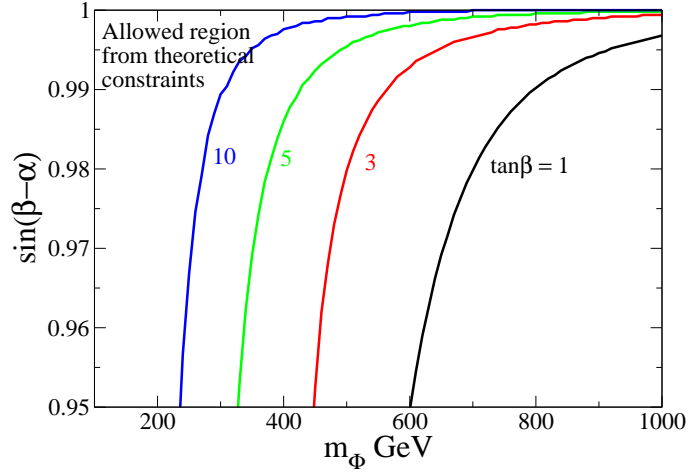} 
\caption{The upper limit on the mass of additional Higgs bosons $m_\Phi(\equiv m_{H^\pm}=m_A=m_H)$ as a function of $\sin(\beta-\alpha)$ for 
each fixed value of $\tan\beta$ in the case of  $\cos(\beta-\alpha)<0$. 
The left regions from each curve are allowed by the constraints of vacuum stability and unitarity. }
\label{th}
%========================================
\end{figure}

In Fig.~\ref{th}, we show the allowed parameter region on the $m_\Phi$-$\sin(\beta-\alpha)$ plane $(m_\Phi\equiv m_{H^\pm}=m_A=m_H)$ from the constraints of vacuum stability and unitarity.
It is seen that a large mass of additional Higgs bosons is allowed in a case with $\sin(\beta-\alpha)\simeq 1$. 
As another view of this figure, we can extract the scale of the mass of the second Higgs boson from the precise measurement of $\kappa_V^{}$ using Eq.~(\ref{scaling3}). 
For example, if 1\% deviation in the $hVV$ coupling is found at future collider experiments, then the second Higgs boson should exist below about 800 GeV. 

\subsection{The oblique parameters}

The $S$, $T$ and $U$ parameters proposed by Peskin and Takeuchi~\cite{Peskin-Takeuchi} are modified in the THDM from those predicted in the SM 
due to the additional Higgs boson loop contributions and modified values of the SM-like Higgs boson coupling constants~\cite{delrho_THDM}. 
We define the differences of $S$, $T$ and $U$ parameters as $\Delta S= S_{\text{THDM}}-S_{\text{SM}}$,
$\Delta T = T_{\text{THDM}}-T_{\text{SM}}$ and $\Delta U = U_{\text{THDM}}-U_{\text{SM}}$. These are calculated in terms of $x$ defined in Eq.~(\ref{xpara}) as 
\begin{align}
\Delta S & = \frac{1}{4\pi} \Bigg\{ F_5'(m_Z^2;m_H,m_A)-\frac{1}{3}\ln m_{H^\pm}^2 \notag\\
&\quad \quad\quad  +x^2\Big[F_\Delta'\Big(\frac{m_A}{m_h},\frac{m_Z}{m_h}\Big)-F_\Delta'\Big(\frac{m_A}{m_H},\frac{m_Z}{m_H}\Big)+G_\Delta'\Big(\frac{m_H}{m_Z},\frac{m_h}{m_Z}\Big)  \Big]\Bigg\}
+\mathcal{O}(x^3), \\
\Delta T & = 
\frac{1}{4\pi e^2v^2} \Bigg\{
F_5(0;m_A,m_{H^\pm})+m_H^2F_\Delta\left(\frac{m_{H^\pm}}{m_H},\frac{m_A}{m_H}\right)\notag\\
&\quad\quad +x^2\Big[m_H^2F_\Delta\left(\frac{m_A}{m_H},\frac{m_{H^\pm}}{m_H}\right)+m_h^2F_\Delta\left(\frac{m_{H^\pm}}{m_h},\frac{m_A}{m_h}\right)
+m_W^2F_\Delta\left(\frac{m_{H}}{m_W},\frac{m_h}{m_W}\right)\notag\\
&\quad\quad +m_Z^2F_\Delta\left(\frac{m_h}{m_Z},\frac{m_H}{m_Z}\right)+4m_W^2 G_\Delta\Big(\frac{m_H}{m_W},\frac{m_h}{m_W}\Big)-4m_Z^2G_\Delta\Big(\frac{m_H}{m_Z},\frac{m_h}{m_Z}\Big)\Big] \Bigg\}+\mathcal{O}(x^3),  \\
\Delta U & = 
\frac{1}{4\pi}  \Bigg\{
F_\Delta'\Big(\frac{m_A}{m_{H^\pm}},\frac{m_H}{m_{H^\pm}}\Big)-\frac{1}{3}\ln m_{H^\pm}^2-F_5'(m_Z^2;m_A,m_H)\notag\\
& \quad \quad\quad +x^2\Big[
F_\Delta'\Big(\frac{m_A}{m_H},\frac{m_Z}{m_H}\Big)
-F_\Delta'\Big(\frac{m_A}{m_h},\frac{m_Z}{m_h}\Big)
+F_\Delta'\Big(\frac{m_{H^\pm}}{m_h},\frac{m_W}{m_h}\Big)
-F_\Delta'\Big(\frac{m_{H^\pm}}{m_H},\frac{m_W}{m_H}\Big) \notag\\
&\quad \quad\quad +G_\Delta'\Big(\frac{m_H}{m_Z},\frac{m_h}{m_Z}\Big)-G_\Delta'\Big(\frac{m_H}{m_W},\frac{m_h}{m_W}\Big) \Bigg\}+\mathcal{O}(x^3), 
\end{align}
where $F_5'(m_V^2;m_1,m_2)=[F_5(m_V^2;m_1,m_2)-F_5(0;m_1,m_2)]/m_V^2$. 
The loop functions are given by  
\begin{align}
F_5(p^2,m_1,m_2)&= \int_0^1 dx \left[(2x-1)(m_1^2-m_2^2)+p^2(2x-1)^2\right]\ln\Delta_B,\\ 
F_\Delta(x_1,x_2)&=\frac{1}{2}(x_1^2-x_2^2)+\frac{x_1^2}{1-x_1^2}\ln x_1^2-\frac{x_2^2}{1-x_2^2}\ln x_2^2, \\
G_\Delta(x_1,x_2)&=\frac{1}{2}\ln \frac{x_1^2}{x_2^2}-\frac{1+x_1^2}{2(1-x_1^2)}\ln x_1^2-\frac{1+x_2^2}{2(1-x_2^2)}\ln x_2^2, \\
F_\Delta'(x_1,x_2)& = \frac{1}{3}\left[\frac{2(x_1^2-x_2^2)(1-x_1^2x_2^2)}{(1-x_1^2)^2(1-x_2^2)^2}-\frac{x_1^4(x_1^2-3)}{(1-x_1^2)^3}\ln x_1^2+\frac{x_2^4(x_2^2-3)}{(1-x_2^2)^3}\ln x_2^2\right], \\
G_\Delta'(x_1,x_2)&=2\left[-\frac{1-x_1^4+2x_1^2\ln x_1^2}{(1-x_1^2)^3}+\frac{1-x_2^4+2x_2^2\ln x_1^2}{(1-x_2^2)^3}\right],  
\end{align} 
where
\begin{align}
\Delta_B  &= -x(1-x)p^2 + xm_1^2 + (1-x)m_2^2.
\end{align}
In the case of $p^2=0$, the $F_5$ function is expressed by 
\begin{align}
F_5(0;m_1,m_2)&=\frac{1}{2}(m_1^2+m_2^2)+\frac{2m_1^2m_2^2}{m_1^2-m_2^2}\ln\frac{m_2}{m_1}, 
\end{align}
which gives zero in the case of $m_1=m_2$. 
Therefore, it is seen that $\Delta T$ becomes zero when $x=0$ and $m_A=m_{H^\pm}$ or $x=0$ and $m_H=m_{H^\pm}$ is taken. 

\subsection{Flavour Constraints}

The mass of $H^\pm$ can be constrained from various $B$ physics processes, because contributions from the SM $W$-boson mediation are replaced 
by $H^\pm$. 
In most of the cases, the constraint from the $b\to s\gamma$ process provides 
the most stringent lower limit on $m_{H^\pm}$~\cite{bsg,Misiak}. 
In Ref.~\cite{Misiak}, the branching ratio of $\bar{B} \to X_s \gamma$ has been calculated at the next-to-next-to-leading order in the 
Type-I and Type-II THDMs. 
A lower bound has been found to be $m_{H^\pm} \gtrsim 380$ GeV at 95\% confidence level (CL) in the Type-II THDM with $\tan\beta\gtrsim 2$.
A stronger bound for $m_{H^\pm}$ is obtained for smaller values of $\tan\beta$. 
On the other hand, 
in the Type-I THDM, the bound from $b\to s\gamma$ is important in the case with low $\tan\beta$; e.g., 
$m_{H^\pm}\lesssim$ 200 (800) GeV is excluded at 95\% CL in the case of $\tan\beta=2~(1)$. 
When we consider the case with $\tan\beta\gtrsim 2.5$,  
the bound on $m_{H^\pm}$ is weaker than the lower bound from the direct search at LEP, namely, about 80 GeV~\cite{PDG}. 
The similar bounds as those given in the Type-II and Type-I THDMs
can be obtained in the Type-Y and Type-X THDMs, respectively, 
because of the same structure of quark Yukawa interactions. 

For a large $\tan\beta$ case, 
bounds from $B\to \tau\nu$~\cite{Btaunu,Maria_Btaunu}, $\tau\to \mu\nu\bar{\nu}$~\cite{Maria_Btaunu,Maria_Tau} 
and the muon anomalous magnetic moment~\cite{Haber_g2,Maria_g2} can be more important as compared to the bound from $b\to s\gamma$ in the Type-II THDM. 
For example, the lower limit on $m_{H^\pm}$ to be about 400 GeV is given at 95\% CL in the case of $\tan\beta \gtrsim 50$ in the Type-II THDM~\cite{Maria_Btaunu}. 

For a small $\tan\beta$ case, the $B^0$-$\bar{B}^0$ mixing is getting important to obtain a severe constraint on $m_{H^\pm}$ in the THDMs. 
In the case of $\tan\beta=1$, $m_{H^\pm}\lesssim 500$ GeV is exluded at 95\% CL in all the types of THDMs~\cite{Stal}. 
This gives the stronger (weaker) bound than that from $b\to s\gamma$ in the Type-II and Type-Y (Type-I and Type-X) THDMs.

\subsection{Direct searches for additional Higgs bosons at the LHC (7-8 TeV)}

The neutral Higgs bosons in the MSSM have been searched in the $\tau^+\tau^-$ decay mode
in the gluon fusion and bottom quark associated productions~\cite{tautau-ATLAS,tautau-CMS} using data with 7 TeV and 8 TeV of the collision energy and  
4.9 fb$^{-1}$ and 19.7 fb$^{-1}$ of the integrated luminosity, respectively.
Because the production cross section of the CP-odd Higgs boson from the bottom quark associated production is proportional to $\tan^2\beta$, 
high-$\tan\beta$ regions can be excluded by this process. 
For example, $\tan\beta\gtrsim 10$ and $\tan\beta\gtrsim 40$ have been excluded at 95\% CL for the fixed value of the mass of the CP-odd Higgs boson to be 
300 GeV and 800 GeV, respectively~\cite{tautau-CMS}. 
We can obtain a similar bound on $\tan\beta$ for a fixed value of $m_A$ 
in the Type-II THDM, because the structure of the Yukawa interaction is the same as that in the MSSM. 
Although the $Hf\bar{f}$ coupling constant can be different in the Type-II THDM and the MSSM, we can achieve 
a similar value by taking $\sin(\beta-\alpha)\simeq 1$, especially for the case with a rather large mass of the CP-odd Higgs boson in the MSSM. 

When $\sin(\beta-\alpha)\neq 1$ is given, $H\to W^+W^-/ZZ$ decays can open in addition to the decay modes into a fermion pair. 
The search for the $H\to WW \to e\nu\mu\nu$ signal has been performed~\cite{HWW-ATLAS} in the range of $135~\text{GeV}<m_H<300~\text{GeV}$
using data with 8 TeV of the collision energy and 13 fb$^{-1}$ of the integrated luminosity. 
The bound is presented in the $m_H$-$\cos\alpha$ plane for each fixed value of $\tan\beta$ in the Type-I and Type-II THDMs. 
In the Type-I THDM with $\tan\beta > 1$, the strongest lower limit on $m_H$ is given to be about 220 GeV at 95\% CL. 
On the other hand, in the Type-II THDM, similar bounds have been given as in the Type-I THDM. 
However, for a case with large $\tan\beta$, the excluded regions are shrinked due to an enhancement of fermonic decay modes such as $H\to b\bar{b}$. 

In Ref.~\cite{HA-THDM-CMS}, $H\to hh$ and $A\to Zh$ decays have been searched in the THDMs with data of the collision energy to be 8 TeV and 
the integrated luminosity to be 19.5 fb$^{-1}$. 
Multi-lepton and di-photon final states have been used for this search.  
The upper limit on the cross section times branching ratio has been presented for each of the processes $gg\to H\to hh$ and $gg\to A\to Zh$
; e.g., the upper limit of 8 (4) pb is given for the case of $m_H=260$ (360) GeV in the $H\to hh$ decay, while 
that of 1.6 (1.0) pb is given for the case of $m_H=260$ (360) GeV in the $H\to Zh$ decay. 
These bounds can be translated into the excluded regions on the
$\cos(\beta-\alpha)$-$\tan\beta$ plane for given values of $m_H$ depending on the type of Yukawa interaction.  

\subsection{Measurements of the Higgs boson couplings at LHC (7-8 TeV), and future collider experiments}

Both the ATLAS and CMS Collaborations have provided scaling factors for the Higgs boson couplings 
extracted from combined data of Higgs boson searches with $\sqrt{s}=$ 7 and 8 TeV and 25 fb$^{-1}$ of the integrated luminosity~\cite{ATLAS_Coupling1,ATLAS_Coupling2,ATLAS_Coupling3,CMS_Coupling0,CMS_Coupling}. 
Under assumptions of the universal scaling factors for fermions and vector bosons; i.e., 
$\kappa_F^{} = \kappa_t=\kappa_b=\kappa_\tau$ and $\kappa_V^{} = \kappa_W^{}=\kappa_Z^{}$, 
current data gives 
\begin{align}
&\kappa_V^{} = 1.15\pm 0.08, \quad \kappa_F^{} = 0.99^{+0.08}_{-0.15},\quad \text{ATLAS~\cite{ATLAS_Coupling2}}, \label{hc1}\\
&\kappa_V^{} = 1.01\pm 0.07, \quad \kappa_F^{} = 0.87^{+0.14}_{-0.13},\quad \text{CMS~\cite{CMS_Coupling}}, \label{hc2}
\end{align}
from the two parameters ($\kappa_F^{}$ and $\kappa_V^{}$) fit analysis based on Ref.~\cite{Handbook}. 
The scaling factors for the loop induced Higgs boson couplings $\kappa_g$ and $\kappa_\gamma$ have also been 
measured under the assumptions of $\kappa_F=\kappa_V=1$, 
\begin{align}
&\kappa_g = 1.08^{+0.15}_{-0.13},\quad  \kappa_\gamma = 1.19^{+0.15}_{-0.12},\quad \text{ATLAS~\cite{ATLAS_Coupling2}}, \label{hc3}\\
&\kappa_g = 0.89^{+0.11}_{-0.10},\quad  \kappa_\gamma = 1.14^{+0.12}_{-0.13},\quad \text{CMS~\cite{CMS_Coupling}}, \label{hc4}
\end{align}
from the two parameters ($\kappa_g^{}$ and $\kappa_\gamma$) fit analysis
%\footnote{The scalling factors provided in Eqs.~(\ref{hc1})-(\ref{hc4}) are given by assuming no contributions from new particles beyond the SM to the total width of the Higgs boson.}  
based on Ref.~\cite{Handbook}. 
We can see that all the SM predictions ($\kappa_X^{}=1$) are included within the 2-$\sigma$ uncertainty of the measured scaling factors, where  
the current 1-$\sigma$ uncertainties of the scaling factors are typically of ${\cal O}(10\%)$.

These scaling factors are expected to be measured more precisely at future collider experiments 
such as the HL-LHC and the ILC.
In TABLE~\ref{Tab:Sensitivity}, expected accuracies of the measurement for the scaling factors are listed at the LHC and at the ILC with 
several collision energies and integrated luminosities. 

%%%%%%%%%%
\begin{table}
\centering
\begin{tabular}{lcccccc}\hline\hline
Facility                     &   LHC            &   HL-LHC         &   ILC500  & ILC500-up & ILC1000 &   ILC1000-up \\% &   CLIC          & TLEP (4 IPs) \\
$\sqrt{s}$ (GeV)                 &   14,000         &   14,000         &   250/500 & 250/500 & 250/500/1000 & 250/500/1000 \\% &  350/1400/3000  &  240/350    \\
$\int{\cal L}dt$ (fb$^{-1}$) & 300/expt         & 3000/expt        &   250+500 & 1150+1600 & 250+500+1000 & 1150+1600+2500  \\% & 500+1500+2000  & 10,000+2600  \\
\hline 
 $\kappa_{\gamma}$ &  $5-7$\%      & $2-5$\%       & 8.3\% & 4.4\% & 3.8\%  & 2.3\%            \\% & $-$/5.5/$<$5.5\% & 1.45\% \\         
 $\kappa_g$           &  $6-8$\%       & $3-5$\%       & 2.0\% & 1.1\% & 1.1\%  & 0.67\%            \\% & 3.6/0.79/0.56\% & 0.79\% \\
 $\kappa_W$           & $4-6$\% & $2-5$\% & 0.39\% & 0.21\% & 0.21\% & 0.2\%         \\% & 1.5/0.15/0.11\% & 0.10\%\\
 $\kappa_Z$           & $4-6$\% & $2-4$\% & 0.49\% & 0.24\% & 0.50\% & 0.3\%        \\% & 0.49/0.33/0.24\%& 0.05\%\\ 

 $\kappa_{E}^{}$      & $6-8$\%       &  $2-5$\%      & 1.9\%  & 0.98\% & 1.3\%  & 0.72\%            \\% & 3.5/1.4/$<$1.3\% & 0.51\%\\
 $\kappa_D =\kappa_b$ & $10-13$\%        &  $4-7$\%       & 0.93\% & 0.60\% & 0.51\% & 0.4\%           \\% & 1.7/0.32/0.19\% & 0.39\% \\
 $\kappa_U =\kappa_t$ & $14-15$\%        &  $7-10$\%      & 2.5\%  &  1.3\% & 1.3\%   & 0.9\%            \\% & 3.1/1.0/0.7\% & 0.69\% \\ 
%\hline
\hline\hline
 \end{tabular}
 \caption{Expected precision on the Higgs boson couplings and total width at the 1-$\sigma$ level
   from a constrained 7-parameter fit quoted from Table 1-20 in Ref.~\cite{snowmass}.  }
 \label{Tab:Sensitivity}
\end{table}

\section{Renormalization} \label{renormalization}

We discuss the renormalization of the Higgs boson couplings, i.e, $hZZ$, $hWW$, $hf\bar{f}$ and $hhh$ at the one-loop level. 
In previous works, each part of the renormalized Higgs boson couplings has been calculated. 
The one-loop corrected $hZZ$ and $hhh$ couplings have been evaluated in Ref.~\cite{KOSY} in the Type-II THDM, and 
the $hf\bar{f}$ couplings have been calculated in Ref.~\cite{KKY} in the four types of THDMs. 

We perform renormalization calculations based on the on-shell scheme which has been applied in Ref.~\cite{KOSY}\footnote{For the determination of the counter term for $M^2$, 
the minimal subtraction scheme has been applied. }. 
However, it has been pointed out that there remains gauge dependence in the determination of the counter term of $\beta$ in Ref.~\cite{gauge_depend}. 
We thus construct a new renormalization scheme for $\beta$ to get rid of the gauge dependence. 
As pointed out later in the paper, the gauge dependence is not completely removed, 
but shifted to a sector which does not contribute to the investigated couplings.

First, we prepare a set of independent counter terms by shifting all the relevant bare parameters in the Lagrangian. 
We then give the renormalized one- and two-point functions which are written in terms of the contributions from 1PI diagrams and counter terms. 
After that, we set the same number of renormalization conditions as the number of independent counter terms to determine them. 

\subsection{Parameter shift and renormalized functions}

We first perform the parameter shift of the electroweak sector and Yukawa sector as the following
\begin{align}
m_V^2 &\to m_V^2 +\delta m_V^2~~(V=W,Z),\quad 
\alpha_{\text{em}} \to \alpha_{\text{em}}  +\delta \alpha_{\text{em}} ,\notag\\
m_f &\to m_f+\delta m_f,\quad T_{1,2}\to \delta T_{1,2},\notag\\
m_{\varphi}^2 &\to m_{\varphi}^2+\delta m_{\varphi}^2,\quad
\alpha\to \alpha+\delta\alpha,\quad
\beta\to \beta+\delta\beta,\quad
M^2\to M^2+\delta M^2,
\end{align}
where $\varphi=H^\pm,~A,~H$ and $h$.
The wave functions for the SM gauge bosons $B_\mu$ and $W_\mu^a$ and the SM left (right) handed fermions $\psi_L$ ($\psi_R$) are shifted as 
\begin{align}
B_\mu\to & \left(1+\frac{1}{2}\delta Z_B\right)B_\mu,\quad 
W_\mu^a \to  \left(1+\frac{1}{2}\delta Z_W\right)W_\mu^a, 
\quad \psi_{L/R} \to  \left(1+\frac{1}{2}\delta Z_{L/R}^f\right)\psi_{L/R}. 
\end{align}

We can then write down the renormalized two point functions for each particle. 
In the following, $\hat{\Pi}_{XY}(p^2)$ and $\Pi_{XY}^{\text{1PI}}(p^2)$ respectively denote the renormalized two point functions 
and the 1PI diagram contributions for fields $X$ and $Y$ with the external momentum $p_\mu$. 
The analytic formulae for the 1PI diagram contributions are given in Appendix~C. 
For the gauge boson two point functions $W^+W^-$, $ZZ$, $\gamma\gamma$ and the $Z$-$\gamma$ mixing, we have   
\begin{align}
\hat{\Pi }_{WW}(p^2)&=\Pi_{WW}^{\text{1PI}}(p^2)-\delta m_W^2+\delta Z_{W}(p^2-m_W^2),\\
\hat{\Pi }_{ZZ}(p^2)&=\Pi_{ZZ}^{\text{1PI}}(p^2)-\delta m_Z^2+\delta Z_{Z}(p^2-m_Z^2),\\
\hat{\Pi}_{\gamma\gamma}(p^2)&=\Pi_{\gamma\gamma}^{\text{1PI}}(p^2)+p^2\delta Z_{\gamma}, \label{Pigg} \\
\hat{\Pi}_{Z\gamma}(p^2)&=\Pi_{Z\gamma}^{\text{1PI}}(p^2)-\delta Z_{Z\gamma}\left(p^2-\frac{1}{2}m_Z^2\right)-m_Z^2\frac{\delta s_W^2}{2s_Wc_W}, 
\end{align}
where
\begin{align}
&\left(\begin{array}{c}
\delta Z_Z\\
\delta Z_\gamma
\end{array}\right)
=\left(
\begin{array}{cc}
c_W^2 & s_W^2\\
s_W^2 & c_W^2
\end{array}\right)
\left(\begin{array}{c}
\delta Z_W\\
\delta Z_B
\end{array}\right),\quad 
\frac{\delta s_W^2}{s_W^2} = \frac{c_W^2}{s_W^2}\left(\frac{\delta m_Z^2}{m_Z^2}-\frac{\delta m_W^2}{m_W^2}\right),\notag\\
&\delta Z_{Z\gamma}=c_Ws_W(\delta Z_W-\delta Z_B)=\frac{c_Ws_W}{c_W^2-s_W^2}(\delta Z_Z-\delta Z_\gamma). 
\label{count_depend}
\end{align}
The renormalized fermion two point function is expressed by the following two parts: 
\begin{align}
\hat{\Pi}_{ff}(p^2) &= \hat{\Pi}_{ff,V}(p^2) +\hat{\Pi}_{ff,A}(p^2), 
\end{align}
where
\begin{align}
\hat{\Pi}_{ff,V}(p^2) &= 
p\hspace{-2mm}/\left[\Pi_{ff,V}^{\text{1PI}}(p^2) + \delta Z_V^f\right] 
+m_f \left[\Pi_{ff,S}^{\text{1PI}}(p^2)-\delta Z_V^f-\frac{\delta m_f}{m_f}\right], \notag\\
\hat{\Pi}_{ff,A}(p^2) &= - p\hspace{-2mm}/\gamma_5\left[\Pi_{ff,A}^{\text{1PI}}(p^2) + \delta Z_A^f\right],  \label{fermi_2p}
\end{align}
with
\begin{align}
&\delta Z_V^f = \frac{\delta Z_L^f+\delta Z_R^f}{2},\quad  \delta Z_A^f = \frac{\delta Z_L^f -\delta Z_R^f}{2}. 
\end{align}
In Eq.~(\ref{fermi_2p}), $\Pi_{ff,V}^{\text{1PI}}$, $\Pi_{ff,A}^{\text{1PI}}$ and $\Pi_{ff,S}^{\text{1PI}}$ 
are the vector, axial vector and scalar parts of the 1PI diagram contributions at the one-loop level, respectively.

For the scalar sector, we first define shifts in the weak eigenbasis of the scalar fields: 
\begin{align}
\begin{pmatrix}
h_1 \\ 
h_2
\end{pmatrix} \to \tilde{Z}_{\text{even}}
\begin{pmatrix}
h_1 \\ 
h_2
\end{pmatrix},~
\begin{pmatrix}
z_1 \\ 
z_2
\end{pmatrix} \to \tilde{Z}_{\text{odd}}
\begin{pmatrix}
z_1 \\ 
z_2
\end{pmatrix},~
\begin{pmatrix}
w_1^\pm \\ 
w_2^\pm
\end{pmatrix} \to \tilde{Z}_{\pm}
\begin{pmatrix}
w_1^\pm \\ 
w_2^\pm
\end{pmatrix},  
\end{align}
where $\tilde{Z}_{\text{even}}$, $\tilde{Z}_{\text{odd}}$ and $\tilde{Z}_{\pm}$ are arbitrary real $2\times 2$ matrices. 
We then express shifts of the scalar fields in the mass eigenbasis as 
\begin{align}\hspace{-5mm}
\begin{pmatrix}
H \\
h
\end{pmatrix}
\to R(-\delta \alpha)Z_{\text{even}}
\begin{pmatrix}
H \\
h
\end{pmatrix},\,
\begin{pmatrix}
G^0 \\
A
\end{pmatrix}
\to R(-\delta \beta)Z_{\text{odd}}
\begin{pmatrix}
G^0 \\
A
\end{pmatrix},\,
\begin{pmatrix}
G^\pm \\
H^\pm
\end{pmatrix}
\to R(-\delta \beta)Z_{\pm}
\begin{pmatrix}
G^\pm \\
H^\pm
\end{pmatrix}, 
\end{align} 
where we introduce $Z_\text{even} \equiv R(-\alpha)\tilde{Z}_{\text{even}}R(\alpha)$ and 
$Z_{\text{odd}/\pm} \equiv R(-\beta)\tilde{Z}_{\text{odd}/\pm}R(\beta)$. 
We define the matrix elements of them as follows: 
\begin{align}
Z_\text{even} = 
\begin{pmatrix}
1+\frac{1}{2}Z_{H} & \delta C_{Hh} \\
\delta C_{hH} & 1+\frac{1}{2}Z_h
\end{pmatrix},~
Z_\text{odd} = 
\begin{pmatrix}
1+\frac{1}{2}Z_{G} & \delta C_{GA} \\
\delta C_{AG} & 1+\frac{1}{2}Z_A
\end{pmatrix},~
Z_\pm = 
\begin{pmatrix}
1+\frac{1}{2}Z_{G^\pm} & \delta C_{G^+H^-} \\
\delta C_{H^+G^-} & 1+\frac{1}{2}Z_{H^\pm}
\end{pmatrix}.   \label{wave}
\end{align}
We note that in Ref.~\cite{KOSY}, the above matrices are chosen to be a symmetric form; i.e., $\delta C_{Hh}=\delta C_{hH}$, 
$\delta C_{GA}=\delta C_{AG}$ and $\delta C_{G^+H^-}=\delta C_{H^+G^-}$.  
In this paper, we do not take the symmetric form, and we use the additional degrees of freedom 
to remove the gauge dependence in the renormalization of $\delta \beta$
as it will be discussed in Sec. III-D. 
Finally, we can express the shifts of the scalar fields by 
\begin{align}
&\left(\begin{array}{c}
H\\
h
\end{array}\right) 
\to 
\begin{pmatrix}
1+\frac{1}{2}\delta Z_H & \delta C_{Hh}+\delta\alpha\\
\delta C_{hH}-\delta\alpha  & 1+\frac{1}{2}\delta Z_h
\end{pmatrix}
\left(\begin{array}{c}
H\\
h
\end{array}\right),\notag\\
&\left(\begin{array}{c}
G^0\\
A
\end{array}\right) \to 
\begin{pmatrix}
1+\frac{1}{2}\delta Z_{G^0} & \delta C_{GA}+\delta\beta\\
\delta C_{AG}-\delta\beta  & 1+\frac{1}{2}\delta Z_{A}
\end{pmatrix}
\left(\begin{array}{c}
G^0\\
A
\end{array}\right),\notag\\
&\left(\begin{array}{c}
G^\pm\\
H^\pm
\end{array}\right) \to
\begin{pmatrix}
1+\frac{1}{2}\delta Z_{H^+} & \delta C_{G^+H^-}+\delta\beta\\
\delta C_{H^+G^-}-\delta\beta & 1+\frac{1}{2}\delta Z_{H^\pm}
\end{pmatrix}
\left(\begin{array}{c}
G^\pm\\
H^\pm
\end{array}\right).  
\label{shift}
\end{align}

For the scalar sector, we have the renormalized one-point function for $h$ and $H$ as 
\begin{align}
\hat{T}_h = \delta T_h + \Gamma_h^{\rm{1PI}},\quad 
\hat{T}_H = \delta T_H + \Gamma_H^{\rm{1PI}}, 
\end{align}
where
\begin{align}
\left(\begin{array}{c}
\delta T_1\\
\delta T_2
\end{array}\right)=
R(\alpha)\left(\begin{array}{c}
\delta T_H\\
\delta T_h
\end{array}\right).
\end{align}
The renormalized two-point functions are expressed as
\begin{align}
\hat{\Pi}_{hh}(p^2)&=\tilde{\Pi}_{hh}^{\text{1PI}}(p^2)+\left[(p^2-m_h^2)\delta Z_h-\delta m_h^2\right],\\
\hat{\Pi}_{HH}(p^2)&=\tilde{\Pi}_{HH}^{\text{1PI}}(p^2)+\left[(p^2-m_H^2)\delta Z_H-\delta m_H^2\right],\\
\hat{\Pi}_{AA}(p^2)&=\tilde{\Pi}_{AA}^{\text{1PI}}(p^2)+\left[(p^2-m_A^2)\delta Z_A-\delta m_A^2\right],\\
\hat{\Pi}_{H^+H^-}(p^2)&=\tilde{\Pi}_{H^+H^-}^{\rm{1PI}}(p^2)+\left[(p^2-m_{H^\pm}^2)\delta Z_{H^\pm}
-\delta m_{H^\pm}^2\right], 
\end{align}
and those of the scalar mixings are given by 
\begin{align}
\hat{\Pi}_{Hh}(p^2)&=\tilde{\Pi}_{Hh}^{\text{1PI}}(p^2)
+p^2(\delta C_{hH}+\delta C_{Hh})+m_h^2(\delta \alpha -\delta C_{hH})-m_H^2(\delta \alpha +\delta C_{Hh}),\\
\hat{\Pi}_{AG}(p^2)&=\tilde{\Pi}_{AG}^{\text{1PI}}(p^2)+p^2(\delta C_{AG}+\delta C_{GA})+m_A^2(\delta\beta-\delta C_{AG}), \\
\hat{\Pi}_{H^+G^-}(p^2)&=\tilde{\Pi}_{H^+ G^-}^{\rm{1PI}}(p^2)+p^2(\delta C_{H^+G^-}+\delta C_{G^+H^-})+m_{H^\pm}^2(\delta\beta-\delta C_{H^+G^-}), 
\end{align}
where
\begin{align}
\tilde{\Pi}_{hh}^{\text{1PI}}(p^2)&=\Pi_{hh}^{\text{1PI}}(p^2)+\frac{s_\alpha^2\delta T_1}{c_\beta v}+\frac{c_\alpha^2\delta T_2}{s_\beta v}, \\
\tilde{\Pi}_{HH}^{\text{1PI}}(p^2)&=\Pi_{HH}^{\text{1PI}}(p^2)+\frac{c_\alpha^2\delta T_1}{c_\beta v}+\frac{s_\alpha^2\delta T_2}{s_\beta v}, \\
\tilde{\Pi}_{AA}^{\text{1PI}}(p^2)&=\Pi_{AA}^{\text{1PI}}(p^2)+\frac{s_\beta^2\delta T_1}{c_\beta v}+\frac{c_\beta^2\delta T_2}{s_\beta v}, \\
\tilde{\Pi}_{H^+H^-}^{\rm{1PI}}(p^2)&=\Pi_{H^+H^-}^{\rm{1PI}}(p^2)+\frac{s_\beta^2\delta T_1}{c_\beta v}+\frac{c_\beta^2\delta T_2}{s_\beta v}, \\
\tilde{\Pi}_{Hh}^{\text{1PI}}(p^2)&=\Pi_{Hh}^{\text{1PI}}(p^2)-s_\alpha c_\alpha\left(\frac{\delta T_1}{c_\beta v}-\frac{\delta T_2}{s_\beta v}\right), \\
\tilde{\Pi}_{AG}^{\text{1PI}}(p^2)
&=\Pi_{AG}^{\text{1PI}}(p^2)+\frac{1}{v}\left[\sin(\beta-\alpha)T_H -\cos(\beta-\alpha)T_h  \right], \\
\tilde{\Pi}_{H^+G^-}^{\text{1PI}}(p^2)
&=\Pi_{H^+G^-}^{\text{1PI}}(p^2)+\frac{1}{v}\left[\sin(\beta-\alpha)T_H -\cos(\beta-\alpha)T_h  \right]. 
\end{align}

\subsection{Renormalization conditions in the electroweak gauge sector}

The renormalization of the electroweak parameters can be done in the same way as in the SM, because 
the number of parameters to describe the electroweak observables are the same in the THDM. 
This nature is also applied to models based on the $SU(2)_L\times U(1)_Y$ gauge symmetry with $\rho =1$ at the tree level\footnote{When 
we discuss models without $\rho=1$ at the tree level such as models with isosipin triplet scalar fields, 
one additional input parameter is required to express the electroweak sector. 
Therefore, we need an additional renormalization condition to determine the extra counter term associated with the parameter. 
In the model with a $Y=0$ Higgs triplet field, 
the renormalization of electroweak parameters has been discussed in Refs.~\cite{Blank_Hollik,Chen-Dawson-Jackson}. 
Furthermore, in the model with a $Y=1$ Higgs triplet field, that has also been discussed in Refs.~\cite{Kanemura-Yagyu,AKKY_Full}. }. 

We apply the electroweak on-shell scheme based on Ref.~\cite{Hollik} to our model. 
There are five counter terms in the electroweak sector; i.e., 
$\delta m_W^2$, $\delta m_Z^2$, $\delta \alpha_\text{em}$, $\delta Z_W$ and $\delta Z_B$. 
Therefore, we need the following five renormalization conditions to determine them: 
\begin{align}
&\text{Re}\hat{\Pi}_{WW}(m_W^2) = 0,\quad \text{Re}\hat{\Pi}_{ZZ}(m_Z^2)=0,\label{rc1}\\
&\frac{d}{dp^2}\hat{\Pi}_{\gamma\gamma}(p^2)\Big|_{p^2=0} =0,\quad
\hat{\Pi}_{Z\gamma}(0) = 0,\label{rc3}\\
&\hat{\Gamma}_\mu^{\gamma ee }(q^2=0,~p_1\hspace{-3mm}/=p_2\hspace{-3mm}/=m_e) = ie\gamma_\mu \label{rc4}, 
\end{align} 
where $\hat{\Gamma}_\mu^{\gamma ee}$ is the renormalized photon-electron-positron vertex. 
From the above conditions, we obtain 
\begin{align}
&\delta m_W^2 = \text{Re}\Pi_{WW}^{\text{1PI}}(m_W^2),\quad  
\delta m_Z^2  = \text{Re}\Pi_{ZZ}^{\text{1PI}}(m_Z^2),\quad 
\frac{\delta \alpha_{\text{em}}}{\alpha_{\text{em}}}
=\Pi_{\gamma\gamma}^{\text{1PI}}(0)^\prime-\frac{2s_W}{c_W}\frac{\Pi_{Z\gamma}^{\text{1PI}}(0)}{m_Z^2}, \\
&\delta Z_\gamma =  -\Pi_{\gamma\gamma}^{\text{1PI}}(0)^\prime, \quad
\delta Z_{Z\gamma} = -\frac{2}{m_Z^2}\Pi_{Z\gamma}^{\text{1PI}}(0)+\frac{\delta s_W^2}{s_Wc_W}, \label{zzz}
\end{align} 
where $\Pi_{\gamma\gamma}^{\text{1PI}}(0)^\prime= \frac{d}{dp^2}\Pi_{\gamma\gamma}^{\text{1PI}}(p^2)\Big|_{p^2 =0}$. 
The other counter terms are also determined by 
\begin{align}
\delta Z_Z &= -\Pi_{\gamma\gamma}^\text{1PI}(0)^\prime-\frac{2(c_W^2-s_W^2)}{c_Ws_W}\frac{\Pi_{Z\gamma}^{\text{1PI}}(0)}{m_Z^2}
+\frac{c_W^2-s_W^2}{c_W^2}\frac{\delta s_W^2}{s_W^2},\\
\delta Z_W &=  -\Pi_{\gamma\gamma}^{\text{1PI}}(0)^\prime
-\frac{2c_W}{s_W}\frac{\Pi_{Z\gamma}^{\text{1PI}}(0)}{m_Z^2}
+\frac{\delta s_W^2}{s_W^2},\\
\frac{\delta s_W^2}{s_W^2} & 
=\frac{c_W^2}{s_W^2}\left[\frac{\Pi_{ZZ}^{\text{1PI}}(m_Z^2)}{m_Z^2}-\frac{\Pi_{WW}^{\text{1PI}}(m_W^2)}{m_W^2}\right]. \label{delZzg}
\end{align}
The counter term for the VEV $\delta v$ is also obtained through the tree level relation:
\begin{align}
v^2 = \frac{m_W^2s_W^2}{\pi\alpha_{\text{em}}}, 
\end{align}
as
\begin{align}
\frac{\delta v}{v}&= \frac{1}{2}
\left[
\frac{s_W^2-c_W^2}{s_W^2}\frac{\Pi_{WW}^{\text{1PI}}(m_W^2)}{m_W^2}+\frac{c_W^2}{s_W^2}\frac{\Pi_{ZZ}^{\text{1PI}}(m_Z^2)}{m_Z^2}
-\Pi_{\gamma\gamma}^{\text{1PI}}(0)^\prime+\frac{2s_W}{c_W}\frac{\Pi_{Z\gamma}^{\text{1PI}}(0)}{m_Z^2}\right].  \label{del_vev} 
\end{align}

We here note that the fermion-loop contribution to $\Pi_{\gamma\gamma}^{\text{1PI}}(0)^\prime$ is given by 
\begin{align}
\Pi_{\gamma\gamma}^{\text{1PI}}(0)^\prime = \sum_f \frac{\alpha_{\text{em}}}{3\pi}N_c^fQ_f^2 (\Delta -\ln m_f^2), 
\end{align}
where $Q_f$ is the electric charge of a fermion $f$, 
$N_c^f$ is the color factor: $N_c^f=3~(1)$ for $f$ being quarks (leptons), 
and $\Delta$ is the divergent part of the loop integral as defined in Eq.~(\ref{div}) in Appendix~B.
In order to avoid to input the light quark masses, we can use the following relation obtained 
from Eqs.~(\ref{Pigg}) and (\ref{zzz}) 
\begin{align}
\Pi_{\gamma\gamma}^{\text{1PI}}(0)^\prime = \frac{1}{m_Z^2}\left[\Pi_{\gamma\gamma}^{\text{1PI}}(m_Z^2)-\hat{\Pi}_{\gamma\gamma}(m_Z^2) \right]
 = \frac{1}{m_Z^2}\Pi_{\gamma\gamma}^{\text{1PI}}(m_Z^2)  +\Delta \alpha_{\text{em}}, 
\end{align}
where $\Delta\alpha_{\text{em}}$ is the shift of the structure constant that we can quote the experimental value. 
In the right hand side of the above equation, the light fermion mass dependence in 
$\Pi_{\gamma\gamma}^{\text{1PI}}(m_Z^2)/m_Z^2$ is of order $m_f^2/m_Z^2$, so that we can neglect it.

\subsection{Renormalization conditions in the Yukawa sector}

In the Yukawa sector, there are three counter terms $\delta m_f$, $\delta Z_V^f$ and $\delta Z_A^f$. 
To determine them, we impose the following three conditions for the fermion two point functions~\cite{KKY}: 
\begin{align}
&\hat{\Pi}_{ff,V}(m_f^2) = 0, \quad
\frac{d}{dp\hspace{-2mm}/}\hat{\Pi}_{ff,V}(p^2)\Big|_{p^2 =m_f^2} = 0, \quad 
\frac{d}{dp\hspace{-2mm}/}\hat{\Pi}_{ff,A}(p^2)\Big|_{p^2 =m_f^2} = 0,
\label{rc_f2}
\end{align}
we obtain 
\begin{align}
\frac{\delta m_f}{m_f} &= \Pi_{ff,V}^{\text{1PI}}(m_f^2) +\Pi_{ff,S}^{\text{1PI}}(m_f^2),\notag\\
\delta Z_V^f
&=-\Pi_{ff,V}^{\text{1PI}}(m_f^2) -2m_f^2\left[\frac{d}{dp^2}\Pi_{ff,V}^{\text{1PI}}(p^2)\Big|_{p^2=m_f^2}+\frac{d}{dp^2}\Pi_{ff,S}^{\text{1PI}}(p^2)\Big|_{p^2=m_f^2}\right], \notag\\
\delta Z_A^f &= - \Pi_{ff,A}^{\text{1PI}}(m_f^2)+2m_f^2\frac{d}{dp^2}\Pi_{ff,A}^{\text{1PI}}(p^2)\Big|_{p^2=m_f^2}. 
\label{del_zvf}
\end{align}

\subsection{Renormalization conditions in the Higgs potential}

There are totally 21 counter terms in the Higgs potential, namely, the counter terms
for two tadpoles $\delta T_h$ and $\delta T_H$, four mass parameters $\delta m_\varphi^2$ ($\varphi=H^\pm,~A,~H$ and $h$), two mixing angles $\delta \alpha$ 
and $\delta \beta$, four wave function factors $\delta Z_\varphi$, 
six wave function mixing factors $\delta C_{ij}$, and $\delta M^2$
\footnote{In addition to them, there are two more counter terms $\delta Z_{G^\pm}$ and 
$\delta Z_{G^0}$. However, they do not enter the following discussion. }. 
First, we impose two tadpole conditions at the one-loop level, i.e.,
\begin{align}
\hat{T}_h=\hat{T}_H=0. 
\end{align}
We then obtain 
\begin{align}
\delta T_h = -\Gamma_h^{\rm{1PI}},\quad  \delta T_H = -\Gamma_H^{\rm{1PI}}. 
\end{align}

Second, eight on-shell conditions for the two-point functions: 
\begin{align}
&\hat{\Pi}_{\varphi\varphi}(m_\varphi^2)=0,  \label{rc1_s}\\
&\frac{d}{dp^2}\hat{\Pi}_{\varphi\varphi}(p^2)\big|_{p^2=m_\varphi^2}=0,\quad \text{for}~~\varphi=H^\pm,~A,~H\text{ and }h, 
\label{rc2_s}
\end{align}
which determine the following eight counter terms 
\begin{align}
\delta m_h^2&=\Pi_{hh}^{\text{1PI}}(m_h^2)+\frac{s_\alpha^2\delta T_1}{c_\beta v}+\frac{c_\alpha^2\delta T_2}{s_\beta v},\\
\delta m_H^2&=\Pi_{HH}^{\text{1PI}}(m_H^2)+\frac{c_\alpha^2\delta T_1}{c_\beta v}+\frac{s_\alpha^2\delta T_2}{s_\beta v},\\
\delta m_A^2&=\Pi_{AA}^{\text{1PI}}(m_A^2)+\frac{s_\beta^2\delta T_1}{c_\beta v}+\frac{c_\beta^2\delta T_2}{s_\beta v},\\
\delta m_{H^\pm}^2&=\Pi_{H^+H^-}^{\rm{1PI}}(m_{H^\pm}^2)+\frac{s_\beta^2\delta T_1}{c_\beta v}+\frac{c_\beta^2\delta T_2}{s_\beta v}, 
\end{align}
and 
\begin{align}
\delta Z_\varphi &=-\frac{d}{d p^2}\Pi_{\varphi\varphi}^{\text{1PI}}(p^2)\Big|_{p^2=m_\varphi^2}. \label{Zphi}
\end{align}

Three counter terms $\delta\alpha$, $\delta C_{hH}$ and $\delta C_{Hh}$ related to the mixing between the CP-even scalar states are determined 
by imposing the following three conditions
\begin{align}
&\hat{\Pi}_{Hh}(m_h^2)=\hat{\Pi}_{Hh}(m_H^2)=0, \quad \delta C_{hH}=\delta C_{Hh}\equiv\delta C_h. 
\end{align}
They give
\begin{align}
&\delta\alpha = \frac{1}{2(m_H^2-m_h^2)}\left[\Pi_{Hh}^{\text{1PI}}(m_h^2)+\Pi_{Hh}^{\text{1PI}}(m_H^2)-2s_\alpha c_\alpha
\left(\frac{\delta T_1}{c_\beta v}-\frac{\delta T_2}{s_\beta v}\right)\right],\\
&\delta C_h = \frac{1}{2(m_H^2-m_h^2)}\left[\Pi_{Hh}^{\text{1PI}}(m_h^2)-\Pi_{Hh}^{\text{1PI}}(m_H^2)\right]. 
\end{align}

Three counter terms $\delta\beta$, $\delta C_{AG}$ and $\delta C_{GA}$ related to the mixing between the CP-odd scalar states are determined 
by three conditions. 
Similar to the CP-even sector, we first impose the following two conditions as
\begin{align}
\hat{\Pi}_{AG}(0)=\hat{\Pi}_{AG}(m_A^2)=0. \label{xxxx}
\end{align}
We then obtain  
\begin{align}
&\delta\beta-\delta C_{AG}=-\frac{1}{m_A^2}\tilde{\Pi}_{AG}^{\text{1PI}}(0),  
\quad \delta \beta + \delta C_{GA}  = -\frac{1}{m_A^2}\tilde{\Pi}_{AG}^{\text{1PI}}(m_A^2). \label{delca}
\end{align}
In order to determine three counter terms, we need to impose one more renormalization condition in addition to that given in Eq.~(\ref{xxxx}). 
This third condition can be used to remove the gauge dependence in $\delta \beta$ which was already mentioned in the beginning of this section. 
To define such a condition, we separate $\tilde{\Pi}_{AG}^{\text{1PI}}(p^2)$ into the gauge dependent (G.D.) part and the gauge independent (G.I.) part as 
\begin{align}
\tilde{\Pi}_{AG}^{\text{1PI}}(p^2) = \tilde{\Pi}_{AG}^{\text{1PI}}(p^2)\big|_{\text{G.D.}}
+\tilde{\Pi}_{AG}^{\text{1PI}}(p^2)\big|_{\text{G.I.}}. 
\end{align}
Then, we imposed the third condition as
\begin{align}
&\delta\beta =  -\frac{1}{2m_A^2}\tilde{\Pi}_{AG}^{\text{1PI}}(m_A^2)\big|_{\text{G.I.}}. \label{abc}
\end{align}
Using Eq.~(\ref{delca}), the remaining two counter terms are also determined:
\begin{align}
&\delta C_{AG} = -\frac{1}{2m_A^2}\left[\tilde{\Pi}_{AG}^{\text{1PI}}(m_A^2)\big|_{\text{G.I.}}-2\tilde{\Pi}_{AG}^{\text{1PI}}(0)\big|_{\text{G.D.}}\right],   \label{12}\\
&\delta C_{GA} = -\frac{1}{2m_A^2}\left[\tilde{\Pi}_{AG}^{\text{1PI}}(m_A^2)\big|_{\text{G.I.}}+2\tilde{\Pi}_{AG}^{\text{1PI}}(m_A^2)\big|_{\text{G.D.}}\right]. \label{13} 
\end{align}
We note that in $\tilde{\Pi}_{AG}^{\text{1PI}}(0)$ only the G.D. part is survived; i.e., $\tilde{\Pi}_{AG}^{\text{1PI}}(0)=\tilde{\Pi}_{AG}^{\text{1PI}}(0)\big|_{\text{G.D.}}$. 
As it can be seen in Eqs. (\ref{12}) and (\ref{13}), there still remains the gauge dependence in $\delta C_{AG}$ and $\delta C_{GA}$. 
However, they do not appear in the following calculations for the renormalization of the  Higgs boson couplings. 
Instead of applying the above renormalization scheme for $\delta \beta$, 
we can apply the $\overline{\text{MS}}$ scheme in which the gauge dependence can also be removed at the one-loop level 
as discueed in Ref.~\cite{gauge_depend}. 
In the following discussion, we apply the renormalized $\tan\beta$ determined by Eq.~(\ref{abc}).

The above $A$-$G^0$ mixing can be replaced by the mixing between $A$ and the physical $Z$ boson by the help of 
the Ward-Takahashi identity; i.e., the condition $\hat{\Pi}_{AG}(m_A^2)=0$ 
is equivalent to that of vanishing renormalized $A$-$Z$ mixing; i.e., $\hat{\Pi}_{ZA}(m_A^2)=0$, 
which can be defined in the following way. 
The $Z$-$A$ mixing is obtained from the kinetic term: 
\begin{align}
\mathcal{L}_{\text{kin}} = m_Z(\partial_\mu G^0)Z^\mu + \cdots \to  m_Z(\delta\beta + \delta C_{GA})(\partial_\mu A^0)Z^\mu + \cdots .
\end{align}
The renormalized $Z$-$A$ mixing $\hat{\Pi}_{ZA}^\mu \equiv -ip^\mu\hat{\Pi}_{ZA}(p^2)$ is then expressed by 
\begin{align}
\hat{\Pi}_{ZA}(p^2)=
 m_Z (\delta\beta + \delta C_{GA}) + \Pi_{ZA}^{\text{1PI}}(p^2),
\end{align}
where $p^\mu$ is the incoming momentum of $A$.
The 1PI diagram contribution to the $Z$-$A$ mixing $\Pi_{ZA}^{\text{1PI}}(p^2)$ is given in Appendix. 
Because of the relation $\tilde{\Pi}_{AG}^{\text{1PI}}(m_A^2)/m_A^2=\Pi_{ZA}^{\text{1PI}}(m_A^2)/m_Z$, 
the condition $\hat{\Pi}_{AG}(m_A^2)=0$ can be replaced by  $\hat{\Pi}_{ZA}(m_A^2)=0$. 
Therefore, Eq.~(\ref{abc}) is rewritten as 
\begin{align}
\delta\beta =  -\frac{1}{2m_Z}\Pi_{ZA}^{\text{1PI}}(m_A^2)\big|_{\text{G.I.}}.    \label{mod_beta2}
\end{align}
We note that the numerical difference between in our scheme and in the previous scheme applied in Ref.~\cite{KOSY}
is negligibly small as long as we discuss the case with $\sin(\beta-\alpha)\simeq 1$ or $x\ll 1$.

Two counter terms $\delta C_{H^+G^-}$ and $\delta C_{G^+H^-}$ 
for the mixing between the singly-charged scalar states are determined by requiring the vanishment of the mixing between $G^\pm$ and $H^\pm$ at $p^2=0$ and $p^2=m_{H^\pm}^2$: 
\begin{align}
&\hat{\Pi}_{H^+G^-}(0)=\hat{\Pi}_{H^+G^-}(m_{H^\pm}^2)=0. 
\end{align}
We obtain 
\begin{align}
\delta C_{H^+G^-} = \delta \beta-\frac{1}{m_{H^\pm}^2}\tilde{\Pi}_{H^+G^-}^{\text{1PI}}(0), \quad 
\delta C_{G^+H^-} = -\delta\beta-\frac{1}{m_{H^\pm}^2}\tilde{\Pi}_{H^+G^-}^{\text{1PI}}(m_{H^\pm}^2). 
\end{align}

Until here, we did not discuss the determination of $\delta M^2$. 
As adopted in Ref.~\cite{KOSY}, we apply the minimal subtraction scheme for $\delta M^2$, where 
it is determined so as to absorb only the divergent part in the $hhh$ vertex at the one-loop level, that is 
\begin{align}
\frac{\delta M^2}{M^2} = \frac{1}{16\pi^2 v^2}
\Big[2\sum_f N_c^f m_f^2 \xi_f^2 +4M^2 -2m_{H^\pm}^2 -m_A^2 + \frac{\sin2\alpha}{\sin2\beta}(m_H^2-m_h^2)-3(2m_W^2+m_Z^2)\Big]\Delta. 
\end{align}

\begin{table}[t]
\begin{center}
{\renewcommand\arraystretch{1.5}
\begin{tabular}{c||ccc}\hline\hline
&$\delta\xi_h^u$ &$\delta\xi_h^d$&$\delta\xi_h^e$\\\hline
Type-I  &$-\frac{c_\alpha}{s_\beta}(\cot\beta\delta \beta+\tan\alpha\delta\alpha )$
&$-\frac{c_\alpha}{s_\beta}(\cot\beta\delta \beta+\tan\alpha\delta\alpha )$&$-\frac{c_\alpha}{s_\beta}(\cot\beta\delta \beta+\tan\alpha\delta\alpha )$ \\\hline
Type-II &$-\frac{c_\alpha}{s_\beta}(\cot\beta\delta \beta+\tan\alpha\delta\alpha )$&
$-\frac{s_\alpha}{c_\beta}(\tan\beta \delta\beta+\cot\alpha \delta \alpha)$
&$-\frac{s_\alpha}{c_\beta}(\tan\beta \delta\beta+\cot\alpha \delta \alpha)$
\\\hline
Type-X  &$-\frac{c_\alpha}{s_\beta}(\cot\beta\delta \beta+\tan\alpha\delta\alpha )$&$-\frac{c_\alpha}{s_\beta}(\cot\beta\delta \beta+\tan\alpha\delta\alpha )$&$-\frac{s_\alpha}{c_\beta}(\tan\beta \delta\beta+\cot\alpha \delta \alpha)$\\\hline
Type-Y  &$-\frac{c_\alpha}{s_\beta}(\cot\beta\delta \beta+\tan\alpha\delta\alpha )$&$-\frac{s_\alpha}{c_\beta}(\tan\beta \delta\beta+\cot\alpha \delta \alpha)$
&$-\frac{c_\alpha}{s_\beta}(\cot\beta\delta \beta+\tan\alpha\delta\alpha )$\\\hline\hline
\end{tabular}}
\caption{The counter term for the mixing factors in Yukawa interactions.}
\label{delxi}
\end{center}
\end{table}

\section{One-loop corrected Higgs boson couplings}\label{reno_couplings}

\subsection{Analytic expressions}

In the previous section, all the counter terms are determined by the set of renormalization conditions. 
Now, we can evaluate the one-loop corrected Higgs boson couplings $hWW$, $hZZ$, $hf\bar{f}$ and $hhh$. 
In addition to the above couplings, we also give formulae for the loop induced decay rates $h\to \gamma\gamma$, $h\to Z\gamma$ and $h\to gg$. 

The renormalized $hVV$, $hf\bar{f}$ and $hhh$ vertices are expressed as 
\begin{align}
\hat{\Gamma}_{hVV}^i(p_1^2,p_2^2,q^2)&=\Gamma^{i,\text{tree}}_{hVV}+\delta \Gamma^i_{hVV}+\Gamma^{i,\text{1PI}}_{hVV}(p_1^2,p_2^2,q^2),\\
\hat{\Gamma}_{hff}^j(p_1^2,p_2^2,q^2)  &=\Gamma_{hff}^{j,\text{tree}}+\delta \Gamma_{hff}^j+\Gamma_{hff}^{j,\text{1PI}}(p_1^2,p_2^2,q^2), \\
\hat{\Gamma}_{hhh}(p_1^2,p_2^2,q^2)&=\Gamma_{hhh}^{\text{tree}}+\delta \Gamma_{hhh}+\Gamma_{hhh}^{\text{1PI}}(p_1^2,p_2^2,q^2), 
\end{align}
where $\Gamma^{\text{tree}}_{hXX}$, $\delta \Gamma_{hXX}$ and $\Gamma^{\text{1PI}}_{hXX}$
are the contributions from the tree level, the counter terms and the 1PI diagrams for the $hXX$ vertices, respectively. 
In the above expressions, $p_1$ and $p_2$ ($q=p_1+p_2$) are the incoming momenta of particle $X$ (outgoing momentum for $h$). 

For the $hVV$ and $hf\bar{f}$ vertices, the indices $i$ and $j$ label the following form factors:
\begin{align}
\hat{\Gamma}_{hVV}^{\mu\nu}&=\hat{\Gamma}_{hVV}^1g^{\mu\nu}+\hat{\Gamma}_{hVV}^2\frac{p_1^\mu p_2^\nu}{m_V^2}
+i\hat{\Gamma}_{hVV}^3\epsilon^{\mu\nu\rho\sigma}\frac{p_{1\rho} p_{2\sigma}}{m_V^2},  \label{form_factor} \\
\hat{\Gamma}_{hff}&=
\hat{\Gamma}_{hff}^S+\gamma_5 \hat{\Gamma}_{hff}^P+p_1\hspace{-3.5mm}/\hspace{2mm}\hat{\Gamma}_{hff}^{V1}
+p_2\hspace{-3.5mm}/\hspace{2mm}\hat{\Gamma}_{hff}^{V2}\notag\\
&\quad +p_1\hspace{-3.5mm}/\hspace{2mm}\gamma_5 \hat{\Gamma}_{hff}^{A1}
+p_2\hspace{-3.5mm}/\hspace{2mm}\gamma_5\hat{\Gamma}_{hff}^{A2}
+p_1\hspace{-3.5mm}/\hspace{2mm}p_2\hspace{-3.5mm}/\hspace{2mm}\hat{\Gamma}_{hff}^{T}
+p_1\hspace{-3.5mm}/\hspace{2mm}p_2\hspace{-3.5mm}/\hspace{2mm}\gamma_5\hat{\Gamma}_{hff}^{PT}.
\end{align}
The tree-level contributions are given as
\begin{align}
&\Gamma^{1,\text{tree}}_{hVV}=\frac{2m_V^2}{v}\sin(\beta-\alpha),\quad  
\Gamma_{hff}^{\text{tree}}=-\frac{m_f}{v}\xi_h^f, \quad \Gamma_{hhh}^{\text{tree}}
=-6\lambda_{hhh}, \notag\\
&\Gamma^{2,\text{tree}}_{hVV}=\Gamma^{3,\text{tree}}_{hVV}=\Gamma^{j,\text{tree}}_{hff}=0 ~~(j\neq S). 
\end{align}
The counter-term contributions are 
\begin{align}
&\delta \Gamma^1_{hVV}=\frac{2m_V^2}{v}\left[\sin(\beta-\alpha)
\left(\frac{\delta m_V^2}{m_V^2} +\delta Z_V+\frac{1}{2}\delta Z_h -\frac{\delta v}{v}\right)
+\cos(\beta-\alpha)(\delta\beta+\delta C_h)\right], \notag\\
&\delta \Gamma_{hff}^S=-\frac{m_f}{v}\xi_h^f\left[\frac{\delta m_f}{m_f}-\frac{\delta v}{v}+\delta Z_V^f+\frac{1}{2}\delta Z_h
+\frac{\delta\xi_h^f}{\xi_h^f}+\frac{\xi_H^f}{\xi_h^f}(\delta C_h+\delta \alpha)\right], \notag\\
&\delta\Gamma_{hhh}=6\left[\delta\lambda_{hhh}+\frac{3}{2}\delta Z_h+\lambda_{Hhh}(\delta\alpha+\delta C_h)\right], \notag\\
&\delta \Gamma^2_{hVV}=\delta \Gamma^3_{hVV}=\delta \Gamma^j_{hff}=0, ~~ (j\neq S), 
\end{align}
where 
\begin{align}
\delta \lambda_{hhh} &=
-\lambda_{hhh}\frac{\delta v}{v}+\frac{1}{v\sin2\beta}\cos^2(\beta-\alpha)\cos(\alpha+\beta)\delta M^2\notag\\
&-\frac{1}{4v\sin2\beta}\left[\cos(3\alpha-\beta)+3\cos(\alpha+\beta)\right]\delta m_h^2\notag\\
&+\frac{1}{2v}\cos(\beta-\alpha)\left[3\frac{\sin2\alpha}{\sin2\beta}(m_h^2-M^2)+M^2\right]\delta \alpha\notag\\
&+\frac{1}{4v\sin^22\beta}\cos(\beta-\alpha)[(4+4\cos2\alpha\cos2\beta-2\sin2\alpha\sin2\beta)m_h^2 \notag\\
&\hspace{3cm}-(5-\cos4\beta+4\cos2\alpha\cos2\beta-2\sin2\alpha\sin2\beta)M^2 ]\delta\beta . \label{del_lamhhh}
\end{align}
The counter terms $\delta \xi_h^f$ appearing in the Yukawa couplings are expressed in terms of $\delta \beta$ and $\delta \alpha$ as listed in Table~\ref{delxi}. 
We define the renormalized scaling factors in the following way: 
\begin{subequations}
\begin{align}
\hat{\kappa}_V^{} &= \frac{\hat{\Gamma}_{hVV}^1(m_V^2,m_h^2,q^2)_{\text{THDM}}}{\hat{\Gamma}_{hVV}^1(m_V^2,m_h^2,q^2)_{\text{SM}}},\\
\hat{\kappa}_f^{} &= \frac{\hat{\Gamma}^S_{hff}(m_f^2,m_f^2,q^2)_{\text{THDM}}}{\hat{\Gamma}^S_{hff}(m_f^2,m_f^2,q^2)_{\text{SM}}},\\
\hat{\kappa}_h^{} &= \frac{\hat{\Gamma}_{hhh}(m_h^2,m_h^2,q^2)_{\text{THDM}}}{\hat{\Gamma}_{hhh}(m_h^2,m_h^2,q^2)_{\text{SM}}}.
\end{align}
\label{BottomUp:kappa_hat}
\end{subequations}
The momentum $q^2$ is fixed to be $(m_V^{}+m_h)^2$, $m_h^2$ and $(2m_h)^2$ for $\hat{\kappa}_V^{}$, $\hat{\kappa}_f^{}$
and $\hat{\kappa}_h^{}$, respectively, in the following discussion.

The deviations in the renormalized Higgs boson couplings are approximately expressed by keeping 
the non-decoupling effects of extra Higgs bosons and top and bottom masses dependence ($m_A\simeq m_H$ is assumed) as 
\begin{align}
\Delta\hat{\kappa}_{V} &\simeq -\frac{1}{2}\,x^2 -\frac{1}{16\pi^2}\frac{1}{6}\sum_{\Phi=A,H,H^\pm}c_\Phi^{}\frac{m_\Phi^2}{v^2}\left(1-\frac{M^2}{m_\Phi^2}\right)^2, \label{kv} \\
\Delta\hat{\kappa}_{\tau} &\simeq \Delta \hat{\kappa}_V +\xi_e \,x , \label{ktau}\\
\Delta\hat{\kappa}_{c} &\simeq \Delta\hat{\kappa}_V +\xi_u \,x , \label{kc}\\
\Delta\hat{\kappa}_b &\simeq \Delta \hat{\kappa}_V+\xi_d \,x
-\frac{1}{16\pi^2}\xi_u\xi_d \frac{4m_t^2}{v^2}\left[1-\frac{M^2}{m_{H^\pm}^2}
+\frac{m_t^2}{m_{H^\pm}^2}\left(1+\ln\frac{m_t^2}{m_{H^\pm}^2}\right) \right]\notag\\
&-\frac{1}{16\pi^2}\frac{1}{3}\xi_d^2\sum_{\Phi=A,H,H^\pm}\frac{m_b^4}{v^2m_{\Phi}^2}, \label{kb} \\
\Delta\hat{\kappa}_t &\simeq \Delta\hat{\kappa}_V+\xi_u \,x -\frac{1}{16\pi^2}\frac{1}{3}\left[
\xi_u^2\sum_{\Phi=A,H,H^\pm}\frac{m_t^4}{v^2m_{\Phi}^2}+\xi_d^2\frac{m_b^2m_t^2}{v^2m_{H^\pm}^2}\right], \\
\Delta\hat{\kappa}_{h} &\simeq \left(\frac{3}{2}-\frac{2M^2}{m_h^2}\right)\,x^2 
+\frac{1}{16\pi^2}\sum_{\Phi=A,H,H^\pm}c_\Phi^{} \frac{4}{3} \frac{m_\Phi^4}{m_h^2 v^2}\left(1-\frac{M^2}{m_\Phi^2}\right)^3, 
\end{align} 
where $c_\Phi^{}=2~(1)$ for $\Phi=H^\pm~(H,~A)$. 
We can see that there appears the term 
$m_\Phi^2/v^2\left(1-M^2/m_\Phi^2\right)^2$
in $\Delta\hat{\kappa}_V$ which 
comes from the counter term $\delta Z_h$; i.e., the derivative of the $h$ two point function given in Eq.~(\ref{Zphi}). 
When we consider the case with $M^2 \lesssim v^2$, this term gives the quadratic power like dependence of the mass of additional Higgs bosons.  
This corresponds to the case where the masses of the additional Higgs bosons, which is expressed schematically as  
$m_\Phi^2= \lambda_i v^2+M^2$, mostly come from the Higgs VEV $v$. 
In such a situation, it is known that the decoupling theorem does not work.  
On the other hand, if we consider the case of $M^2\gg v^2$, the amount of $\Delta\hat{\kappa}_f$ is reduced as $1/m_\Phi^2$ according 
to the decoupling theorem.    
The same contribution from $\delta Z_h$ is also seen in $\Delta\hat{\kappa}_f$ ($f=\tau,c,b,t$) through the term $\Delta \hat{\kappa}_V$. 
Notice here that 
there are additional terms proportional to the top or bottom quark masses in $\Delta\hat{\kappa}_b$ and $\Delta\hat{\kappa}_t$. 
Apart from $\Delta\hat{\kappa}_V$ and $\Delta\hat{\kappa}_f$, let us discuss the expression of $\Delta\hat{\kappa}_h$. 
There appears the term $m_\Phi^4/(m_h^2 v^2)\left(1-M^2/m_\Phi^2\right)^3$ which comes from the additional Higgs boson loop contributions to the 
1PI $hhh$ diagrams.
When we consider the non-decoupling case; i.e., $M^2\lesssim v^2$, it gives the quartic power like dependence of $m_\Phi$. 
Similar to the case in $\Delta\kappa_V$, this effect is decoupled by $1/m_\Phi^2$ when $M^2\gg v^2$ is taken.  

Similarly, the decay rates of $h\to \gamma\gamma$ and $h\to gg$ are expressed in terms of $x$ $(x\ll 1)$ as  
\begin{align}
\Gamma(h\to \gamma\gamma)& \simeq \frac{G_F\alpha_{\text{em}}^2 m_h^3}{128\sqrt{2}\pi^3}\left|
-\frac{1}{3}\left(1-\frac{M^2}{m_{H^\pm}^2}\right)+\sum_fQ_fN_c^f(1+\xi_f \,x-\frac{x^2}{2})I_F+(1-\frac{x^2}{2})I_W\right|^2, \label{hgg_x}\\
\Gamma(h\to gg)
&\simeq 
\frac{G_F\alpha_{s}^2 m_h^3}{64\sqrt{2}\pi^3}\left|\sum_q(1+\xi_q \,x-\frac{x^2}{2})I_F \right|^2, 
\end{align} 
where $I_F$ and $I_W$ are the loop functions. 
The exact expressions for the decay rates for $h\to \gamma\gamma$, $h\to Z\gamma$ and $h\to gg$ 
are given in Eqs.~(\ref{hgamgam_full}), (\ref{hZgam_full}) and (\ref{hgg_full}) in Appendix C, respectively. 
In Eq.~(\ref{hgg_x}), the first term in $\Gamma(h\to \gamma\gamma)$ proportional to $(1-M^2/m_{H^\pm}^2)$ is the charged Higgs boson loop contribution. 
When we take the limit of $M^2\to 0$, this term approaches to the constant $-1/3$. 
This can also be understood as the consequence of the non-decoupling effect of the charged Higgs boson loop contribution, but it is not
like the quartic (quadratic) power like dependence as seen in 
$\Delta\hat{\kappa}_h$ ($\Delta\hat{\kappa}_V$ and  $\Delta\hat{\kappa}_f$).

 %% Numerically, the loop functions are evaluated as 
 %% \begin{align}
 %% I_t \simeq -1.38,\quad I_b \simeq 8.04\times 10^{-2}-0.113\,i,\quad I_W \simeq 8.33. 
 %% \end{align}

\subsection{Numerical evaluations}

\begin{figure}[t]
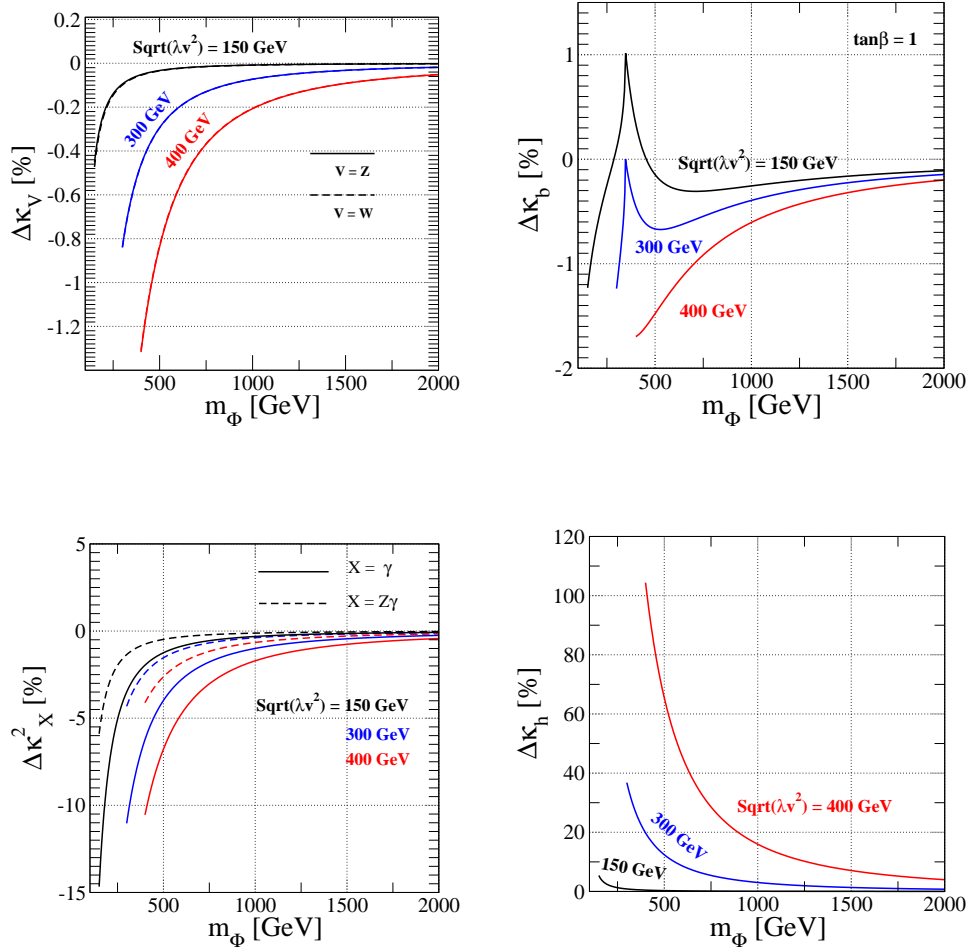

%========================================
\centering
\includegraphics[width=6cm]{delhat_kv_mphi.eps}	\hspace{5mm}
\includegraphics[width=6cm]{delhat_kb_mphi.eps}\\\vspace{13mm}
\includegraphics[width=6cm]{delR_mphi.eps}  \hspace{5mm}
\includegraphics[width=6cm]{delhat_kh_mphi.eps}
\caption{Deviations in the scaling factors for $hVV$ (upper left), $hb\bar{b}$ (upper right), $h\gamma\gamma/hZ\gamma$ (bottom left) and $hhh$
(bottom right) at the one-loop level as a function of $m_\Phi^{}(=m_{H^\pm}^{}=m_A^{}=m_H^{})$ in the case of $\sin(\beta-\alpha)=1$ and $\tan\beta=1$
The black, blue and red curves respectively show the cases of $\sqrt{\lambda v^2} (=\sqrt{m_\Phi^2-M^2})=150$, 300 and 400 GeV.  
}
\label{Fig:oneloop1}
%========================================
\end{figure}

\begin{figure}[t]
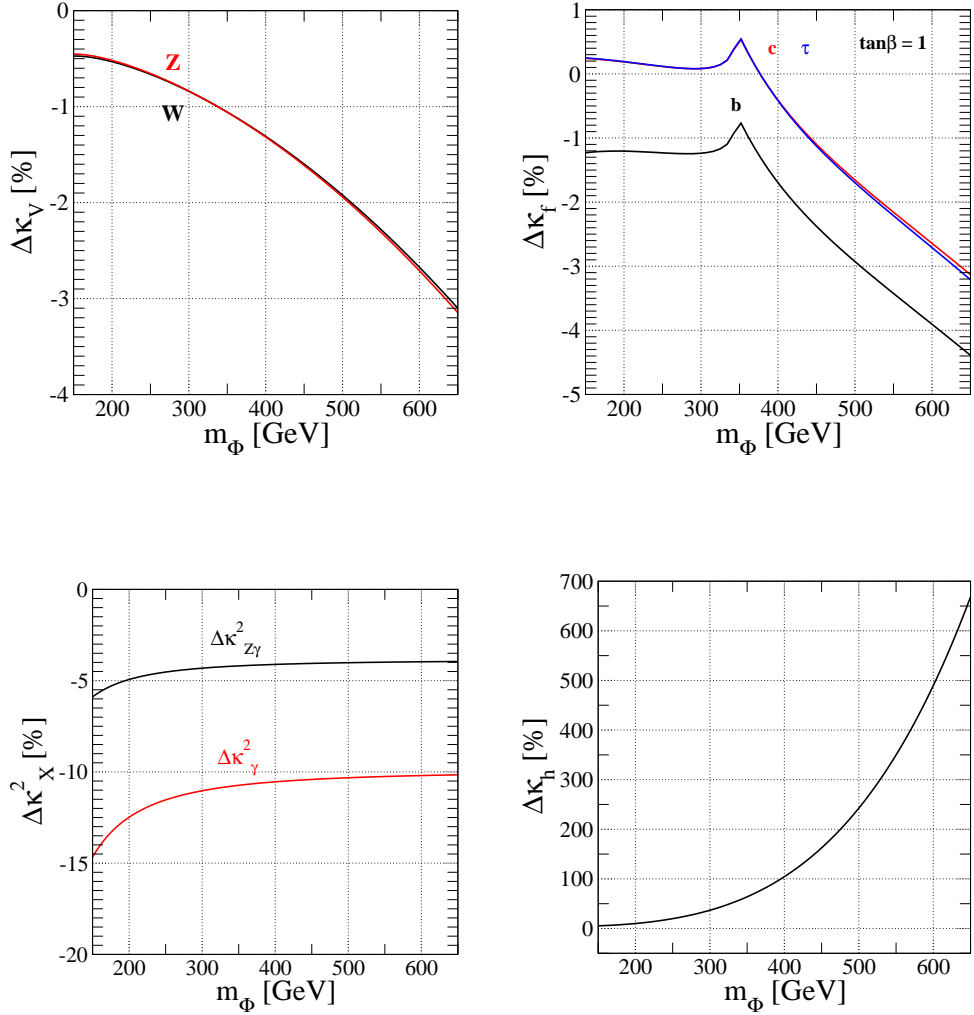

%========================================
\centering
\includegraphics[width=6cm]{delhat_kv_M0.eps}\hspace{7mm}
\includegraphics[width=6cm]{delhat_kf_M0.eps}\\\vspace{15mm}
\includegraphics[width=6cm]{delR_M0.eps}\hspace{7mm} 
\includegraphics[width=6cm]{delhat_kh_M0.eps}
\caption{Deviations in the scaling factors for $hVV$ (upper left), $hf\bar{f}$ (upper right), $h\gamma\gamma/hZ\gamma$ (bottom left) and $hhh$
(bottom right) at the one-loop level as a function of $m_\Phi^{}(=m_{H^\pm}^{}=m_A^{}=m_H^{})$ in the case of $M^2=0$, $\sin(\beta-\alpha)=1$ and $\tan\beta=1$. }
\label{Fig:oneloop2}
%========================================
\end{figure}

In the following, we show numerical results for the Higgs boson couplings at the one-loop level. 
 We use the following inputs~\cite{PDG}: 
 \begin{align}
 &m_Z=91.1875~\text{GeV},~G_F=1.16639\times 10^{-5}~\text{GeV}^{-2},~\alpha_{\text{em}}^{-1}=137.035989,~\Delta\alpha_{\text{em}}=0.06635, \notag\\
 &m_t=173.07~\text{GeV},~m_b=4.66~\text{GeV},~m_c=1.275~\text{GeV},~m_\tau=1.77684~\text{GeV}, \notag\\
 & m_h=126~\text{GeV}. 
 \end{align}

We first show the case of the SM-like limit $x=0$. In this case, the deviations in the Higgs boson couplings 
purely comes from the additional Higgs boson loop effects. 
We note that the $\tan\beta$ dependence in the renormalized scaling factors appears only in $\hat{\kappa}_f$. 
We take all the masses of additional Higgs bosons to be the same; i.e., $m_{H^\pm}=m_A=m_H~(\equiv m_\Phi)$ for simplicity. 

In Fig.~\ref{Fig:oneloop1}, we show the decoupling behavior of additional Higgs boson loop contributions to the Higgs boson couplings. 
The upper-left, upper-right, lower-left and lower-right panels respectively show $\Delta\hat{\kappa}_V$, $\Delta\hat{\kappa}_b$, 
$\Delta\kappa_{\gamma/Z\gamma}^2$ and $\Delta\hat{\kappa}_h$ 
as a function of $m_\Phi$ for several fixed values of $\sqrt{\lambda v^2}~(=\sqrt{m_\Phi^2-M^2})$ in the case of $\tan\beta=1$. 
We can see that all the deviations approach to zero in the large mass region due to the decoupling theorem~\cite{decoupling}. 

In Fig.~\ref{Fig:oneloop2}, we show 
the deviation in the Higgs boson couplings $\Delta\hat{\kappa}_V^{}$ (upper-left), $\Delta\hat{\kappa}e_f$ (upper-right), 
$\Delta\kappa_{\gamma/Z\gamma}^2$ (lower-left) and $\Delta\hat{\kappa}_h$ (lower-right) as a function of $m_\Phi^{}$. 
We take $M^2=0$ and $\tan\beta=1$ for all panels.
In this case, the magnitude of deviations increase when $m_\Phi$ becomes larger due to the non-decouipling effect of 
the extra Higgs boson loops except for $\Delta\kappa_{\gamma/Z\gamma}^2$.

\section{Determination of inner parameters from the Higgs boson coupling measurements}\label{determination}
 
In this section, we investigate how we can fingerprint the THDMs using the one-loop corrected Higgs boson couplings 
and also future precision measurements of these couplings at 
the HL-LHC and the ILC. 
We carefully see how the tree level analysis for the model discrimination discussed in Sec.~II or in Ref.~\cite{fingerprint}
can be improved by the analysis with radiative corrections. 
Furthermore, we demonstrate how the inner parameters such as $x$, $\tan\beta$ and masses of additional Higgs bosons 
can be extracted from the measurement of the couplings for the Higgs boson $h$. 
In our analysis below, we assume that the deviations in scale factors of the Higgs boson couplings are
measured as expected in Table~\ref{benchmark}. 
We also assume that the SM values of these coupling constants are well predicted without large uncertainties 
which mainly come from QCD corrections\footnote{According to Refs.~\cite{QCD_hbb,QCD_Peskin}, 
the current uncertainty of the bottom Yukawa coupling $hb\bar{b}$ due to the QCD corrections is 0.77\% in the SM. 
This uncertainty could be reduced in future studies using the lattice calculation up to 0.10\%~\cite{QCD_Peskin} which is 
better than the expected accuracy of the measurement of the $hb\bar{b}$ coupling at the ILC1000-up as listed in Table \ref{Tab:Sensitivity} (0.4\%). 
}.

\begin{table}[t]
\begin{center}
{\renewcommand\arraystretch{1.5}
\begin{tabular}{|c||c|c|c|c|c|c}\hline\hline
  & Set A  & Set B & Set C &  Set D & Set E   \\\hline
$\Delta \kappa_V$     & $-2\%$ & $-2\%$ & $-2\%$ & $-1\%$ & $-0.4\%$  \\
$\Delta \kappa_\tau$  & $+18\%$ &$+10\%$ & $+5\%$ & $+18\%$ & +18\%  \\
$\Delta\kappa_b$     & +18\% & +10\% & +5\% & +18\% & +18\%  \\\hline\hline
\end{tabular}}
\caption{Benchmark sets for the central values of measured scaling factors for the $hVV$, $hb\bar{b}$ and $h\tau\tau$ couplings. 
The expected 1-$\sigma$ uncertainties for each scaling factor at the HL-LHC and the ILC 500 are shown in Eq.~(\ref{error}). }
\label{benchmark}
\end{center}
\end{table}

Let us suppose that $\Delta\kappa_V$, $\Delta\kappa_\tau$ and $\Delta\kappa_b$ are measured at the HL-LHC and the ILC500. 
We consider five benchmark sets for the central values of $(\Delta\kappa_V^{},\Delta\kappa_\tau,\Delta\kappa_b)$ as 
listed in Table~\ref{benchmark}. 
Set A is the typical case where Yukawa couplings deviate from the SM values rather significantly (18\%) with 
a relatively large deviation in the $hVV$ couplings ($-2\%$). 
Set B and Set C correspond to the cases with smaller deviations in Yukawa couplings with the same deviation in gauge couplings as Set A. 
Set D and Set E do to the cases with smaller deviations in gauge couplings with fixing the same deviation in Yukawa couplings as Set A. 
According to Table~\ref{Tab:Sensitivity}, 
the 1-$\sigma$ uncertainty for these scaling factors are given as
\begin{align}
&[\sigma(\kappa_V), \sigma(\kappa_b), \sigma(\kappa_\tau)] = [2\%,~4\%,~2\%], ~~~~~~~~~\text{for HL-LHC} , \notag\\
&[\sigma(\kappa_V), \sigma(\kappa_b), \sigma(\kappa_\tau)] = [0.4\%,~0.9\%,~1.9\%],~~ \text{for ILC500}.  \label{error}
\end{align}
From the tree level analysis in Fig.~\ref{FP_tree}, these benchmark sets indicate that 
the Higgs sector is the THDM with the Type-II (Type-I) Yukawa interaction assuming $x\simeq\cos(\beta-\alpha)<0$ ($x>0$).  
In order to further discriminate Type-I or Type-II, we need additional information to determine the sign of $x$
such as the measurement of $\Delta\kappa_c$, namely, 
if $\Delta \kappa_c$ is given to be a negative (positive) value, 
then we can completely determine the Yukawa interaction to be Type-II (Type-I). 
In the following, we consider the case of $\Delta\kappa_c<0$, so that we assume the case of the Type-II THDM.   

For all Set A to Set E, we survey parameter regions in which values of $\kappa$'s are predicted around the central values 
within the 1-$\sigma$ uncertainty expressed 
in Eq.~(\ref{error}) by scanning the inner parameters $x$, $\tan\beta$, $m_\Phi^{}~(=m_{H^\pm}^{}=m_A^{}=m_H^{})$ and $M^2$ 
in the Type-II THDM. 
We also take into account the constraints from vacuum stability and perturbative unitarity in order to constrain the parameter space. 
The scanned regions for $\tan\beta$ and $m_\Phi^{}$ are taken as $\tan\beta \geq 1$ and $m_\Phi^{}\geq$ 300 GeV, respectively. 
Values of the other parameters $M^2$ and $x$ are scanned over ranges which are enough wide 
to obtain the maximally allowed parameter spaces.   

\begin{figure}[t]
%========================================
\centering
\includegraphics[width=4.3cm]{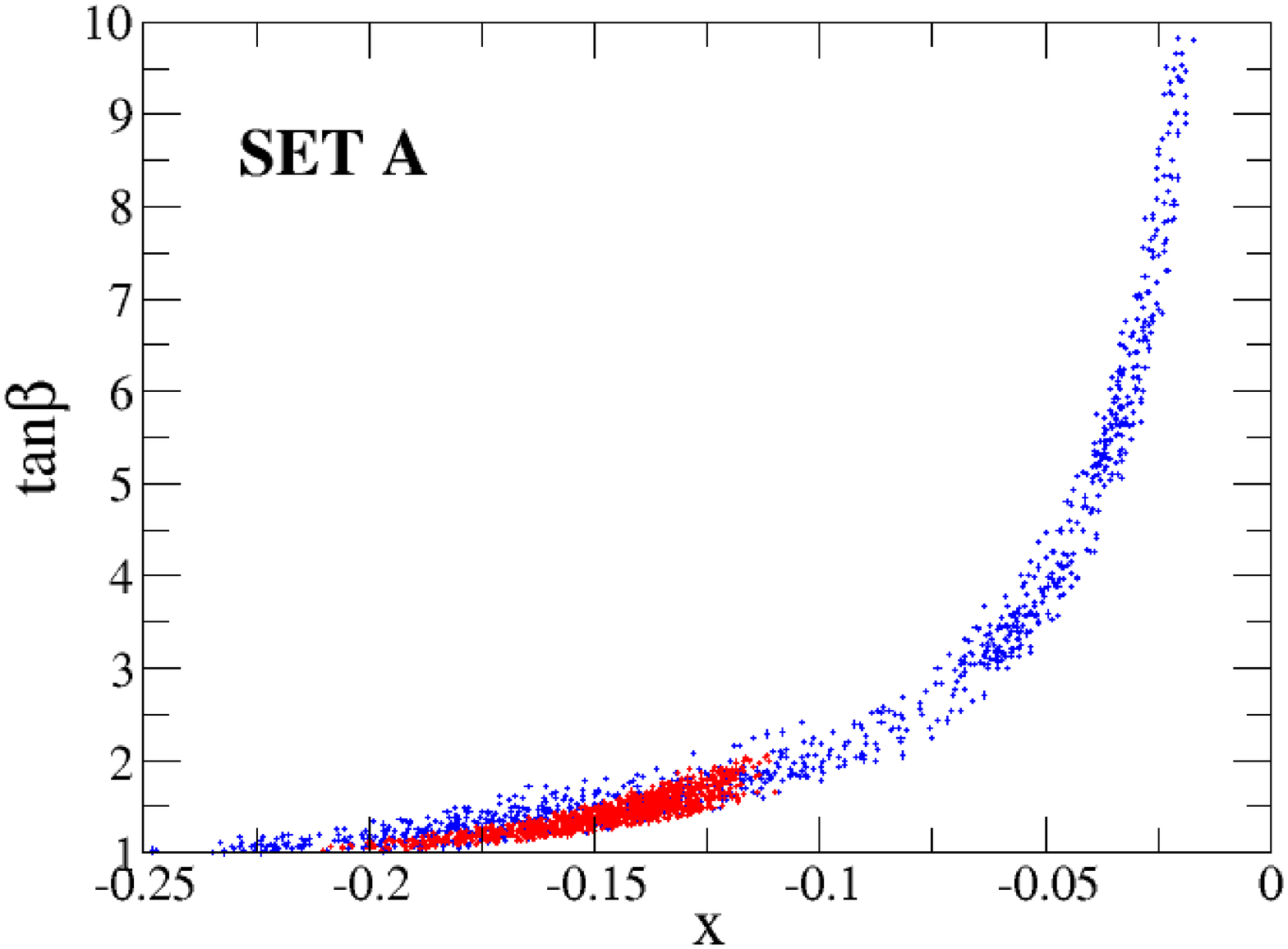}	\hspace{-5mm}
\includegraphics[width=4.3cm]{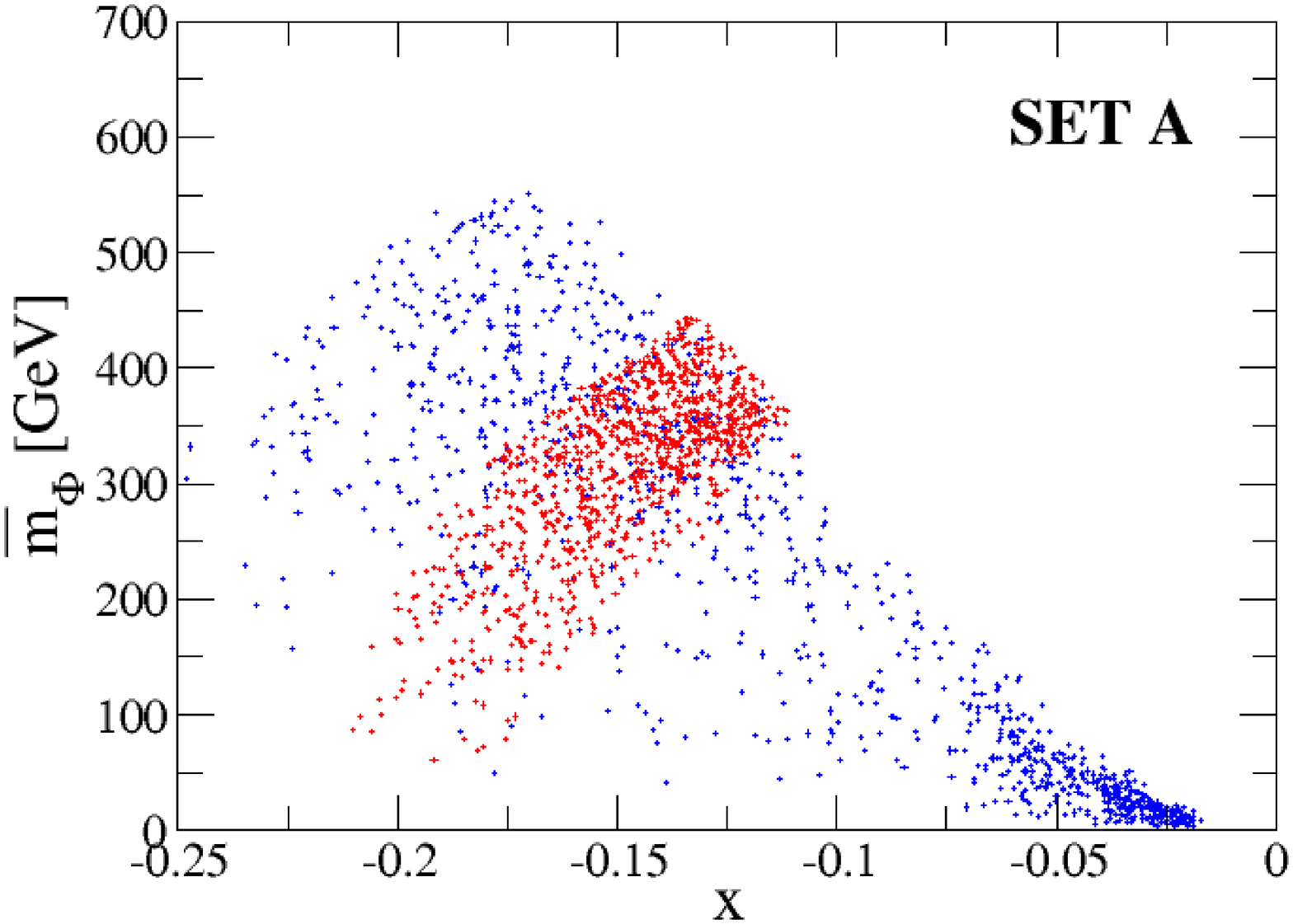}  \hspace{-5mm}
\includegraphics[width=4.3cm]{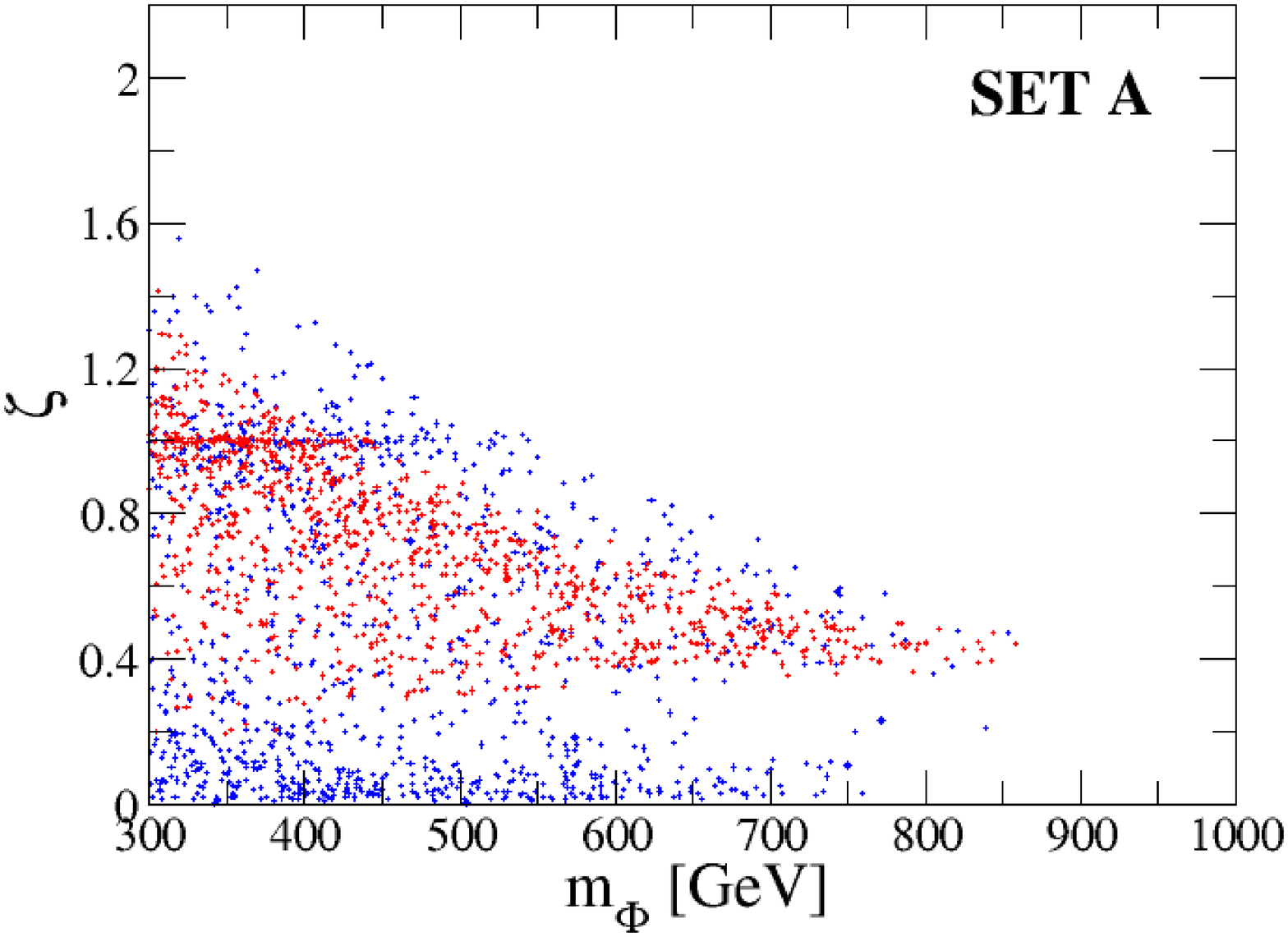}  \hspace{-5mm}  
\includegraphics[width=4.3cm]{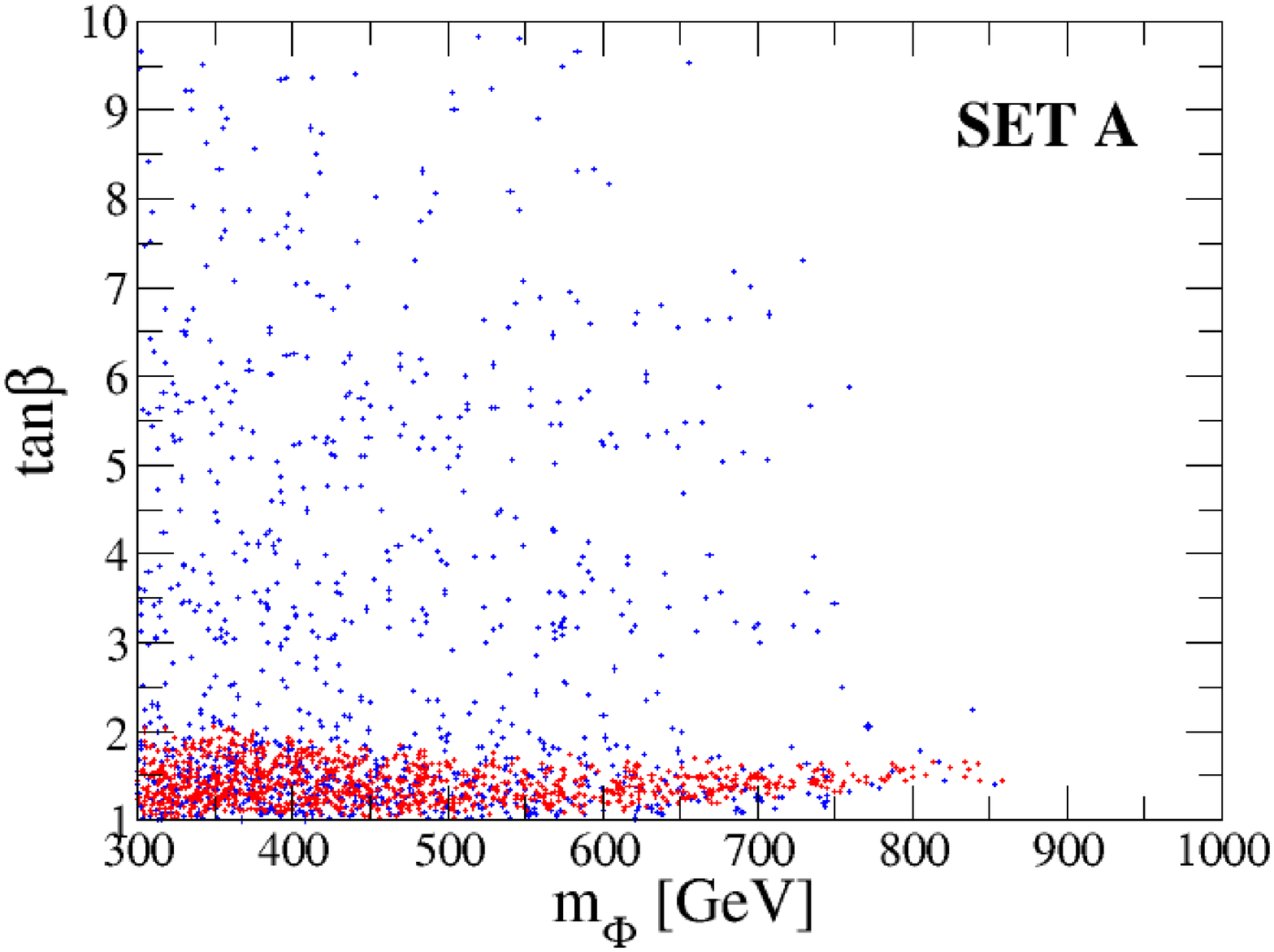} \\ 
\includegraphics[width=4.3cm]{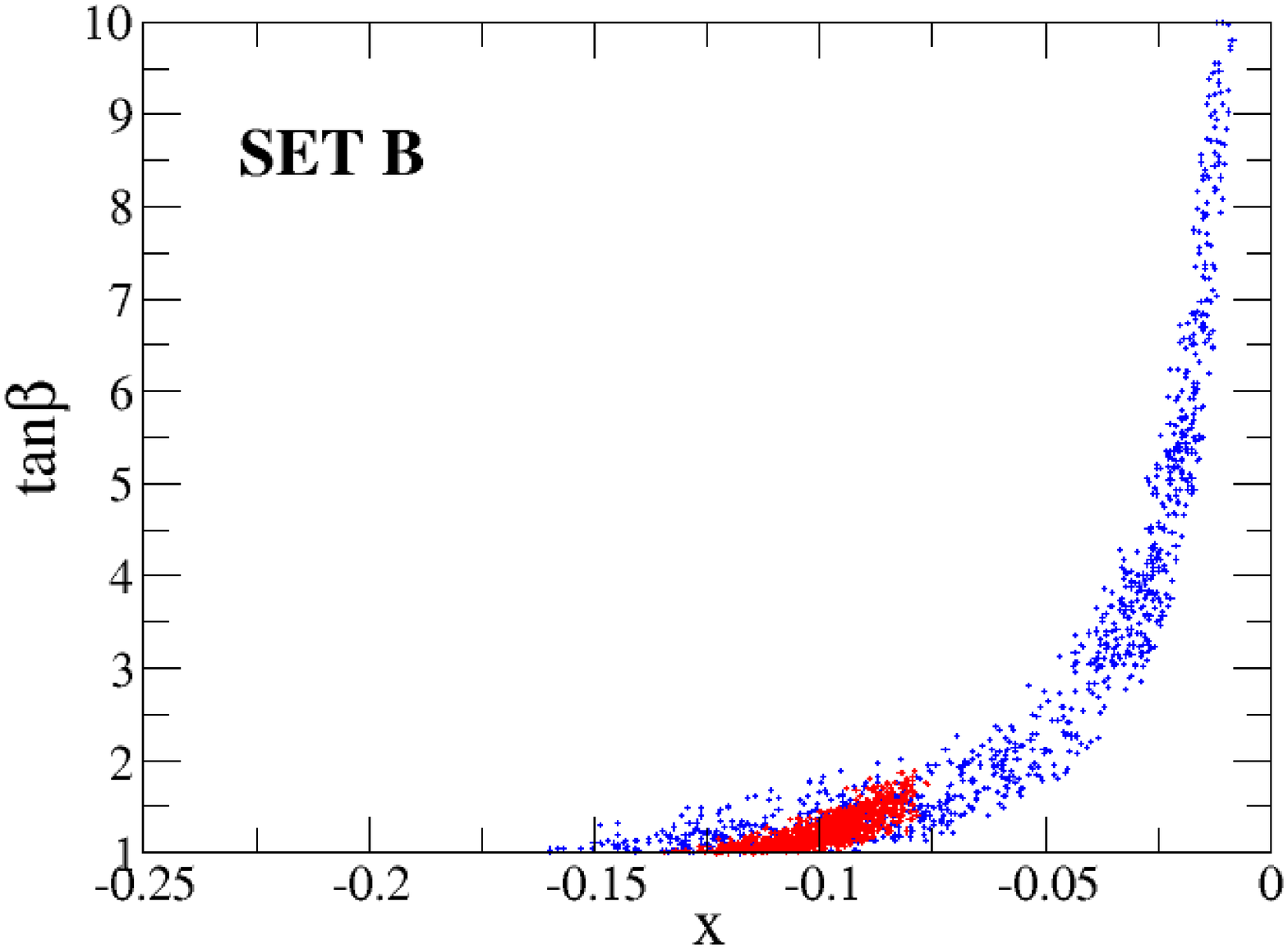}	\hspace{-5mm}
\includegraphics[width=4.3cm]{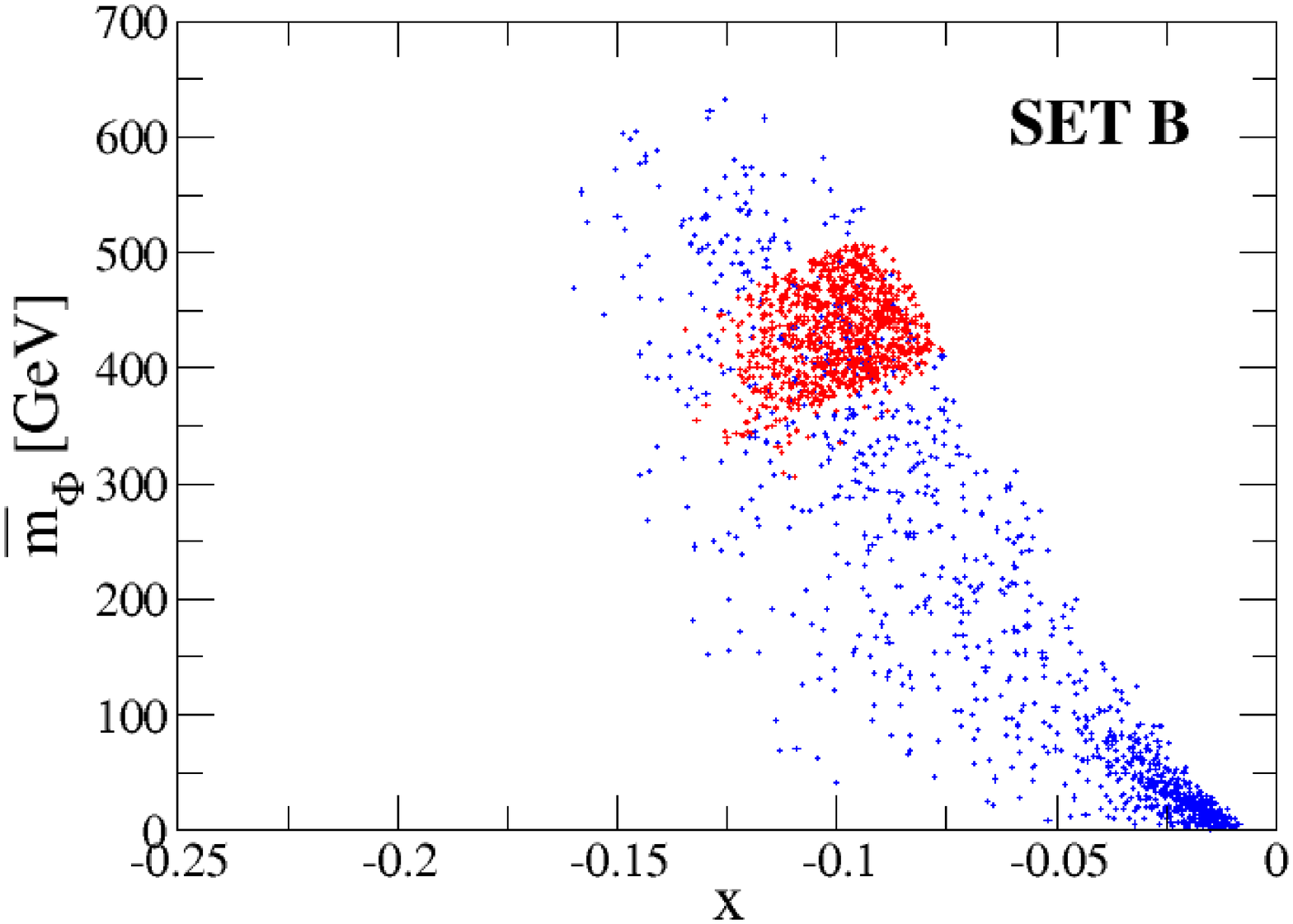}  \hspace{-5mm}
\includegraphics[width=4.3cm]{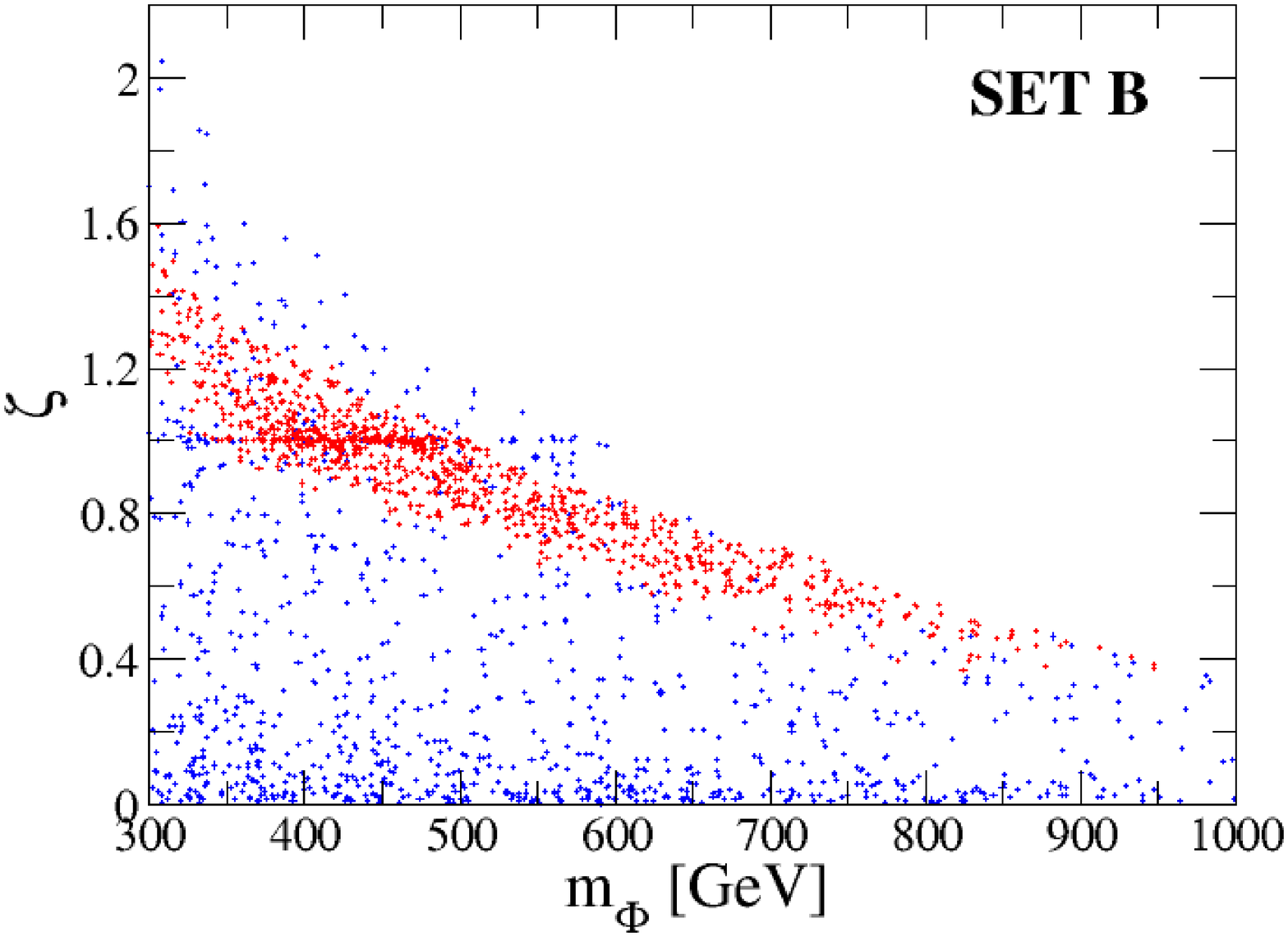}  \hspace{-5mm}
\includegraphics[width=4.3cm]{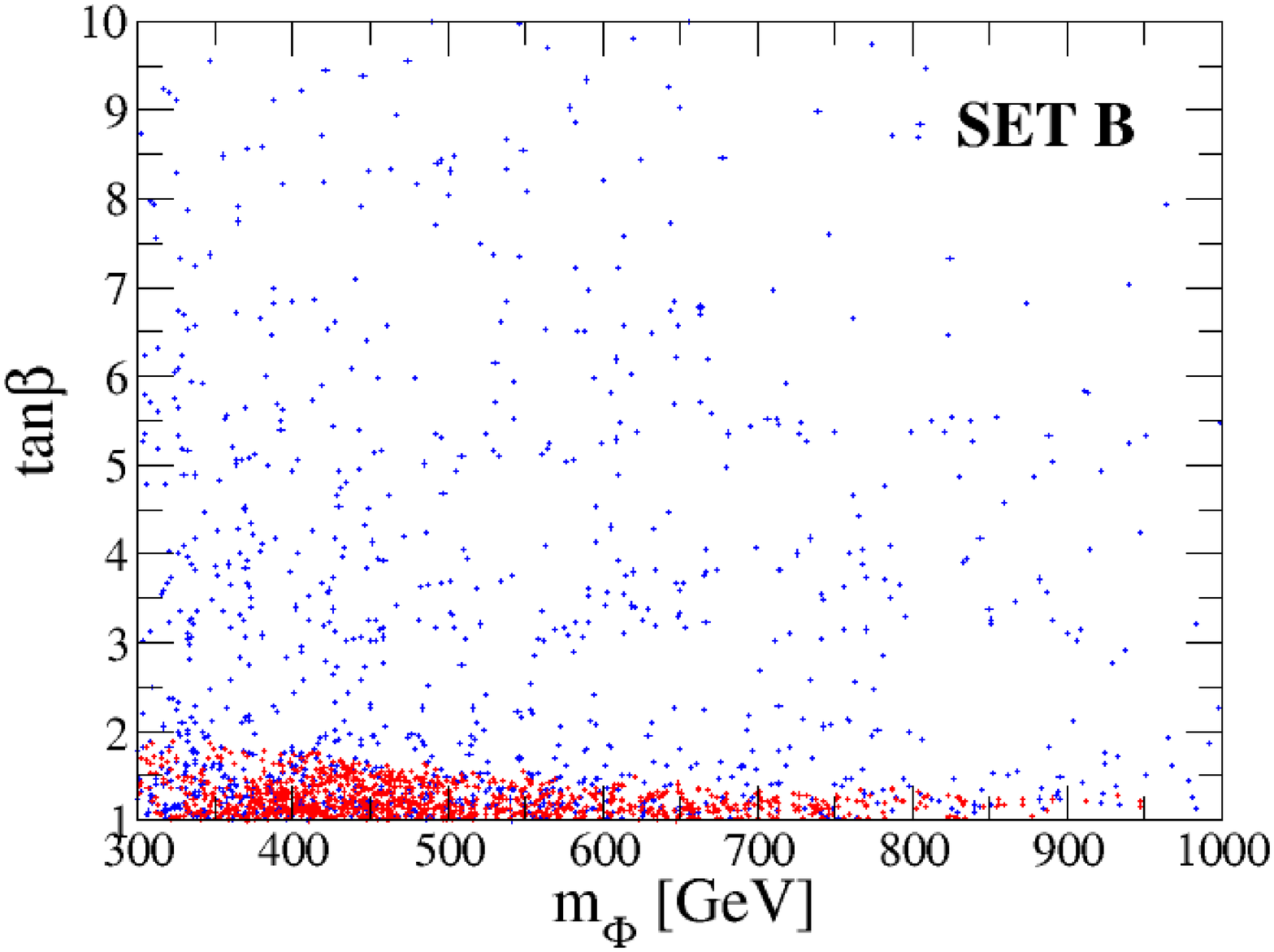} \\ 
\includegraphics[width=4.3cm]{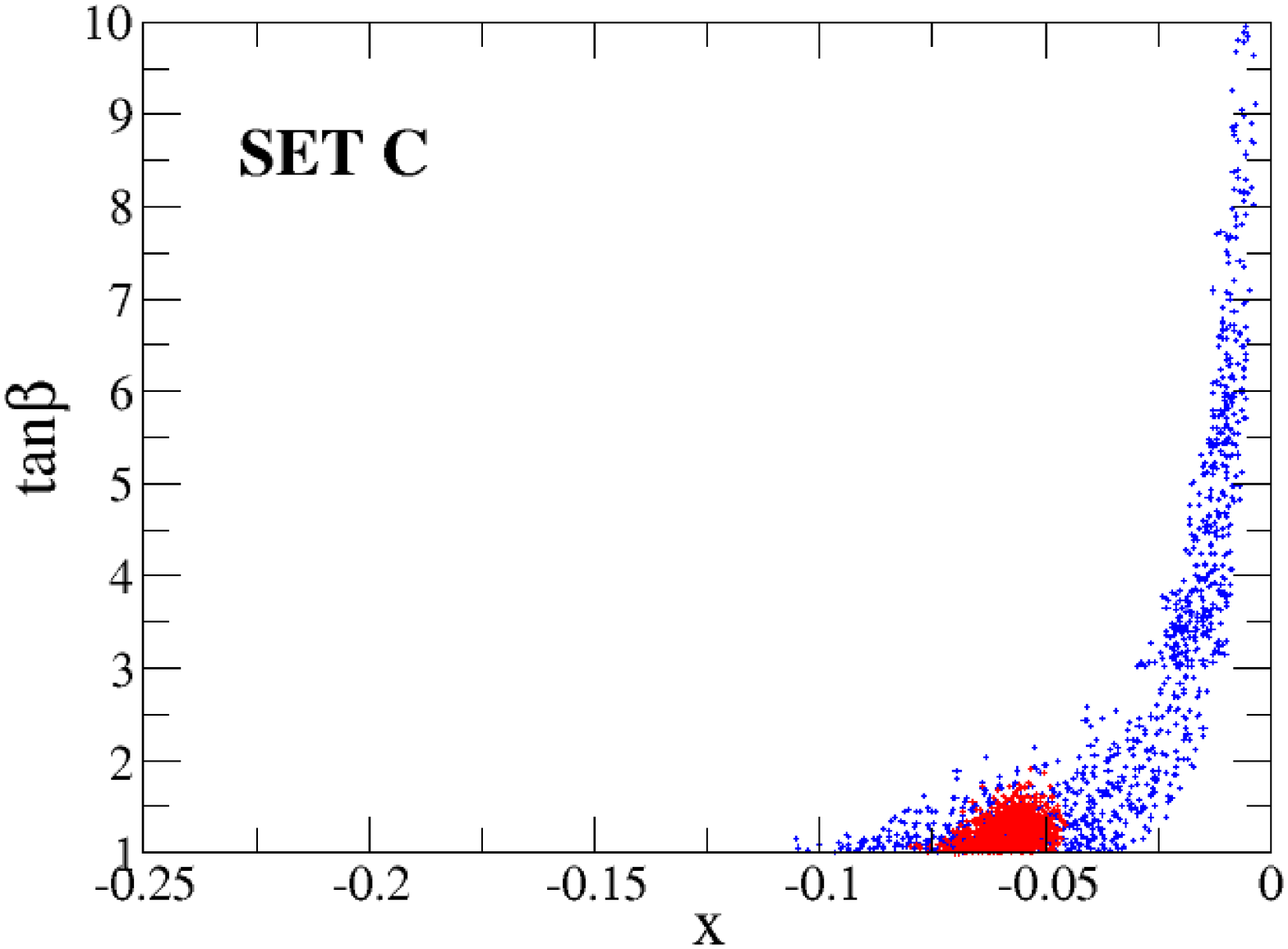}	\hspace{-5mm}
\includegraphics[width=4.3cm]{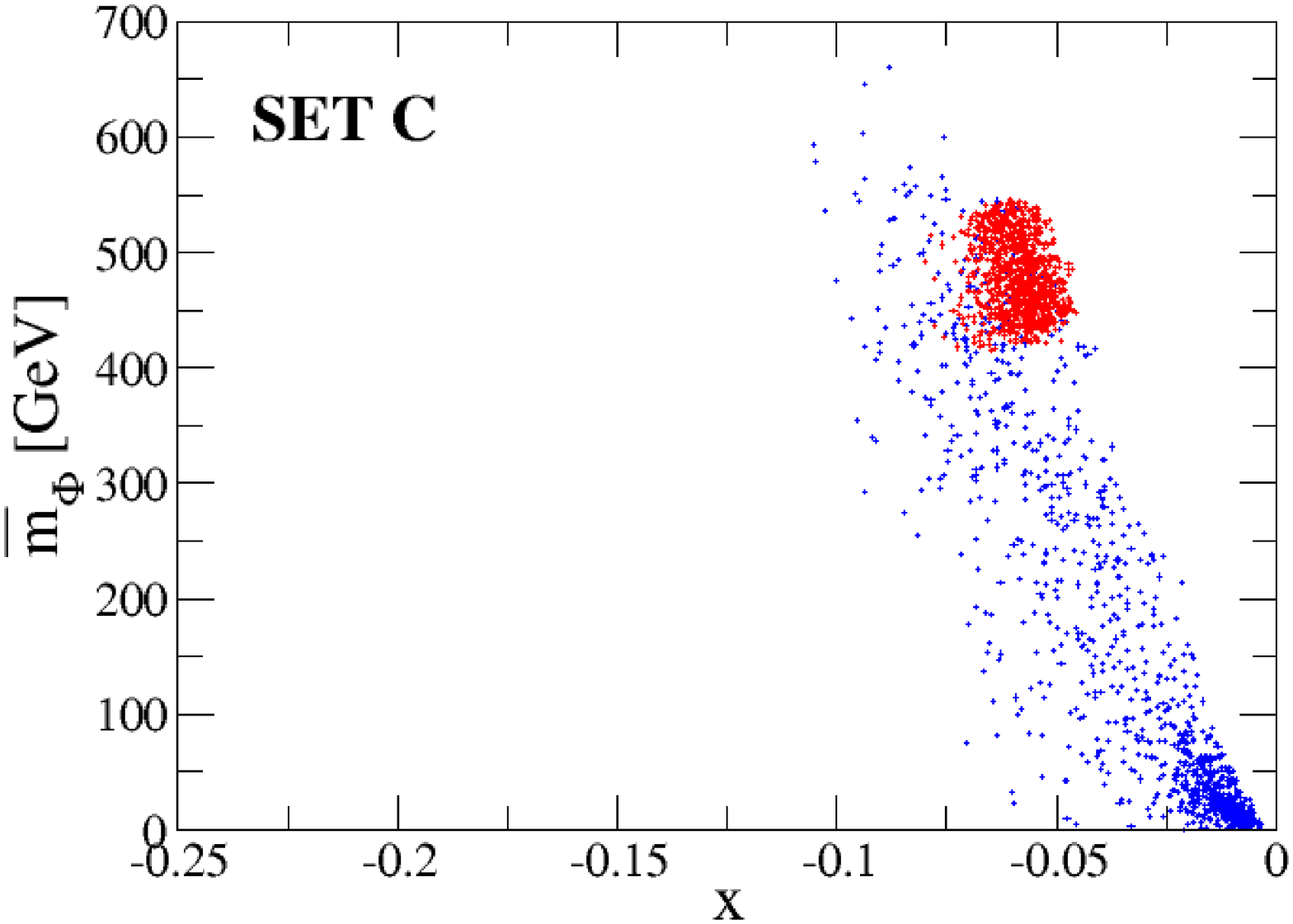}  \hspace{-5mm}
\includegraphics[width=4.3cm]{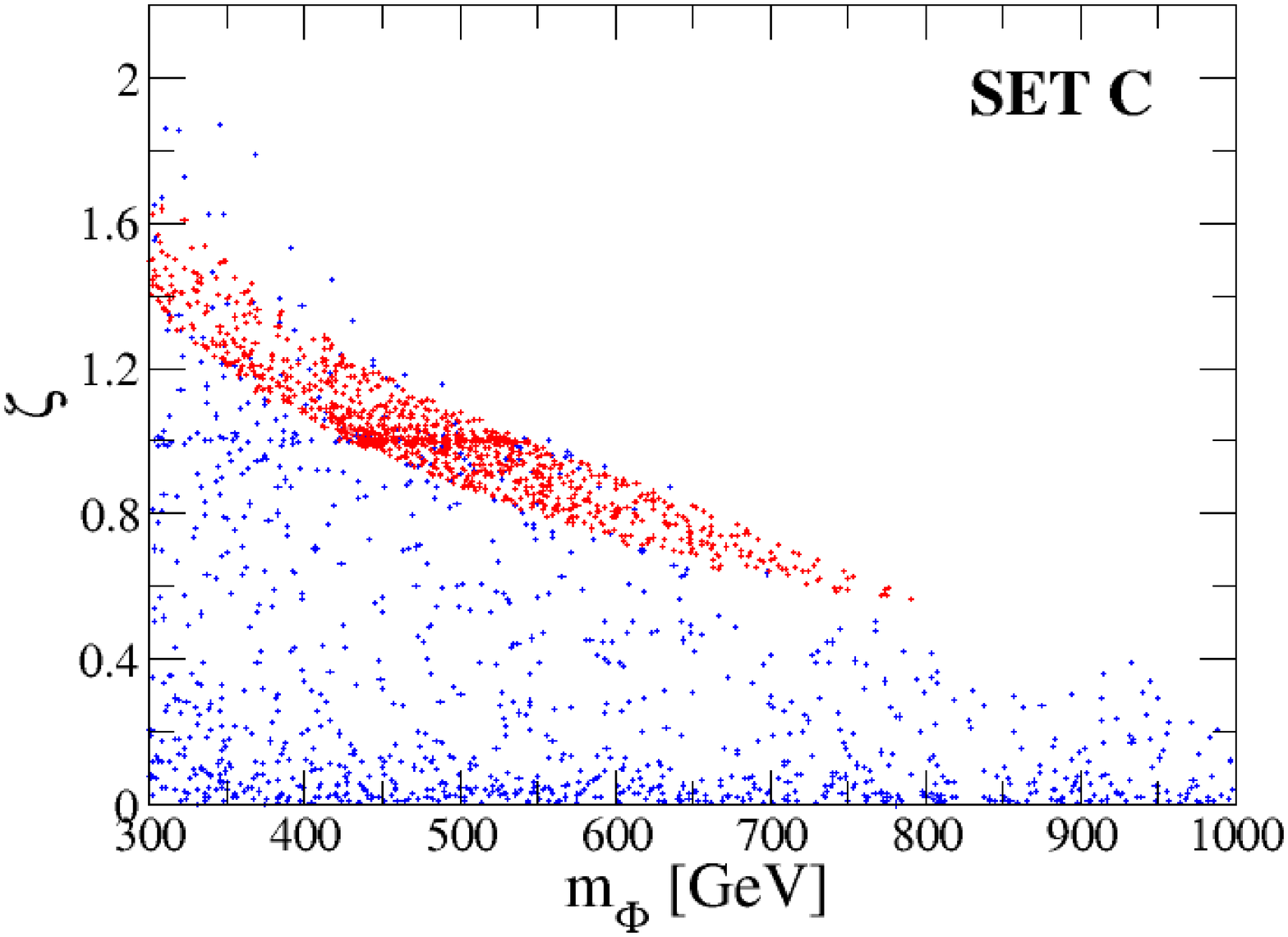}  \hspace{-5mm}  
\includegraphics[width=4.3cm]{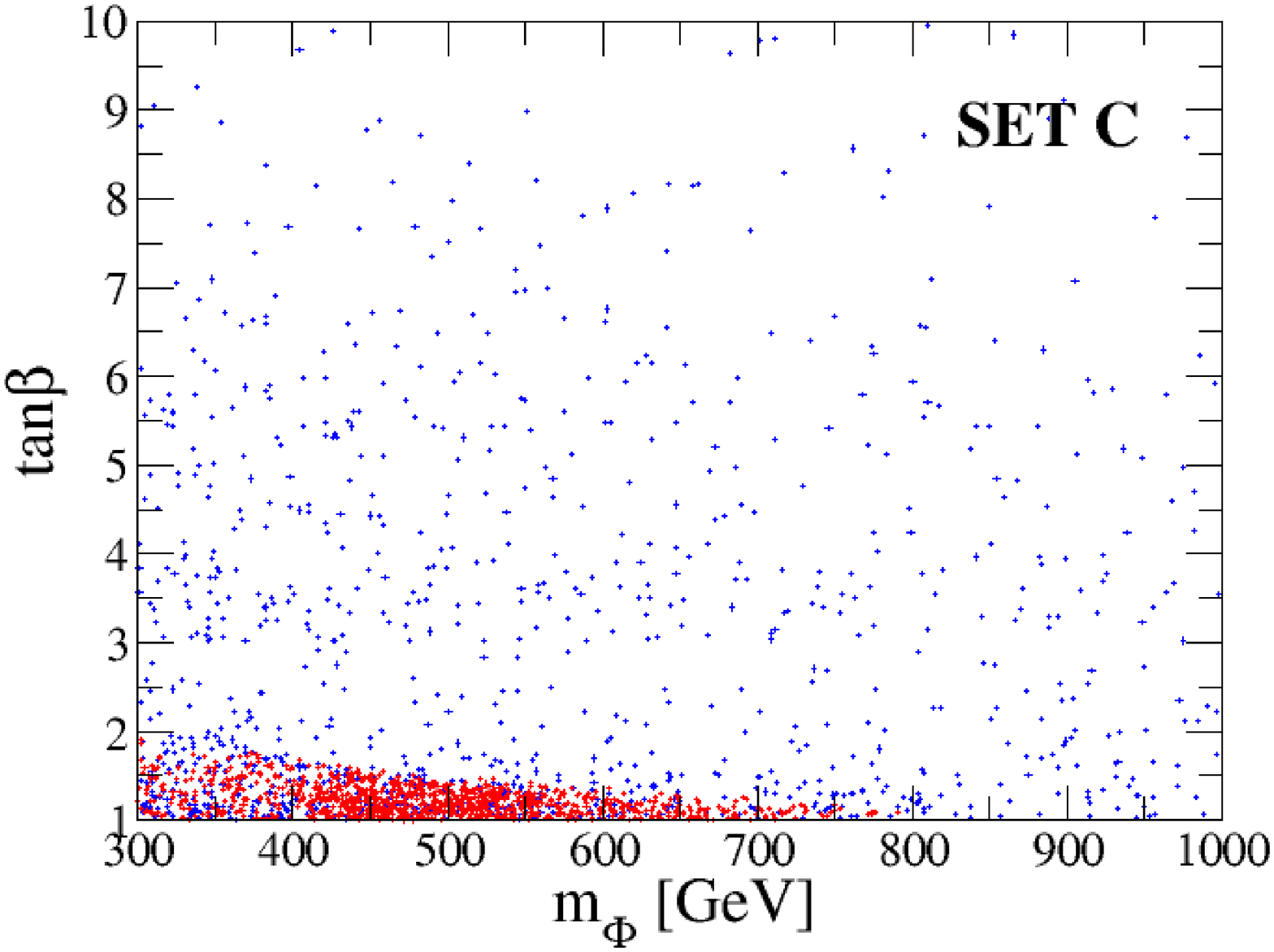} \\ 
\includegraphics[width=4.3cm]{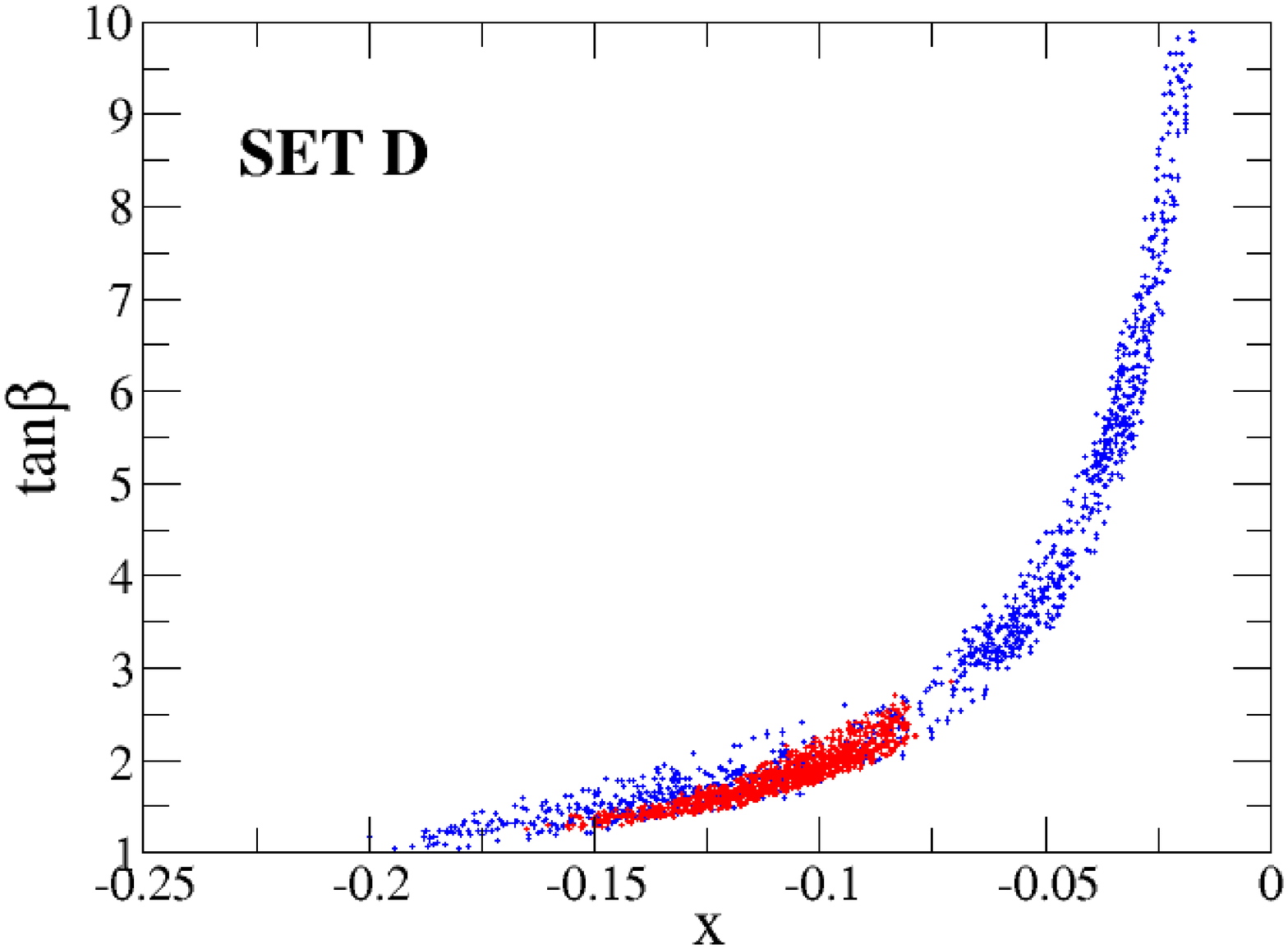}	\hspace{-5mm}
\includegraphics[width=4.3cm]{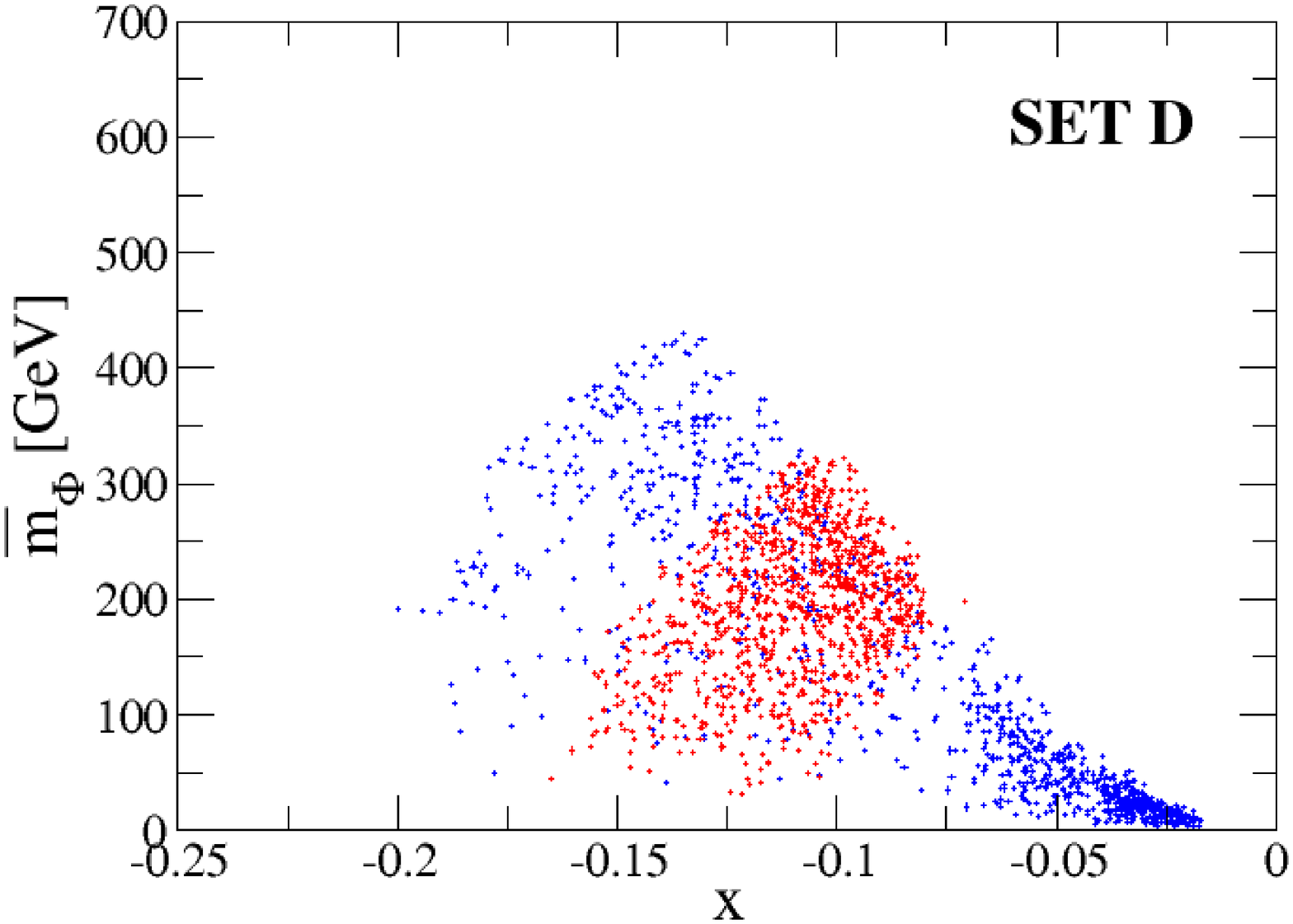}  \hspace{-5mm}
\includegraphics[width=4.3cm]{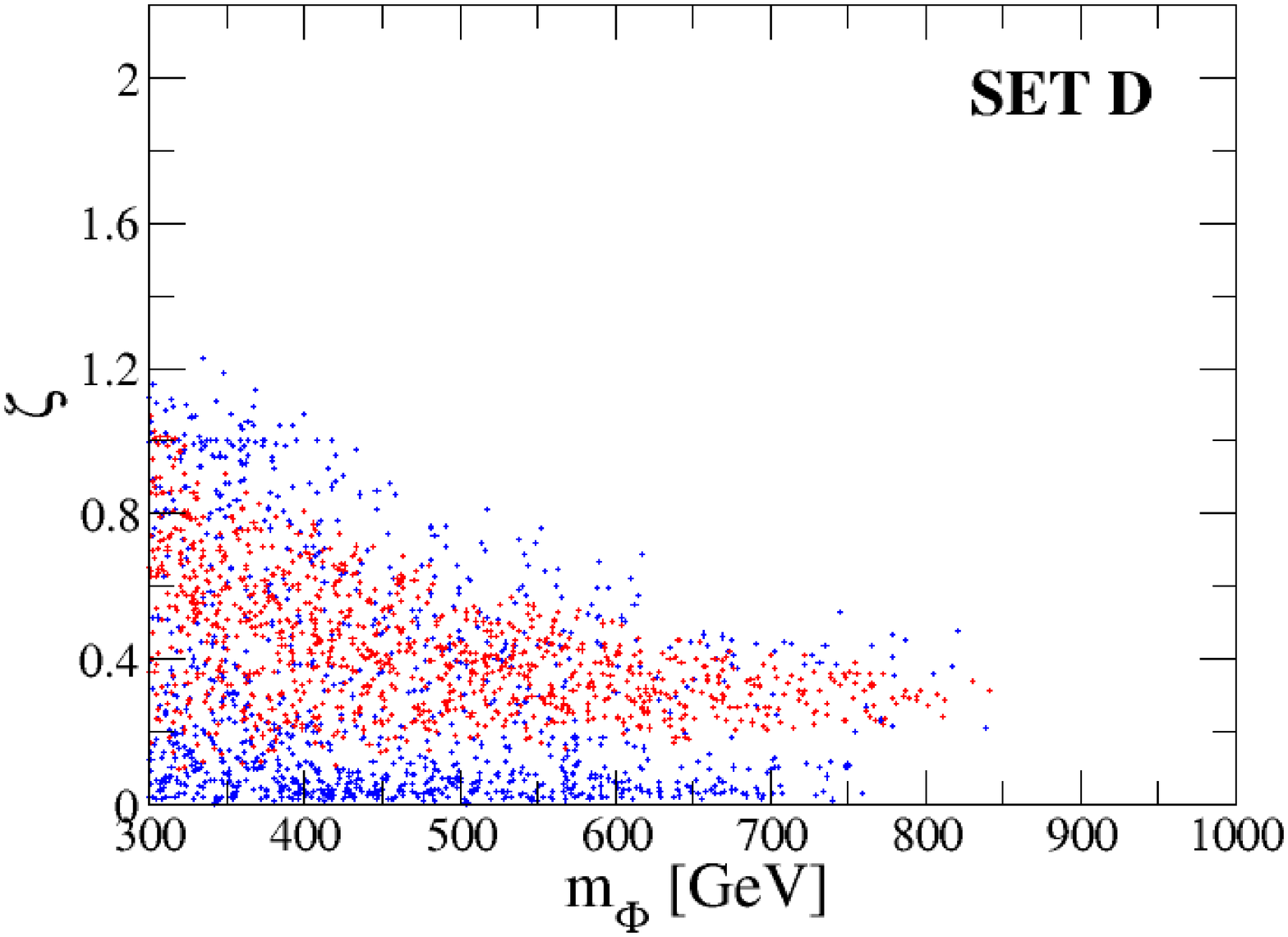}  \hspace{-5mm} 
\includegraphics[width=4.3cm]{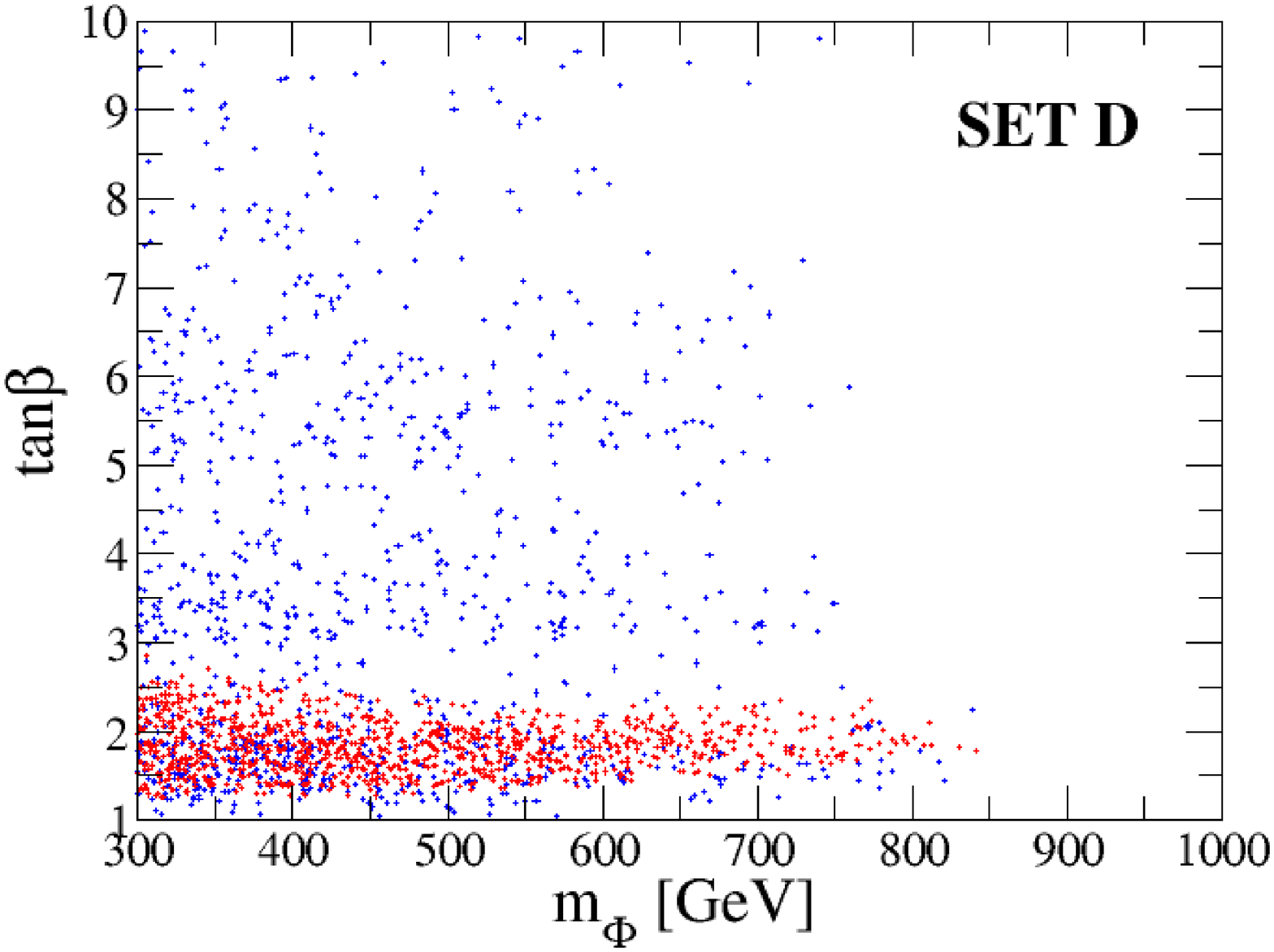} \\ 
\includegraphics[width=4.3cm]{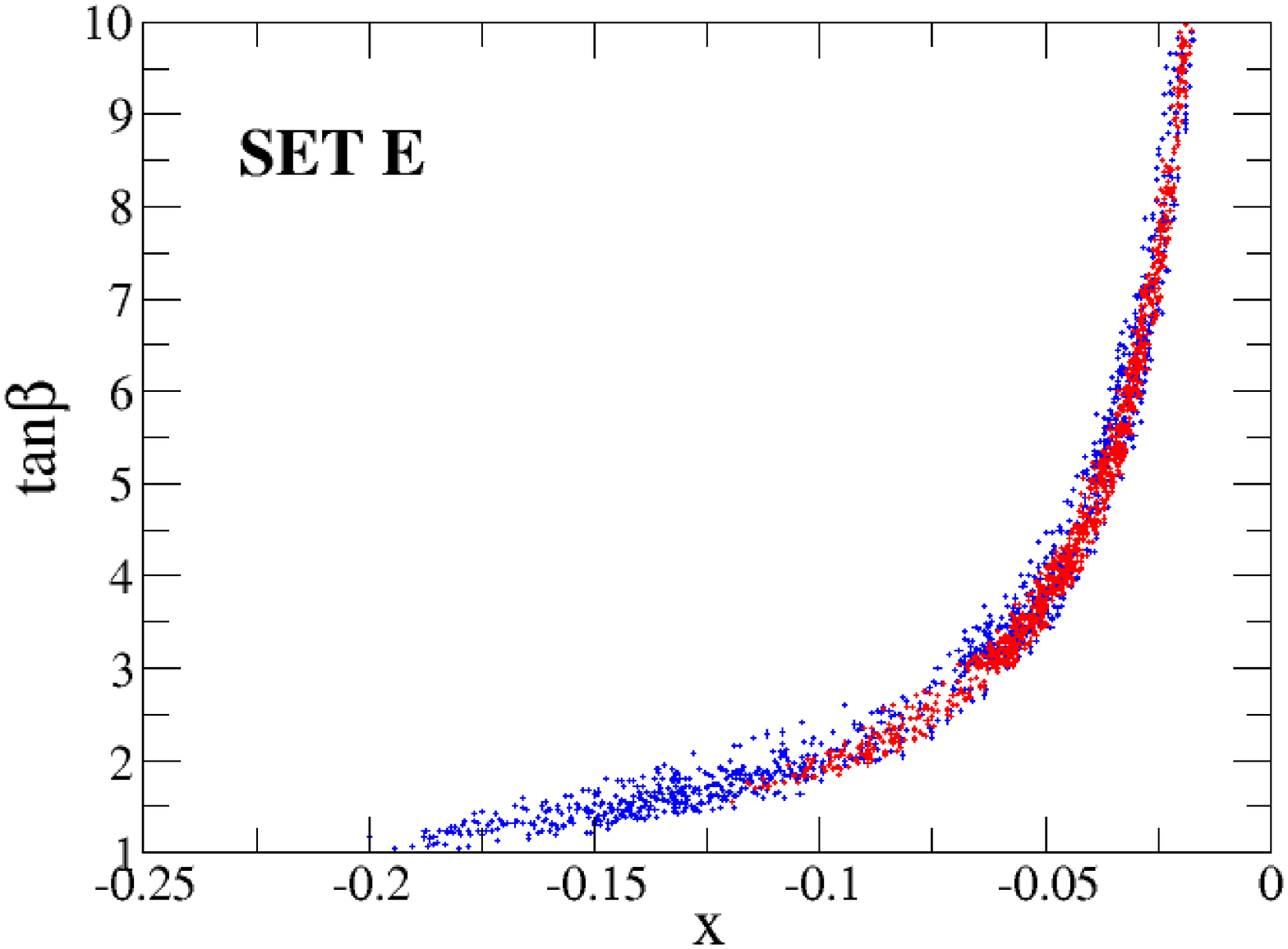}	\hspace{-5mm}
\includegraphics[width=4.3cm]{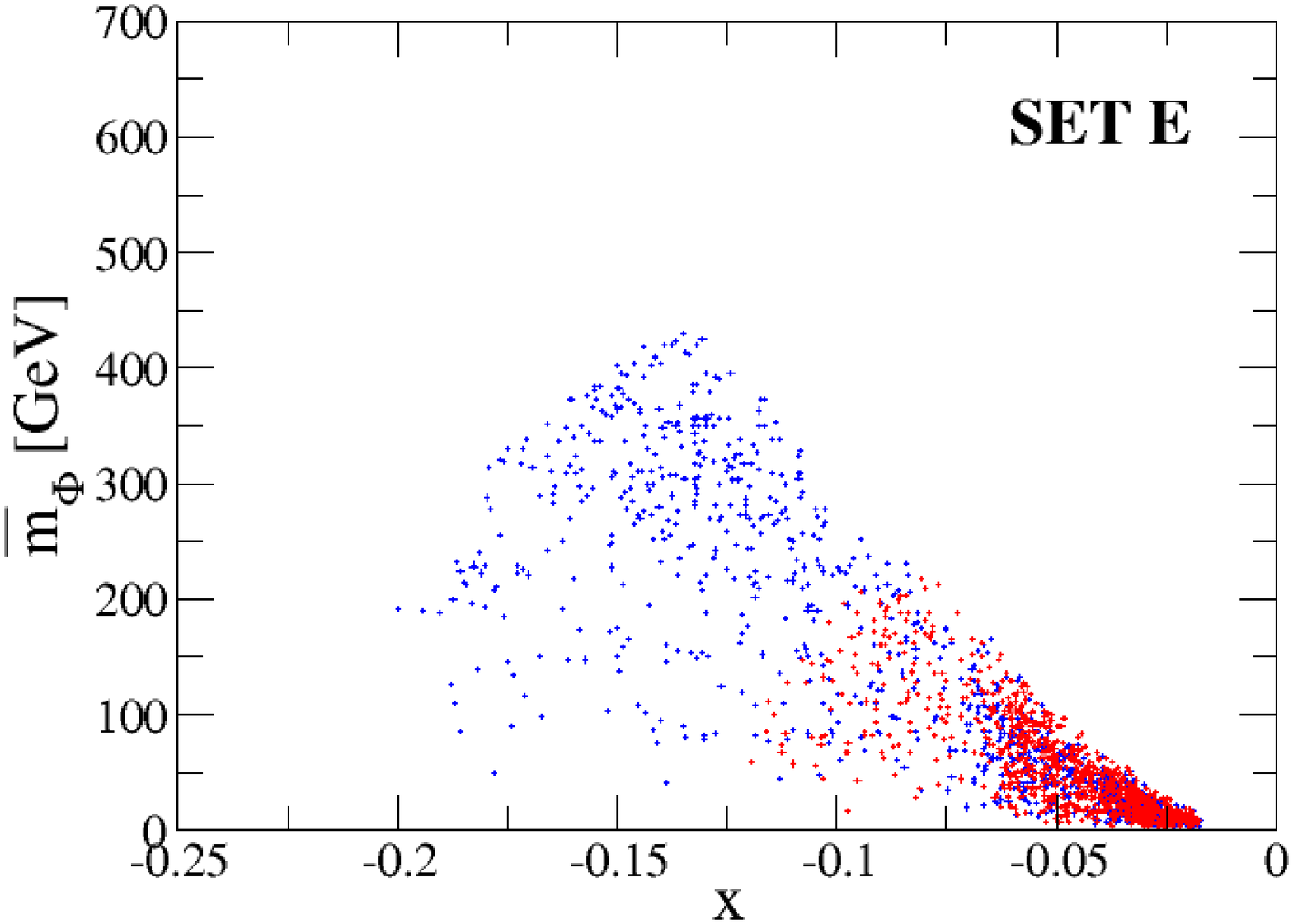}  \hspace{-5mm}
\includegraphics[width=4.3cm]{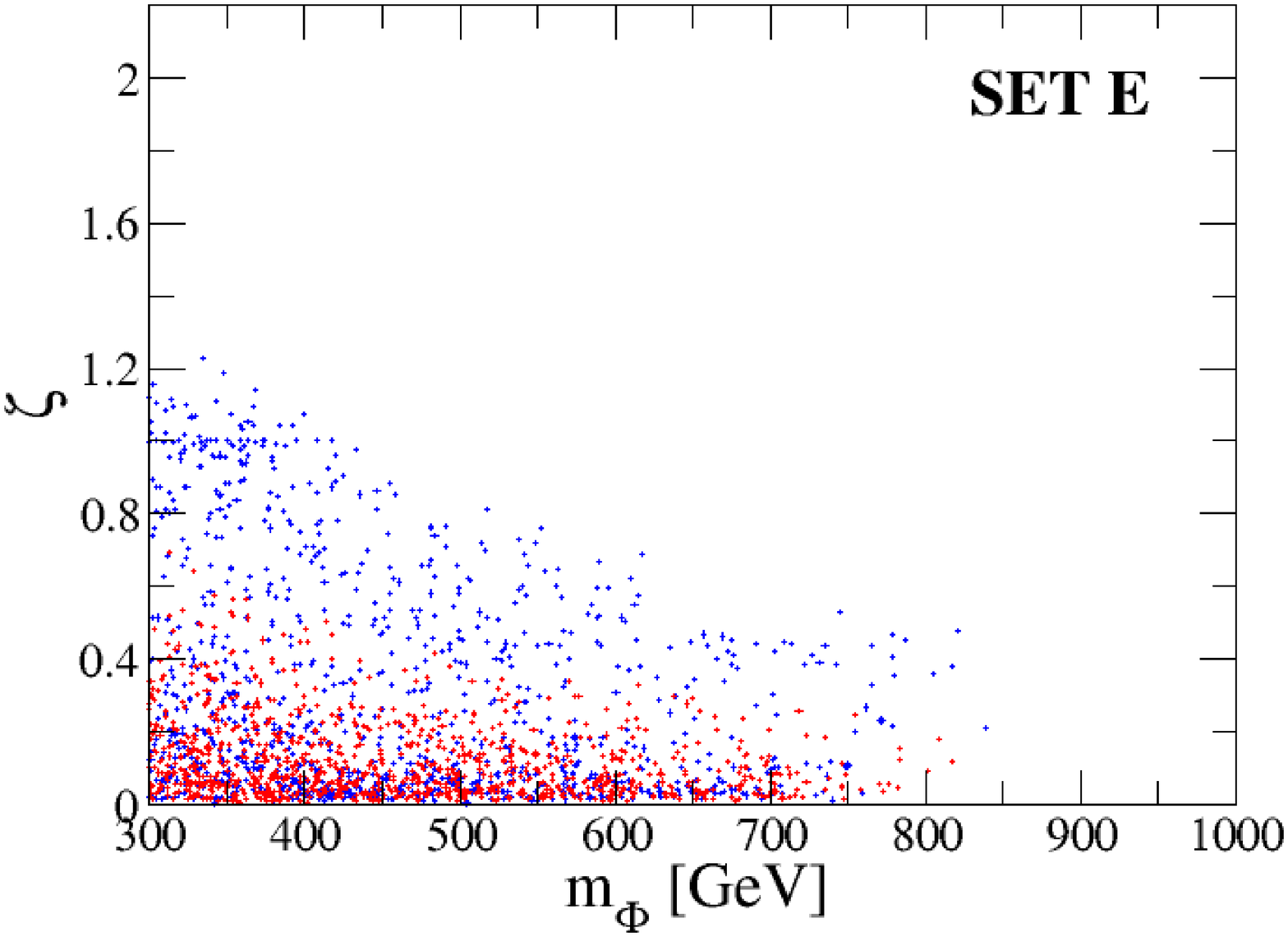}  \hspace{-5mm}  
\includegraphics[width=4.3cm]{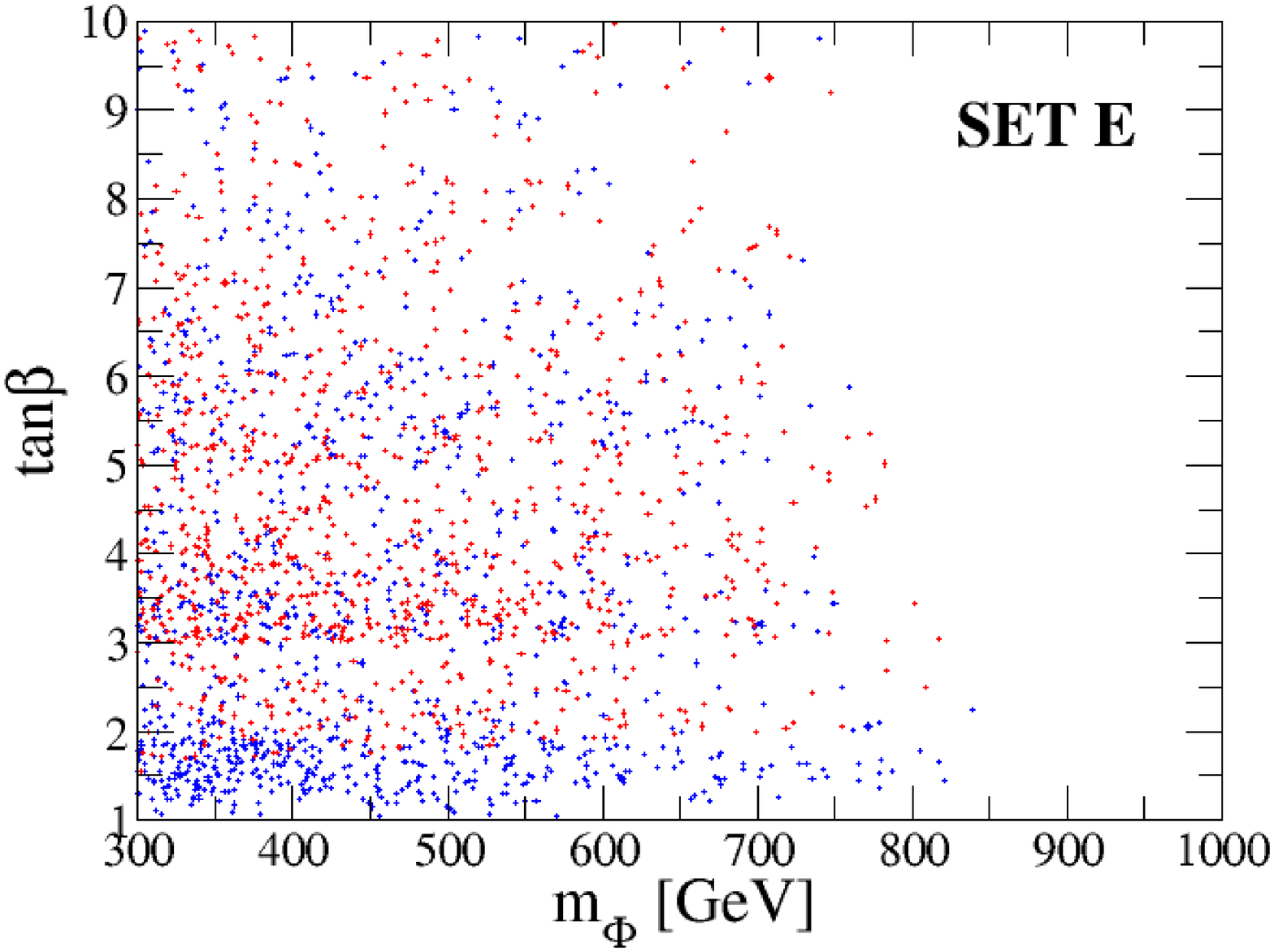} 
\caption{Scatter plots for Set A, B, C, D and E from upper to bottom panels. 
The cyan and red points satisfy the benchmark sets within the 1-$\sigma$ uncertainty at the HL-LHC and ILC500 given in Eq.~(\ref{error}), respectively. 
For the panels shown in the second and the third columns, the vertical axis $\bar{m}_\Phi$ and $\zeta$ are respectively defined by 
$\bar{m}_\Phi \equiv  m_\Phi (1-M^2/m_\Phi^2)$ and $\zeta \equiv 1-M^2/m_\Phi^2$. }
\label{scatter1}
%========================================
\end{figure}

%% ($-0.2,1.0$)  
%% ($-0.2,0.61$) 
%% ($-0.2,0.35$) 
%% ($-0.14,1.4$) 
%% ($-0.089,2.1$)

In Fig.~\ref{scatter1}, we show the allowed parameter regions on the
$x$-$\tan\beta$, $x$-$\bar{m}_\Phi^{}$, $m_\Phi^{}$-$\zeta$ and $m_\Phi^{}$-$\tan\beta$ planes from the left to right panels, where 
we define 
\begin{align}
\zeta \equiv 1-M^2/m_\Phi^2, \quad \bar{m}_\Phi^{} \equiv  m_\Phi^{}\zeta. 
\end{align} 
The parameters $x$ and $\bar{m}_\Phi^{}$ give deviations of the Higgs boson couplings by the mixing effect and the loop effect, 
respectively. 
Notice that the scale of $\bar{m}_\Phi^{}$ corresponds to the mass of the extra Higgs boson when $M^2=0$. 
The physics meaning of $\zeta$ is to measure the magnitude of non-decouplingness of the loop effects of 
extra Higgs bosons. If $\zeta$ is unity, we have $M^2=0$,  
while if $\zeta<1$ with nonzero value of $M^2$ ($>0$), the mass of the extra Higgs bosons partially comes from $M^2$
so that the non-decouplingness is smaller. 
The central values of $\Delta\kappa$'s are chosen from Set A, B, C, D and E from the upper to bottom panels. 
The blue and red points correspond to the region within 
the 1-$\sigma$ uncertainty at the HL-LHC and ILC500, respectively, from the central value in Table~\ref{benchmark}. 

%\noindent
%[Set A]

For Set A in Fig.~\ref{scatter1}, let us first explain the behavior of the red points on the $x$-$\tan\beta$ plane.  
In this case, $-2.4\%<\Delta\kappa_V^{}<-1.6\%$ is allowed at the ILC500, which can be explained by 
taking $-0.22\lesssim x\lesssim -0.18$ at the tree level from the expression of $\Delta\kappa_V\simeq -x^2/2$. 
At the same time, both $\Delta\kappa_\tau$ and $\Delta\kappa_b$ are approximately given by $-x\tan\beta$ 
in the Type-II THDM at the tree level, so that $\tan\beta$ is determined by a fixed value of $x$ from $\tan\beta \simeq -\Delta\kappa_{\tau/b}/x$, which is around unity 
if we take the central value of $\Delta\kappa_V$ and $\Delta\kappa_{\tau/b}$.  
In fact, by looking at the top-left panel in Fig.~\ref{scatter1}, 
the above mentioned values of $x$ and $\tan\beta$ are allowed. 
However, the actual allowed region 
of $x$ 
inclucing radiative corrections 
is about from $-0.22$ to $-0.12$ which is wider than the allowed region 
estimated at the tree level. 
This can be understood by taking into account the additional Higgs boson loop contributions to $\kappa_V^{}$ at the
one-loop level.  
The approximate formula for $\Delta\hat{\kappa}_V^{}$ is given in Eq.~(\ref{kv}), where 
the second term in the right hand side corresponds to the one-loop contribution. 
The point here is that the sign of one-loop effect is negative, and it is proportional to the factor $\zeta^2$.
Therefore, the allowed region above $x\simeq -0.18$ is explained from the one-loop contribution with a non-zero value of $\zeta$. 
On the other hand, the one-loop correction to $\kappa_\tau$ is given by the same form as for $\kappa_V^{}$ as given in Eq.~(\ref{ktau}), 
so that the difference $\Delta\hat{\kappa}_\tau -\Delta\hat{\kappa}_V^{}$ is approximately given by the same form $-x\tan\beta$ 
as that given at the tree level. 
Now from the measurement, since the difference is determined with the uncertainty, $-x\tan\beta$ is also fixed at the one-loop level. 
We thus can understand the shape of the allowed region of this plot.  
Although for $\Delta\hat{\kappa}_b$ the top quark, the bottom quark and $H^\pm$ loop diagrams give 
an additional contribution as shown in Eq.~(\ref{kb}), this is not so significant in the scanned regions. 
As a consequence for Set A, when the measurement at the ILC500 is assumed, 
the allowed value of $x$ and $\tan\beta$ can be determined to be about from 
$-0.22$ to $-0.12$ and from 1 to 2, respectively. 
On the other hand at the HL-LHC, $\Delta\kappa_V=0$ is included within the 1-$\sigma$ uncertainty.  
Thus, $x\simeq 0$ is still allowed, so that 
the value of $\tan\beta$ is not determined at all because of the relation
$\tan\beta \simeq -\Delta\kappa_{\tau/b}/x$. 
In addition, we can only extract the lower limit of $x$ to be about $-0.22$.  

Next, we discuss the behavior of the second panel for Set A in Fig.~\ref{scatter1}. 
As we mentioned in the above, 
the vertical axis $\bar{m}_\Phi^{}$ measures the size of one-loop contribution to the deviation in the Higgs boson couplings.  
At the ILC500, in the region with $x\simeq -0.20$, the value of $\bar{m}_\Phi^{}$ is determined to be a smaller value, but 
$\bar{m}_\Phi^{}\simeq 0$ is not included because of the constraint from vacuum stability. 
This can be understood that the deviation from the tree level mixing is dominant in this case. 
On the other hand, when the value of $x$ approaches to zero, a sizable value of $\bar{m}_\Phi^{}$ is extracted, in which 
the deviation driven by the one-loop contribution becomes more important to compensate the reduced contribution from the 
tree level mixing. 
In addition, the upper limit of $\bar{m}_\Phi^{}$ to be about 450 GeV is determined by the constraint from perturbative unitarity.  
At the HL-LHC, although the blue plots are spread over the region with $x\simeq 0$ as we observed in the $x$-$\tan\beta$ plot, 
the upper and lower limit of $\bar{m}_\Phi$ is given by the constraint from unitarity and vacuum stability, respectively. 

The third panel for Set A in Fig.~\ref{scatter1} shows the allowed region on the $m_\Phi^{}$-$\zeta$ plane, where $\zeta$ is the parameter indicating the 
non-decouplingness of the extra Higgs bosons. 
For Set A, the allowed regions for ILC500 are shown by the red points while those for HL-LHC by the blue points. 
There are upper and lower bounds for $\zeta$ for each value of $m_\Phi^{}$. 
They are crossed at around $m_\Phi^{}=$ 850 GeV which corresponds to the upper bound of the mass of extra Higgs boson. 
The region of $\zeta$ is from 0.2 to 1.4 at $m_\Phi^{}=300$ GeV. 
The region of $\zeta>1$ corresponds to $M^2<0$, where non-decoupling effects are effectively large.
The exclusion of $\zeta <0.2$ means that there must be some non-decoupling loop effects of extra Higgs bosons in order to explain 
this benchmark point. 
At the HL-LHC, the similar behavior can be observed. However, $\zeta=0$ is still allowed, so that we cannot say something about 
the non-decoupling effect. 

The last panel for Set A in Fig.~\ref{scatter1} shows the allowed regions on the $m_\Phi^{}$-$\tan\beta$ plane. 
At the ILC500, $\tan\beta$ can be determined to be less than 2, and the upper bound of the mass of the extra Higgs bosons 
are obtained to be less 850 GeV, 
while at the HL-LHC, $\tan\beta$ is undetermined and only the upper bound of the mass of the extra Higgs bosons is obtained.    

The panels shown in the second and third rows in Fig.~\ref{scatter1} display the allowed parameter regions for Set~B and Set~C, respectively, where 
the central value of $\Delta\kappa_\tau(=\Delta \kappa_b)$ is taken to be smaller than that of Set A, while $\Delta\kappa_V^{}$ is taken to be the same. 
By looking at the panels for the $x$-$\tan\beta$ plane, we can see that a smaller value of $|x|$ is preferred as compared to the case for Set A. 
Furthermore, a smaller value of $\tan\beta$ is favored in addition to a smaller value of $|x|$ as seen in the result at the ILC500. 
These tendencies can be understood in such a way that   
the deviations in Yuakwa couplings are proportional to $-x\tan\beta$ at the tree level. 
Because of the smaller value of $|x|$, the deviation in $\kappa_V^{}$ cannot be explained only from the tree level contribution, so that the 
one-loop effect is necessary to compensate the tree level contribution. 
That is the reason why the red points in the second and the third panels for Set B and Set C 
are given in the upper region which does not include $\bar{m}_\Phi^{}\simeq 0$ and $\zeta\simeq 0$. 
Therefore, the non-decoupling effect can be extracted at the ILC500 for these two benchmark sets.  
From the results of ILC500, the upper limit on $m_\Phi^{}$ is extracted to be about 950 GeV and 800 GeV for Set~B and Set~C, respectively. 

The panels shown in the fourth and fifth rows in Fig.~\ref{scatter1} display the allowed parameter regions for Set~D and Set~E, respectively, where 
the central value of $\Delta\kappa_V^{}$ is taken to be smaller than that of Set A, while $\Delta\kappa_\tau~(=\Delta\kappa_b)$ is taken to be the same. 
From the red points in the left panels, it is seen that the 
values of smaller $|x|$ and larger $\tan\beta$ are allowed, which can be explained by the tree level 
formulae of $\Delta\kappa_V^{}=-x^2/2$ and $\Delta \kappa_{\tau/b}=-x\tan\beta$.
For Set~E unlike the other benchmark sets, values of $x$ and $\tan\beta$ are not well determined even at the ILC500,
because $\Delta\kappa_V^{}\simeq 0$ is included within the 1-$\sigma$ uncertainty of ILC500. 
The extraction for $\bar{m}_\Phi^{}$, $\zeta$ and $m_\Phi^{}$ is done from the ILC500 as 
$50 \lesssim \bar{m}_\Phi^{}\lesssim 300$ GeV, $0.1 \lesssim \zeta \lesssim 1.1$ GeV and $m_\Phi^{}<850$ GeV for Set D and 
$0 \lesssim \bar{m}_\Phi^{}\lesssim 200$ GeV, $0 \lesssim \zeta \lesssim 0.7$ GeV and $m_\Phi^{}<800$ GeV for Set E.

\begin{figure}[t]
%========================================
\centering
\includegraphics[width=4.3cm]{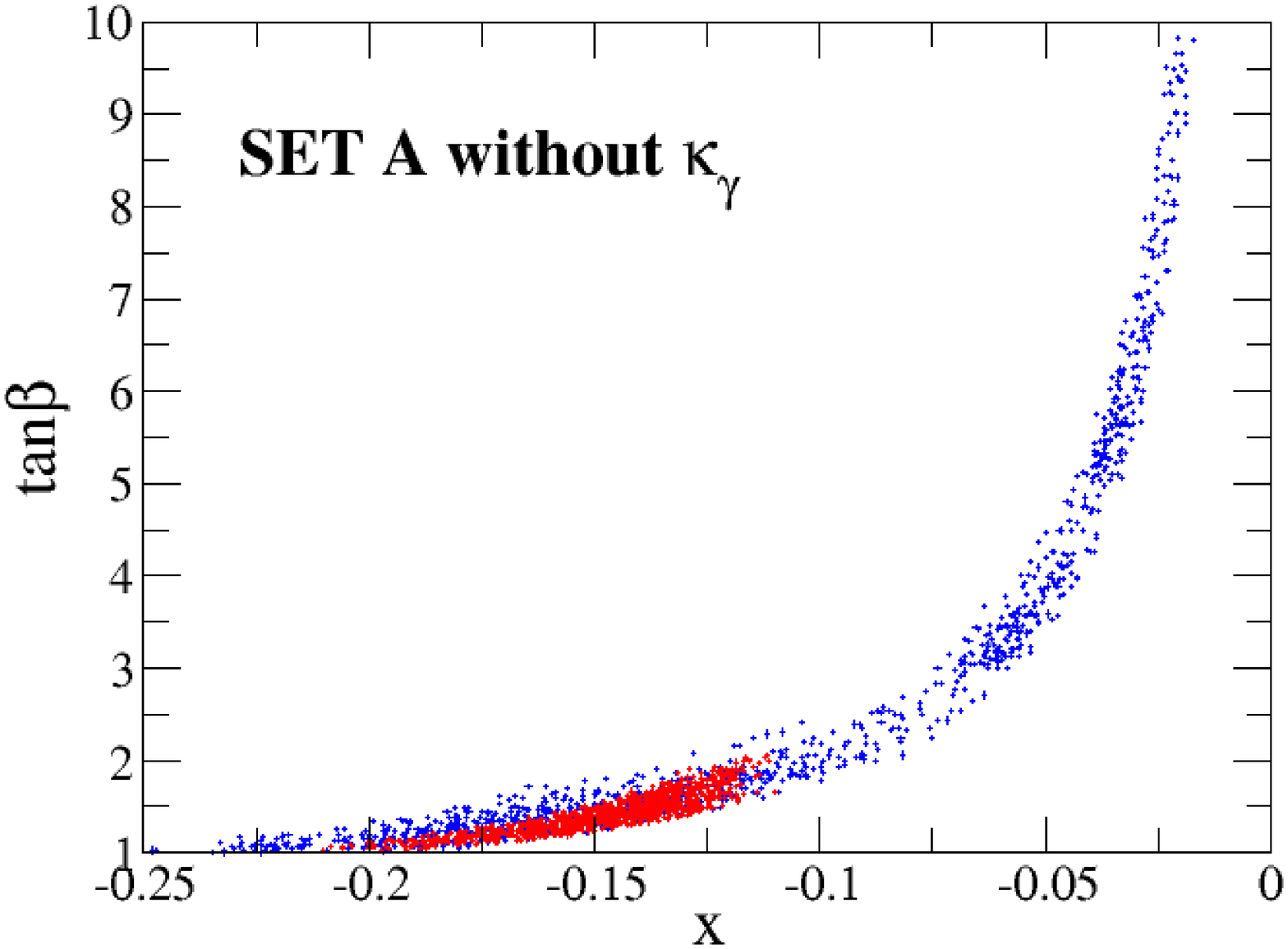}	\hspace{-5mm}
\includegraphics[width=4.3cm]{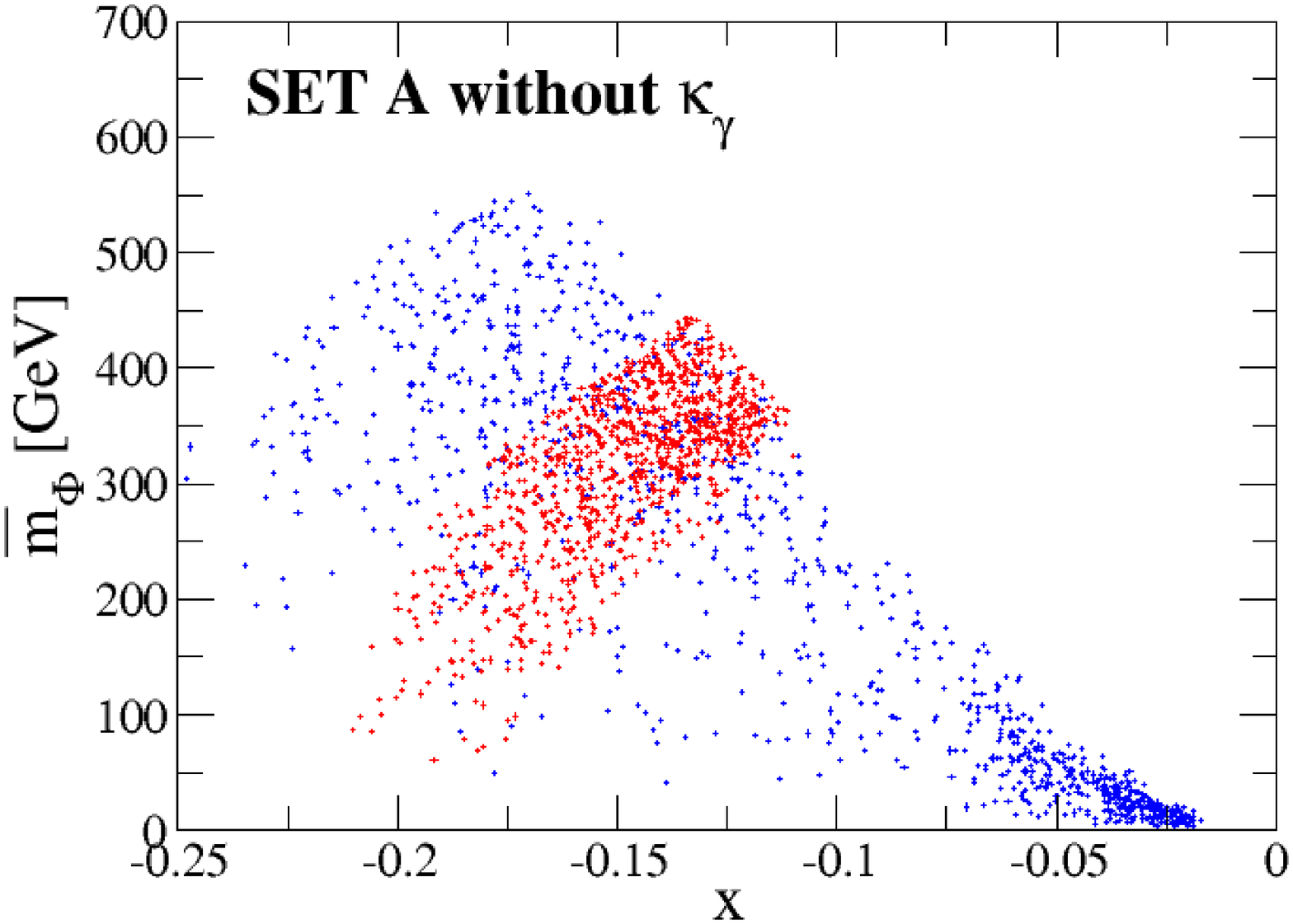}  \hspace{-5mm}
\includegraphics[width=4.3cm]{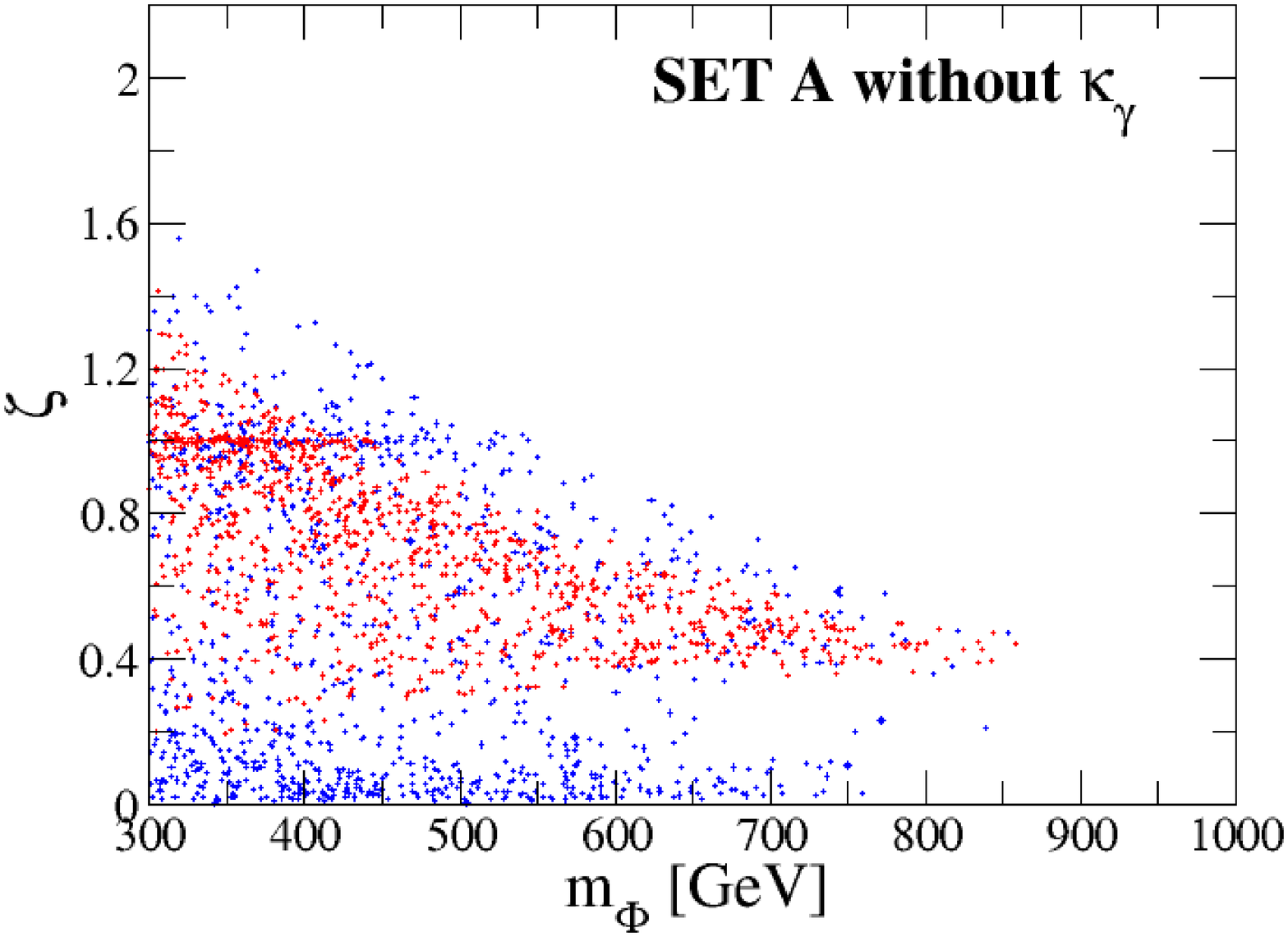}  \hspace{-5mm}  
\includegraphics[width=4.3cm]{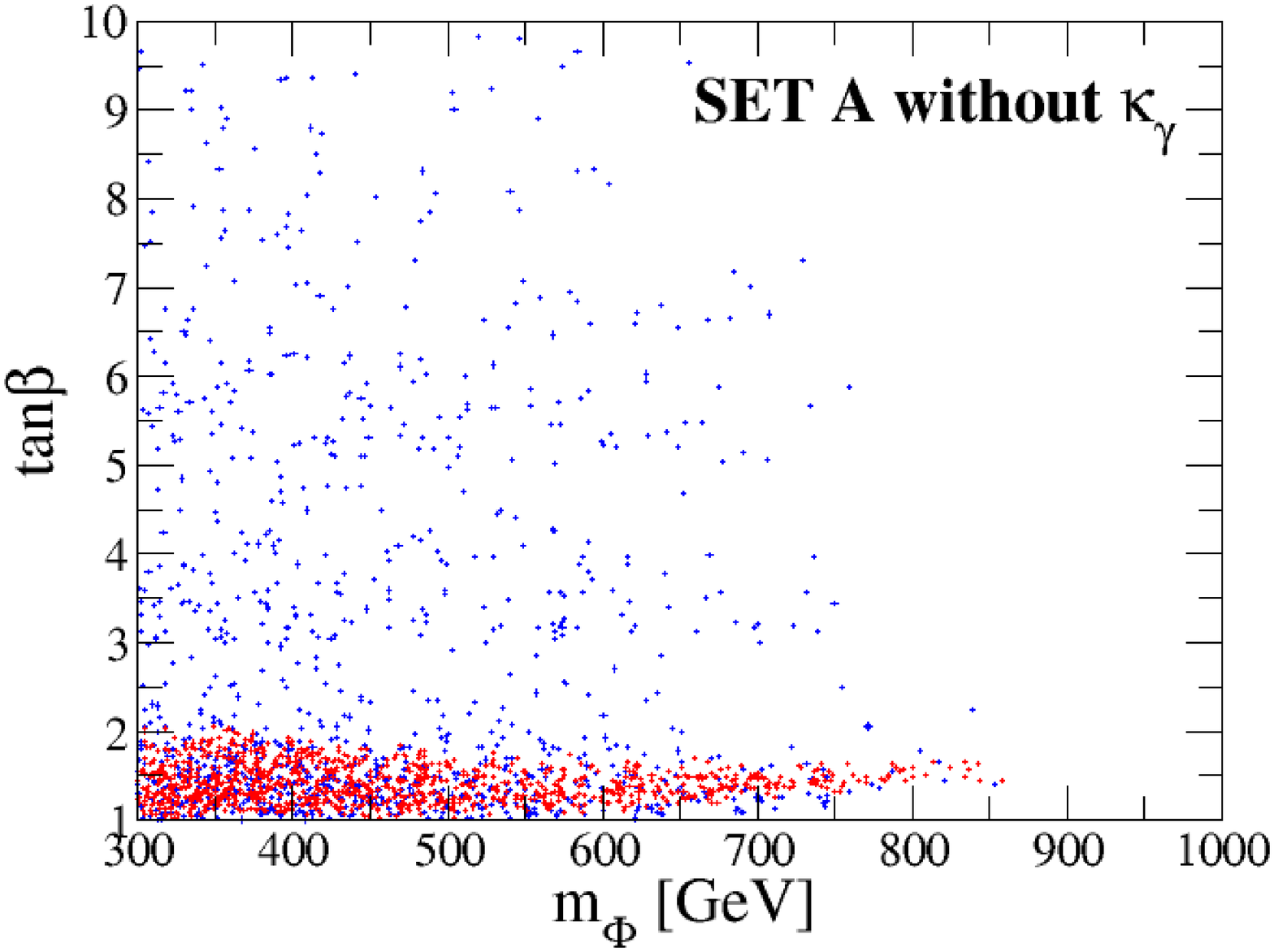} \\ 
\includegraphics[width=4.3cm]{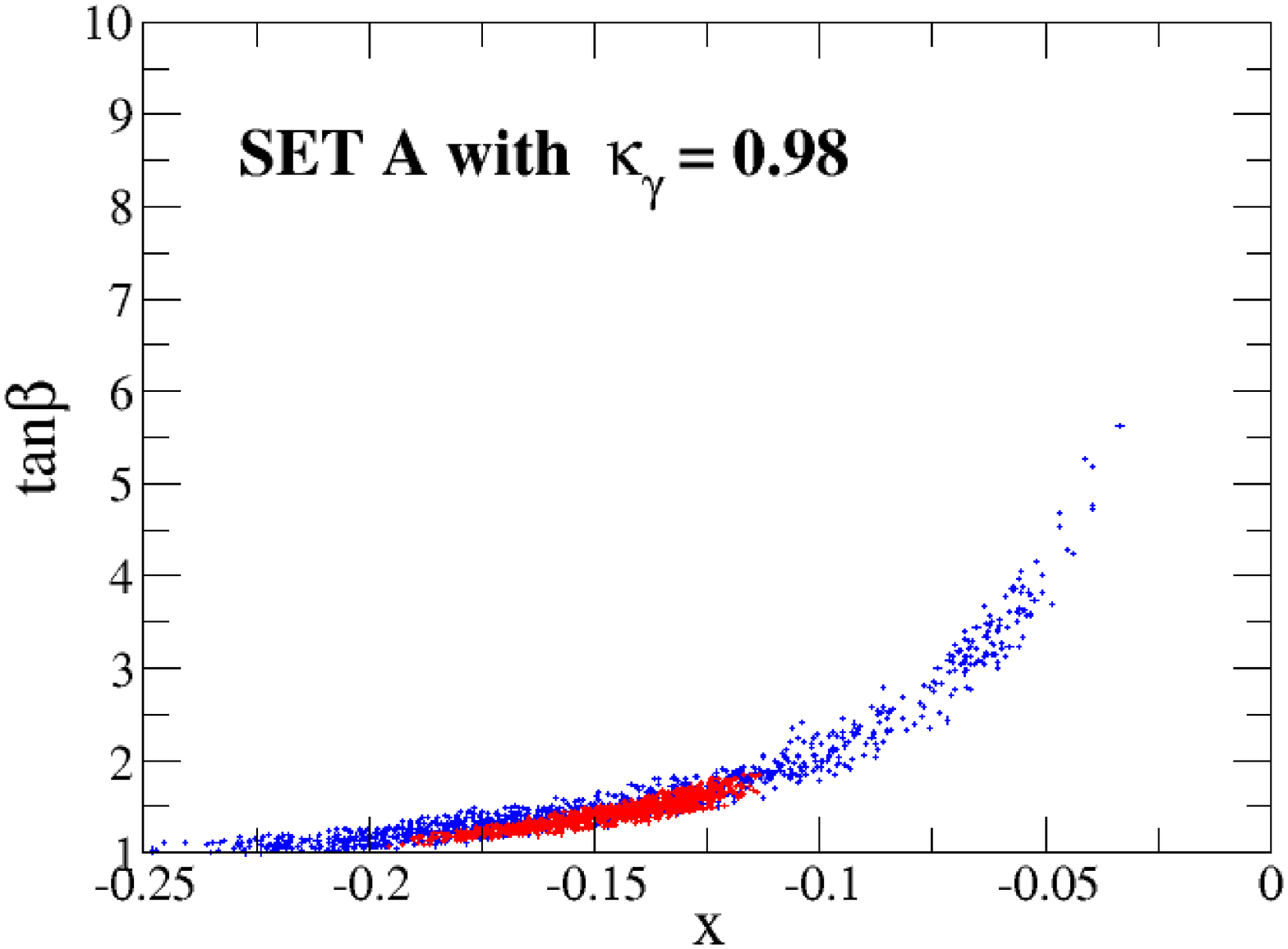}	\hspace{-5mm}
\includegraphics[width=4.3cm]{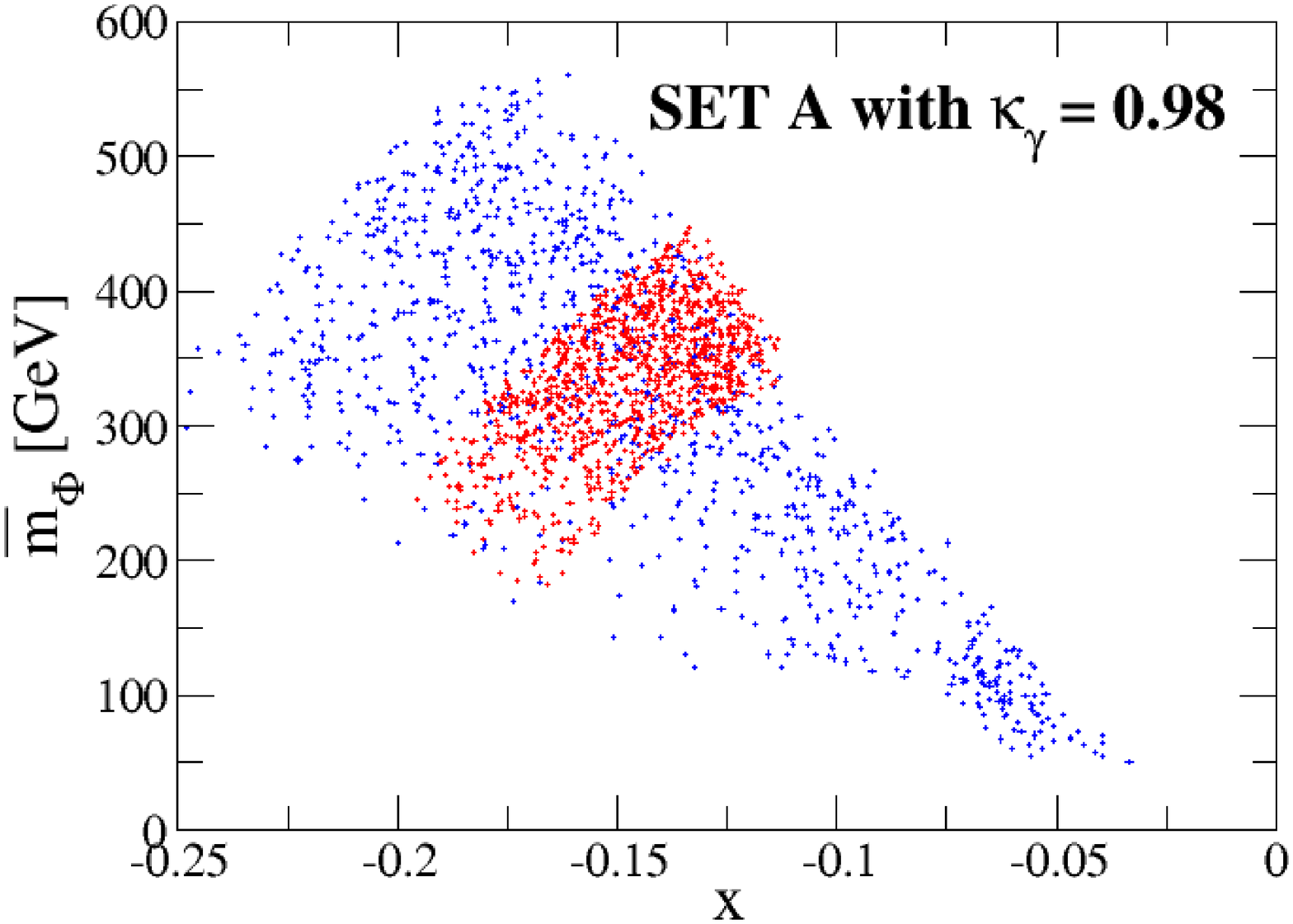}  \hspace{-5mm}
\includegraphics[width=4.3cm]{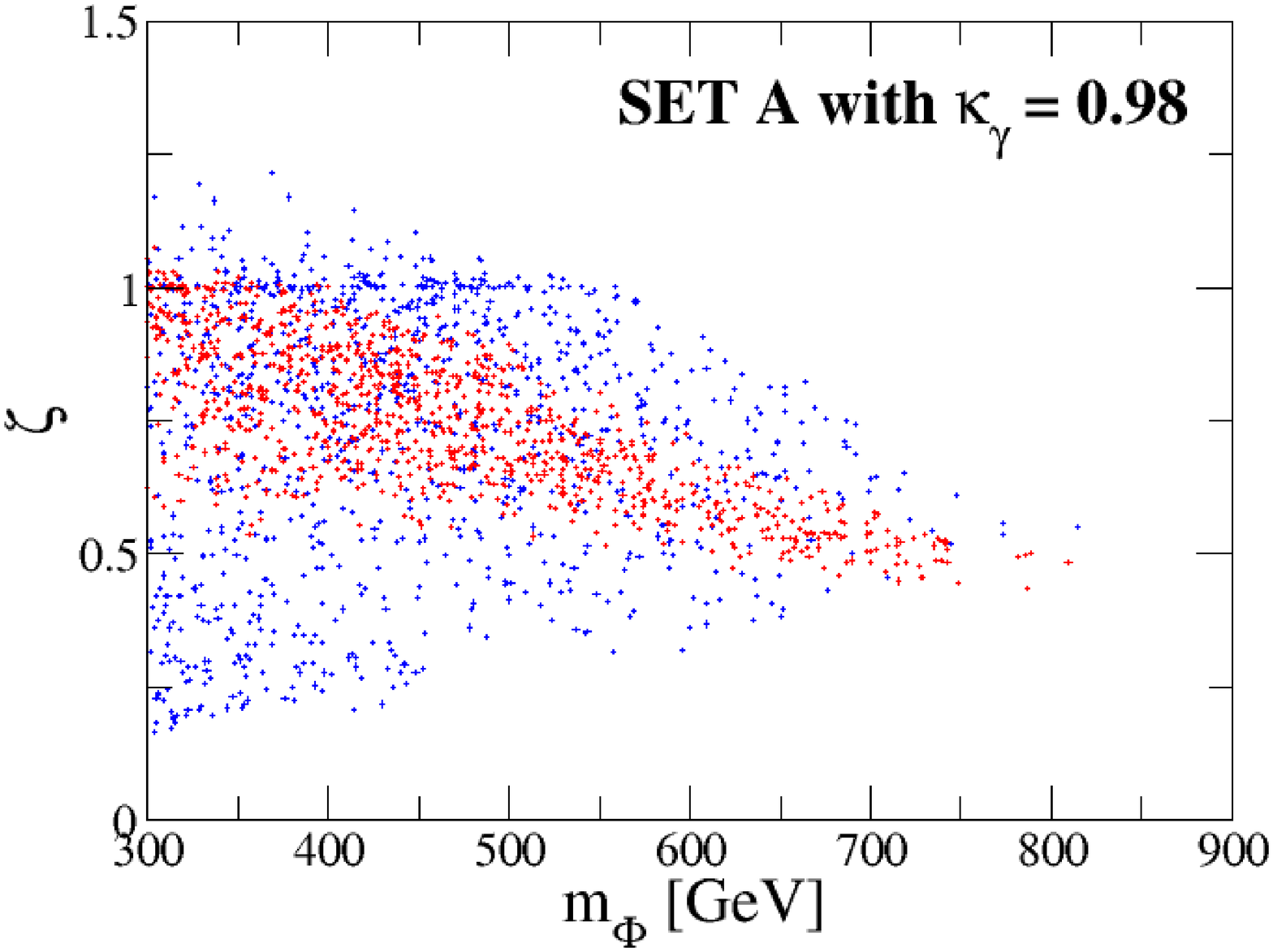}   \hspace{-5mm} 
\includegraphics[width=4.3cm]{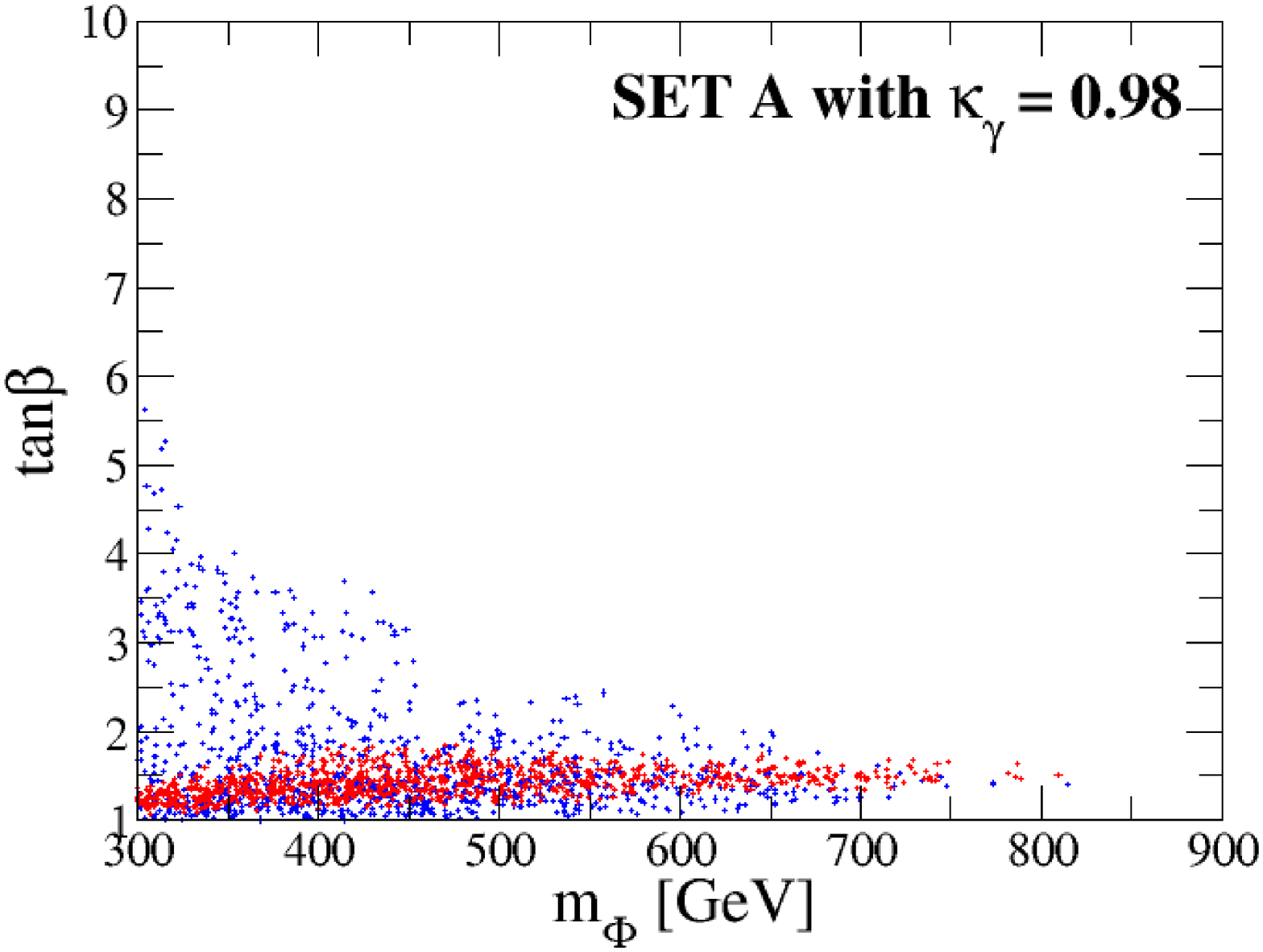} \\ 
\includegraphics[width=4.3cm]{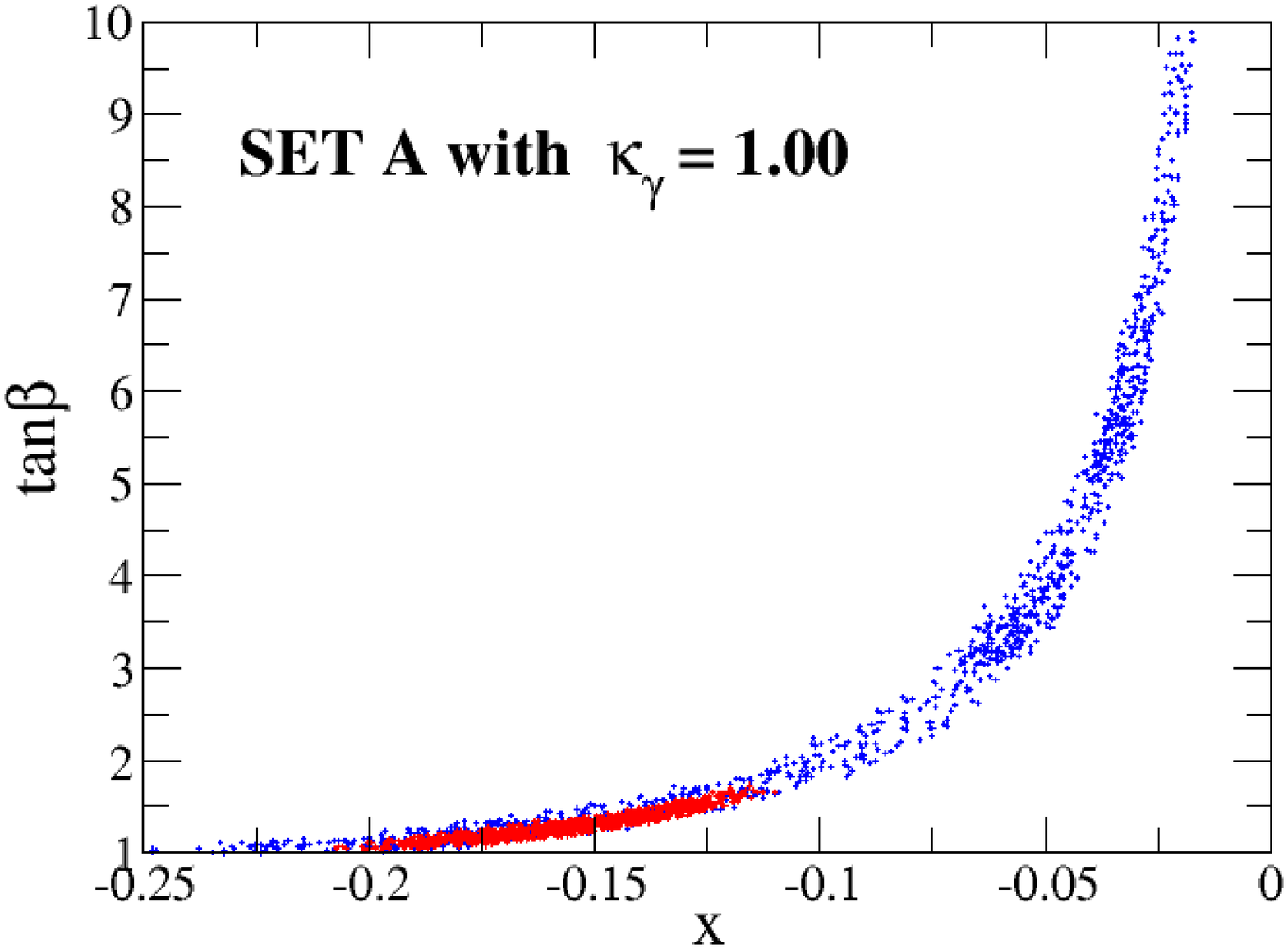}	\hspace{-5mm}
\includegraphics[width=4.3cm]{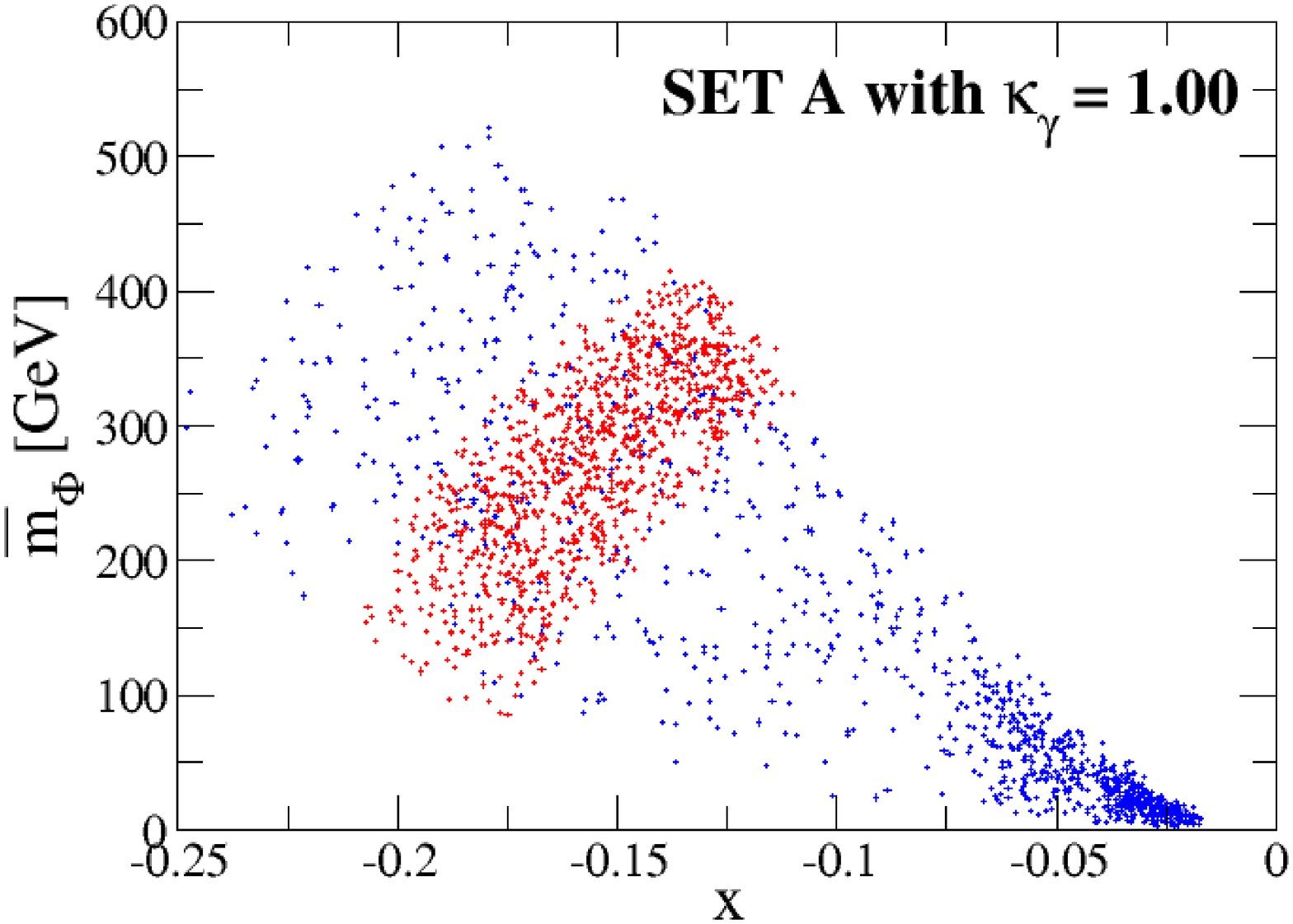}  \hspace{-5mm}
\includegraphics[width=4.3cm]{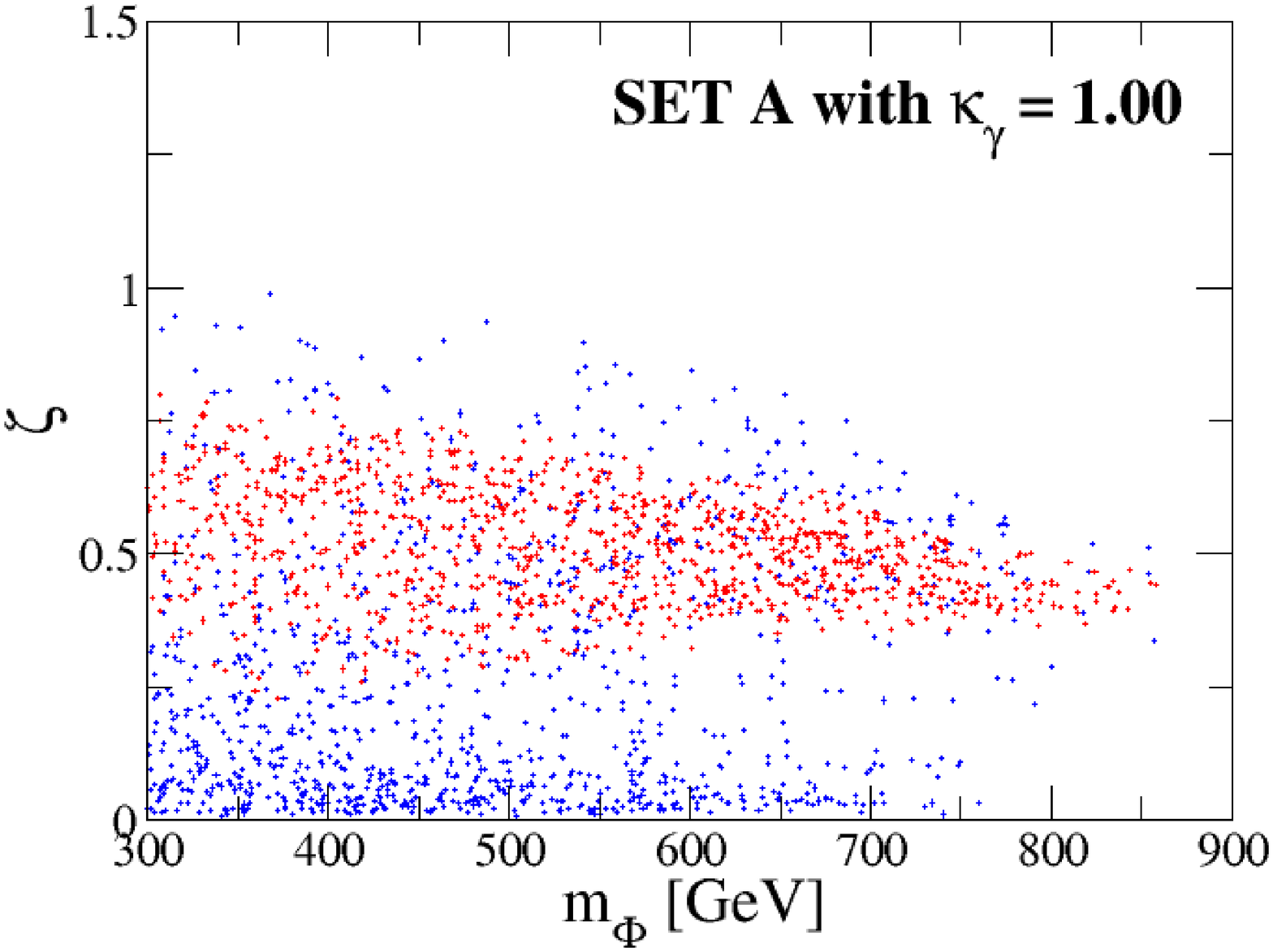}  \hspace{-5mm} 
\includegraphics[width=4.3cm]{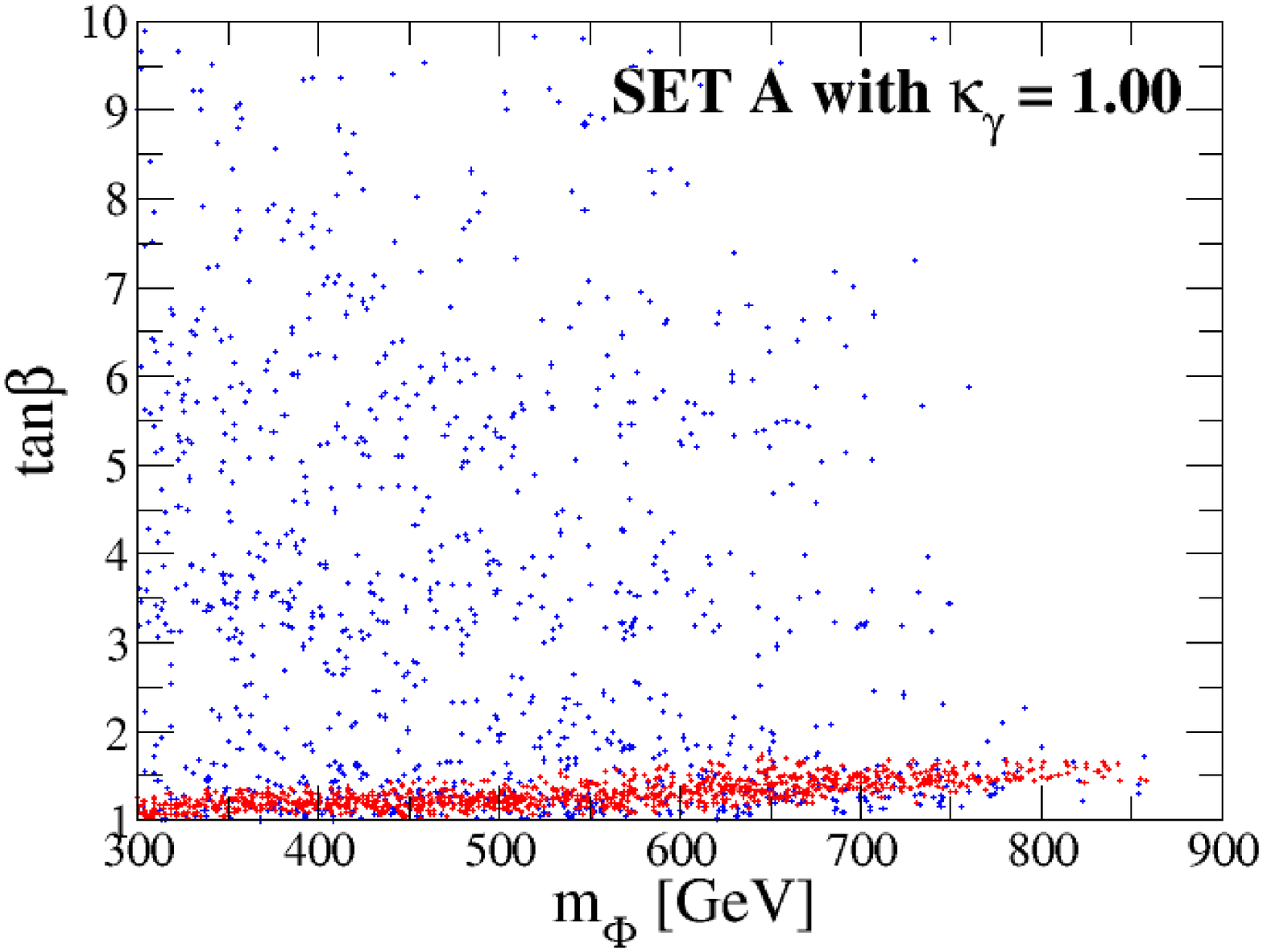} \\ 
\includegraphics[width=4.3cm]{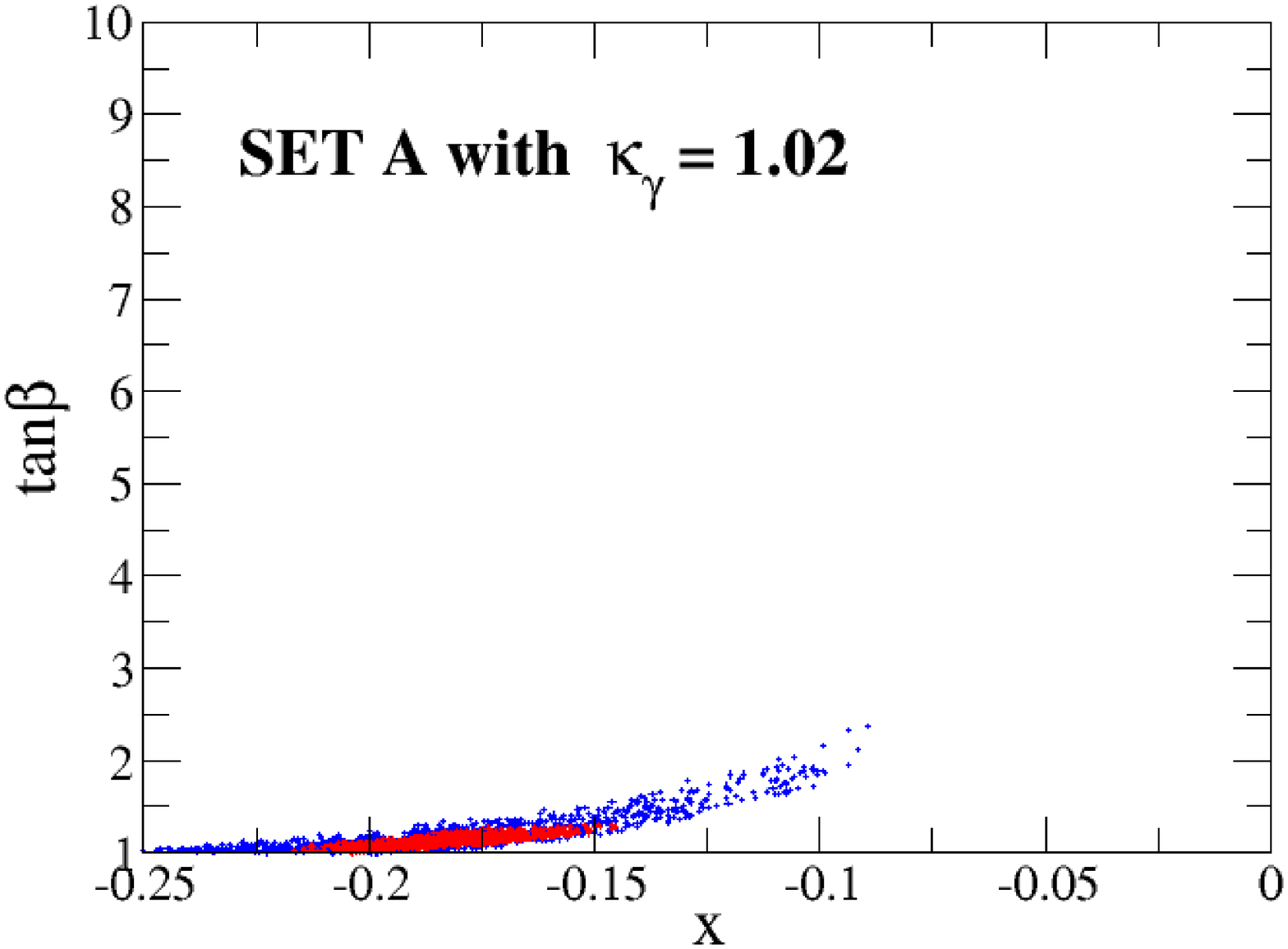}	 \hspace{-5mm}
\includegraphics[width=4.3cm]{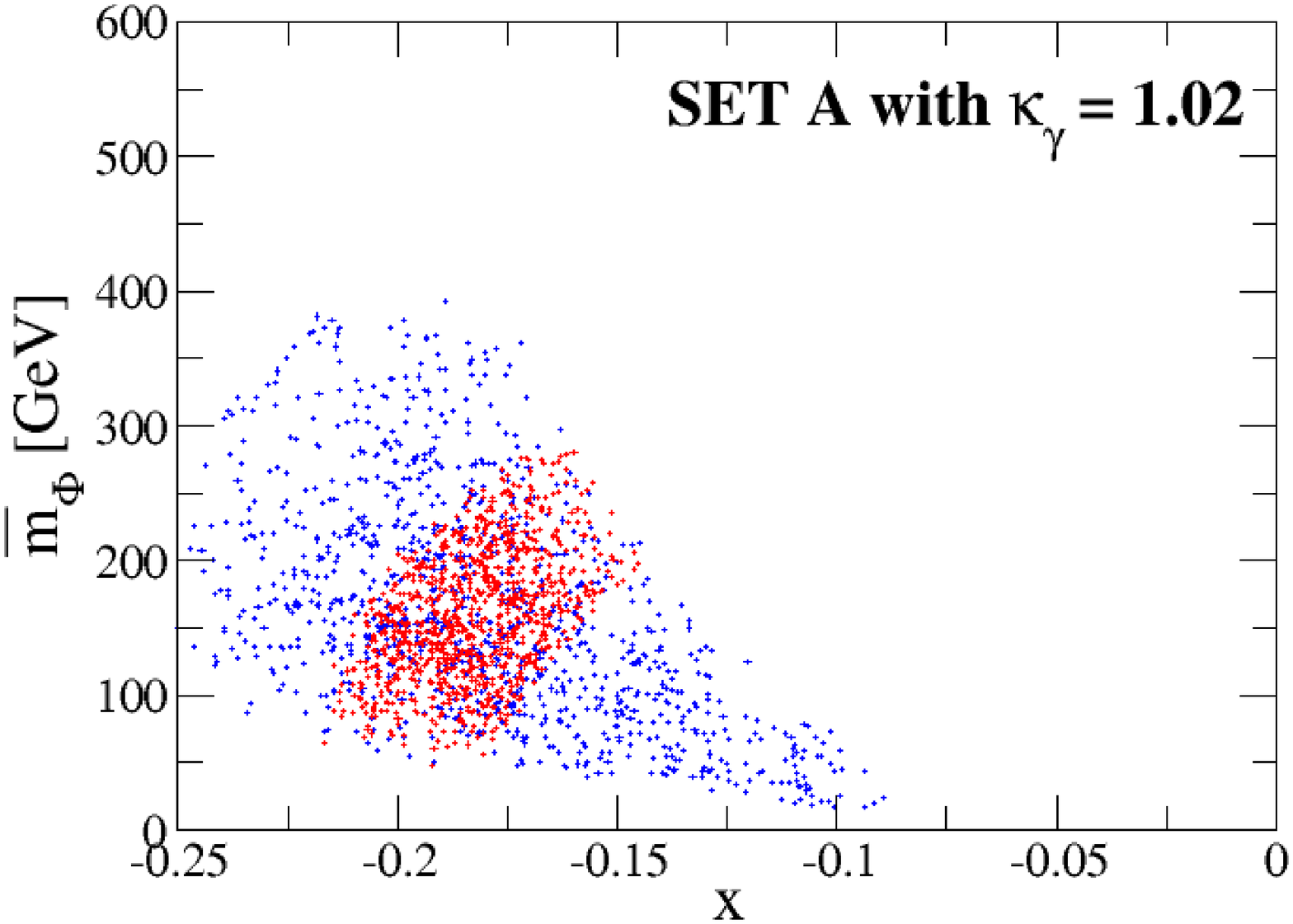}  \hspace{-5mm}
\includegraphics[width=4.3cm]{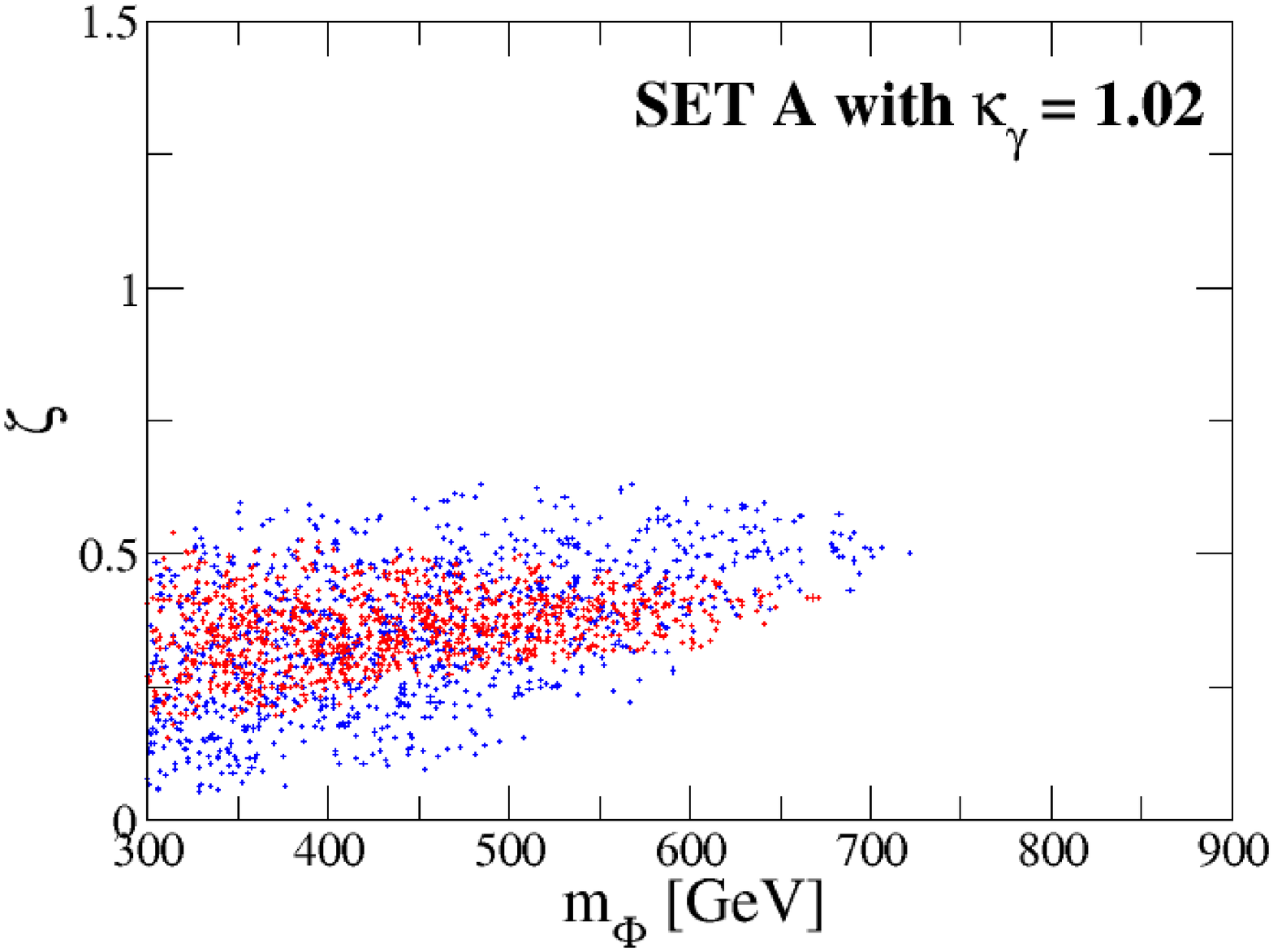}  \hspace{-5mm} 
\includegraphics[width=4.3cm]{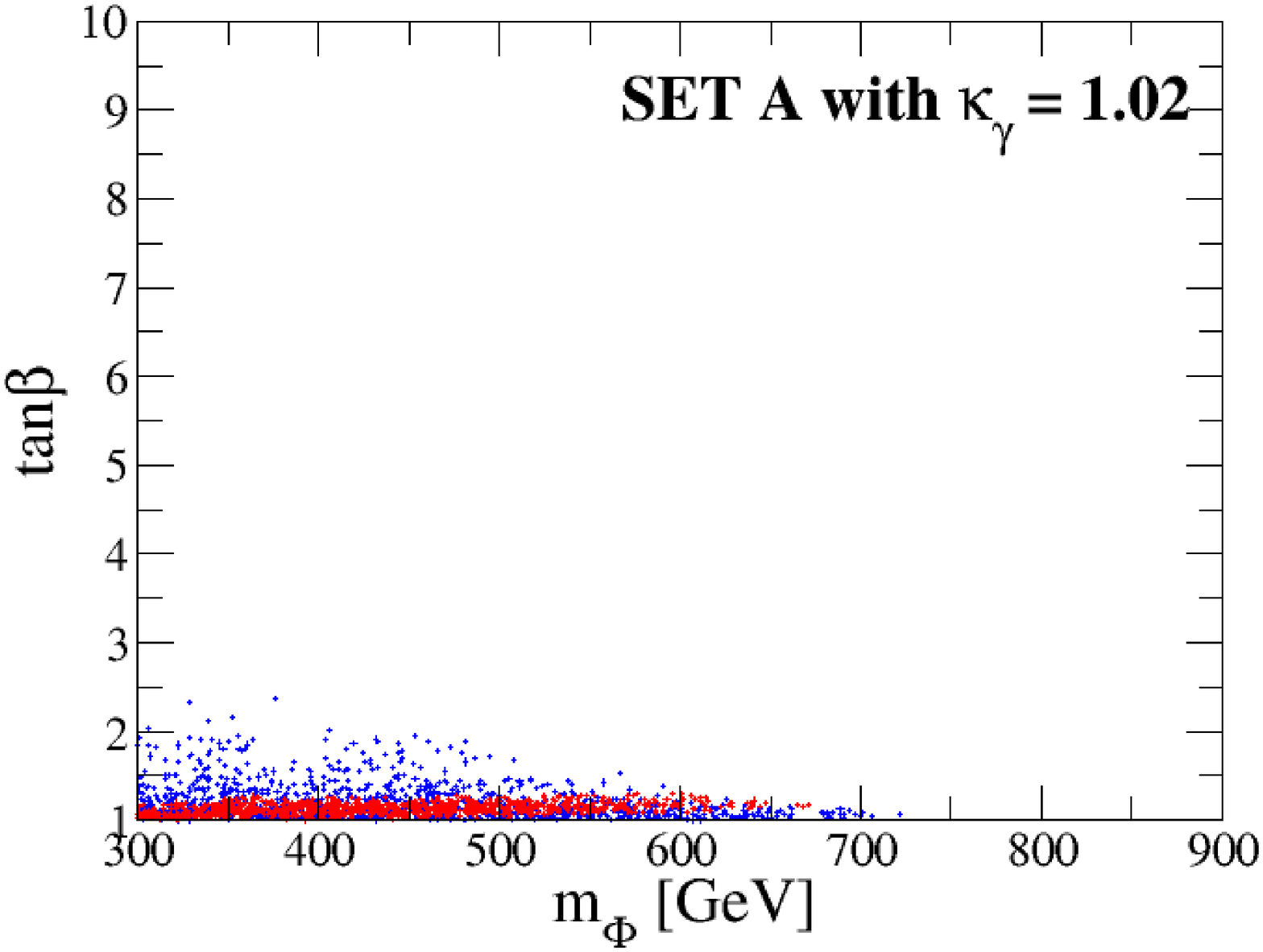}  \\ 
\caption{Scatter plots for Set A with the additional constraint from $\kappa_\gamma=0.98,~1.00$ and 1.02 for 
upper, center and bottom panels. 
The 1-$\sigma$ uncertainty of $\kappa_\gamma$ is assumed to be $2\%$ as expected at the HL-LHC. 
The cyan and red points satisfy the benchmark sets within the 1-sigma uncertainty at the HL-LHC and ILC500 given in Eq.~(\ref{error}), respectively. 
For the panels shown in the second and the third columns, the vertical axis $\bar{m}_\Phi$ and $\zeta$ are respectively defined by 
$\bar{m}_\Phi \equiv  m_\Phi (1-M^2/m_\Phi^2)$ and $\zeta \equiv 1-M^2/m_\Phi^2$.
}
\label{scatter2}
%========================================
\end{figure}

Up to now, we have discussed the extraction of the inner parameters from the three experimental inputs; i.e., $\Delta\kappa_V^{}$, $\Delta\kappa_\tau$ and $\Delta\kappa_b$. 
In Fig.~\ref{scatter2}, we show how the extraction can be improved by adding information of $\kappa_\gamma$ in addition to the above three inputs. 
The panels shown in the first row 
are the same as those shown in the first row in Fig.~\ref{scatter1}, which are displayed in order to compare the results with $\kappa_\gamma$. 
The panels displayed in the second, third and fourth rows respectively show the allowed region for Set~A with the central value of $\kappa_\gamma$ of 
0.98, 1.00 and 1.02 within the 1-$\sigma$ uncertainty of $\pm 2\%$ as expected at the HL-LHC (see Table~\ref{Tab:Sensitivity}). 
Because the accuracy of the measurement of $\kappa_\gamma$ at the ILC500 is not better than that of the best value at the HL-LHC, $2\%$, 
we also use $2\%$ for the analysis at the ILC500. 
As we see Eq.~(\ref{hgg_x}), the $H^\pm$ loop contribution to the decay rate of the $h\to \gamma\gamma$ mode gives a different dependence of the non-decouplingness 
from that in $\Delta\hat{\kappa}_V^{}$ and $\Delta\hat{\kappa}_f$, which is not proportional to $\bar{m}_\Phi^{}$, but proportional to $\zeta$, 
so that the non-decouplingness $\zeta$ can be expected to be extracted more precisely depending on the measured value of $\kappa_\gamma$. 
In fact, we can observe that $\zeta$ is determined more precisely to be $0.5 \lesssim \zeta \lesssim 1.0$, 
$0.25 \lesssim \zeta \lesssim 1.1$ and $0.2 \lesssim \zeta \lesssim 0.5$ at the ILC500
for the cases with the central value of $\kappa_\gamma=0.98$, $\kappa_\gamma=1.00$ and $\kappa_\gamma=1.02$, respectively, as 
compared to the case without $\kappa_\gamma$ ($0.2 \lesssim \zeta \lesssim 1.2$). 
The determination of $\bar{m}_\Phi$ is also improved, because $\bar{m}_\Phi$ is given as a function of $\zeta$.  
We note that smaller values of $\zeta$ and $\bar{m}_\Phi^{}$  are favored in the case of the larger central value of $\kappa_\gamma$, because 
the $H^\pm$ loop effect gives a destructive contribution to the W boson loop contribution.

\begin{figure}[t]
%========================================
\centering
\includegraphics[width=4.3cm]{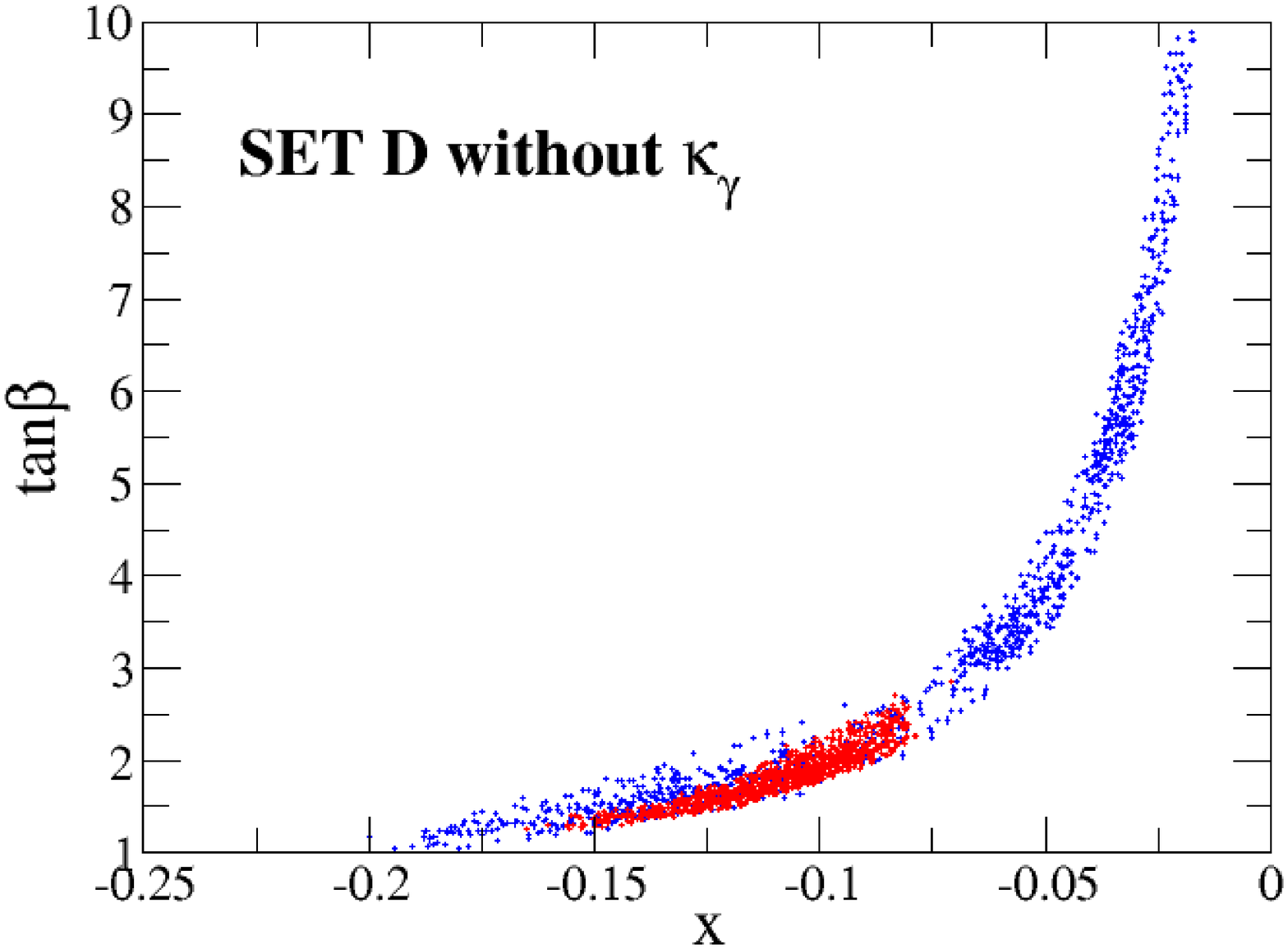}	\hspace{-5mm}
\includegraphics[width=4.3cm]{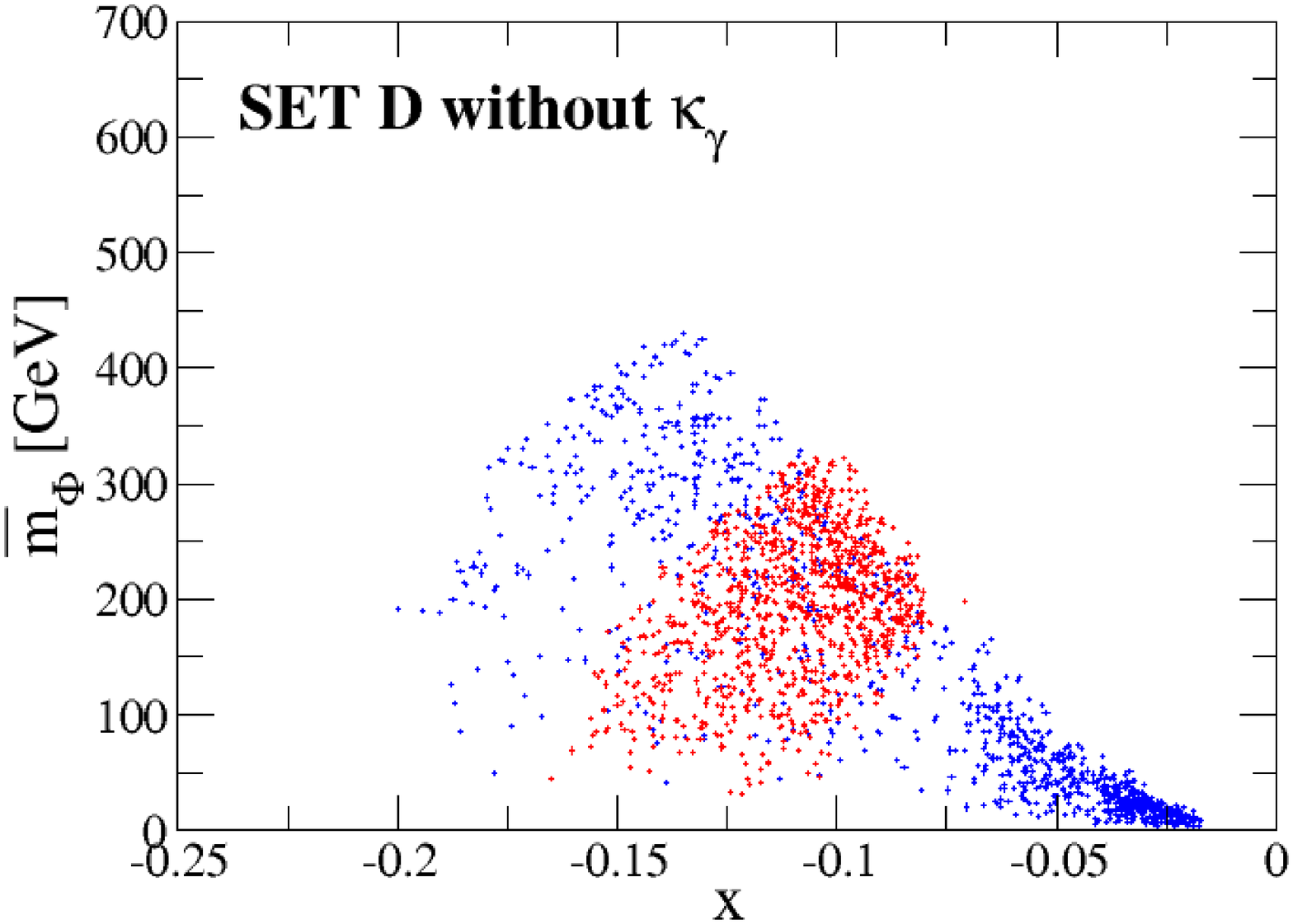}  \hspace{-5mm}
\includegraphics[width=4.3cm]{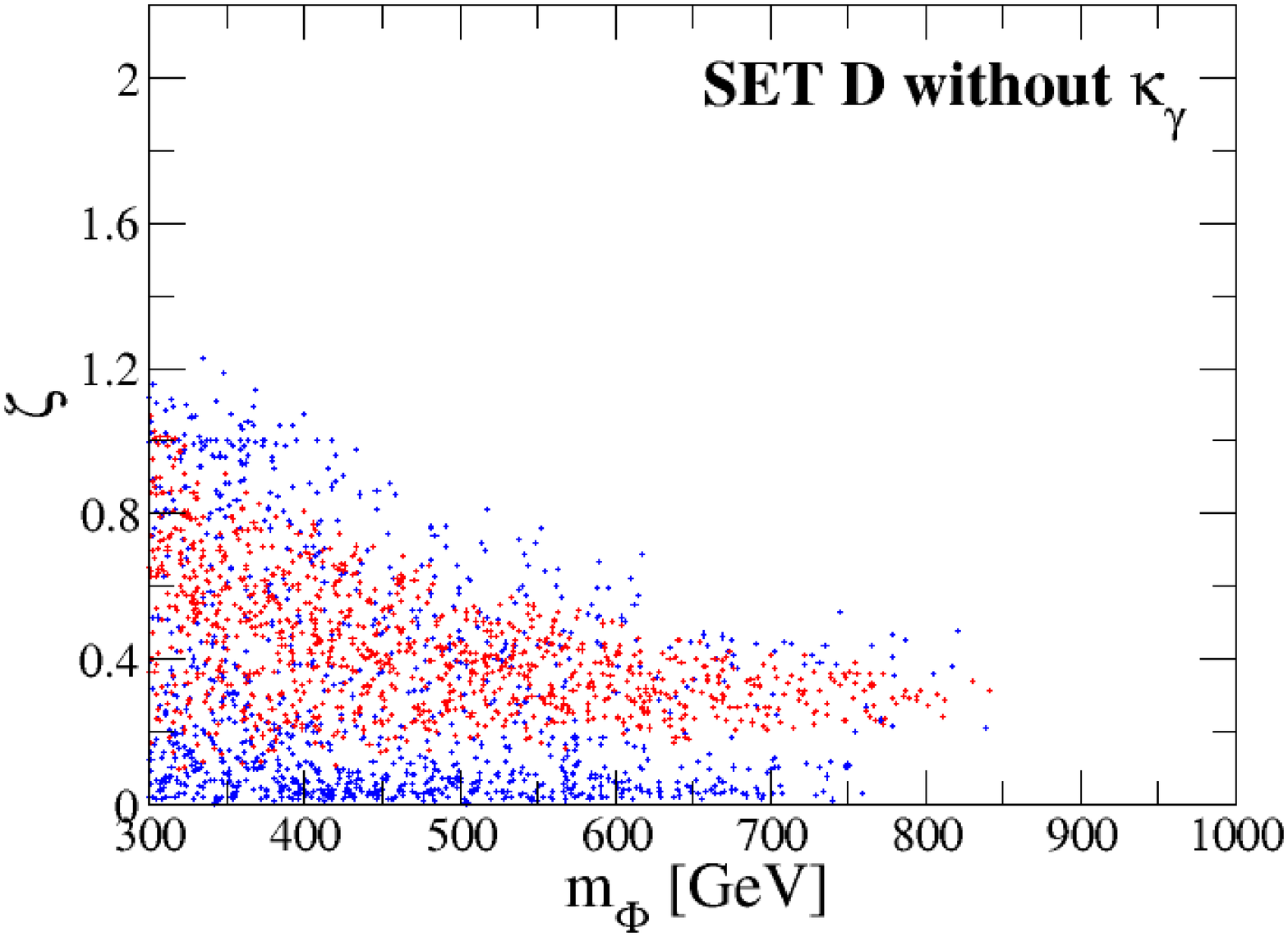}  \hspace{-5mm} 
\includegraphics[width=4.3cm]{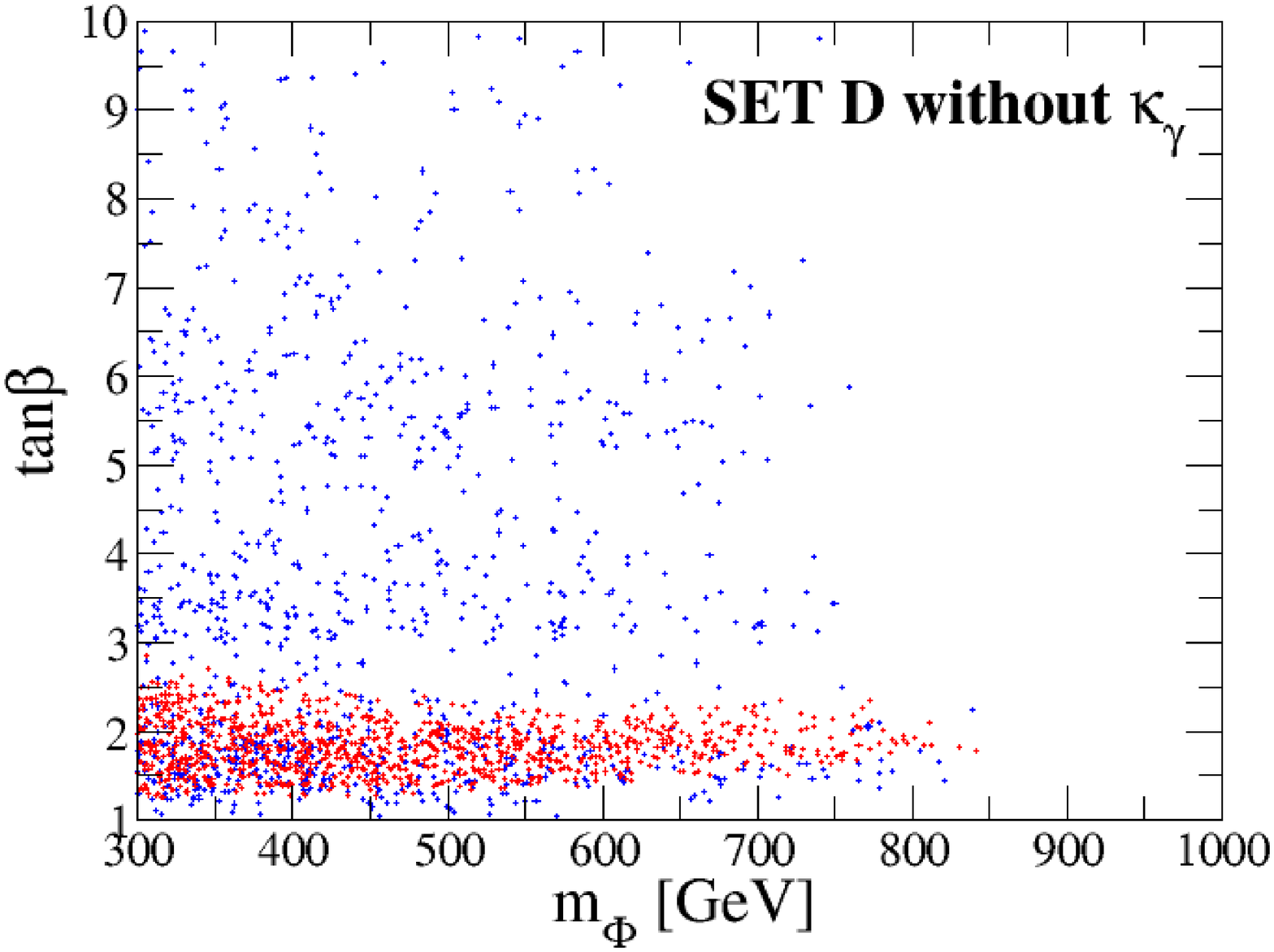}  \\ 
\includegraphics[width=4.3cm]{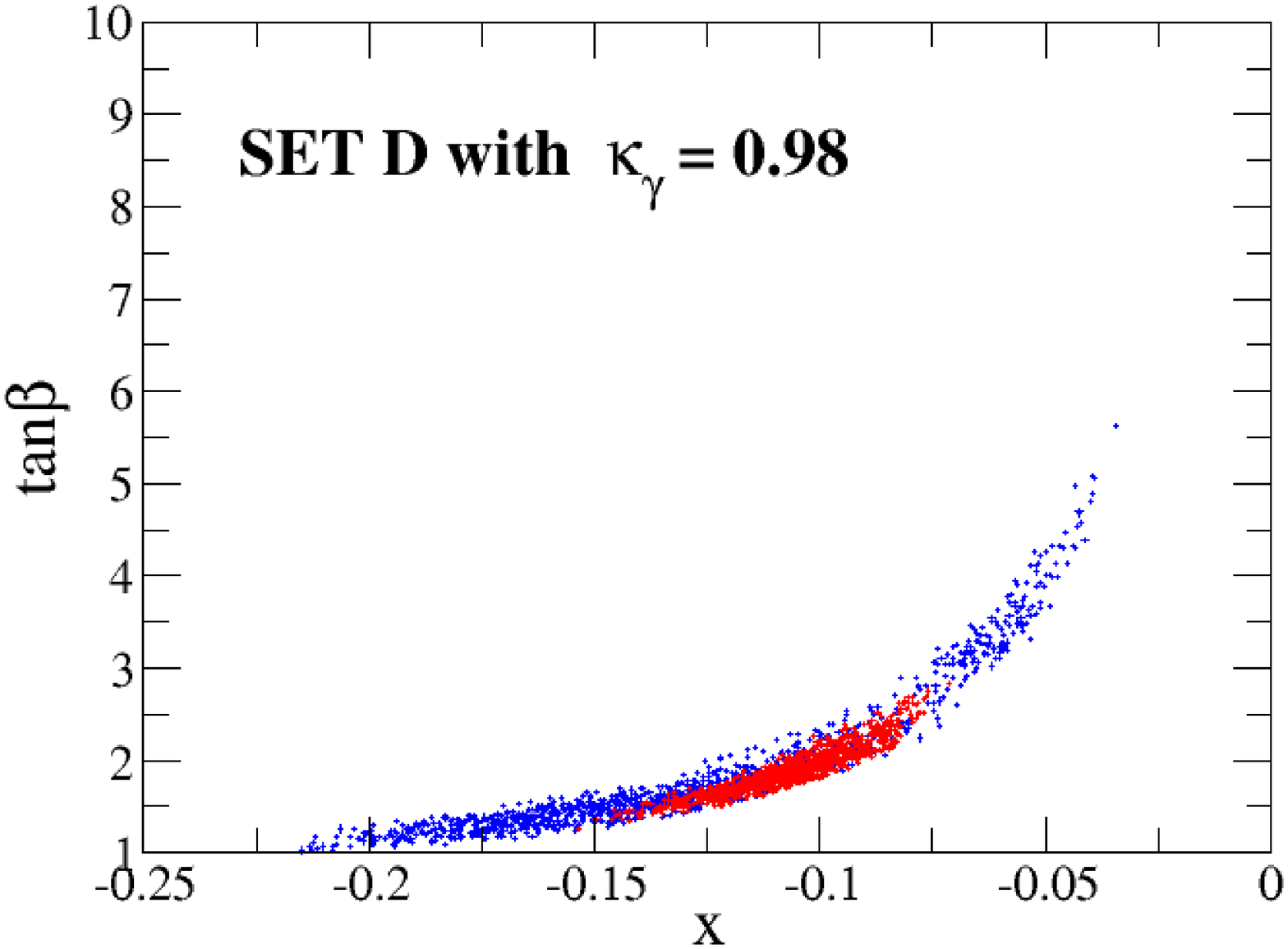}	 \hspace{-5mm}
\includegraphics[width=4.3cm]{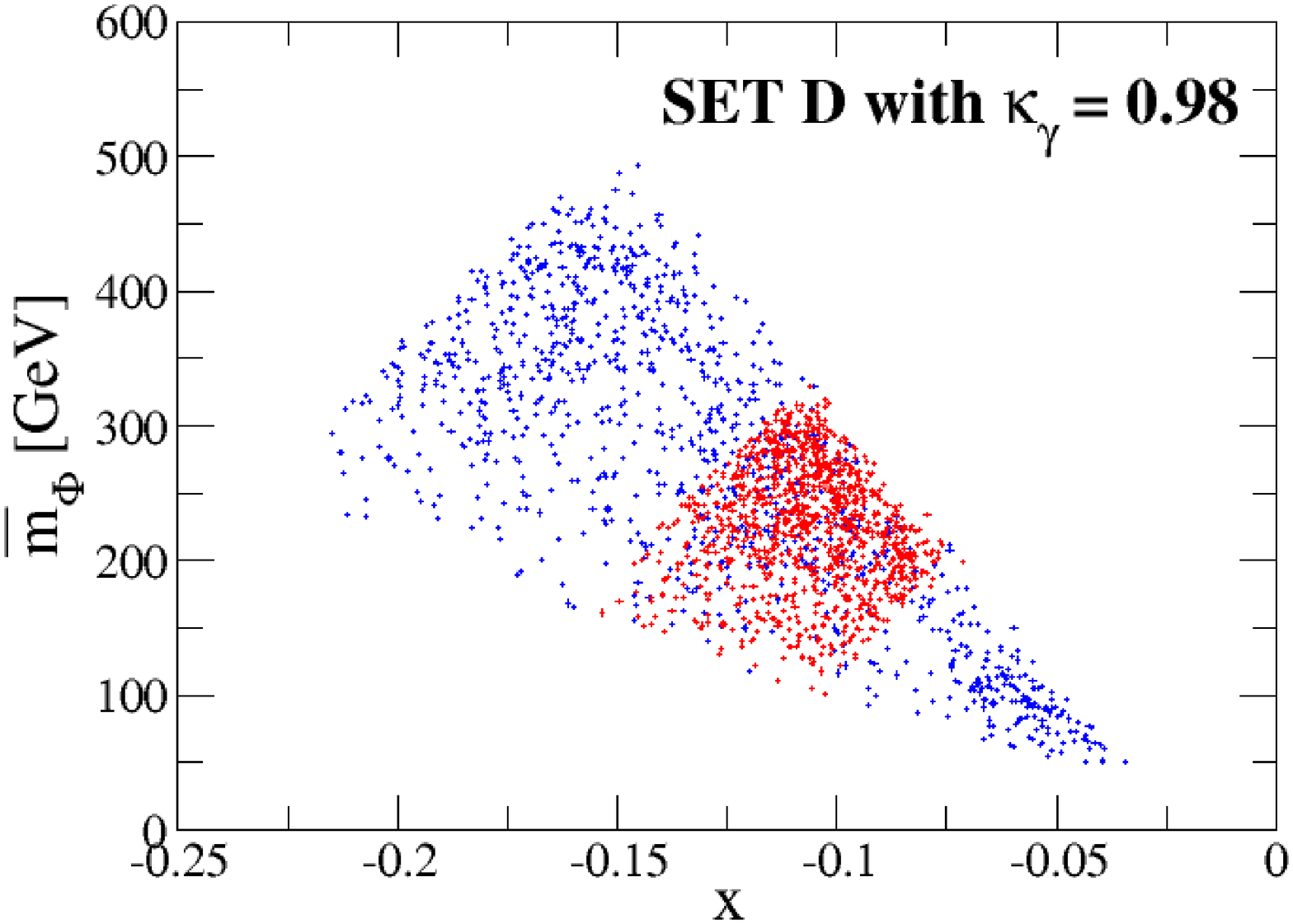}  \hspace{-5mm}
\includegraphics[width=4.3cm]{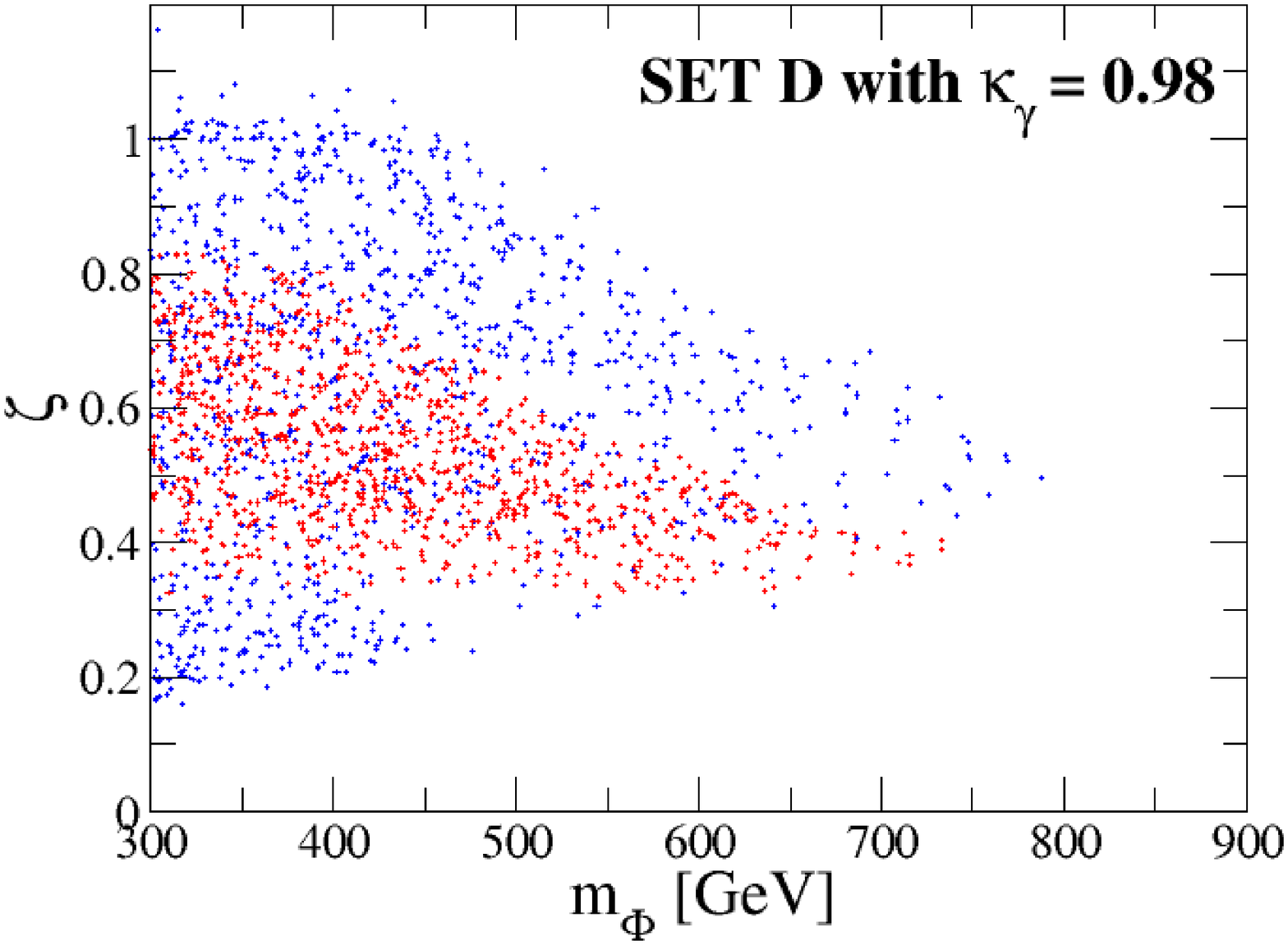}  \hspace{-5mm} 
\includegraphics[width=4.3cm]{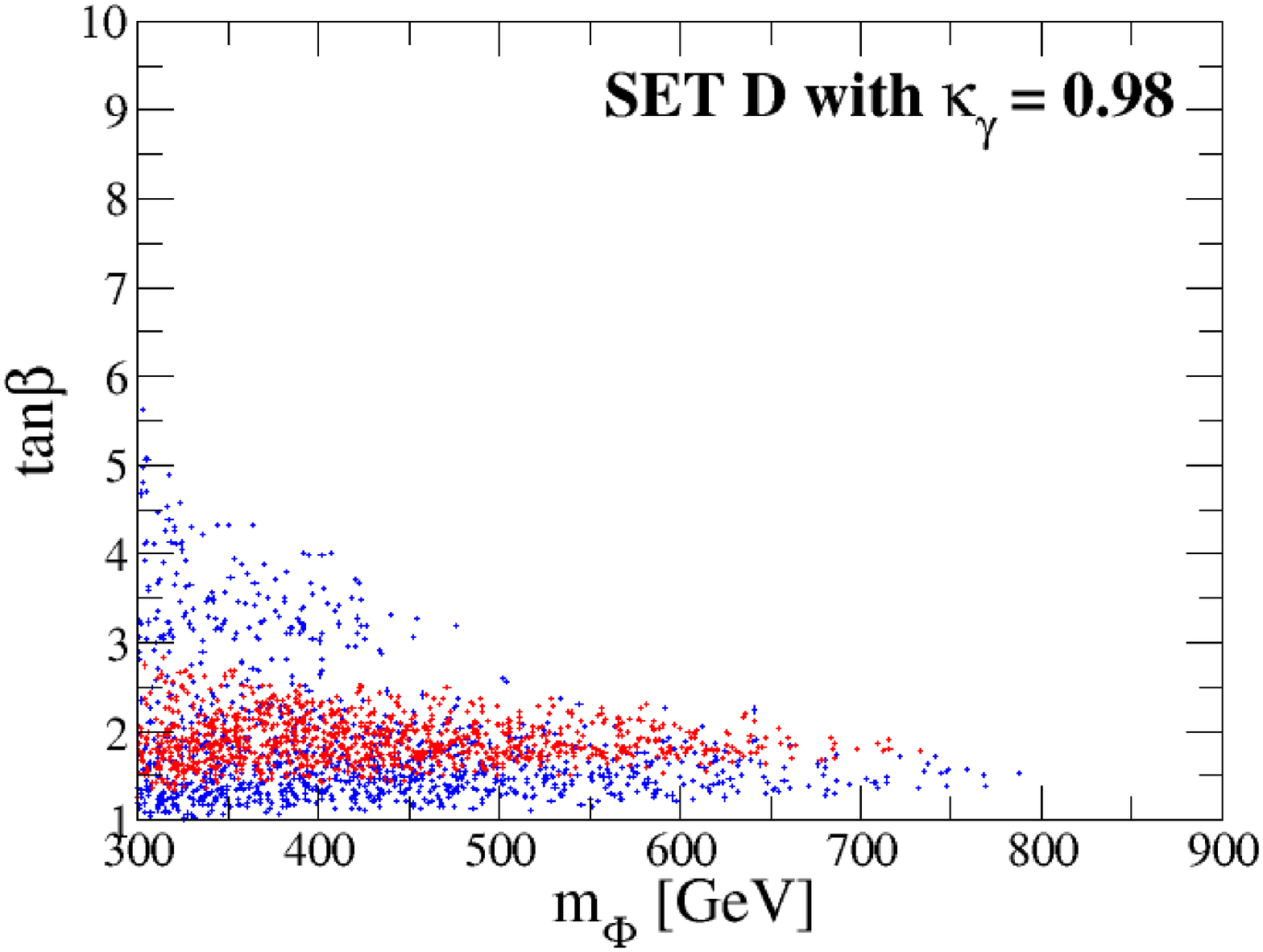}  \\ 
\includegraphics[width=4.3cm]{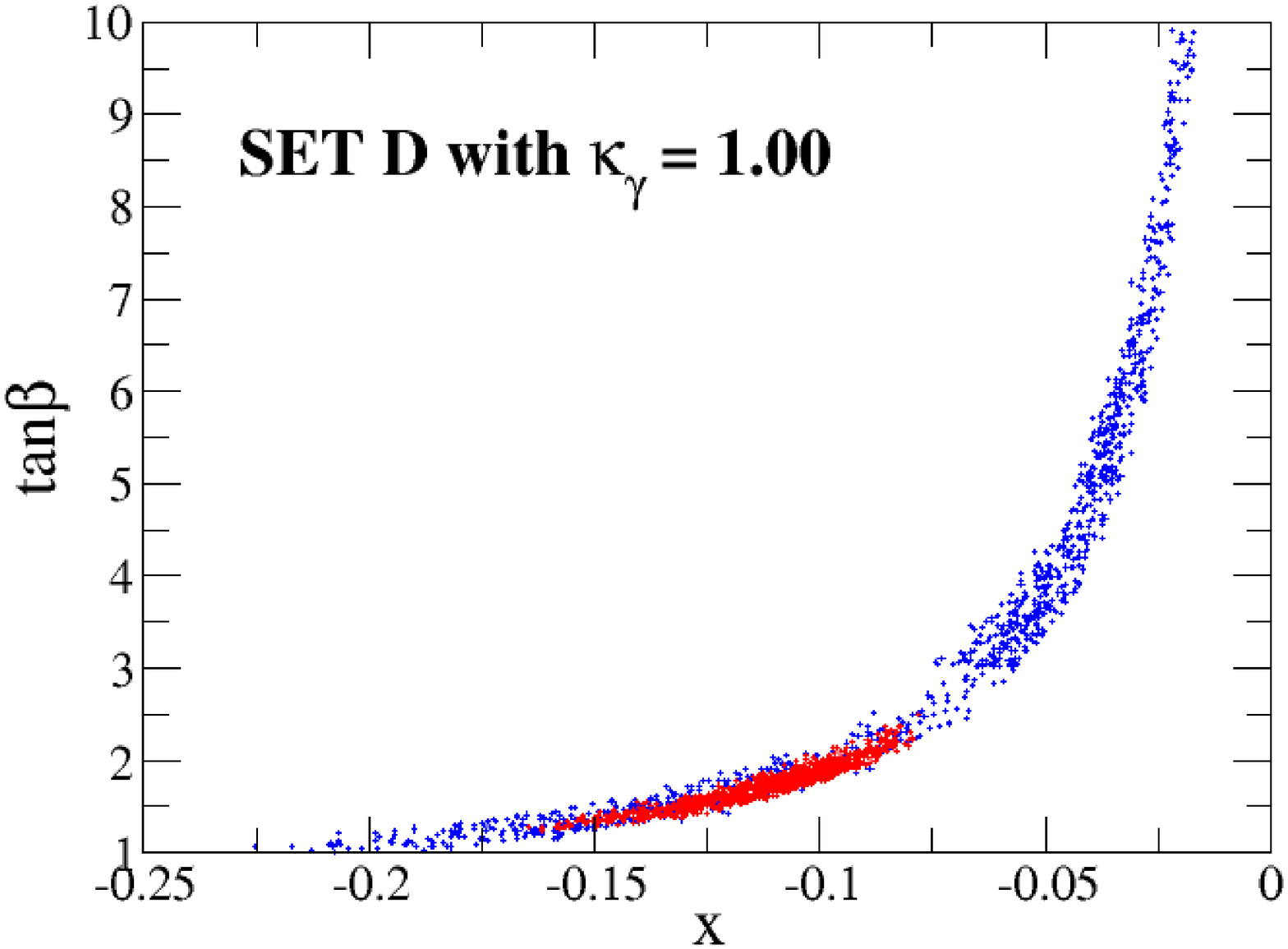}	 \hspace{-5mm}
\includegraphics[width=4.3cm]{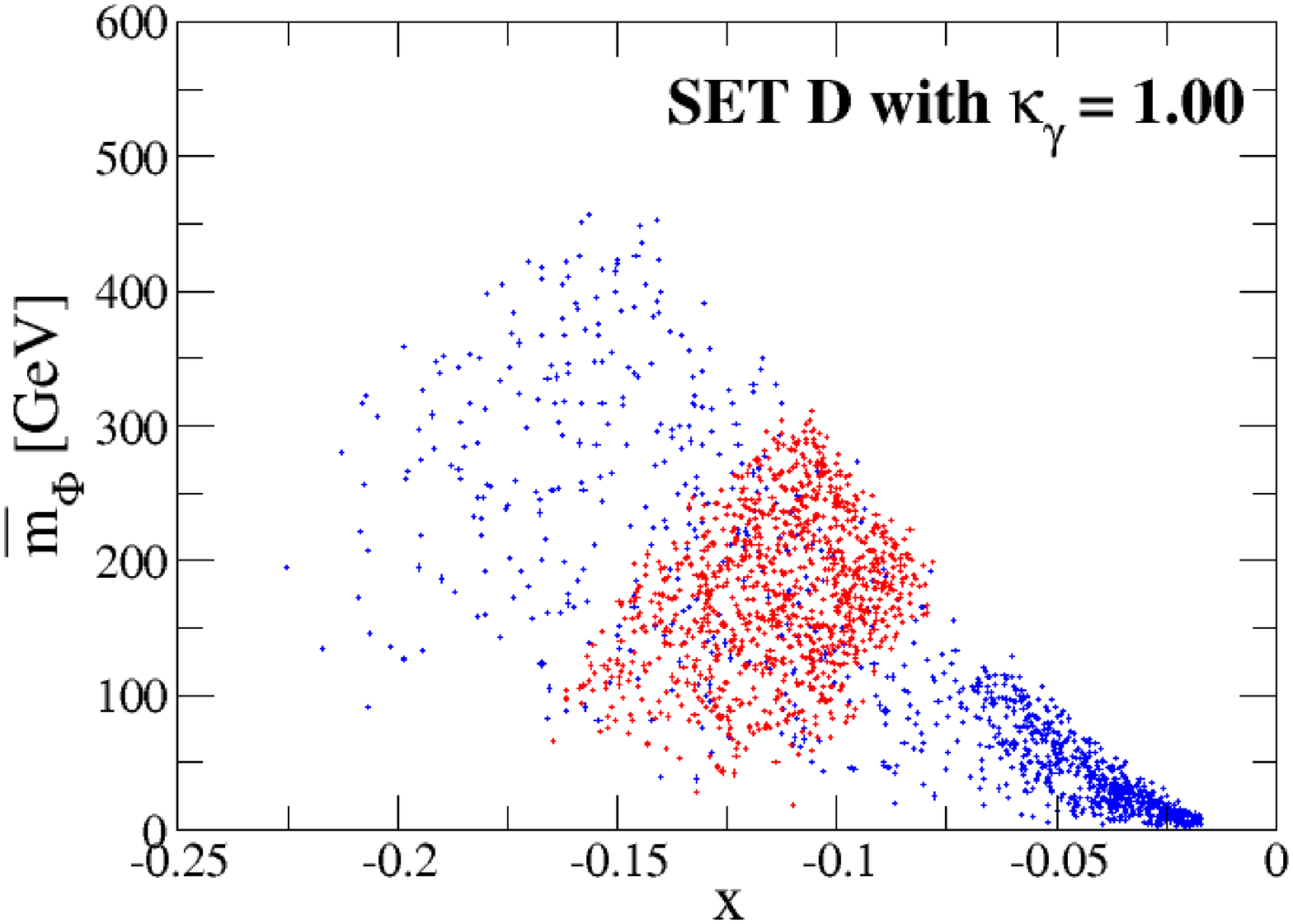}  \hspace{-5mm}
\includegraphics[width=4.3cm]{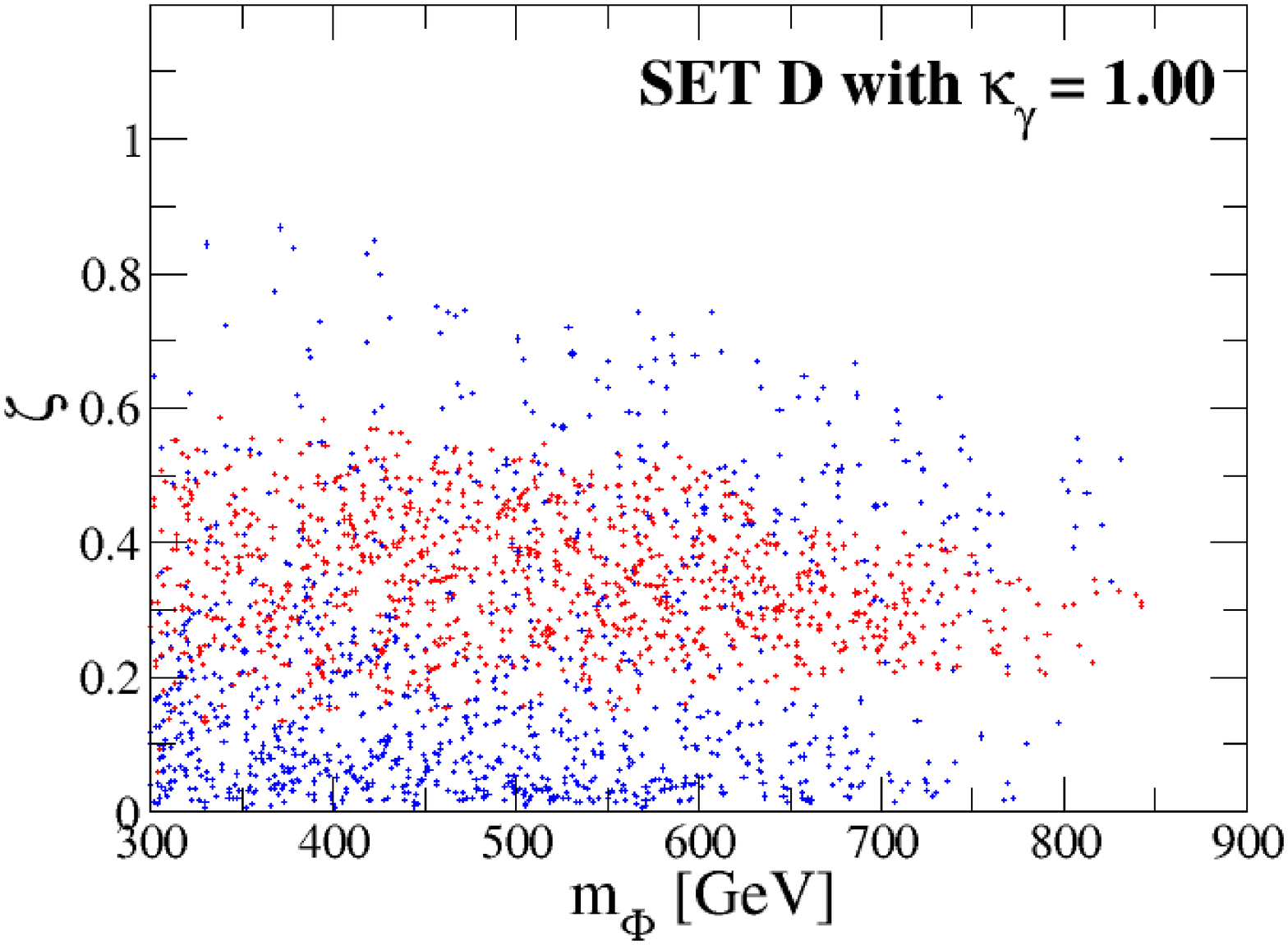}  \hspace{-5mm}
\includegraphics[width=4.3cm]{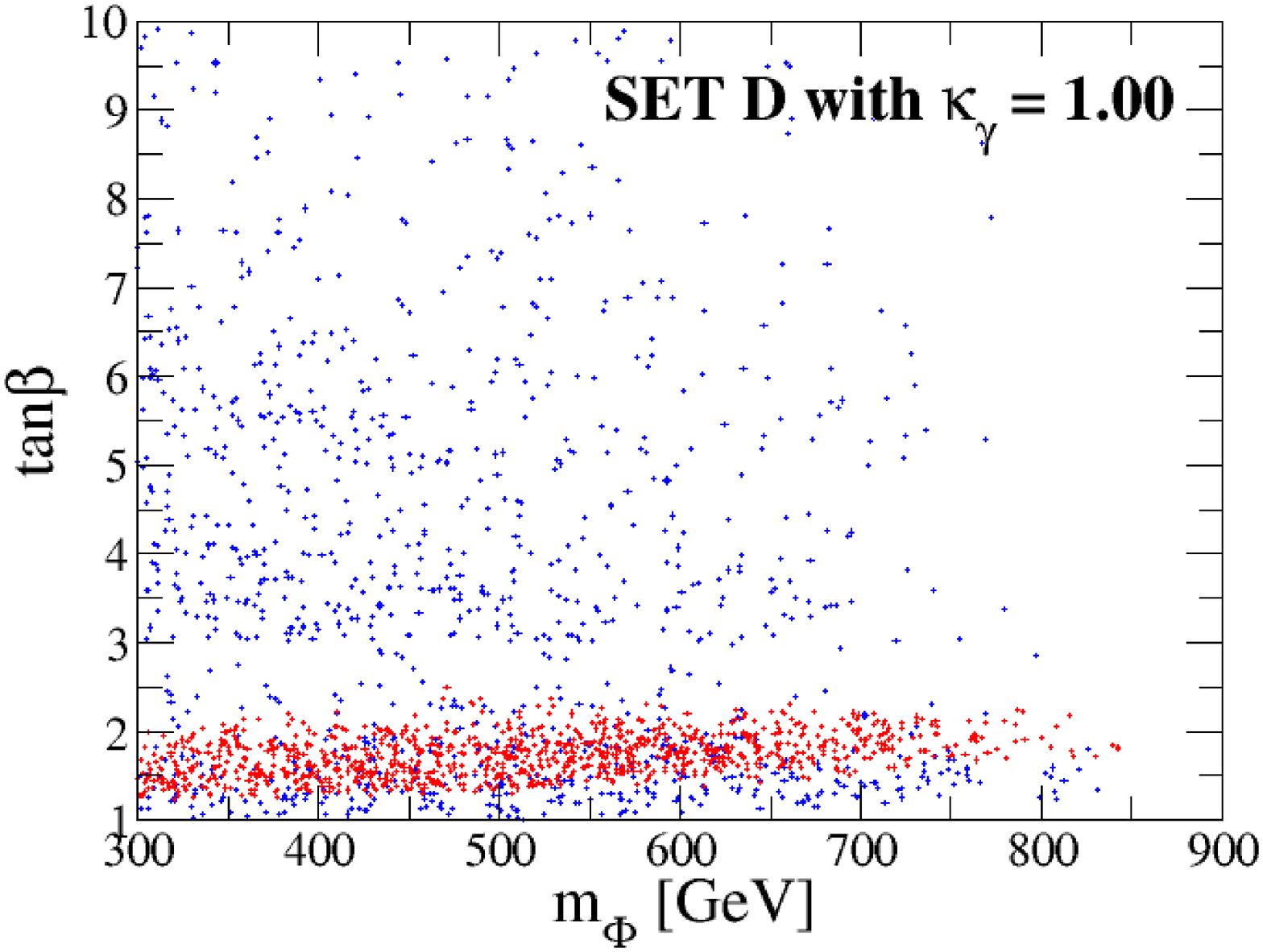} \\
\includegraphics[width=4.3cm]{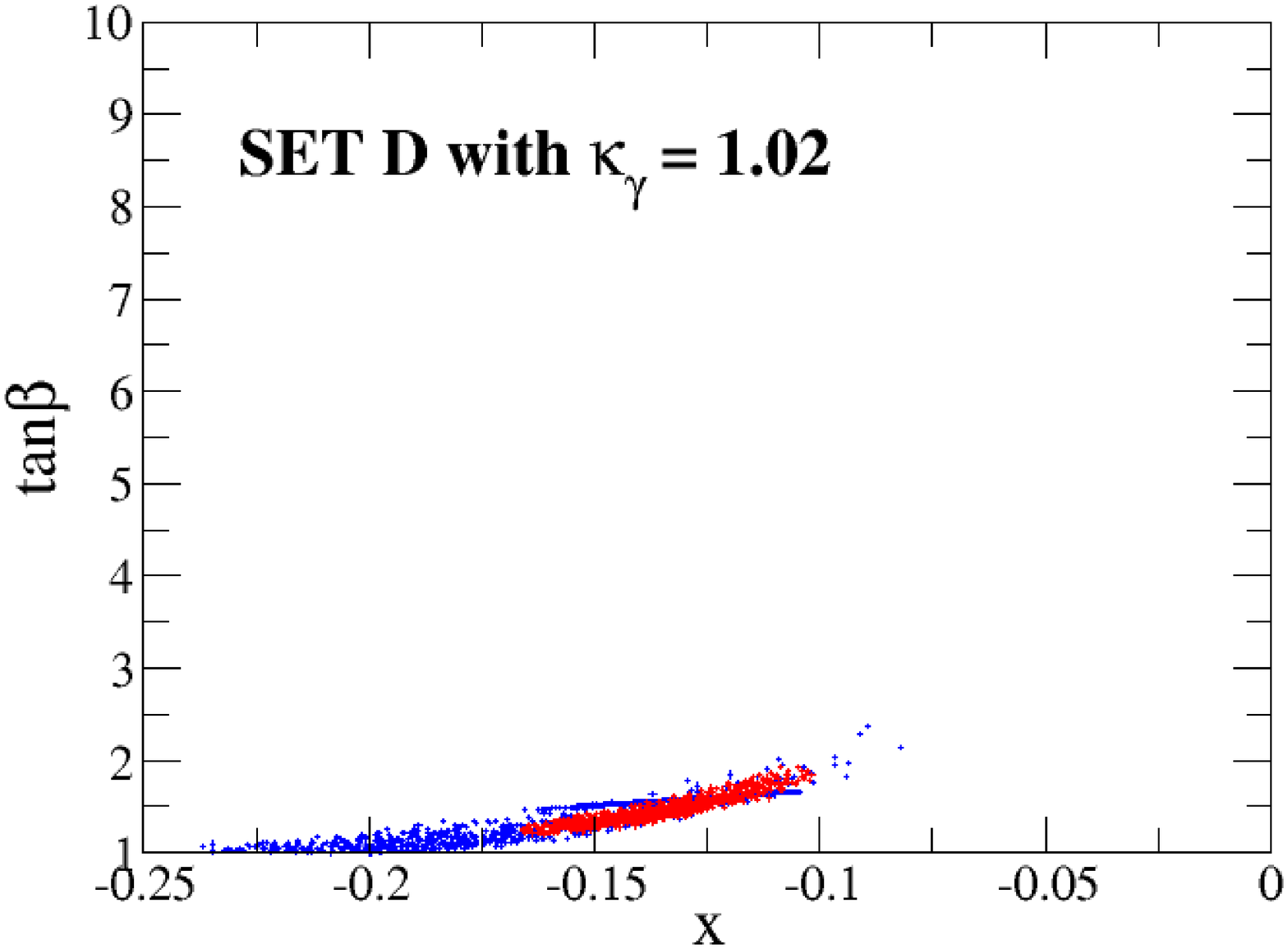}	\hspace{-5mm}
\includegraphics[width=4.3cm]{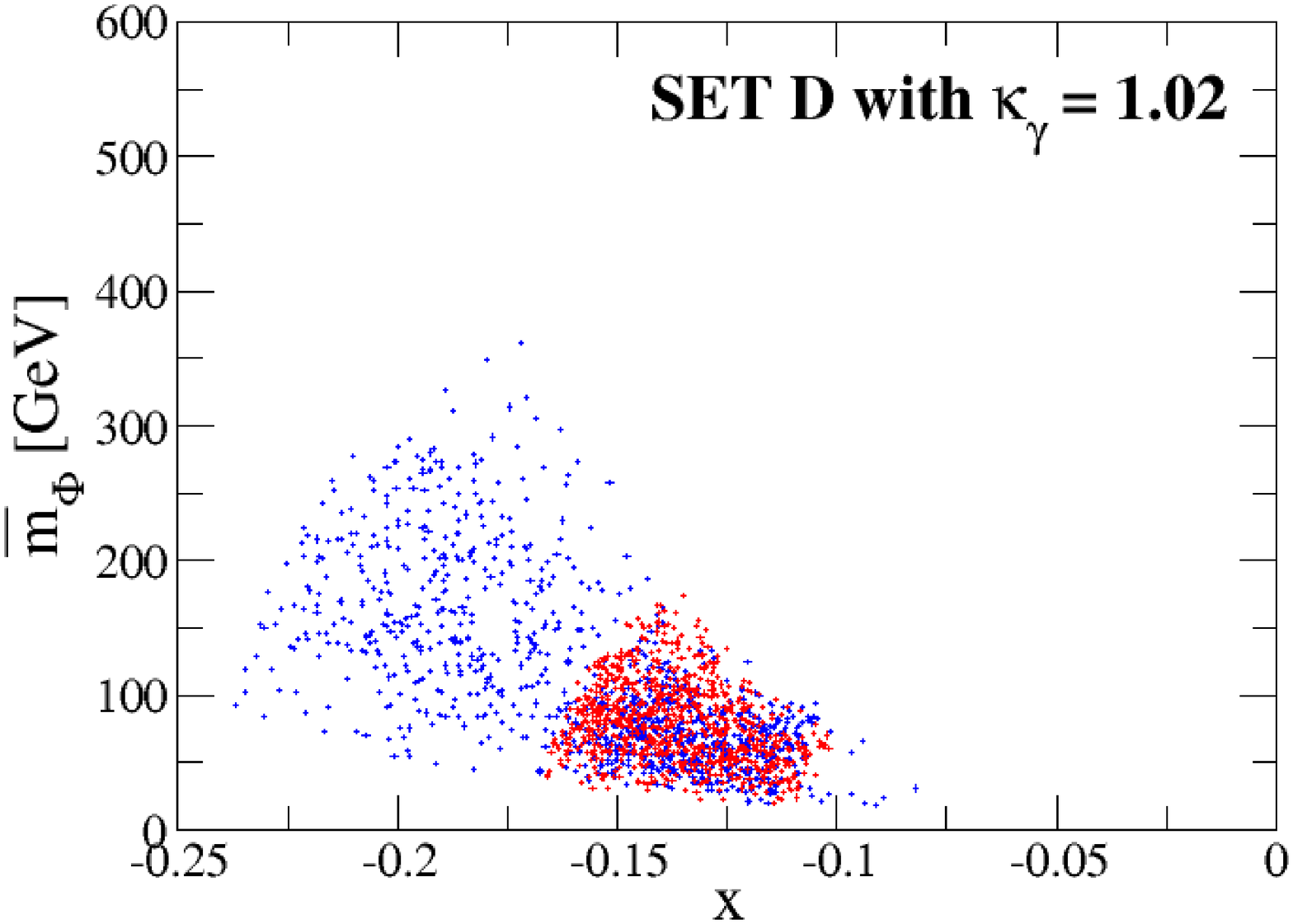}  \hspace{-5mm}
\includegraphics[width=4.3cm]{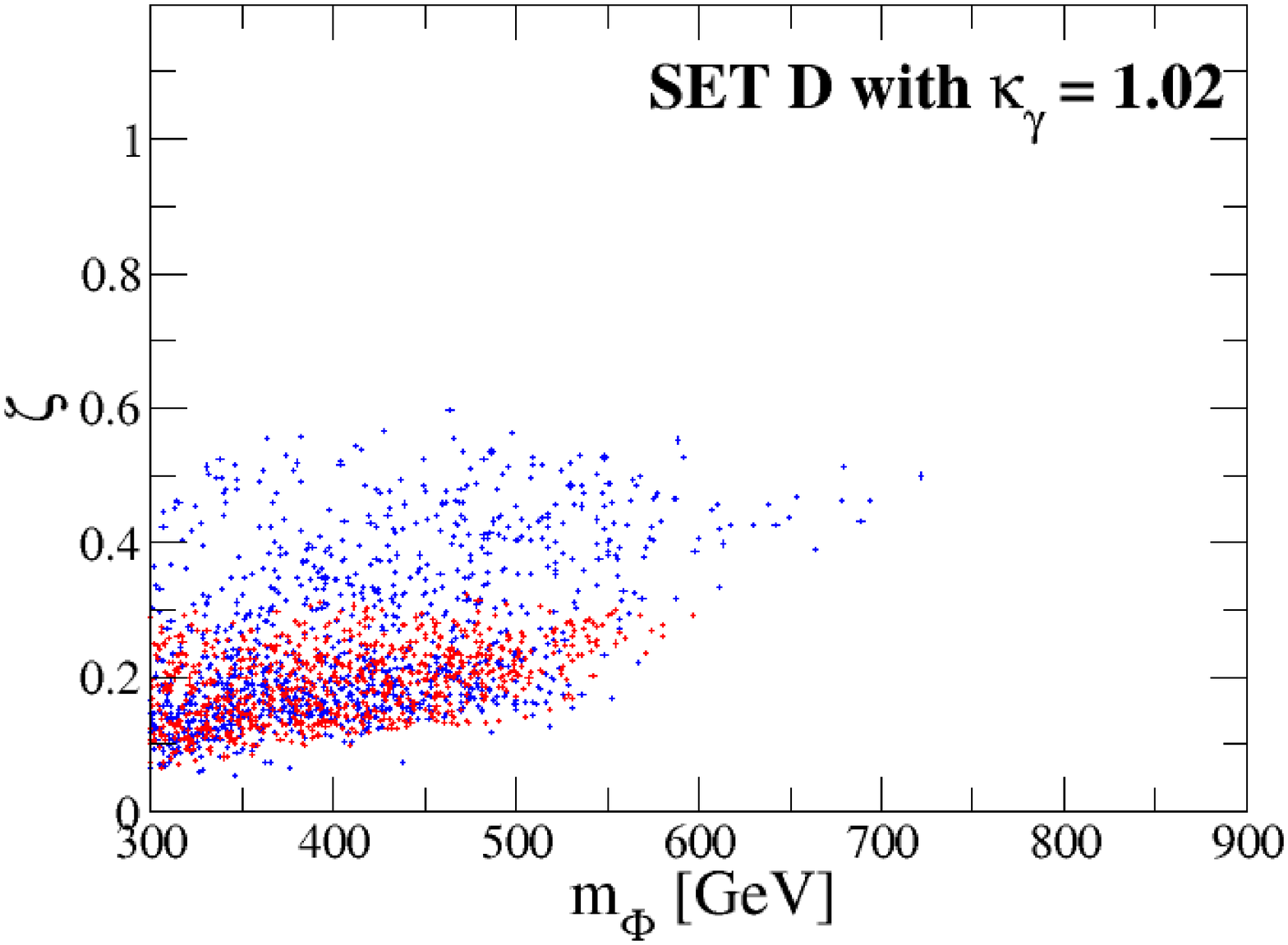}  \hspace{-5mm} 
\includegraphics[width=4.3cm]{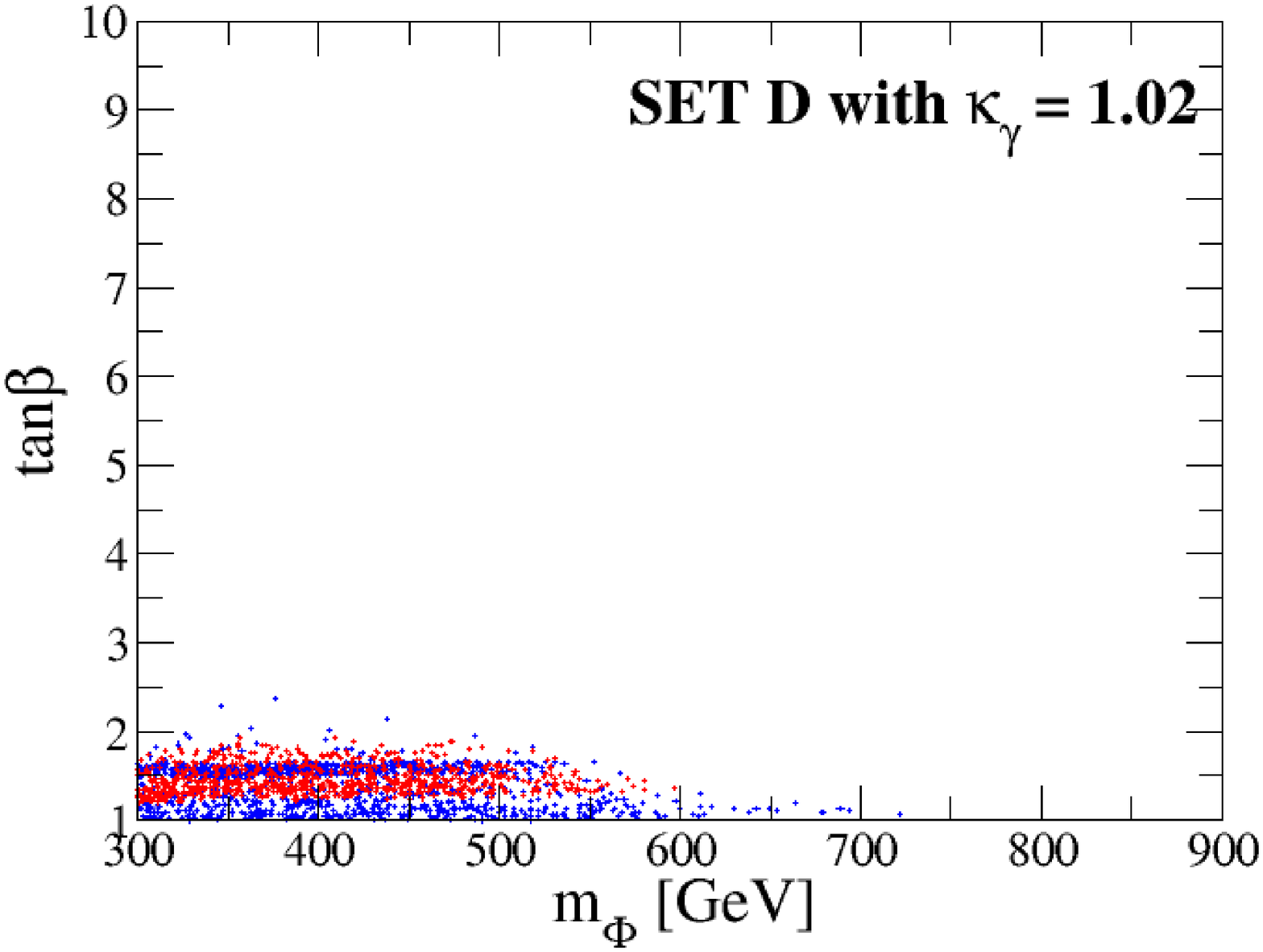} 
\caption{Scatter plots for Set D with the additional constraint from $\kappa_\gamma=0.98,~1.00$ and 1.02 for 
upper, center and bottom panels. 
The 1-$\sigma$ uncertainty of $\kappa_\gamma$ is assumed to be $2\%$ as expected at the HL-LHC. 
The cyan and red points satisfy the benchmark sets within the 1-sigma uncertainty at the HL-LHC and ILC500 given in Eq.~(\ref{error}), respectively. 
For the panels shown in the second and the third columns, the vertical axis $\bar{m}_\Phi$ and $\zeta$ are respectively defined by 
$\bar{m}_\Phi \equiv  m_\Phi (1-M^2/m_\Phi^2)$ and $\zeta \equiv 1-M^2/m_\Phi^2$.
}
\label{scatter3}
%========================================
\end{figure}

In Fig.~\ref{scatter3}, we also show the allowed parameter region with additional information of $\kappa_\gamma$ for Set D. 
Similar to the results in the previous figure, $\zeta$ and $\bar{m}_\Phi^{}$ are well extracted as compared to the case without $\kappa_\gamma$ displayed in the first 
row in Fig.~\ref{scatter3}. 
For example, $\zeta$ is determined to be $0.3\lesssim \zeta \lesssim 0.8$, $0.1\lesssim \zeta \lesssim 0.6$ and $0.1\lesssim \zeta \lesssim 0.6$ 
for the cases with the central value of $\kappa_\gamma=0.98$, $\kappa_\gamma=1.00$ and $\kappa_\gamma=1.02$, respectively.

\section{Discussions and conclusions}

We have calculated radiative corrections to a full set of 
coupling constants for the Higgs boson $h$ at the one-loop level in the THDMs with the four types of Yukawa interactions under the 
softly-broken discrete $Z_2$ symmetry.
These couplings are evaluated in the on-shell scheme, in which the gauge dependence in the mixing parameter which appears in the previous calculation 
is consistently avoided. 
We have shown the details of our one-loop calculations, and 
have presented the complete set of the analytic formulae of the renormalized couplings. 
We then have numerically demonstrated how the inner parameters of the THDM can be extracted by the future precision measurements of these couplings at the HL-LHC and 
the ILC. 

We have found that the inner parameters of the THDM can be determined 
to a considerable extent as long as  $\kappa_V^{}$ will be  measured with the deviation about 1\%. 
The extraction of the inner parameters using the ILC500 
is much better than that using the HL-LHC. 
That is mainly due to the good accuracy of the $hVV$ coupling measurement at the ILC500 whose uncertainty is expected to be less than 1\%. 
Although we have only demonstrated the results for Set A to Set E assuming the true Higgs sector is of the Type-II THDM, 
the similar analysis can be performed straightforwardly in the other types of THDM or the other extended Higgs sectors, and 
the extraction of inner parameters is expected to be attained as well in these models. 
Our study given in this paper shows that the numerical evaluation of the Higgs boson couplings at the one-loop level in extended Higgs sectors is 
essentially important to indirectly determine the structure of the Higgs sector by using the future precision data. 
In addition, it also shows that in addition to the HL-LHC where especially $h\gamma\gamma$ can be measured precisely  
future lepton colliders such as the ILC are absolutely necessary for our purpose of 
determining the structure of the Higgs sector from the measurement of the coupling constants of the discovered Higgs boson $h$. 

Although we have discussed fingerprinting by using $\kappa_V^{}$, $\kappa_\tau$, $\kappa_b$ and $\kappa_\gamma$, 
the information of $\kappa_c$, $\kappa_t$ and $\kappa_h$ is also important to determine the Higgs sector more deeply. 
In particular, the measurement of the top Yukawa coupling is important not only to determine the nature of the top quark, the heaviest matter particle, 
but also to test the new physics scenarios based on the composite models. 
The measurement of the $hhh$ coupling is essentially important not only to determine the nature of the Higgs potential but also 
to test, for instance, the new physics models with strongly first order phase transition. 
Although at the HL-LHC the cross section of the double Higgs production process is expected to be measured at a few times 10\%
it seems to be hopeless to extract the information of the $hhh$ coupling sufficiently accurately. 
On the other hand, at the ILC with $\sqrt{s}=1$ TeV the $hhh$ coupling can be measured with the 13\% accuracy~\cite{ILC_white,13per}, 
which is sufficient precision to test the strong first order phase transition which is required for successful electroweak baryogenesis.  

We conclude that the combination of the future data for all kinds of the couplings for the Higgs boson $h$
and their theory predictions with radiative corrections 
in various extended Higgs sectors is a promissing way to determine the structure of the Higgs sector and further 
to access new physics beyond the SM, even if a new particle was not directly discovered in the future experiments.\\

\noindent
$Acknowledgments$

S.K. was supported in part by Grant-in-Aid for Scientific Research, Japan
Society for the Promotion of Science (JSPS), Nos. 22244031 and 24340046, and The Ministry of Education, Culture, 
Sports, Science and Technology (MEXT), No. 23104006.
M.K. was supported in part by JSPS, No. 25$\cdot$10031.
K.Y. is supported by JSPS postdoctoral fellowships for research abroad.
%supported in part by the National Science Council of R.O.C. under Grant No. NSC-101-2811-M-008-014.

\begin{appendix}

\section{Higgs boson couplings}

From the Higgs kinetic term, we obtain the two types of the trilinear couplings; i.e., 
Gauge-Gauge-Scalar, Gauge-Scalar-Scalar, and quartic Gauge-Gauge-Scalar-Scalar type couplings. 
These couplings can be expressed as 
\begin{align}
\mathcal{L}=&+g_{\phi V_1V_2}^{}g^{\mu\nu}\phi V_{1\mu} V_{2\nu}  + g_{\phi_1\phi_2 V}(\partial^\mu\phi_1\phi_2-\phi_1\partial^\mu\phi_2) V_\mu+
g_{\phi_1\phi_2 V_1V_2}g^{\mu\nu}\phi_1\phi_2V_{1\mu} V_{2\nu} + \cdots. 
\end{align}
The coefficients $g_{\phi V_1 V_2}$, $g_{\phi_1\phi_2 V}$ and $g_{\phi_1\phi_2 V_1V_2}$ are listed in Table~\ref{FR_GS}, 
where we use $g_Z^{}=g/c_W^{}$ in this table and below.  
Throughout Appendix, we use the shortened notation of the mixing angles, $s_{\beta-\alpha}=\sin(\beta-\alpha)$ and $c_{\beta-\alpha}=\cos(\beta-\alpha)$. 

\begin{table}[h]
\begin{center}
{\renewcommand\arraystretch{1.2}
\begin{tabular}{c|c}\hline\hline
Vertices& $g_{\phi V_1 V_2}$ \\\hline
$hW^+_\mu W^-_\nu$ & $\frac{g^2}{2}vs_{\beta-\alpha}$\\\hline
$HW^+_\mu W^-_\nu$ & $\frac{g^2}{2}vc_{\beta-\alpha}$\\\hline
$hZ_\mu Z_\nu$ & $\frac{g_Z^2}{4}vs_{\beta-\alpha}$\\\hline
$HZ_\mu Z_\nu$ & $\frac{g_Z^2}{4}vc_{\beta-\alpha}$\\\hline
$G^\pm Z_\mu W_\nu^\mp$ & $-\frac{gg_Z}{2}vs_W^2$ \\\hline
$G^\pm A_\mu W_\nu^\mp$ & $\frac{eg}{2}v$ \\\hline\hline
\end{tabular}}
\caption{The Gauge-Gauge-Scalar vertices. }
\end{center}
\end{table}

\begin{table}[t]
{\renewcommand\arraystretch{1.2}
\begin{tabular}{c|c}\hline\hline
Vertices&$g_{\phi_1\phi_2 V}$ \\\hline
$hG^\pm W^\mp_\mu$ & $\mp i\frac{g}{2}s_{\beta-\alpha} $\\\hline
$HG^\pm W^\mp_\mu$ & $\mp i\frac{g}{2}c_{\beta-\alpha} $\\\hline
$G^0G^\pm W^\mp_\mu$ & $-\frac{g}{2}$\\\hline
$hH^\pm W^\mp_\mu$ & $\mp i\frac{g}{2}c_{\beta-\alpha} $\\\hline
$HH^\pm W^\mp_\mu$ & $\pm i\frac{g}{2}s_{\beta-\alpha} $\\\hline
$AH^\pm W^\mp_\mu$ & $-\frac{g}{2}$\\\hline
$G^+G^-Z_\mu$ & $i\frac{g_Z}{2}c_{2W}$\\\hline
$H^+H^-Z_\mu$ & $i\frac{g_Z}{2}c_{2W}$\\\hline
$hG^0Z_\mu$ & $-\frac{g_Z}{2}s_{\beta-\alpha}$\\\hline
$hAZ_\mu$ & $-\frac{g_Z}{2}c_{\beta-\alpha}$\\\hline
$HG^0Z_\mu$ & $-\frac{g_Z}{2}c_{\beta-\alpha}$\\\hline
$HAZ_\mu$ & $\frac{g_Z}{2}s_{\beta-\alpha}$\\\hline
$G^+G^-A_\mu$ & $ie$\\\hline
$H^+H^-A_\mu$ & $ie$\\\hline\hline
\end{tabular}\hspace{10mm}
\begin{tabular}{l|l|l|l}\hline\hline
Vertices&$g_{\phi_1\phi_2 V_1V_2}$&Vertices&$g_{\phi_1\phi_2 V_1V_2}$ \\\hline
$hhW^+_\mu W^-_\nu$ & $\frac{g^2}{4}$&$G^\pm G^0 W_\mu^\mp Z_\nu$&$\pm i\frac{gg_Z}{2}s_W^2$\\\hline
$HHW^+_\mu W^-_\nu$ & $\frac{g^2}{4}$&$H^\pm A W_\mu^\mp Z_\nu$&$\pm i\frac{gg_Z}{2}s_W^2$\\\hline
$AAW^+_\mu W^-_\nu$ & $\frac{g^2}{4}$&$G^\pm H W_\mu^\mp Z_\nu$&$-\frac{gg_Z}{2}s_W^2c_{\beta-\alpha}$\\\hline
$G^0G^0W^+_\mu W^-_\nu$ & $\frac{g^2}{4}$&$H^\pm h W_\mu^\mp  Z_\nu$&$-\frac{gg_Z}{2}s_W^2c_{\beta-\alpha}$\\\hline
$G^+G^- W^+_\mu W^-_\nu$ & $\frac{g^2}{2}$&$G^\pm h W_\mu^\mp Z_\nu$&$-\frac{gg_Z}{2}s_W^2s_{\beta-\alpha}$\\\hline
$H^+H^- W^+_\mu W^-_\nu$ & $\frac{g^2}{2}$&$H^\pm H W_\mu^\mp Z_\nu$&$\frac{gg_Z}{2}s_W^2s_{\beta-\alpha}$\\\hline
$hh Z_\mu Z_\nu$ & $\frac{g_Z^2}{8}$&$H^\pm A W_\mu^\mp A_\nu$  &$\mp \frac{eg}{2}$\\\hline
$HH Z_\mu Z_\nu$ & $\frac{g_Z^2}{8}$&$G^\pm G^0 W_\mu^\mp A_\nu$ &$\mp \frac{eg}{2}$\\\hline
$AA Z_\mu Z_\nu$ & $\frac{g_Z^2}{8}$&$H^\pm h W_\mu^\mp A_\nu$  &$\frac{eg}{2}c_{\beta-\alpha}$\\\hline
$G^0G^0 Z_\mu Z_\nu$ & $\frac{g_Z^2}{8}$&$G^\pm H W_\mu^\mp  A_\nu$  &$\frac{eg}{2}c_{\beta-\alpha}$\\\hline
$G^+G^- Z_\mu Z_\nu$ & $\frac{g_Z^2}{4}c_{2W}^2$&$G^+G^- A_\mu Z_\nu$ & $eg_Zc_{2W}$ \\\hline
$H^+H^- Z_\mu Z_\nu$ & $\frac{g_Z^2}{4}c_{2W}^2$&$H^+H^- A_\mu Z_\nu$ & $eg_Zc_{2W}$ \\\hline
$G^+G^- A_\mu A_\nu$ & $e^2$&$G^\pm h W_\mu^\mp  A_\nu$ & $\frac{eg}{2}s_{\beta-\alpha}$\\\hline 
$H^+H^- A_\mu A_\nu$ & $e^2$&$H^\pm H W_\mu^\mp  A_\nu$  & $-\frac{eg}{2}s_{\beta-\alpha}$\\\hline\hline
\end{tabular}}
\caption{The Scalar-Scalar-Gauge and Scalar-Scalar-Gauge-Gauge type vertices and those coefficients}
\label{FR_GS}
\end{table}

From the Higgs potential, we obtain the scalar trilinear and the scalar quartic couplings. 
When we use the following notation for these couplings
\begin{align}
\mathcal{L}=+\lambda_{\phi_i\phi_j\phi_k}\phi_i\phi_j\phi_k
+\lambda_{\phi_i\phi_j\phi_k\phi_l}\phi_i\phi_j\phi_k\phi_l + \cdots. 
\end{align}
These coefficients are given by
\begin{align}
\lambda_{H^+H^-h }&=\frac{1}{v}\left[(2M^2-2m_{H^\pm}^2-m_h^2)s_{\beta-\alpha}+2(M^2-m_h^2)\cot2\beta c_{\beta-\alpha}  \right],\\
\lambda_{AAh}&=\frac{1}{2v}\left[(2M^2-2m_A^2-m_h^2)s_{\beta-\alpha}+2(M^2-m_h^2)\cot2\beta c_{\beta-\alpha}  \right],\\
\lambda_{HHh}&=\frac{s_{\beta-\alpha}}{2v}\Big[(2M^2-2m_H^2-m_h^2)s_{\beta-\alpha}^2+2(3M^2-2m_H^2-m_h^2)\cot2\beta s_{\beta-\alpha}c_{\beta-\alpha}\notag\\
&\quad\quad \quad\quad  -(4M^2-2m_H^2-m_h^2)c_{\beta-\alpha}^2 \Big],\\
\lambda_{hhh}
&=-\frac{m_h^2}{2v}s_{\beta-\alpha} +\frac{M^2-m_h^2}{v}s_{\beta-\alpha}c_{\beta-\alpha}^2+\frac{M^2-m_h^2}{2v}c_{\beta-\alpha}^3(\cot\beta-\tan\beta),\label{lam_hhh}\\
\lambda_{GGh}&= -\frac{m_h^2}{2v}s_{\beta-\alpha}, \\
\lambda_{H^\pm G^\mp h}&= -\frac{1}{v}(m_h^2-m_{H^\pm}^2)c_{\beta-\alpha}, \\
\lambda_{AGh}&= -\frac{1}{v}(m_h^2-m_A^2)c_{\beta-\alpha}, \\
\lambda_{H^+H^-H}&=
-\frac{1}{v}\Big[2(M^2-m_H^2)\cot2\beta s_{\beta-\alpha}+(2m_{H^\pm}^2+m_H^2-2M^2)c_{\beta-\alpha}\Big],\\
\lambda_{AAH}&=
-\frac{1}{2v}\Big[2(M^2-m_H^2)\cot2\beta s_{\beta-\alpha}+(2m_A^2+m_H^2-2M^2)c_{\beta-\alpha}\Big],\\
\lambda_{HHH}&=
-\frac{1}{2v}\Big[2(M^2-m_H^2)\cot2\beta s_{\beta-\alpha}^3-2(M^2-m_H^2)c_{\beta-\alpha}s_{\beta-\alpha}^2+m_H^2c_{\beta-\alpha}\Big],\\
\lambda_{GGH}&= -\frac{m_H^2}{2v}c_{\beta-\alpha}, \\
\lambda_{H^\pm G^\mp H}&= \frac{1}{v}(m_H^2-m_{H^\pm}^2)s_{\beta-\alpha}, \\
\lambda_{AGH}&= \frac{1}{v}(m_H^2-m_A^2)s_{\beta-\alpha}, \\
\lambda_{Hhh}&=-\frac{c_{\beta-\alpha}}{2v\sin2\beta}\left[(2m_h^2+m_H^2-3M^2)\sin2\alpha +M^2\sin2\beta \right], \\  \label{lam_bhhh}
\lambda_{H^\pm G^\mp A}&= \pm \frac{i}{v}(m_A^2-m_{H^\pm}^2). 
\end{align}
The four point couplings are given by
\begin{align}
\lambda_{H^+H^-AG} &= - \frac{1}{v}(\lambda_{H^+H^-H}s_{\beta-\alpha}-\lambda_{H^+H^-h}c_{\beta-\alpha}), \\
\lambda_{G^+G^-AG} &= - \frac{1}{v}(\lambda_{G^+G^-H}s_{\beta-\alpha}-\lambda_{G^+G^-h}c_{\beta-\alpha}), \\
\lambda_{AAAG} &= - \frac{1}{v}(\lambda_{AAH}s_{\beta-\alpha}-\lambda_{AAh}c_{\beta-\alpha}), \\
\lambda_{AGGG} &= - \frac{1}{v}(\lambda_{GGH}s_{\beta-\alpha}-\lambda_{GGh}c_{\beta-\alpha}). 
\end{align}

%%%%%%%%%%%%%%%%%%%%%%%%%%%%%%%%%%%%%%%%%%%%%%%%%%%%%%%%%%%

\section{Loop Functions}

The Passarino-Veltman functions~\cite{Ref:PV} are quite useful to systematically express the one-loop functions. 
First, we define $A$, $B$ and $C$ functions: 
\begin{align}
\frac{i}{16\pi^2}A(m_1)&=\mu^{4-D}\int\frac{d^Dk}{(2\pi)^D}\frac{1}{N_1},\\
\frac{i}{16\pi^2}[B_0,B^\mu,B^{\mu\nu}](p_1^2;m_1,m_2)&=\mu^{4-D}\int\frac{d^Dk}{(2\pi)^D}\frac{[1,k^\mu,k^\mu k^\nu]}{N_1N_2},\\
\frac{i}{16\pi^2}[C_0,C^\mu,C^{\mu\nu}](p_1^2,p_2^2,(p_1+p_2)^2;m_1,m_2,m_3)&=\mu^{4-D}\int\frac{d^Dk}{(2\pi)^D}\frac{[1,k^\mu,k^\mu k^\nu]}{N_1N_2N_3}, 
\end{align}
where $D=4-2\epsilon$, and $\mu$ is a dimensionful parameter to keep the mass dimension four in the $k$-integral. 
The propagators are defined by 
\begin{align}
&N_1=k^2-m_1^2+i\varepsilon,\quad N_2=(k+p_1)^2-m_2^2+i\varepsilon,\quad N_3=(k+p_1+p_2)^2-m_3^2+i\varepsilon. 
\end{align}
The vector and the tensor functions for $B$ and $C$ are
expressed in terms of the following scalar functions:
\begin{align}
B^\mu&=p_1^\mu B_1,\label{b1}\\
B^{\mu\nu}&=p_1^\mu p_1^\nu B_{21}+g^{\mu\nu}B_{22},\label{b2} \\
C^\mu &=p_1^\mu C_{11}+p_2^\mu C_{12},\\
C^{\mu\nu} &=p_1^\mu p_1^\nu C_{21}+p_2^\mu p_2^\nu C_{22}
+(p_1^\mu p_2^\nu+p_1^\nu p_2^\mu) C_{23}+g^{\mu\nu}C_{24}.    \label{cmunu} 
\end{align} 
By counting the mass demension of the above functions, we can find that the divergent part is contained in 
$A$, $B_0$, $B_1$, $B_{21}$, $B_{22}$ and $C_{24}$. 
All the scalar functions are expressed by the divergent part and finite part as 
\begin{align}
A(m)& = m^2\left(\Delta +1-\ln m^2\right), \\
B_0&=\Delta -\int_0^1 dx \ln \Delta_B, \\
B_1 &=-\frac{\Delta}{2}+\int_0^1 dx (1-x)\ln \Delta_B, \\
B_{21}&=\frac{\Delta}{3} - \int_0^1 dx (1-x)^2 \ln\Delta_B, \\
B_{22}&=
\frac{1}{4}\left(m_1^2+m_2^2-\frac{p^2}{3}\right)\Delta
+\frac{1}{4}\left(m_1^2+m_2^2-\frac{p^2}{3}\right)
-\frac{1}{2}\int_0^1 dx \Delta_B\ln \Delta_B, \\
%%%%%%%%%
C_0&=-\int_0^1 dx \int_0^1dy\frac{y}{\Delta_C},\\ 
C_{11}&=-\int_0^1 dx \int_0^1dy\frac{y(xy-1)}{\Delta_C},\\
C_{12}&=-\int_0^1 dx \int_0^1dy\frac{y(y-1)}{\Delta_C}, \\
C_{21}&=-\int_0^1 dx \int_0^1dy\frac{y(1-xy)^2}{\Delta_C},\\
C_{22}&=-\int_0^1 dx \int_0^1dy\frac{y(1-y)^2}{\Delta_C},\\
C_{23}&=-\int_0^1 dx \int_0^1dy\frac{y(1-xy)(1-y)}{\Delta_C}, \\ 
C_{24}&=\frac{\Delta}{4} -\frac{1}{2}\int_0^1 dx \int_0^1dy \,y\ln \Delta_C, 
\end{align}
where
\begin{align}
\Delta_B  &= -x(1-x)p^2 + xm_1^2 + (1-x)m_2^2, \\
\Delta_C &= y^2(p_1x+p_2)^2+y[x(p_2^2-q^2+m_1^2-m_2^2)+m_2^2-m_3^2-p_2^2]+m_3^2, 
\end{align}
and the divergent part $\Delta$ is given by  
\begin{align}
\Delta \equiv \frac{1}{\epsilon}-\gamma_E+\ln 4\pi+\ln\mu^2, \label{div}
\end{align} 
with $\gamma_E^{}$ being the Euler constant. 
It is convenient to define the following functions~\cite{hhkm}: 
\begin{align}
B_2(p^2,m_1,m_2)&=B_{21}(p^2,m_1,m_2), \\
B_3(p^2,m_1,m_2)&=-B_1(p^2,m_1,m_2)-B_{21}(p^2,m_1,m_2),\label{b3-0}\\
B_4(p^2,m_1,m_2)&=-m_1^2B_1(p^2,m_2,m_1)-m_2^2B_1(p^2,m_1,m_2),\\
B_5(p^2,m_1,m_2)&=A(m_1)+A(m_2)-4B_{22}(p^2,m_1,m_2). 
\end{align}

\section{1PI diagrams}

In this section, we give the analytic expressions for the 1PI diagram contributions to one, two and three point functions by using the 
Passarino-Veltman functions defined in the previous section. 
We calculate 1PI diagrams in the t'~Hooft-Feynman gauge in which the masses of 
Nambu-Goldstone bosons $m_{G^\pm}$ and $m_{G^0}$ and those of Fadeev-Popov ghosts $m_{c^\pm}$ and $m_{c_Z}$ are the same as 
corresponding masses of the gauge bosons; i.e., 
$m_{G^\pm}=m_{c^\pm}=m_W$ and $m_{G^0}=m_{c_Z}=m_Z$. 
1PI diagrams with bosonic external lines are separately calculated by the fermion-loop and boson-loop contritbutions. 
We denote the fermionic- and bosonic-loop contributions by the subscript of $F$ and $B$, respectively.  
Throughout this section, we use the shortened notation of the Passarino-Veltman functions~\cite{Ref:PV} as 
\begin{align}
A(X)=\frac{i}{16\pi^2}A(m_X^{}) \\
B_{i,\,ij}(p^2;X,Y)=\frac{i}{16\pi^2}B_{i,\,ij}(p^2;m_X^{},m_Y^{}),  \\
C_{i,\,ij}(X,Y,Z) =\frac{i}{16\pi^2}C_{i,\,ij}(p_1^2,p_2^2,(p_1+p_2)^2;m_X^{},m_Y^{},m_Z^{}). 
\end{align}

\subsection{One-point functions}

The 1PI tadpole diagrams for $h$ and $H$ are calculated by 
\begin{align}
T_{h,F}^{\text{1PI}}&=-\sum_f\frac{4m_f^2 }{v}N_c^f\xi_h^fA(f),\\
T_{H,F}^{\text{1PI}}&=-\sum_f\frac{4m_f^2 }{v}N_c^f\xi_H^fA(f), \\
T_{h,B}^{\text{1PI}}&= 
s_{\beta-\alpha}\left[3gm_WA(W)+\frac{3}{2}g_Zm_ZA(Z)-2gm_W^3-g_Zm_Z^3\right]\notag\\
&-\lambda_{H^{+}H^{-}h}A({H^\pm})
-\lambda_{AAh}A(A)
-\lambda_{HHh}A(H)-3\lambda_{hhh}A(h)\notag\\
&-\lambda_{G^+G^-h}A({G^\pm})-\lambda_{G^0G^0h}A({G^0}),\\
T_{H,B}^{\text{1PI}}&=c_{\beta-\alpha}^{}\left[3gm_WA(W)+\frac{3}{2}g_Zm_ZA(Z)-2gm_W^3-g_Zm_Z^3\right]\notag\\
&-\lambda_{H^{+}H^{-}H}A({H^\pm})
-\lambda_{AAH}A(A)-3\lambda_{HHH}A(H)-\lambda_{Hhh}A(h)\notag\\
&-\lambda_{G^+G^-H}A({G^\pm})-\lambda_{G^0G^0H}A({G^0}). 
\end{align}

\subsection{Two-point functions}

The 1PI diagram contributions to the scalar boson two point functions are calculated as  
\begin{align}
\Pi_{hh}^{\text{1PI}}(p^2)_F&=-\sum_f\frac{4m_f^2 N_c^f}{v^2}(\xi_h^f)^2
\left[A(f)+\left(2m_f^2-\frac{p^2}{2}\right)B_0(p^2;f,f)\right],\\
\Pi_{HH}^{\text{1PI}}(p^2)_F&=-\sum_f\frac{4m_f^2 N_c^f}{v^2}(\xi_H^f)^2
\left[A(f)+\left(2m_f^2-\frac{p^2}{2}\right)B_0(p^2;f,f)\right],\\
\Pi_{Hh}^{\text{1PI}}(p^2)_F&=-\sum_f\frac{4m_f^2 N_c^f}{v^2}\xi_h^f\xi_H^f
\left[A(f)+\left(2m_f^2-\frac{p^2}{2}\right)B_0(p^2;f,f)\right],\\
\Pi_{AA}^{\text{1PI}}(p^2)_F&=-\sum_f\frac{4m_f^2 N_c^f}{v^2}\xi_f^2
\left[A(f)-\frac{p^2}{2}B_0(p^2;f,f)\right],\\
\Pi_{AG}^{\text{1PI}}(p^2)_F&=-\sum_f\frac{4m_f^2 N_c^f}{v^2}\xi_f
\left[A(f)-\frac{p^2}{2}B_0(p^2;f,f)\right],
\end{align}
\begin{align}
\Pi_{hh}^{\text{1PI}}(p^2)_B&
=g^2\sin^2(\beta-\alpha)(3m_W^2-p^2)B_0(p^2;W,W)+\frac{g^2}{2}\left[4-\sin^2(\beta-\alpha)\right]A(W)\notag\\
&+\frac{g_Z^2}{2}\sin^2(\beta-\alpha)(3m_Z^2-p^2)B_0(p^2;Z,Z)+\frac{g_Z^2}{4}\left[4-\sin^2(\beta-\alpha)\right]A(Z)\notag\\
&-\frac{g^2}{2}\cos^2(\beta-\alpha)\left[2A({W})-A({H^\pm})+(2m_{H^\pm}^2-m_W^2+2p^2)B_0(p^2;W,{H^\pm})\right]\notag\\
&-\frac{g_Z^2}{4}\cos^2(\beta-\alpha)\left[2A({Z})-A(A)+(2m_A^2-m_Z^2+2p^2)B_0(p^2;Z,A)\right]\notag\\
&-\left[\sin^2(\beta-\alpha)+\frac{1}{2}\right](2g^2m_W^2+g_Z^2m_Z^2),\notag\\
&-2\lambda_{H^{+}H^{-}hh}A({H^\pm})
-2\lambda_{AAhh}A(A)-2\lambda_{HHhh}A(H)-12\lambda_{hhhh}A(h)\notag\\
&-2\lambda_{G^+G^-hh}A({G^\pm})-2\lambda_{G^0G^0hh}A({G^0})
\notag\\
&+\lambda_{H^{+}H^{-}h}^2B_0(p^2;{H^\pm},{H^\pm})
+\lambda_{G^{+}G^{-}h}^2B_0(p^2;{G^\pm},{G^\pm})+2\lambda_{H^{+}G^{-}h}^2B_0(p^2;{H^\pm},{G^\pm})\notag\\
&+2\lambda_{AAh}^2B_0(p^2;A,A)+2\lambda_{G^0G^0h}^2B_0(p^2;{G^0},{G^0})
+\lambda_{AG^0h}^2B_0(p^2;A,{G^0})\notag\\
&+2\lambda_{HHh}^2B_0(p^2;H,H)
+18\lambda_{hhh}^2B_0(p^2;h,h)
+4\lambda_{Hhh}^2B_0(p^2;h,H),
\end{align}
\begin{align}
\Pi_{HH}^{\text{1PI}}(p^2)_B&
=g^2\cos^2(\beta-\alpha)(3m_W^2-p^2)B_0(p^2;W,W)+\frac{g^2}{2}[4-\cos^2(\beta-\alpha)]A(W)\notag\\
&+\frac{g_Z^2}{2}\cos^2(\beta-\alpha)(3m_Z^2-p^2)B_0(p^2;Z,Z)+\frac{g_Z^2}{4}\left[4-\cos^2(\beta-\alpha)\right]A(Z)\notag\\
&-\frac{g^2}{2}\sin^2(\beta-\alpha)\left[2A({W})-A({H^\pm})+(2m_{H^\pm}^2-m_W^2+2p^2)B_0(p^2;W,{H^\pm})\right]\notag\\
&-\frac{g_Z^2}{4}\sin^2(\beta-\alpha)\left[2A({Z})-A(A)+(2m_A^2-m_Z^2+2p^2)B_0(p^2;Z,A)\right]\notag\\
&-\left[\cos^2(\beta-\alpha)+\frac{1}{2}\right](2g^2m_W^2+g_Z^2m_Z^2),\notag\\
&-2\lambda_{H^{+}H^{-}HH}A({H^\pm})
 -2\lambda_{AAHH}A(A)-12\lambda_{HHHH}A(H)-2\lambda_{HHhh}A(h)\notag\\
&-2\lambda_{G^+G^-HH}A({G^\pm})-2\lambda_{G^0G^0HH}A({G^0})\notag\\
&+\lambda_{H^{+}H^{-}H}^2B_0(p^2;{H^\pm},{H^\pm})
+\lambda_{G^{+}G^{-}H}^2B_0(p^2;{G^\pm},{G^\pm})+2\lambda_{H^{+}G^{-}H}^2B_0(p^2;{H^\pm},{G^\pm})\notag\\
&+2\lambda_{AAH}^2B_0(p^2;A,A)+2\lambda_{G^0G^0H}^2B_0(p^2;{G^0},{G^0})
+\lambda_{AG^0H}^2B_0(p^2;A,{G^0})\notag\\
&+18\lambda_{HHH}^2B_0(p^2;H,H)
+2\lambda_{Hhh}^2B_0(p^2;h,h)
+4\lambda_{HHh}^2B_0(p^2;h,H),
\end{align}
\begin{align}
\Pi_{Hh}^{\text{1PI}}(p^2)_B&
=s_{\beta-\alpha}c_{\beta-\alpha}^{}\notag\\
&\times\Big\{ g^2(3m_W^2-p^2)B_0(p^2;W,W)-\frac{g^2}{2}A(W)\notag\\
&+\frac{g_Z^2}{2}(3m_Z^2-p^2)B_0(p^2;Z,Z)-\frac{g_Z^2}{4}A(Z)\notag\\
&+\frac{g^2}{2}[2A({W})-A(H^\pm)+(2m_{H^\pm}^2-m_W^2+2p^2)B_0(p^2;W,{H^\pm})]\notag\\
&+\frac{g_Z^2}{4}[2A({Z})-A(A)+(2m_A^2-m_Z^2+2p^2)B_0(p^2;Z,A)]-(2g^2m_W^2+g_Z^2m_Z^2)\Big\}\notag\\
&-\lambda_{H^+H^-Hh}A(H^\pm)
 -\lambda_{AAHh}A(A)-3\lambda_{HHHh}A(H)-3\lambda_{Hhhh}A(h)\notag\\
&-\lambda_{G^+G^-Hh}A(G^\pm)-\lambda_{G^0G^0Hh}A(G^0)\notag\\
&+\lambda_{H^{+}H^{-}h}\lambda_{H^{+}H^{-}H}B_0(p^2;H^\pm,H^\pm)
 +\lambda_{G^{+}G^{-}h}\lambda_{G^{+}G^{-}H}B_0(p^2;G^\pm,G^\pm)\notag\\
&+2\lambda_{H^{+}G^{-}h}\lambda_{H^{+}G^{-}H}B_0(p^2;H^\pm,G^\pm)\notag\\
&+2\lambda_{AAh}\lambda_{AAH}B_0(p^2;A,A)+2\lambda_{hG^0G^0}\lambda_{G^0G^0H}B_0(p^2;G^0,G^0)\notag\\
&+\lambda_{AG^0h}\lambda_{AG^0H}B_0(p^2;A,{G^0})+6\lambda_{HHh}\lambda_{HHH}B_0(p^2;H,H)\notag\\
&+6\lambda_{hhh}\lambda_{Hhh}B_0(p^2;h,h)+4\lambda_{Hhh}\lambda_{HHh}B_0(p^2;H,h),
\end{align}
\begin{align}
\Pi_{AA}^{\text{1PI}}(p^2)_B&=
2g^2A(W)+g_Z^2A(Z)-\frac{1}{2}(2g^2m_W^2+g_Z^2m_Z^2)\notag\\
&-\frac{g^2}{2}\left[2A({W})-A({H^\pm})+(2m_{H^\pm}^2-m_W^2+2p^2)B_0(p^2;W,{H^\pm})\right]\notag\\
&-\frac{g_Z^2}{4}\cos^2(\beta-\alpha)\left[2A({Z})-A({h})+(2m_{h}^2-m_Z^2+2p^2)B_0(p^2;Z,h)\right]\notag\\
&-\frac{g_Z^2}{4}\sin^2(\beta-\alpha)\left[2A({Z})-A(H)+(2m_{H}^2-m_Z^2+2p^2)B_0(p^2;Z,H)\right], \notag\\
&-2\lambda_{H^{+}H^{-}AA}A({H^\pm})-12\lambda_{AAAA}A(A)-2\lambda_{AAHH}A(H)-2\lambda_{AAhh}A(h)\notag\\
&-2\lambda_{G^+G^-AA}A({G^\pm})-2\lambda_{AAG^0G^0}A({G^0})\notag\\
&+2|\lambda_{H^+G^-A}|^2 B_0(p^2;{H^\pm},{G^\pm})
+4\lambda_{AAh}^2B_0(p^2;A,h)\notag\\
&+4\lambda_{AAH}^2B_0(p^2;A,H)
+\lambda_{AG^0h}^2B_0(p^2;h,{G^0})+\lambda_{AG^0H}^2B_0(p^2;H,{G^0}),
\end{align}
\begin{align}
\Pi_{AG}^{\text{1PI}}(p^2)_B&=s_{\beta-\alpha}c_{\beta-\alpha}^{}\notag\\
&\times \Big\{
  \frac{g_Z^2}{4}\left[2A(Z)-A(H)+(2m_H^2-m_Z^2+2p^2)B_0(p^2;Z,H)\right] \notag\\
&-\frac{g_Z^2}{4}[2A(Z)-A(h)+(2m_h^2-m_Z^2+2p^2)B_0(p^2;Z,h)] \Big\} \notag\\
&-\lambda_{H^{+}H^{-}AG^0}A({H^\pm})
-3\lambda_{AAAG^0}A(A)-\lambda_{AG^0HH}A(H)-\lambda_{AG^0hh}A(h)\notag\\
&-\lambda_{G^+G^-AG^0}A({G^\pm})-3\lambda_{AG^0G^0G^0}A({G^0})\notag\\
&+2\lambda_{AAh}\lambda_{AG^0h}B_0(p^2;A,h)+2\lambda_{AAH}\lambda_{AG^0H}B_0(p^2;A,H)\notag\\
&+2\lambda_{AG^0h}\lambda_{G^0G^0h}B_0(p^2;{G^0},h)+2\lambda_{AG^0H}\lambda_{G^0G^0H}B_0(p^2;{G^0},H).
\end{align}

The $Z$-$A$ mixing is given by 
\begin{align}
\Pi_{ZA}(p^2)_F& = \sum_f\frac{2m_f^2}{v^2}m_Z  N_c^f\xi_f B_0(p^2;f,f), \\ 
\Pi_{ZA}(p^2)_B& = m_Z\Big [ 
\frac{2\lambda_{AAH}}{v}s_{\beta-\alpha}(2B_1+B_0)(p^2; A,H)
-\frac{2\lambda_{AAh}}{v}c_{\beta-\alpha}^{}(2B_1+B_0)(p^2; A,h)\notag\\
&-\frac{\lambda_{AGH}}{v}c_{\beta-\alpha}^{}(2B_1+B_0)(p^2; {G^0},H)
-\frac{\lambda_{AGh}}{v}s_{\beta-\alpha}(2B_1+B_0)(p^2; {G^0},h)\notag\\
&-\frac{g_Z^2}{2}s_{\beta-\alpha}c_{\beta-\alpha}^{}(B_1-B_0)(p^2; H,Z)
+\frac{g_Z^2}{2}s_{\beta-\alpha}c_{\beta-\alpha}^{}(B_1-B_0)(p^2; h,Z)\Big], 
\end{align}
The G.I. part appearing in Eq.~(\ref{mod_beta2}) is given by 
\begin{align}
\Pi_{ZA}(p^2)\big|_{\text{G.I.}}& = \Pi_{ZA}(p^2)_F\notag\\
&+\frac{2m_Z}{v}\Big [ 
\lambda_{AAH}s_{\beta-\alpha}(2B_1+B_0)(p^2; A,H)
-\lambda_{AAh}c_{\beta-\alpha}^{}(2B_1+B_0)(p^2; A,h)\Big]. \label{gipart}
\end{align}

The 1PI diagram contributions to the gauge boson two point functions are calculated as   
\begin{align}
\Pi_{WW}^{\text{1PI}}(p^2)_F&=     \sum_{f,f'}     g^2N_c^f\Big(2p^2B_3-B_4\Big)(p^2;f,{f'}),\\
\Pi_{\gamma\gamma}^{\text{1PI}}(p^2)_F&= \sum_f8e^2Q_f^2N_c^fp^2B_3(p^2;f,f),\\
\Pi_{Z\gamma}^{\text{1PI}}(p^2)_F&=     \sum_feg_Z^{}N_c^f\Big[2p^2(2I_fQ_f-4s_W^2Q_f^2)B_3\Big](p^2;f,f), \\
\Pi_{ZZ}^{\text{1PI}}(p^2)_F&=         \sum_fg_Z^2N_c^f\Big[2p^2(4s_W^4Q_f^2-4s_W^2Q_fI_f+2I_f^2)B_3-2I_f^2f^2B_0\Big](p^2;f,f),
\end{align}
\begin{align}
\Pi_{WW}^{\text{1PI}}(p^2)_B
&=g^2\Bigg\{\frac{1}{4}B_5(p^2;A,{H^\pm})+\frac{1}{4}\sin^2(\beta-\alpha)B_5(p^2;H,{H^\pm})\notag\\
&\quad\quad+\frac{1}{4}\cos^2(\beta-\alpha)B_5(p^2;h,{H^\pm})\notag\\
&\quad\quad+\sin^2(\beta-\alpha)\left(m_W^2B_0+\frac{1}{4}B_5\right)(p^2;h,W)\notag\\
&\quad\quad+\cos^2(\beta-\alpha)\left(m_W^2B_0+\frac{1}{4}B_5\right)(p^2;H,W)\notag\\
&\quad\quad+\left[\left(\frac{1}{4}+2c_W^2\right)B_5+(m_W^2-4s_W^2m_W^2+m_Z^2-8p^2c_W^2)B_0\right](p^2;Z,W)\notag\\
&\quad\quad+2s_W^2\Big[B_5+(2m_W^2-4p^2)B_0\Big](p^2;0,W)-\frac{2}{3}p^2   \Bigg\},\\
\Pi_{\gamma\gamma}^{\text{1PI}}(p^2)_B
&=e^2B_5(p^2;{H^\pm},{H^\pm})
-e^2p^2\left[12B_3+5B_0(p^2;W,W)+\frac{2}{3}\right],\\
\Pi_{Z\gamma}^{\text{1PI}}(p^2)_B
&=\frac{eg_Z}{2}B_5(p^2;{H^\pm},{H^\pm})-eg_Zp^2
\left(10B_3+\frac{11}{2}B_0+\frac{2}{3}\right)(p^2;W,W)\notag\\
&-\frac{s_W}{c_W}\Pi_{\gamma\gamma}^{\text{1PI}}(p^2)_B,\\
\Pi_{ZZ}^{\text{1PI}}(p^2)_B&=
g_Z^2\Bigg\{
\frac{1}{4}B_5(p^2;{H^\pm},{H^\pm})+\frac{1}{4}\sin^2(\beta-\alpha)B_5(p^2;H,A)\notag\\
&+\frac{1}{4}\cos^2(\beta-\alpha)B_5(p^2;h,A)] \notag\\
&+\sin^2(\beta-\alpha)\left(m_Z^2B_0+\frac{1}{4}B_5\right)(p^2;h,Z)\notag\\
&+\cos^2(\beta-\alpha)\left(m_Z^2B_0+\frac{1}{4}B_5\right)(p^2;H,Z)\notag\\
&
+\Big[(2m_W^2-\frac{23}{4}p^2)B_0-9p^2B_3\Big](p^2;W,W)-\frac{2}{3}p^2\Bigg\}\notag\\
&-\frac{2s_W}{c_W}\Pi_{Z\gamma}^{\text{1PI}}(p^2)_B
-\frac{s_W^2}{c_W^2}\Pi_{\gamma\gamma}^{\text{1PI}}(p^2)_B, 
\end{align}
where the fermion-loop contributions are the same as those in the SM. 

The fermion two point functions can be decomposed into the following three parts
\begin{align}
\Pi_{ff}^{\text{1PI}}(p^2) = 
p\hspace{-2mm}/\Pi_{ff,V}^{\text{1PI}}(p^2) - p\hspace{-2mm}/\gamma_5\Pi_{ff,A}^{\text{1PI}}(p^2)  
+m_f\Pi_{ff,S}^{\text{1PI}}(p^2). 
\end{align}
Each part is caluclated as
\begin{align}
\Pi_{ff,V}^{\text{1PI}}(p^2)&=-e^2Q_f^2(2B_1+1)(p^2;f,\gamma)
-g_Z^2(v_f^2+a_f^2)(2B_1+1)(p^2;f,Z)\notag\\
&-\frac{g^2}{4}(2B_1+1)(p^2;{f'},W)\notag\\
&-\frac{m_f^2}{v^2}
\Big[(\xi_h^f)^2B_1(p^2;f,h)+(\xi_H^f)^2B_1(p^2;f,H)+\xi_f^2B_1(p^2;f,A)
+B_1(p^2;f,{G^0})\Big]\notag\\
&-\frac{m_f^2+m_{f'}^2}{v^2}B_1(p^2;{f'},{G^\pm})
-\frac{m_f^2\xi_f^2+m_{f'}^2\xi_{f'}^2}{v^2}B_1(p^2;{f'},{H^\pm}),\notag\\
\Pi_{ff,A}^{\text{1PI}}(p^2)&=
-2g_Z^2v_fa_f(2B_1+1)(p^2;f,Z)
-\frac{g^2}{4}(2B_1+1)(p^2;{f'},W)\notag\\
&+\frac{m_f^2-m_{f'}^2}{v^2}B_1(p^2;{f'},{G^\pm})
+\frac{m_f^2\xi_f^2-m_{f'}^2\xi_{f'}^2}{v^2}B_1(p^2;{f'},{H^\pm}),\notag\\
\Pi_{ff,S}^{\text{1PI}}(p^2)&=-2e^2Q_f^2(2B_0-1)(p^2;f,\gamma)
-2g_Z^2(v_f^2-a_f^2)(2B_0-1)(p^2;f,Z)\notag\\
&+\frac{m_f^2}{v^2}
\Big[(\xi_h^f)^2B_0(p^2;f,h)+(\xi_H^f)^2B_0(p^2;f,H)-\xi_f^2B_0(p^2;f,A)-B_0(p^2;f,{G^0})\Big]\notag\\
&-2\frac{m_{f'}^2}{v^2}\left[B_0(p^2;{f'},{G^\pm})
+\xi_f \xi_{f'}B_0(p^2;{f'},{H^\pm})\right], 
\end{align}
where $v_f$ and $a_f$ are the coefficient of the vector coupling and axial vector coupling of $Zf\bar{f}$ vertex given as
\begin{align}
v_f = \frac{I_f}{2}-s_W^2Q_f,\quad a_f = \frac{I_f}{2}. 
\end{align}

\subsection{Three-point functions}

In this subsection, we give analytic expressions for the 1PI diagram contributions to the three point functions. 
The assignment for external momentum is taken in such a way that 
$p_1$ and $(p_2)$ is the incoming momnetum of $h$ $(h)$, $V$ ($V$) and $f$ ($\bar{f}$) for the $hhh$, $hVV$ and $hf\bar{f}$ vertices, respectively, and 
$q=p_1+p_2$ is the outgoing momentum of $h$ for all the above vertices. 

First, the 1PI diagrams for the $hhh$ coupling is calculated as
\begin{align}
&\Gamma_{hhh}^{\text{1PI}}(p_1^2,p_2^2,q^2)_F=
-\sum_f\frac{8m_f^4 N_c^f}{v^3}
(\xi_h^f)^3
\Big[B_0(p_1^2,f,f)+B_0(p_2^2,f,f)+B_0(q^2,f,f)\notag\\
&\quad\quad\quad\quad+(4m_f^2-q^2+p_1\cdot p_2)C_0(f,f,f)\Big],
\end{align}

\begin{align}
&\Gamma_{hhh}^{\text{1PI}}(p_1^2,p_2^2,q^2)_B=
\frac{g^3}{2}m_W^3s_{\beta-\alpha}^3\Big[16C_0(W,W,W)-C_0(c^\pm,c^\pm,c^\pm)\Big]\notag\\
&-\frac{g^3}{2}m_Ws_{\beta-\alpha}\Big[s_{\beta-\alpha}^2C_{hhh}^{SVV}({G^\pm},W,W)+c^2_{\beta-\alpha}C_{hhh}^{SVV}({H^\pm},W,W) \Big]\notag\\
&+\frac{g_Z^3}{4}m_Z^3s_{\beta-\alpha}^3\Big[16C_0(Z,Z,Z)-C_0(c_Z,c_Z,c_Z)\Big]
-\frac{g_Z^3m_Z}{4}s_{\beta-\alpha}\Big[s_{\beta-\alpha}^2C_{hhh}^{SVV}({G^0},Z,Z)c^2_{\beta-\alpha}C_{hhh}^{SVV}(A,Z,Z)\Big]\notag\\
&+\frac{g^2}{2}\lambda_{G^+G^-h}s_{\beta-\alpha}^2C_{hhh}^{VSS}(W,G^\pm,G^\pm)
+\frac{g^2}{2}\lambda_{H^+H^-h}c^2_{\beta-\alpha}C_{hhh}^{VSS}(W,H^\pm,H^\pm)
\notag\\
&+\frac{g^2}{2}\lambda_{H^+G^-h}s_{\beta-\alpha}c_{\beta-\alpha}^{}
[C_{hhh}^{VSS}(W,{G^\pm},{H^\pm})+C_{hhh}^{VSS}(W,H^\pm,G^\pm)]\notag\\
&+\frac{g_Z^2}{2}\lambda_{G^0G^0h}s_{\beta-\alpha}^2
C_{hhh}^{VSS}(Z,G^0,G^0)
+\frac{g_Z^2}{2}\lambda_{AAh}c^2_{\beta-\alpha}
C_{hhh}^{VSS}(Z,A,A)\notag\\
&+\frac{g_Z^2}{4}\lambda_{AG^0h}s_{\beta-\alpha}c_{\beta-\alpha}^{}
[C_{hhh}^{VSS}(Z,{A},{G^0})+C_{hhh}^{VSS}(Z,G^0,A)]\notag\\
&+2g^3m_Ws_{\beta-\alpha}[B_0(p_1^2,W,W)+B_0(p_2^2,W,W)+B_0(q^2,W,W)]-3g^3m_Ws_{\beta-\alpha}\notag\\
&+g_Z^3m_Zs_{\beta-\alpha}[B_0(p_1^2,Z,Z)+B_0(p_2^2,Z,Z)+B_0(q^2,Z,Z)]-\frac{3}{2}g_Z^3m_Zs_{\beta-\alpha}\notag\\
&+2\lambda_{H^{+}H^{-}h}\lambda_{H^{+}H^{-}hh}[B_0(p_1^2,H^\pm,H^\pm)
+B_0(p_2^2,H^\pm,H^\pm)+B_0(q^2,H^\pm,H^\pm)]\notag\\
&+2\lambda_{hG^{+}G^{-}}\lambda_{hhG^{+}G^{-}}[B_0(p_1^2,G^\pm,G^\pm)
+B_0(p_2^2,G^\pm,G^\pm)+B_0(q^2,G^\pm,G^\pm)]\notag\\
&+4\lambda_{H^{+}G^{-}h}\lambda_{H^{+}G^{-}hh}[B_0(p_1^2,H^\pm,G^\pm)
+B_0(p_2^2,H^\pm,G^\pm)+B_0(q^2,H^\pm,G^\pm)]\notag\\
&+4\lambda_{AAh}\lambda_{AAhh}
[B_0(p_1^2,{A},{A})+B_0(p_2^2,{A},{A})+B_0(q^2,{A},{A})]\notag\\
&+4\lambda_{G^0G^0h}\lambda_{G^0G^0hh}
[B_0(p_1^2,{G^0},{G^0})+B_0(p_2^2,{G^0},{G^0})+B_0(q^2,{G^0},{G^0})]\notag\\
&+2\lambda_{AG^0h}\lambda_{AG^0hh}
[B_0(p_1^2,{A},{G^0})+B_0(p_2^2,A,G^0)+B_0(q^2,{A},{G^0})]\notag\\
&+4\lambda_{HHh}\lambda_{HHhh}
[B_0(p_1^2,{H},{H})+B_0(p_2^2,H,H)+B_0(q^2,{H},{H})]\notag\\
&+12\lambda_{Hhh}\lambda_{Hhhh}[B_0(p_1^2,h,H)+B_0(p_2^2,h,H)+B_0(q^2,h,H)]\notag\\
&+72\lambda_{hhh}\lambda_{hhhh}[B_0(p_1^2,h,h)+B_0(p_2^2,h,h)+B_0(q^2,h,h)] \notag\\
%%%
&-2\lambda_{H^+H^-h}^3C_0({H^\pm},{H^\pm},{H^\pm})
-2\lambda_{G^{+}G^{-}h}^3C_0({G^\pm},{G^\pm},{G^\pm})-8\lambda_{G^0G^0h}^3C_0({G^0},{G^0},{G^0})\notag\\
&-8\lambda_{AAh}^3C_0(A,A,A)
-8\lambda_{HHh}^3C_0(H,H,H)
-216\lambda_{hhh}^3C_0(h,h,h)\notag\\
&-2\lambda_{H^+H^-h}\lambda_{H^+G^-h}^2
[C_0({G^\pm},{H^\pm},{H^\pm})+C_0({H^\pm},{G^\pm},{H^\pm})+C_0({H^\pm},{H^\pm},{G^\pm})]\notag\\
&-2\lambda_{G^+G^-h}\lambda_{H^+G^-h}^2
[C_0({H^\pm},{G^\pm},{G^\pm})+C_0({G^\pm},{H^\pm},{W})+C_0({G^\pm},{G^\pm},{H^\pm})]\notag\\
&-2\lambda_{AAh}\lambda_{AG^0h}^2
[C_0({G^0},A,A)+C_0(A,{G^0},A)+C_0(A,A,{G^0})]\notag\\
&-2\lambda_{G^0G^0h}\lambda_{AG^0h}^2
[C_0(A,{G^0},{G^0})+C_0({G^0},A,{G^0})+C_0({G^0},{G^0},A)]\notag\\
&-8\lambda_{HHh}\lambda_{Hhh}^2
[C_0(h,H,H)+C_0(H,H,h)+C_0(H,h,H)]\notag\\
&-24\lambda_{hhh}\lambda_{Hhh}^2
[C_0(h,h,H)+C_0(H,h,h)+C_0(h,H,h)],
\end{align}
where
\begin{align}
&C_{hhh}^{SVV}(X,Y,Z)\equiv 
\Big[p_1^2C_{21}+p_2^2C_{22}+2p_1p_2C_{23}+4C_{24}-\frac{1}{2}
-(q+p_1)(p_1C_{11}+p_2C_{12})+qp_1C_0
\Big](X,Y,Z)\notag\\
&+\Big[p_1^2C_{21}+p_2^2C_{22}+2p_1p_2C_{23}+4C_{24}-\frac{1}{2}
+(3p_1-p_2)(p_1C_{11}+p_2C_{12})+2p_1(p_1-p_2)C_0
\Big](Z,X,Y)\notag\\
&+\Big[p_1^2C_{21}+p_2^2C_{22}+2p_1p_2C_{23}+4C_{24}-\frac{1}{2}
+(3p_1+4p_2)(p_1C_{11}+p_2C_{12})+2q(q+p_2)C_0
\Big](Y,Z,X),\notag\\
&C_{hhh}^{VSS}(X,Y,Z)\equiv \notag\\
&\Big[p_1^2C_{21}+p_2^2C_{22}+2p_1p_2C_{23}+4C_{24}-\frac{1}{2}
+(4p_1+2p_2)(p_1C_{11}+p_2C_{12})+4p_1\cdot q C_0
\Big](X,Y,Z)\notag\\
&\hspace{-5mm}+\Big[p_1^2C_{21}+p_2^2C_{22}+2p_1p_2C_{23}+4C_{24}-\frac{1}{2}
+2p_2(p_1C_{11}+p_2C_{12})-p_1(p_1+2p_2)C_0
\Big](Z,X,Y)\notag\\
&\hspace{-5mm}+\Big[p_1^2C_{21}+p_2^2C_{22}+2p_1p_2C_{23}+4C_{24}-\frac{1}{2}
-2p_2(p_1C_{11}+p_2C_{12})-q(p_1-p_2)C_0
\Big](Y,Z,X).
\end{align}

The $hf\bar{f}$ vertex can be decomposed into the following 8 form factors
\begin{align}
&\Gamma_{hff}^{\text{1PI}}(p_1^2,p_2^2,q^2) = \notag\\
&F_{hff}^S+\gamma_5 F_{hff}^P+p_1\hspace{-3.5mm}/\hspace{2mm}F_{hff}^{V1}
+p_2\hspace{-3.5mm}/\hspace{2mm}F_{hff}^{V2}+p_1\hspace{-3.5mm}/\hspace{2mm}\gamma_5 F_{hff}^{A1}
+p_2\hspace{-3.5mm}/\hspace{2mm}\gamma_5F_{hff}^{A2}
+p_1\hspace{-3.5mm}/\hspace{2mm}p_2\hspace{-3.5mm}/\hspace{2mm}F_{hff}^{T}
+p_1\hspace{-3.5mm}/\hspace{2mm}p_2\hspace{-3.5mm}/\hspace{2mm}\gamma_5F_{hff}^{PT}. 
\end{align}
Each form factor can be calculated by
\begin{align}
&\left(\frac{m_f}{v}\right)^{-1}F^S_{hff}=-2g_Z^4v^2(v_f^2-a_f^2)s_{\beta-\alpha}C_0(Z,f,Z)\notag\\
&-4\xi_h^f\Big\{ 
e^2Q_f^2
[m_f^2C_0+p_1^2(C_{11}+C_{21})+p_2^2(C_{12}+C_{22})+p_1\cdot p_2(2C_{23}-C_0)+4C_{24}-1](f,\gamma,f)
\notag\\
&+g_Z^2(v_f^2-a_f^2)[m_f^2C_0+p_1^2(C_{11}+C_{21})+p_2^2(C_{12}+C_{22})+p_1\cdot p_2(2C_{23}-C_0)+4C_{24}-1](f,Z,f) \Big\}
\notag\\
&+\xi_h^f
\frac{m_f^2}{v^2}\Big[
(\xi_h^f)^2C_{hff}^{FSF}(f,h,f)
+(\xi_H^f)^2C_{hff}^{FSF}(f,H,f) -C_{hff}^{FSF}(f,G^0,f)-\xi_f^2C_{hff}^{FSF}(f,A,f) \Big]  \notag\\
&
-\xi_h^{f'}\frac{2m_{f'}^2}{v^2}\Big[C_{hff}^{FSF}(f',G^\pm,f')
+\xi_f\xi_{f'}C_{hff}^{FSF}(f',H^\pm,f')\Big]\notag\\
&-\frac{m_f^2}{v}\Big\{6(\xi_h^f)^2\lambda_{hhh} C_0(h,f,h)+2(\xi_H^f)^2\lambda_{HHh}C_0(H,f,H)
+2\xi_h^f\xi_H^f\lambda_{Hhh}[C_0(h,f,H)+C_0(H,f,h)]\notag\\
&\quad\quad\quad-2\lambda_{G^0G^0h}C_0(G^0,f,G^0)-2\xi_f^2\lambda_{AAh}C_0(A,f,A)
-\xi_f\lambda_{AG^0h}[C_0(A,f,{G^0})+C_0(G^0,f,A)]  \Big\}\notag\\
&+\frac{2m_{f'}^2}{v}\Big\{
\lambda_{G^+G^-h}C_0(G^\pm,f',G^\pm)+\xi_f\xi_{f'}\lambda_{H^+H^-h}C_0(H^\pm,f',H^\pm)\notag\\
&\quad\quad\quad\quad+\frac{1}{2}\lambda_{H^+G^-h}(\xi_f+\xi_{f'})[C_0({G^\pm},{f'},{H^\pm})+C_0({H^\pm},{f'},{G^\pm})]\Big\}\notag\\
&-\frac{g^2}{4}s_{\beta-\alpha}\Big[
C_{hff}^{VFS}(W,{f'},{G^\pm})+C_{hff}^{SFV}(G^\pm,f',W)\Big]\notag\\
&-\frac{g^2}{4}\xi_fc_{\beta-\alpha}^{}\Big[C_{hff}^{VFS}(W,f',H^\pm)+C_{hff}^{SFV}(H^\pm,f',W)\Big]\notag\\
&-\frac{g_Z^2}{8}s_{\beta-\alpha}\Big[C_{hff}^{VFS}(Z,f,G^0)
+C_{hff}^{SFV}({G^0},f,Z) \Big]\notag\\
&-\frac{g_Z^2}{8}\xi_fc_{\beta-\alpha}^{}\Big[C_{hff}^{VFS}(Z,f,A)+C_{hff}^{SFV}(A,f,Z)\Big], 
\end{align}
\begin{align}
&\left(\frac{m_f}{v}\right)^{-1}  F^P_{hff}= 
\lambda_{H^+G^-h}\frac{m_{f'}^2}{v}(\xi_{f'}-\xi_{f})[C_0({G^\pm},{f'},{H^\pm})-C_0({H^\pm},{f'},{G^\pm})]\notag\\
&-\frac{g^2}{4}s_{\beta-\alpha}^{}\Big[C_{hff}^{VFS}(W,f',G^\pm)-C_{hff}^{SFV}({G^\pm},{f'},W)\Big]\notag\\
&-\frac{g^2}{4}\xi_f c_{\beta-\alpha}^{}\Big[ C_{hff}^{VFS}(W,{f'},{H^\pm})-C_{hff}^{SFV}({H^\pm},{f'},W)\Big]  \notag\\
&-g_Z^2v_fI_f  s_{\beta-\alpha}^{} \Big[C_{hff}^{VFS}(Z,f,{G^0})-C_{hff}^{SFV}({G^0},f,Z) \Big] \notag\\
&-g_Z^2v_fI_f\xi_fc_{\beta-\alpha}^{}\Big[C_{hff}^{VFS}(Z,f,A)-C_{hff}^{SFV}(A,f,Z)\Big],
\end{align}
\begin{align}
&  F^{V1}_{hff}=
\frac{2m_f^2}{v}\xi_h^f \Big[ 
 g_Z^2(v_f^2+a_f^2)(C_0+2C_{11})(f,Z,f)
+e^2Q_f^2(C_0+2C_{11})(f,\gamma,f)\Big]\notag\\
&+g^2\frac{m_{f'}^2}{2v}\xi_h^{f'}(C_0+2C_{11})({f'},W,{f'})\notag\\
&-s_{\beta-\alpha}g_Z^4v(v_f^2+a_f^2)(C_0+C_{11})(Z,f,Z)-s_{\beta-\alpha}\frac{g^4}{4}v(C_0+C_{11})(W,{f'},W)\notag\\
&+\xi_h^f\frac{m_f^4}{v^3}\Big[(\xi_h^f)^2(C_0+2C_{11})(f,h,f)+(\xi_H^f)^2(C_0+2C_{11})(f,H,f)\notag\\
&\quad\quad\quad\quad+ (C_0+2C_{11})(f,{G^0},f)+\xi_f^2(C_0+2C_{11})(f,A,f)\Big]\notag\\
&+\frac{m_{f'}^2}{v^3}\xi_h^{f'}\Big[(m_f^2+m_{f'}^2)(C_0+2C_{11})({f'},{G^\pm},{f'})
+(m_f^2\xi_f^2+m_{f'}^2\xi_{f'}^2)(C_0+2C_{11})({f'},{H^\pm},{f'})\Big]\notag\\
&-\frac{m_f^2}{v^2}\Big\{6(\xi_h^f)^2\lambda_{hhh}(C_0+C_{11})(h,f,h)
+2(\xi_H^f)^2\lambda_{HHh}(C_0+C_{11})(H,f,H)\notag\\
&\quad\quad\quad+2\xi_h^f\xi_H^f\lambda_{Hhh}[(C_0+C_{11})(H,f,h)+(C_0+C_{11})(h,f,H)]\notag\\
&\quad\quad\quad+2\lambda_{G^0G^0h}(C_0+C_{11})({G^0},f,{G^0})
+2\xi_f^2\lambda_{AAh}(C_0+C_{11})(A,f,A)\notag\\
&\quad\quad\quad+\xi_f\lambda_{AG^0h}[(C_0+C_{11})(A,f,{G^0})+(C_0+C_{11})({G^0},f,A)]  \Big\}\notag\\
&-\frac{\lambda_{G^+G^-h}}{v^2}(m_f^2+m_{f'}^2)(C_0+C_{11})({G^\pm},{f'},{G^\pm})
 -\frac{\lambda_{H^+H^-h}}{v^2}(m_f^2\xi_f^2+m_{f'}^2\xi_{f'}^2)(C_0+C_{11})({H^\pm},{f'},{H^\pm})\notag\\
&-\frac{\lambda_{H^+G^-h}}{v^2}(m_f^2\xi_f+m_{f'}^2\xi_{f'})[(C_0+C_{11})({G^\pm},{f'},{H^\pm})+(C_0+C_{11})({H^\pm},{f'},{G^\pm})]\notag\\
&-g^2\frac{m_{f'}^2}{4v}\Big[s_{\beta-\alpha}(2C_0+C_{11})(W,{f'},{G^\pm})
+s_{\beta-\alpha}(-C_0+C_{11})({G^\pm},{f'},W)\notag\\
&\quad\quad  \quad \quad-\xi_{f'}c_{\beta-\alpha}^{}(2C_0+C_{11})(W,{f'},{H^\pm})
-\xi_{f'}c_{\beta-\alpha}^{}(-C_0+C_{11})({H^\pm},{f'},W)\Big]\notag\\
&-g_Z^2\frac{m_f^2}{8v}\Big[
s_{\beta-\alpha}(2C_0+C_{11})(Z,f,{G^0})
+s_{\beta-\alpha}(-C_0+C_{11})({G^0},f,Z)\notag\\
&\quad\quad  \quad \quad -\xi_fc_{\beta-\alpha}^{}(2C_0+C_{11})(Z,f,A)-\xi_fc_{\beta-\alpha}^{}(-C_0+C_{11})(A,f,Z)\Big],
\end{align}
\begin{align}
& F^{V2}_{hff}= 
\frac{2m_f^2}{v}\xi_h^f  \Big[ g_Z^2(v_f^2+a_f^2)(C_0+2C_{12})(f,Z,f)
+e^2Q_f^2(C_0+2C_{12})(f,\gamma,f)\Big]\notag\\
&+g^2\frac{m_{f'}^2}{2v}\xi_h^{f'}(C_0+2C_{12})({f'},W,{f'})\notag\\
&-s_{\beta-\alpha}g_Z^4v(v_f^2+a_f^2)C_{12}(Z,f,Z)-s_{\beta-\alpha}\frac{g^4}{4}vC_{12}(W,{f'},W)\notag\\
&+\xi_h^f\frac{m_f^4}{v^3}\Big[ (\xi_h^f)^2 (C_0+2C_{12})(f,h,f) +(\xi_H^f)^2(C_0+2C_{12})(f,H,f) \notag\\
&\quad\quad\quad\quad+(C_0+2C_{12})(f,{G^0},f)+\xi_f^2(C_0+2C_{12})(f,A,f)\Big]\notag\\
&+\xi_h^{f'}\frac{m_{f'}^2}{v^3}\Big[ (m_f^2+m_{f'}^2)(C_0+2C_{12})({f'},{G^\pm},{f'}) 
 +(m_f^2\xi_f^2+m_{f'}^2\xi_{f'}^2)(C_0+2C_{12})({f'},{H^\pm},{f'}) \Big] \notag\\
&-\frac{m_f^2}{v^2}\Big\{ 6(\xi_h^f)^2\lambda_{hhh}C_{12}(h,f,h) +2(\xi_H^f)^2\lambda_{HHh}C_{12}(H,f,H)
+2\xi_h^f\xi_H^f\lambda_{Hhh}[C_{12}(H,f,h)+C_{12}(h,f,H)]\notag\\
&\quad\quad\quad\quad+2\lambda_{G^0G^0h}C_{12}({G^0},f,{G^0})
+2\xi_f^2\lambda_{AAh}C_{12}(A,f,A)
+2\xi_f\lambda_{AG^0h}[C_{12}({G^0},f,A)+C_{12}(A,f,{G^0})]\Big\}
\notag\\
&-\frac{\lambda_{G^+G^-h}}{v^2}(m_f^2+m_{f'}^2)C_{12}({G^\pm},{f'},{G^\pm})
-\frac{\lambda_{H^+H^-h}}{v^2}(m_f^2\xi_f^2+m_{f'}^2\xi_{f'}^2)C_{12}({H^\pm},{f'},{H^\pm})\notag\\
&-\frac{\lambda_{H^+G^-h}}{v^2}(m_f^2\xi_f+m_{f'}^2\xi_{f'})[C_{12}({G^\pm},{f'},{H^\pm})+C_{12}({H^\pm},{f'},{G^\pm})]\notag\\
&-\frac{g^2}{4}\frac{m_{f'}^2}{v}\Big[
s_{\beta-\alpha}(2C_0+C_{12})(W,{f'},{G^\pm})
+s_{\beta-\alpha}(-C_0+C_{12})({G^\pm},{f'},W)\notag\\
&\quad\quad\quad\quad-\xi_{f'}c_{\beta-\alpha}^{}(2C_0+C_{12})(W,{f'},{H^\pm})
-\xi_{f'}c_{\beta-\alpha}^{}(-C_0+C_{12})({H^\pm},{f'},W)\Big] \notag\\
&-\frac{g_Z^2}{8}\frac{m_f^2}{v}\Big[s_{\beta-\alpha}(2C_0+C_{12})(Z,f,{G^0})
+s_{\beta-\alpha}(C_{12}-C_0)({G^0},f,Z)\notag\\
&\quad\quad\quad\quad+\xi_f c_{\beta-\alpha}^{}(2C_0+C_{12})(Z,f,A)
+\xi_fc_{\beta-\alpha}^{}(C_{12}-C_0)(A,f,Z)\Big],
\end{align}
\begin{align}
&   F^{A1}_{hff}= 
-4g_Z^2v_fa_f\frac{m_f^2}{v}\xi_h^f(C_0+2C_{11})(f,Z,f)
-g^2\frac{m_{f'}^2}{2v}\xi_h^{f'}(C_0+2C_{11})({f'},W,{f'})\notag\\
&+2s_{\beta-\alpha}g_Z^4v_fa_fv(C_0+C_{11})(Z,f,Z)
+s_{\beta-\alpha}\frac{g^4}{4}v(C_0+C_{11})(W,{f'},W)\notag\\
&+\frac{m_{f'}^2}{v^3}\xi_h^{f'} \Big[
(m_f^2-m_{f'}^2)(C_0+2C_{11})({f'},{G^\pm},{f'})+(m_f^2\xi_f^2-m_{f'}^2\xi_{f'}^2)(C_0+2C_{11})({f'},{H^\pm},{f'}) \Big]\notag\\
&-\frac{\lambda_{G^+G^-h}}{v^2}(m_f^2-m_{f'}^2)(C_0+C_{11})({G^\pm},{f'},{G^\pm})
-\frac{\lambda_{H^+H^-h}}{v^2}(m_f^2\xi_f^2-m_{f'}^2\xi_{f'}^2)(C_0+C_{11})({H^\pm},{f'},{H^\pm})\notag\\
&-\frac{\lambda_{H^+G^-h}}{v^2}(m_f^2\xi_f-m_{f'}^2\xi_{f'})[(C_0+C_{11})({G^\pm},{f'},{H^\pm})+(C_0+C_{11})({H^\pm},{f'},{G^\pm})]\notag\\
&+\frac{g^2}{4}\frac{m_{f'}^2}{v}\Big[s_{\beta-\alpha}(2C_0+C_{11})(W,f',{G^\pm})+s_{\beta-\alpha}(-C_0+C_{11})(G^\pm,f',W)\notag\\
&\quad\quad\quad\quad -\xi_{f'}c_{\beta-\alpha}^{}(2C_0+C_{11})(W,f',H^\pm)-\xi_{f'}c_{\beta-\alpha}^{}(-C_0+C_{11})({H^\pm},{f'},W)\Big]   \notag\\
&+g_Z^2I_fv_f\frac{m_f^2}{v}\Big[
s_{\beta-\alpha}(2C_0+C_{11})(Z,{f},{G^0})
+s_{\beta-\alpha}(-C_0+C_{11})({G^0},{f},Z)\notag\\
&\quad\quad\quad\quad+\xi_fc_{\beta-\alpha}^{}(2C_0+C_{11})(Z,f,A)+\xi_f c_{\beta-\alpha}^{}(-C_0+C_{11})(A,f,Z)\Big], 
\end{align}
\begin{align}
& F^{A2}_{hff}=
-4\xi_h^fg_Z^2v_fa_f\frac{m_f^2}{v}(C_0+2C_{12})(f,Z,f)-\xi_h^{f'}g^2\frac{m_{f'}^2}{2v}(C_0+2C_{12})({f'},W,{f'})\notag\\
&+2s_{\beta-\alpha}g_Z^4v_fa_fvC_{12}(Z,f,Z)
+s_{\beta-\alpha}\frac{g^4}{4}vC_{12}(W,{f'},W)\notag\\
&+\xi_h^{f'}\frac{m_{f'}^2}{v^3}\Big[(m_f^2-m_{f'}^2)(C_0+2C_{12})({f'},{G^\pm},{f'})
+(m_f^2\xi_f^2-m_{f'}^2\xi_{f'}^2)(C_0+2C_{12})({f'},{H^\pm},{f'})\Big]\notag\\
&-\frac{\lambda_{G^+G^-h}}{v^2}(m_f^2-m_{f'}^2)C_{12}({G^\pm},{f'},{G^\pm})
 -\frac{\lambda_{H^+H^-h}}{v^2}(m_f^2\xi_f^2-m_{f'}^2\xi_{f'}^2)C_{12}({H^\pm},{f'},{H^\pm})\notag\\
&-\frac{\lambda_{H^+G^-h}}{v^2}(m_f^2\xi_f-m_{f'}^2\xi_{f'})[C_{12}({G^\pm},{f'},{H^\pm})+C_{12}({H^\pm},{f'},{G^\pm})]\notag\\
&+\frac{g^2}{4}\frac{m_{f'}^2}{v}\Big[ s_{\beta-\alpha}(2C_0+C_{12})(W,{f'},{G^\pm})
+s_{\beta-\alpha}(-C_0+C_{12})({G^\pm},{f'},W)\notag\\
&\quad\quad\quad\quad-\xi_{f'}c_{\beta-\alpha}^{}(2C_0+C_{12})(W,{f'},{H^\pm})-\xi_{f'}c_{\beta-\alpha}^{}(-C_0+C_{12})({H^\pm},{f'},W)\Big]\notag\\
&+g_Z^2I_fv_f\frac{m_f^2}{v}  \Big[  s_{\beta-\alpha}(2C_0+C_{12})(Z,{f},{G^0})
                                   +s_{\beta-\alpha}(-C_0+C_{12})({G^0},{f},Z)\notag\\
&\quad\quad\quad\quad+\xi_fc_{\beta-\alpha}^{}(2C_0+C_{12})(Z,{f},A)
+\xi_fc_{\beta-\alpha}^{}(-C_0+C_{12})(A,{f},Z)\Big],
\end{align}
\begin{align}
&\left(\frac{m_f}{ v}\right)^{-1}  F^{T}_{hff}=
\xi_h^f\frac{m_f^2}{v^2}\Big[
(\xi_h^f)^2(C_{11}-C_{12})(f,h,f)
+(\xi_H^f)^2(C_{11}-C_{12})(f,H,f)\notag\\
&-(C_{11}-C_{12})(f,{G^0},f)-\xi_f^2(C_{11}-C_{12})(f,A,f)\Big]\notag\\
&-\xi_h^{f'}\frac{2m_{f'}^2}{v^2}\Big[(C_{11}-C_{12})({f'},{G^\pm},{f'})
+\xi_f\xi_{f'}(C_{11}-C_{12})({f'},{H^\pm},{f'})\Big]\notag\\
&-\frac{g^2}{4}\Big[s_{\beta-\alpha}(-2C_0-2C_{11}+C_{12})(W,{f'},{G^\pm})
+s_{\beta-\alpha}(-C_0-C_{11}+2C_{12})({G^\pm},{f'},W)\notag\\
&\quad\quad +\xi_fc_{\beta-\alpha}^{}(-2C_0-2C_{11}+C_{12})(W,{f'},{H^\pm})
+\xi_fc_{\beta-\alpha}^{}(-C_0-C_{11}+2C_{12})({H^\pm},{f'},W)\Big]  \notag\\
&-\frac{g_Z^2}{8}\Big[ s_{\beta-\alpha}^{}(-2C_0-2C_{11}+C_{12})(Z,{f},{G^0})
+s_{\beta-\alpha}^{}(-C_0-C_{11}+2C_{12})({G^0},{f},Z)\notag\\
&\quad\quad +\xi_f c_{\beta-\alpha}^{} (-2C_0-2C_{11}+C_{12})(Z,{f},A)
+\xi_f c_{\beta-\alpha}^{}(-C_0-C_{11}+2C_{12})(A,{f},Z)\Big], 
\end{align}
\begin{align}
& \left(\frac{m_f}{v}\right)^{-1} F^{PT}_{hff}=
\frac{g^2}{4}\Big[s_{\beta-\alpha}^{}(2C_0+2C_{11}-C_{12})(W,{f'},{G^\pm})
-s_{\beta-\alpha}^{}(C_0+C_{11}-2C_{12})({G^\pm},{f'},W)\notag\\
&-\xi_fc_{\beta-\alpha}^{}(-2C_0-2C_{11}+C_{12})(W,{f'},{H^\pm})
-\xi_fc_{\beta-\alpha}^{}(C_0+C_{11}-2C_{12})({H^\pm},{f'},W)\Big]\notag\\
&-g_Z^2I_fv_f\Big[s_{\beta-\alpha}^{}(-2C_0-2C_{11}+C_{12})(Z,{f},{G^0})
                 +s_{\beta-\alpha}^{}(C_0+C_{11}-2C_{12})({G^0},{f},Z)\notag\\
&+\xi_fc_{\beta-\alpha}^{}(-2C_0-2C_{11}+C_{12})(Z,f,A)
+\xi_f c_{\beta-\alpha}^{}(C_0+C_{11}-2C_{12})(A,f,Z)\Big],  
\end{align}
where 
\begin{align}
&C_{hff}^{FSF}(X,Y,Z)\equiv\notag\\
&[m_F^2C_0+p_1^2(C_{11}+C_{21})+p_2^2(C_{12}+C_{22})+2p_1\cdot p_2(C_{12}+C_{23})+4C_{24}](X,Y,Z)-\frac{1}{2},\notag\\
&C_{hff}^{VFS}(X,Y,Z)\equiv \notag\\
&[p_1^2(2C_0+3C_{11}+C_{21})+p_2^2(2C_{12}+C_{22})+2p_1\cdot p_2(2C_0+2C_{11}+C_{12}+C_{23})+4C_{24}](X,Y,Z)-\frac{1}{2},\notag\\
&C_{hff}^{SFV}(X,Y,Z)\equiv 
[p_1^2(C_{21}-C_0)+p_2^2(C_{22}-C_{12})+2p_1\cdot p_2(C_{23}-C_{12})+4C_{24}](X,Y,Z)-\frac{1}{2}. 
\end{align}

The 1PI diagram contributions to the form factors of the $hZZ$ and $hWW$ vertices which are defined in Eq.~(\ref{form_factor}) are 
calculated as 
\begin{align}
&\Gamma^{1,\text{1PI}}_{hZZ}(p_1^2,p_2^2,q^2)_F=\sum_f\frac{16m_f^2m_Z^2N_c^f}{v^3}
\Bigg\{(v_f^2+a_f^2)
\Big[B_0(p_1^2,f,f)+B_0(p_2^2,f,f)+2B_0(q^2,f,f)\notag\\
&+(4m_f^2-p_1^2-p_2^2)C_0(f,f,f)-8C_{24}(f,f,f)\Big]\notag\\
&-(v_f^2-a_f^2)\Big[B_0(p_2^2,f,f)+B_0(p_1^2,f,f)
+(4m_f^2-q^2)C_0(f,f,f)\Big]\Bigg\},\\
&\Gamma^{2,\text{1PI}}_{hZZ}(p_1^2,p_2^2,q^2)_F=-\sum_f\frac{32m_f^2m_Z^4N_c^f}{v^3}\notag\\
&\Big[(v_f^2+a_f^2)
(4C_{23}+3C_{12}+C_{11}+C_0)+(v_f^2-a_f^2)(C_{12}-C_{11})\Big](f,f,f),\\
&\Gamma^{3,\text{1PI}}_{hZZ}(p_1^2,p_2^2,q^2)_F=\sum_f\frac{64m_f^2m_Z^4N_c^f}{v^3}
v_fa_f(C_{11}+C_{12}+C_0)(f,f,f), \\
%%%%%%%%
&\Gamma^{1,\text{1PI}}_{hWW}(p_1^2,p_2^2,q^2)_F=\sum_{f,f'}\frac{4m_W^2m_f^2N_c^f}{v^3}
\Bigg[\frac{1}{2}B_0(p_2^2,f,f')+B_0(q^2,f,f)+\frac{1}{2}B_0(p_1^2,f,f')\notag\\
&-4C_{24}(p_1^2,p_2^2,q^2,f,f',f)
+\frac{1}{2}(2m_f^2+2m_{f'}^2-p_1^2-p_2^2)C_0(f,f',f)\Bigg]+(m_f\leftrightarrow m_{f'}),\\
&\Gamma^{2,\text{1PI}}_{hWW}(p_1^2,p_2^2,q^2)_F=\frac{-4m_W^4m_f^2N_c^f}{v^3}
\left(4C_{23}+3C_{12}+C_{11}+C_0\right)(f,f',f)+(m_f\leftrightarrow m_{f'}),\\
&\Gamma^{3,\text{1PI}}_{hWW}(p_1^2,p_2^2,q^2)_F= \frac{-4m_W^4m_f^2N_c^f}{v^3}\left(C_{11}+C_{12}+C_0\right)(f,f',f)
+(m_f\leftrightarrow m_{f'}), 
\end{align}
 \begin{align}
&\Gamma^{1,\text{1PI}}_{hZZ}(p_1^2,p_2^2,q^2)_B=2g_Z^2\lambda_{G^+G^-h} m_W^2 s_W^4 C_0(G^\pm,W,G^\pm)\notag\\ 
&+ 	 g^3m_W s_{\beta-\alpha}^{}  \Big\{2c_W^2C_{hVV1}^{VVV}(W,W,W)-2c_W^2C_{24}(c^\pm,c^\pm,c^\pm)+s_W^2C_{hVV1}^{SVV}(G^\pm,W,W) +s_W^2C_{hVV1}^{VVS}(W,W,G^\pm)  \notag\\ 
%% !~~~~~~~~~~~ 2-Gauge ~~~~~~~~~~~~~~~~~~~~~~~~~~~~~~!
 	&\quad\quad\quad\quad\quad  -2\frac{s_W^4}{c_W^2}m_W^2s_{\beta-\alpha} C_0(W,G^\pm,W)- (c_W^2-s_W^2)\frac{s_W^2}{c_W^2}[ C_{24}(W,G^\pm,G^\pm)+C_{24}(G^\pm,G^\pm,W) ]\Big\} 	\notag\\
 	&  + \frac{g_Z^3}{2} m_Z s_{\beta-\alpha}\Big\{ -2 m_Z^2 \left[s_{\beta-\alpha}^2 C_0(Z,h,Z)+c_{\beta-\alpha}^2 C_0(Z,H,Z)\right]+ s_{\beta-\alpha}^2 [C_{24}(G^0,h,Z)+C_{24}(Z,h,G^0)]			\notag\\
 	&  +c_{\beta-\alpha}^2 
   \left[C_{24}(A,h,Z)+C_{24}(Z,h,A)+C_{24}(G^0,H,Z)+C_{24}(Z,H,G^0)-C_{24}(A,H,Z)-C_{24}(Z,H,A)\right]   
 	\Big\}\notag\\
%% !~~~~~~~~~~~ 1-Gauge ~~~~~~~~~~~~~~~~~~~~~~~~~~~~~~!
 	& +2g_Z^2 m_Z^2 \Big\{3\lambda_{hhh}s_{\beta-\alpha}^2 C_0(h,Z,h)			
 	 +\lambda_{HHh} c_{\beta-\alpha}^2 C_0(H,Z,H)
 	 +\lambda_{Hhh} s_{\beta-\alpha}^{}c_{\beta-\alpha}^{} [C_0(H,Z,h)+C_0(h,Z,H)]  \Big\}\notag\\
%% !~~~~~~~~~~~ 0-Gauge ~~~~~~~~~~~~~~~~~~~~~~~~~~~~~~!
	& -2g_Z^2(c_W^2-s_W^2)^2 \left[ \lambda_{G^+G^-h} C_{24}(G^\pm,G^\pm,G^\pm) + \lambda_{H^+H^-h} C_{24}(H^\pm,H^\pm,H^\pm)\right]\notag\\
	& -2g_Z^2 s_{\beta-\alpha}^2 \Big[ 3\lambda_{hhh} C_{24}(h,G^0,h)+\lambda_{HHh} C_{24}(H,A,H)+\lambda_{GGh} C_{24}(G^0,h,G^0)+\lambda_{AAh} C_{24}(A,H,A)\Big]\notag\\
        & -2g_Z^2 c_{\beta-\alpha}^2 \Big[ 3\lambda_{hhh} C_{24}(h,A,h)  +\lambda_{HHh} C_{24}(H,G^0,H)+\lambda_{AAh} C_{24}(A,h,A)  +\lambda_{GGh} C_{24}(G^0,H,G^0)\Big]\notag\\
	& -2g_Z^2 s_{\beta-\alpha}^{}c_{\beta-\alpha}^{} \lambda_{Hhh}[ C_{24}(h,G^0,H)+C_{24}(H,G^0,h)-C_{24}(h,A,H)-C_{24}(H,A,h) ]\notag\\
	& -2g_Z^2    s_{\beta-\alpha}^{}c_{\beta-\alpha}^{}  \lambda_{AGh}[ C_{24}(A,h,G^0)+C_{24}(G^0,h,A)-C_{24}(A,H,G^0)-C_{24}(G^0,H,A)]\notag\\
%% !~~~~~~~~~~~ B function-like ~~~~~~~~~~~~~~~~~~~~~~~~~~~~~~!
	& +\frac{g_Z^2}{2}\lambda_{G^+G^-h}  (c_W^2-s_W^2)^2 B_0(q^2,G^\pm ,G^\pm)
	  +\frac{g_Z^2}{2}\lambda_{H^+H^-h}  (c_W^2-s_W^2)^2 B_0(q^2,H^\pm,H^\pm)\notag\\
	& +\frac{g_Z^2}{2}\lambda_{GGh}  B_0(q^2,G^0,G^0)
	 +\frac{g_Z^2}{2}\lambda_{AAh}  B_0(q^2,A,A)			
	 +\frac{g_Z^2}{2}\lambda_{HHh}  B_0(q^2,H,H)	
	 +\frac{3g_Z^2}{2}\lambda_{hhh}  B_0(q^2,h,h)\notag\\		
	& -g^3\frac{s_W^4}{c_W^2}m_W s_{\beta-\alpha}^{} [ B_0(p_2^2,W,G^\pm)+B_0(p_1^2,G^\pm,W) ]
	 -\frac{g_Z^3}{2}m_Z s_{\beta-\alpha}^{}   [ B_0(p_1^2,h,Z)+B_0(p_2^2,h,Z)] \notag\\
	& -6 g^3 c_W^2 m_W  s_{\beta-\alpha}^{} B_0(q^2,W,W) +4 g^3 c_W^2 m_W s_{\beta-\alpha}^{} , 
\end{align}
\begin{align}
&(g_Z^2m_Z^2)^{-1}\Gamma_{hZZ}^{2,\text{1PI}}(p_1^2,p_2^2,q^2)_B
=2gm_Wc_W^4s_{\beta-\alpha}C_{hVV2}^{VVV}(W,W,W)-2gc_W^4m_Ws_{\beta-\alpha}C_{1223}(c^\pm,c^\pm,c^\pm)\notag\\
&+gm_Ws_W^2c_W^2s_{\beta-\alpha}[C_{hVV2}^{SVV}(G^\pm,W,W)+C_{hVV2}^{VVS}(W,W,G^\pm)]\notag\\
&-gm_W(c_W^2-s_W^2)s_W^2 [C_{hVV2}^{SSV}(G^\pm,G^\pm,W)+C_{hVV2}^{VSS}(W,G^\pm,G^\pm)]\notag\\
&+\frac{g_Z}{2}m_Z[C_{hVV2}^{VSS}(Z,h,G^0)+C_{hVV2}^{VSS}(G^0,h,Z)]\notag\\
 & + \frac{g_Z}{2}m_Z s_{\beta-\alpha}^3[ C_{hVV2}^{VSS}(Z,h,G^0) +  C_{hVV2}^{SSV}(G^0,h,Z)]\notag\\
&  + \frac{g_Z}{2}m_Zs_{\beta-\alpha}c_{\beta-\alpha}^2[ C_{hVV2}^{VSS}(Z,h,A)+C_{hVV2}^{VSS}(Z,H,G^0)-C_{hVV2}^{VSS}(Z,H,A) \notag\\ 
&\hspace{24mm}+C_{hVV2}^{SSV}(A,h,Z)+C_{hVV2}^{SSV}(G^0,H,Z)-C_{hVV2}^{SSV}(A,H,Z)] \notag\\
& -2(c_W^2-s_W^2)^2 \left[ \lambda_{G^+G^-h} C_{1223}(G^\pm,G^\pm,G^\pm) + \lambda_{H^+H^-h} C_{1223}(H^\pm,H^\pm,H^\pm)\right]\notag\\
	& -2 s_{\beta-\alpha}^2 \Big[ 3\lambda_{hhh} C_{1223}(h,G^0,h)+\lambda_{HHh} C_{1223}(H,A,H)+\lambda_{GGh} C_{1223}(G^0,h,G^0)+\lambda_{AAh} C_{1223}(A,H,A)\Big]\notag\\
        & -2 c_{\beta-\alpha}^2 \Big[ 3\lambda_{hhh} C_{1223}(h,A,h)  +\lambda_{HHh} C_{1223}(H,G^0,H)+\lambda_{AAh} C_{1223}(A,h,A)  +\lambda_{GGh} C_{1223}(G^0,H,G^0)\Big]\notag\\
	& -2 s_{\beta-\alpha}^{}c_{\beta-\alpha}^{} \lambda_{Hhh}[ C_{1223}(h,G^0,H)+C_{1223}(H,G^0,h)-C_{1223}(h,A,H)-C_{1223}(H,A,h) ]\notag\\
	& -2    s_{\beta-\alpha}^{}c_{\beta-\alpha}^{}  \lambda_{AGh}[ C_{1223}(A,h,G^0)+C_{1223}(G^0,h,A)-C_{1223}(A,H,G^0)-C_{1223}(G^0,H,A)], \\
&\Gamma_{hZZ}^{3,\text{1PI}}(p_1^2,p_2^2,q^2)_B=0, 
\end{align}

\begin{align}
&\Gamma^{1,\text{1PI}}_{hWW}(p_1^2,p_2^2,q^2)_B=\notag\\
%% !~~~~~~~~~~~ 3-Gauge ~~~~~~~~~~~~~~~~~~~~~~~~~~~~~~!
&	 g^3m_W s_{\beta-\alpha}^{}  [ C_{hVV1}^{VVV}(Z,W,Z) + c_W^2C_{hVV1}^{VVV}(W,Z,W) + s_W^2  C_{hVV1}^{VVV}(W,\gamma,W)\notag\\
&\quad\quad\quad\quad\quad\quad	 - C_{24}(c_Z,c^\pm,c_Z) -c_W^2  C_{24}(c^\pm,c_Z,c^\pm)- s_W^2 C_{24}(c^\pm,c_\gamma,c^\pm)  
	         	       ] \notag\\
%% !~~~~~~~~~~~ 2-Gauge ~~~~~~~~~~~~~~~~~~~~~~~~~~~~~~!
& -\frac{g^3}{2}m_W s_W^2 s_{\beta-\alpha}^{}  [C_{hVV1}^{SVV}(G^\pm,Z,W)-C_{hVV1}^{SVV}(G^\pm,\gamma,W)+C_{hVV1}^{VVS}(W,Z,G^\pm)-C_{hVV1}^{VVS}(W,\gamma,G^\pm)] \notag\\
&	 -g^3m_W^3 \frac{s_W^4}{c_W^4} s_{\beta-\alpha}^{} C_0(Z,G^\pm,Z)
 -g^3m_W^3 s_{\beta-\alpha}^3 C_0(W,h,W)
 -g m_W^3  s_{\beta-\alpha}^{} c_{\beta-\alpha}^2  C_0(W,H,W)\notag\\
%!~~~~~~~~~~~ 1-Gauge ~~~~~~~~~~~~~~~~~~~~~~~~~~~~~~!
&	 +g^2\frac{s_W^4}{c_W^2} m_W^2 \lambda_{G^+G^-h} C_0(G^\pm,Z,G^\pm) +s_W^2 m_W^2 \lambda_{G^+G^-h} C_0(G^\pm,\gamma,G^\pm)\notag\\
&	 +6g^2 \lambda_{hhh} m_W^2  s_{\beta-\alpha}^2 C_0(h,W,h)
	 +2g^2 \lambda_{HHh} m_W^2c_{\beta-\alpha}^2  C_0(H,W,H)\notag\\
&	 +2g^2\lambda_{Hhh} m_W^2 c_{\beta-\alpha}^{} s_{\beta-\alpha}^{}   [C_0(h,W,H)+C_0(H,W,h)]\notag\\
&	 +\frac{g^3}{2}m_W s_{\beta-\alpha}\Big\{ s_{\beta-\alpha}^2  [ C_{24}(W,h,G^\pm) + C_{24}(G^\pm,h,W) ]\notag\\
&\quad\quad\quad\quad\quad\quad	 +c_{\beta-\alpha}^2 [C_{24}(W,H,G^\pm) + C_{24}(G^\pm,H,W) +C_{24}(W,h,H^\pm) + C_{24}(H^\pm,h,W)\notag\\
&\quad\quad\quad\quad \quad\quad\quad\quad  \quad\quad     -C_{24}(W,H,H^\pm)-C_{24}(H^\pm,H,W)] \Big\}\notag\\
&	 +\frac{g^3}{2}m_W\frac{s_W^2}{c_W^2}s_{\beta-\alpha}^{}[ C_{24}(G^0,G^\pm,Z) + C_{24}(Z,G^\pm,G^0) ]\notag\\
&	 -g^2\Big[
          \lambda_{G^+G^-h} C_{24}(G^\pm,G^0,G^\pm)+\lambda_{H^+H^-h} C_{24}(H^\pm,A,H^\pm)  \notag \\
&\quad\quad\quad	  +2\lambda_{GGh}C_{24}(G^0,G^\pm,G^0)+2\lambda_{AAh}C_{24}(A,H^\pm,A)\Big]\notag\\
&	 -g^2s_{\beta-\alpha}^2\Big[  6\lambda_{hhh}    C_{24}(h,G^\pm,h)
	                           +2\lambda_{HHh}    C_{24}(H,H^\pm,H) \notag\\
& \quad\quad\quad\quad             +\lambda_{G^+G^-h}  C_{24}(G^\pm,h,G^\pm) 
	                           +\lambda_{H^+H^-h}  C_{24}(H^\pm,H,H^\pm)\Big] \notag\\ 
&	 -g^2c_{\beta-\alpha}^2\Big[6\lambda_{hhh}C_{24}(h,H^\pm,h)
	                         +2\lambda_{HHh}C_{24}(H,G^\pm,H) \notag\\
& \quad\quad\quad\quad           +\lambda_{G^+G^-h}C_{24}(G^\pm,H,G^\pm) 
	                         +\lambda_{H^+H^-h}C_{24}(H^\pm,h,H^\pm)  \Big] \notag\\
&	 -g^2 \lambda_{H^+G^-h}s_{\beta-\alpha}^{} c_{\beta-\alpha}^{}  [ C_{24}(G^\pm,h,H^\pm)+C_{24}(H^\pm,h,G^\pm)-C_{24}(G^\pm,H,H^\pm)-C_{24}(H^\pm,H,G^\pm) ]\notag\\
&	 -2g^2\lambda_{Hhh} s_{\beta-\alpha}^{} c_{\beta-\alpha}^{}   [C_{24}(h,G^\pm,H)+C_{24}(H,G^\pm,h)-C_{24}(h,H^\pm,H)-C_{24}(H,H^\pm,h)] \notag\\
%% !~~~~~~~~~~~ B function-like ~~~~~~~~~~~~~~~~~~~~~~!
&	 -g^3m_Ws_{\beta-\alpha}^{}\Big[3B_0(q^2,W,W) +3B_0(q^2,Z,Z)-4\Big]\notag\\
&	 +\frac{g^2}{2}\lambda_{G^+G^-h} B_0(q^2,G^\pm,G^\pm)
	 +\frac{g^2}{2}\lambda_{GGh}   B_0(q^2,G^0,G^0)
	 +\frac{3g^2}{2}\lambda_{hhh}  B_0(q^2,h,h)\notag\\
&	 +\frac{g^2}{2}\lambda_{H^+H^-h} B_0(q^2,H^\pm,H^\pm)
	 +\frac{g^2}{2}\lambda_{AAh}   B_0(q^2,A,A)
	 +\frac{g^2}{2}\lambda_{HHh} B_0(q^2,H,H)\notag\\
&	 -\frac{g^3}{2}m_Ws_{\beta-\alpha}^{} \Big\{ B_0(p_1^2,W,h) + B_0(p_2^2,W,h) 
	 +\frac{s_W^4}{c_W^2}[ B_0(p_1^2,Z,G^\pm) + B_0(p_2^2,Z,G^\pm) ]\notag\\
&\quad\quad\quad\quad\quad\quad	 + s_W^2 [ B_0(p_1^2,\gamma,G^\pm) + B_0(p_2^2,\gamma,G^\pm) ]
\Big\}, 
\end{align}
%%%
\begin{align}
&(g^2m_W^2)^{-1}\Gamma_{hWW}^{2,\text{1PI}}(p_1^2,p_2^2,q^2)_B 
=\notag\\
&gm_Ws_{\beta-\alpha}\Big[C_{hVV2}^{VVV}(Z,W,Z)
+c_W^2C_{hVV2}^{VVV}(W,Z,W)+s_W^2C_{hVV2}^{VVV}(W,\gamma,W)\notag\\
&\quad\quad\quad\quad
-C_{1223}(c_Z^{},c^\pm,c_Z^{})
 -c_W^2C_{1223}(c^\pm,c_Z^{},c^\pm)-s_W^2C_{1223}(c^\pm,c_\gamma,c^\pm)\Big]\notag\\
&-\frac{g}{2}s_W^2m_Ws_{\beta-\alpha}[
C_{hVV2}^{SVV}(G^\pm,Z,W)-C_{hVV2}^{SVV}(G^\pm,\gamma,W)%
+C_{hVV2}^{VVS}(W,Z,G^\pm)-C_{hVV2}^{VVS}(W,\gamma,G^\pm)]\notag\\
&+\frac{g}{2}m_Ws_{\beta-\alpha}^3\Big[C_{hVV2}^{VSS}(W,h,G^\pm)+C_{hVV2}^{SSV}(G^\pm,h,W)\Big]\notag\\
&+\frac{g}{2}m_Ws_{\beta-\alpha}c_{\beta-\alpha}^2\Big[C_{hVV2}^{VSS}(W,H,G^\pm)+C_{hVV2}^{VSS}(W,h,H^\pm)-C_{hVV2}^{VSS}(W,H,H^\pm)\notag\\
&\quad\quad\quad\quad\quad\quad +C_{hVV2}^{SSV}(G^\pm,H,W)+C_{hVV2}^{SSV}(H^\pm,h,W)-C_{hVV2}^{SSV}(H^\pm,H,W)\Big]\notag\\
&+\frac{g}{2}\frac{s_W^2}{c_W^2}m_Ws_{\beta-\alpha}^3
\Big[C_{hVV2}^{VSS}(Z,G^\pm,G^0)+C_{hVV2}^{SSV}(G^0,G^\pm,Z)\Big] \notag\\
&	 -\Big[
          \lambda_{G^+G^-h} C_{1223}(G^\pm,G^0,G^\pm)+\lambda_{H^+H^-h} C_{1223}(H^\pm,A,H^\pm)  \notag \\
&\quad\quad\quad	  +2\lambda_{GGh}C_{1223}(G^0,G^\pm,G^0)+2\lambda_{AAh}C_{1223}(A,H^\pm,A)\Big]\notag\\
&	 -s_{\beta-\alpha}^2\Big[  6\lambda_{hhh}    C_{1223}(h,G^\pm,h)
	                           +2\lambda_{HHh}    C_{1223}(H,H^\pm,H) \notag\\
& \quad\quad\quad\quad             +\lambda_{G^+G^-h}  C_{1223}(G^\pm,h,G^\pm) 
	                           +\lambda_{H^+H^-h}  C_{1223}(H^\pm,H,H^\pm)\Big] \notag\\ 
&	 -c_{\beta-\alpha}^2\Big[6\lambda_{hhh}C_{1223}(h,H^\pm,h)
	                         +2\lambda_{HHh}C_{1223}(H,G^\pm,H) \notag\\
& \quad\quad\quad\quad           +\lambda_{G^+G^-h}C_{1223}(G^\pm,H,G^\pm) 
	                         +\lambda_{H^+H^-h}C_{1223}(H^\pm,h,H^\pm)  \Big] \notag\\
&	 - \lambda_{H^+G^-h}s_{\beta-\alpha}^{} c_{\beta-\alpha}^{}  [ C_{1223}(G^\pm,h,H^\pm)+C_{1223}(H^\pm,h,G^\pm)-C_{1223}(G^\pm,H,H^\pm)-C_{1223}(H^\pm,H,G^\pm) ]\notag\\
&	 -2\lambda_{Hhh} s_{\beta-\alpha}^{} c_{\beta-\alpha}^{}   [C_{1223}(h,G^\pm,H)+C_{1223}(H,G^\pm,h)-C_{1223}(h,H^\pm,H)-C_{1223}(H,H^\pm,h)], \\ 
&\Gamma_{hWW}^{3,\text{1PI}}(p_1^2,p_2^2,q^2)_B=0, 
\end{align}
where 
\begin{align}
&C_{hVV1}^{VVV}(X,Y,Z)\equiv \notag\\
&\left[18C_{24}+p_1^2(2C_{21}+3C_{11}+C_{0})+p_2^2(2C_{22}+C_{12})+p_1\cdot p_2(4C_{23}+3C_{12}+C_{11}-4C_0)\right]
(X,Y,Z)-3,\notag\\
&C_{hVV1}^{SVV}(X,Y,Z)\equiv\notag\\
&\left[3C_{24}+p_1^2(C_{21}-C_0)+p_2^2(C_{22}-2C_{12}+C_0)+2p_1\cdot p_2 (C_{23}-C_{11})\right](X,Y,Z)-\frac{1}{2},\notag\\
&C_{hVV1}^{VVS}(X,Y,Z)\equiv\notag\\
&\left[3C_{24}+p_1^2(C_{21}+4C_{11}+4C_0)+p_2^2(C_{22}+2C_{12})+2p_1\cdot p_2 (C_{23}+2C_{12}+C_{11}+2C_0)\right](X,Y,Z)-\frac{1}{2}, \notag\\
&C_{hVV2}^{VVV}(X,Y,Z)\equiv \left(10C_{23}+9C_{12}+C_{11}+5C_0\right)(X,Y,Z),\notag\\
&C_{hVV2}^{SVV}(X,Y,Z)\equiv \left(4C_{11}-3C_{12}-C_{23}\right)(X,Y,Z),\notag\\
&C_{hVV2}^{VVS}(X,Y,Z)\equiv \left(2C_{11}-5C_{12}-2C_0-C_{23}\right)(X,Y,Z), \notag\\
&C_{hVV2}^{VSS}(X,Y,Z)\equiv (C_{23}+C_{12}+2C_{11}+2C_0)(X,Y,Z), \notag\\
&C_{hVV2}^{SSV}(X,Y,Z)\equiv (C_{23}-C_{12})(X,Y,Z), \notag\\
&C_{1223}(X,Y,Z)\equiv (C_{12}+C_{23}) (X,Y,Z). 
\end{align}

\subsection{Decay rates for loop induced processes}

The decay rates for the loop induced processes are given by 
\begin{align}
\Gamma(h\to\gamma \gamma)&=\frac{\sqrt{2}G_F\alpha_{\text{em}}^2m_h^3}{256\pi^3 } \Big|s_{\beta-\alpha}
I_V+\sum_fQ_f^2N_c^f\xi_h^f I_F-\frac{\lambda_{H^+H^-h}}{v}I_S \Big|^2, \label{hgamgam_full} \\
\Gamma(h\to Z\gamma)
&=\frac{\sqrt{2}G_F\alpha_{\text{em}}^2m_h^3}{128\pi^3}\left(1-\frac{m_Z^2}{m_h^2}\right)^3\notag\\
&\hspace{8mm}\times \Big|s_{\beta-\alpha} J_V+\sum_fQ_fN_c^fv_fJ_F-\frac{\lambda_{H^+H^-h}}{v}\frac{g_Z}{2}(c_W^2-s_W^2)J_S \Big|^2, \label{hZgam_full} \\
\Gamma(h\to gg)&=\frac{\sqrt{2}G_F\alpha_s^2 m_h^3}{128\pi^3 }\Big|\sum_q\xi_h^qI_F\Big|^2, \label{hgg_full}
\end{align}
%where $g_Z^{}=g/c_W^{}$ and $v_f=I_f/2 -s^2_WQ_f$. 
The loop functions are defined as 
\begin{align}
I_S &= \frac{2v^2}{m_h^2}[1+2m_{H^\pm}^2C_0(0,0,m_h^2,m_{H^\pm},m_{H^\pm},m_{H^\pm})],\\
I_F& = -\frac{8m_f^2}{m_h^2}\left[1+\left(2m_f^2-\frac{m_h^2}{2}\right)C_0(0,0,m_h^2,m_f,m_f,m_f)\right], \\
I_V& = \frac{2m_W^2}{m_h^2}\left[6+\frac{m_h^2}{m_W^2}+(12m_W^2-6m_h^2)C_0(0,0,m_h^2,m_W,m_W,m_W)\right], 
\end{align}
and 
\begin{align}
J_V&=\frac{2m_W^2}{s_Wc_W(m_h^2-m_Z^2)}
\Big\{\left[c_W^2\left(5+\frac{m_h^2}{2m_W^2}\right)-s_W^2\left(1+\frac{m_h^2}{2m_W^2}\right)\right]\notag\\
&\left[1+2m_W^2C_0+\frac{m_Z^2}{m_h^2-m_W^2}(B_0(m_h^2,m_W,m_W)-B_0(m_Z^2,m_W,m_W))\right]\notag\\
&-6c_W^2(m_h^2-m_Z^2)C_0+2s_W^2(m_h^2-m_Z^2)C_0\Big\},\\
J_F&=-\frac{8m_f^2 }{s_Wc_W(m_h^2-m_Z^2)}\Big[1+\frac{1}{2}(4m_f^2-m_h^2+m_Z^2)C_0(0,m_Z^2,m_h^2,m_f,m_f,m_f)\notag\\
&+\frac{m_Z^2}{m_h^2-m_Z^2}(B_0(m_h^2,m_f,m_f)-B_0(m_Z^2,m_f,m_f))\Big],\\
J_S &=\frac{2v^2}{e(m_h^2-m_Z^2)}
\Big\{1+2m_{H^\pm}^2C_0(0,m_Z^2,m_h^2,m_{H^\pm},m_{H^\pm},m_{H^\pm}) \notag\\
&+\frac{m_Z^2}{m_h^2-m_Z^2}\left[B_0(m_h^2,m_{H^\pm},m_{H^\pm})-B_0(m_Z^2,m_{H^\pm},m_{H^\pm})\right]\Big\}. 
\end{align}

\end{appendix}

%\vspace*{-4mm}

\end{document}